\documentclass[a4paper, 11pt]{article}
\usepackage[]{graphicx}\usepackage[]{color}
\makeatletter
\def\maxwidth{ %
  \ifdim\Gin@nat@width>\linewidth
    \linewidth
  \else
    \Gin@nat@width
  \fi
}
\makeatother

\definecolor{fgcolor}{rgb}{0.345, 0.345, 0.345}
\newcommand{\hlnum}[1]{\textcolor[rgb]{0.686,0.059,0.569}{#1}}%
\newcommand{\hlstr}[1]{\textcolor[rgb]{0.192,0.494,0.8}{#1}}%
\newcommand{\hlopt}[1]{\textcolor[rgb]{0,0,0}{#1}}%
\newcommand{\hlstd}[1]{\textcolor[rgb]{0.345,0.345,0.345}{#1}}%
\newcommand{\hlkwa}[1]{\textcolor[rgb]{0.161,0.373,0.58}{\textbf{#1}}}%
\newcommand{\hlkwb}[1]{\textcolor[rgb]{0.69,0.353,0.396}{#1}}%
\newcommand{\hlkwc}[1]{\textcolor[rgb]{0.333,0.667,0.333}{#1}}%
\newcommand{\hlkwd}[1]{\textcolor[rgb]{0.737,0.353,0.396}{\textbf{#1}}}%

\usepackage{framed}
\makeatletter
\newenvironment{kframe}{%
 \def\at@end@of@kframe{}%
 \ifinner\ifhmode%
  \def\at@end@of@kframe{\end{minipage}}%
  \begin{minipage}{\columnwidth}%
 \fi\fi%
 \def\FrameCommand##1{\hskip\@totalleftmargin \hskip-\fboxsep
 \colorbox{shadecolor}{##1}\hskip-\fboxsep
     \hskip-\linewidth \hskip-\@totalleftmargin \hskip\columnwidth}%
 \MakeFramed {\advance\hsize-\width
   \@totalleftmargin\z@ \linewidth\hsize
   \@setminipage}}%
 {\par\unskip\endMakeFramed%
 \at@end@of@kframe}
\makeatother

\definecolor{shadecolor}{rgb}{.97, .97, .97}
\definecolor{messagecolor}{rgb}{0, 0, 0}
\definecolor{warningcolor}{rgb}{1, 0, 1}
\definecolor{errorcolor}{rgb}{1, 0, 0}
\newenvironment{knitrout}{}{} % an empty environment to be redefined in TeX

\usepackage{alltt}

\usepackage[utf8]{inputenc}
\usepackage{authblk}
\usepackage[scale=0.7]{geometry}
\usepackage[T1]{fontenc}
\usepackage{hyperref} 
\usepackage{amssymb}
\usepackage{amsmath,amsthm} 
\usepackage[english]{babel}
\usepackage{graphicx}
\usepackage{natbib} 
\usepackage{caption}
\usepackage{breakcites}
\usepackage{hyperref}
\usepackage{todonotes}
\usepackage{bm}
\usepackage{stmaryrd} 
\usepackage{gensymb}
\usepackage{color}
\usepackage[]{algorithm2e}
\usepackage{ulem} % pour barrer de texte

\usepackage{tikz}
\usepackage{color}
\usepackage{multirow}
\usepackage{ulem} % pour barrer de texte

\usepackage[toc,page]{appendix}
\usepackage{tocloft}
\usepackage{array}

\newcommand{\pkg}[1]{{\fontseries{b}\selectfont #1}}
\newcommand{\proglang}[1]{{\fontseries{b}\selectfont #1}}
\providecommand{\keywords}[1]{\textbf{\textit{Keywords---}} #1}
\newcommand{\code}[1]{\texttt{#1}}

\newcommand{\prior}{\textit{prior }}
\newcommand{\posterior}{\textit{posterior }}

\IfFileExists{upquote.sty}{\usepackage{upquote}}{}
\begin{document}

\title{\pkg{CaliCo}: a \proglang{R} package for Bayesian calibration}
\author[1,2,3]{Mathieu Carmassi}
\author[1]{Pierre Barbillon}
\author[3]{Merlin Keller}
\author[1]{Eric Parent}
\affil[1]{UMR MIA-Paris, AgroParisTech, INRA, Paris }
\affil[2]{EDF R\&D, TREE department, Moret-sur-Loing}
\affil[3]{EDF R\&D, PRISME department, Chatou}

\maketitle

\begin{abstract}
In this article, we present a recently released \proglang{R} package for Bayesian calibration. Many industrial fields are facing unfeasible or costly field experiments. These experiments are replaced with numerical/computer experiments which are realized by running a numerical code. Bayesian calibration intends to estimate, through a posterior distribution, input parameters of the code in order to make the code outputs close to the available experimental data. The code can be time consuming while the Bayesian calibration implies a lot of code calls which makes studies too burdensome. A discrepancy might also appear between the numerical code and the physical system when facing incompatibility between experimental data and numerical code outputs.
The package \pkg{CaliCo} deals with these issues through four statistical models which deal with a time consuming code or not and with discrepancy or not. A guideline for users is provided in order to illustrate the main functions and their arguments. Eventually, a toy example is detailed using \pkg{CaliCo}. This example (based on a real physical system) is in five dimensions and uses simulated data.
\end{abstract}


\section{Introduction}
\label{sec:intro}

Many industrial fields make recourse to numerical experiments to bypass the economic burden of field experiments which are 
expensive to make \citep{santner2013,fang2005}. 
The numerical experiments, based on a physical modeling, might not be accurate enough regarding the real physical system.
Calibration intends to reduce this gap with the help of data from field experiments which are usually scarce.
Calibration is performed with respect to a statistical model that shall take into account additional uncertainties.
There are two kinds of uncertainties. The first comes from the measurement error. Indeed, the sensors, that are recording data from the experimental field, are carrying uncertainty. The second is the code discrepancy or code error which takes into account the mismatch between the code and the physical system \citep{kennedy2001,bayarri2007,higdon2004}. The complexity of numerical codes has also increased for the last few years and has become greedy in computational time \citep{sacks1989}. As calibration requires a lot of code calls, the computational time burden becomes quickly intractable for such methods.\newline

Three packages have been developed for Bayesian calibration. The package \pkg{BACCO} \citep{BACCO} is a bundle of several other packages. It imports \pkg{emulator} \citep{emulator}, \pkg{mvtnorm} \citep{mvtnorm}, \pkg{calibrator} \citep{calibrator} and \pkg{approximator} \citep{approximator}. These packages contain functions that realize Bayesian calibration and also prediction. The statistical model implemented concerns only the case where the numerical code is time consuming and when a code error is added (the model introduced by \citet{kennedy2001}). Moreover, \pkg{BACCO} explores the prior distribution, of the parameters from the statistical model, using analytic or numerical integration. Similarly, another package called \pkg{SAVE} \citep{SAVE} deals with Bayesian calibration through four main functions: \code{SAVE}, \code{bayesfit}, \code{predictcode} and \code{validate}. The function \code{SAVE} creates the statistical model when \code{bayesfit}, \code{predictcode} and \code{validate} realize respectively Bayesian calibration, prediction and validation of a model. Calibration is done in \pkg{SAVE} in a similar way to \pkg{BACCO} because it is based on the same statistical model. Both packages are not flexible with the numerical code and the statistical model. A design of experiments has to be run upstream on the code before running calibration. The package \pkg{RobustCalibration} based on \citet{gu2017} \citep{RobustCalibration} is a package that realizes calibration of inexact mathematical models and implements the discrepancy with a ``scaled Gaussian process''.\newline

\pkg{CaliCo} offers more flexibility on the statistical model choice. If one is interested in calibrating a numerical code inexpensive in computation time, \pkg{CaliCo} allows the user to upload the numerical code in the model and run Bayesian calibration with it. A very intuitive perspective is given to the user by using four functions called \code{model}, \code{prior}, \code{calibrate} and \code{forecast}, that are detailed Section \ref{sec:guidelines}. \pkg{CaliCo} also allows the user to access several \pkg{ggplot2} \citep{ggplot2} graphs very easily and to load each of them to change the layout at one's convenience. At some point, if the code is time consuming, calibration needs a surrogate to emulate it. Usually, a Gaussian process is chosen \citep{sacks1989,cox2001}. Three packages are related to the establishment of a Gaussian process as a surrogate: \pkg{GPfit} \citep{GPfit}, \pkg{DiceKriging} \citep{DiceKriging} and \pkg{RobustGaSP} based on \citet{gu2018} \citep{RobustGaSP}. We use in \pkg{CaliCo}, the package \pkg{DiceDesign} \citep{DiceDesign} to establish design of experiments (DOE). For compatibility matters, we have chosen \pkg{DiceKriging} to generate surrogates. Moreover, the time consuming steps of the Bayesian calibration are coded in \proglang{C++} and linked to \proglang{R} via the package \pkg{Rcpp} \citep{Rcpp}.\newline

In this article, the first part (Section \ref{sec:statisticalBackground}) provides a recall of all the statistical basis for a good understanding of \pkg{CaliCo}. Then, the second part (Section \ref{sec:guidelines}) presents the main functions and functionalities of the package. The last section provides a Bayesian calibration illustration on a toy example. It is based on a physical modeling of a damped harmonic oscillator with five parameters to calibrate.

\section{Statistical background}
\label{sec:statisticalBackground}

A numerical code generally depends on two kinds of inputs: variables and parameters. The variables are input variables (observable and often controllable) which are set during a field experiment and can encompass environmental variables that can be measured. These variables are data necessary to run the numerical code and often enforced to the user. The parameters can generally be interpreted as physical constants required in the physical modeling. They can also encompass the tuning parameters which have no physical interpretation. They have to be set by the user to run the code and chosen carefully to make the code mimic the real physical phenomenon. Let us call $\boldsymbol{x}$ the vector of $d$ input variables and $\mathcal{H}$ the input variable space such that $\boldsymbol{x}\in\mathcal{H}\subset\mathbb{R}^d$. Similarly, $\boldsymbol{\theta}$ stands for the vector of $p$ parameters and $\mathcal{Q}$ for the input space parameter so that $\boldsymbol{\theta}\in\mathcal{Q}\subset\mathbb{R}^p$. Usually, in an industrial context, a quantity of interest $\boldsymbol{y}_{e}$ is measured with the help of sensors which are carrying uncertainty. Let us consider that $n$ observations are available. In what follows, we will denote by $\boldsymbol{X}$ the matrix of the input variables in $\mathcal{M}_{n\times d}(\mathbb{R})$ and $\boldsymbol{x}_i$ the $i^{th}$ row of the matrix $\boldsymbol{X}$ (also representing the vector of input variables corresponding to the $i^{th}$ observation). The physical system observed will be called $\zeta$ and depends only on input variables because parameters intervene only in the numerical code. We use the following model for field data:

\begin{equation}
\forall i \in \llbracket1,\dots,n\rrbracket \quad y_{e}^i=\zeta(\boldsymbol{x}_i) + \epsilon_i
\end{equation}
where $\epsilon_i\overset{iid}{\sim}\mathcal{N}(0,\sigma_e^2)$ is a white Gaussian noise which accounts for the measurement error.
The model for the $\epsilon_i$'s remains the same in the following.  
Since the numerical code is created to mimic the physical system, a first model ($\mathcal{M}_1$) which takes the code into account, can be written straightforwardly as:

\begin{equation}
\mathcal{M}_1:\quad \forall i \in \llbracket1,\dots,n\rrbracket \quad y_{e}^i=f_c(\boldsymbol{x}_i,\boldsymbol{\theta}) + \epsilon_i\,.
\end{equation}
 When the code is time consuming, this model becomes intractable for Bayesian calibration. Indeed, Bayesian calibration generally uses Monte Carlo Markov Chains (MCMC) algorithms which call the numerical code a high number of times. That is why \citet{sacks1989} has introduced a surrogate of the numerical code based on a Gaussian process. Then, \citet{cox2001} introduced this surrogate into a statistical model ($\mathcal{M}_2$) for calibration:

\begin{equation}
\mathcal{M}_2:\quad \forall i \in \llbracket1,\dots,n\rrbracket \quad y_{e}^i=F(\boldsymbol{x}_i,\boldsymbol{\theta}) + \epsilon_i
\end{equation}
where $F(\bullet,\bullet)\sim\mathcal{PG}(m_F(\bullet,\bullet),c_F(\{\bullet,\bullet\},\{\bullet,\bullet\})$ stands for the Gaussian process that emulates the numerical code. To find a suitable Gaussian process that mimics "well enough" the code,
a DOE of $n_d$ points (denoted by $D$) is taken in $\mathcal{H}\times\mathcal{Q}$. Then, the output of the code $\boldsymbol{y}_c=f_c(D)$ can be used for estimating parameters inherent to the Gaussian process.\newline

Some papers advocate to add a code error (also called discrepancy) in the statistical modeling \citep{kennedy2001,bayarri2007,higdon2004}. This term $\delta(\bullet)$ is a realization of a Gaussian process and depends only on $\boldsymbol{x}$ because it aims to find some structural correlation between the points in $\boldsymbol{y}_e$. Then, if the code is not time consuming:

\begin{equation}
\mathcal{M}_3:\quad \forall i \in \llbracket1,\dots,n\rrbracket \quad y_{e}^i=f_c(\boldsymbol{x}_i,\boldsymbol{\theta}) + \delta(\boldsymbol{x}_i) + \epsilon_i
\end{equation}
where  $\delta(\bullet)\sim\mathcal{PG}(m_{\delta}(\bullet),c_{\delta}(\bullet,\bullet))$ stands for the discrepancy. With a time consuming code, the last model ($\mathcal{M}_4$) is close to the one introduced in \citet{kennedy2001}:

\begin{equation}
\mathcal{M}_4:\quad \forall i \in \llbracket1,\dots,n\rrbracket \quad y_{e}^i=F(\boldsymbol{x}_i,\boldsymbol{\theta}) + \delta(\boldsymbol{x}_i) + \epsilon_i
\end{equation}
where  $F(\bullet,\bullet)\sim\mathcal{PG}(m_F(\bullet,\bullet),c_F(\{\bullet,\bullet\},\{\bullet,\bullet\})$ and $\delta(\bullet)\sim\mathcal{PG}(m_{\delta}(\bullet),c_{\delta}(\bullet,\bullet))$. \newline

At every Gaussian processes added in the statistical modeling, it brings new parameters to calibrate (see Appendix \ref{app:GP} for more details on Gaussian processes). In a Bayesian framework, a prior distribution has to be defined for each parameters. In the general framework of $\mathcal{M}_4$ several estimation methods exist. The first is introduced in \citet{higdon2004} and implies that the full likelihood (using all collected data such as $\boldsymbol{y}_e$ and $\boldsymbol{y}_c$) is maximized to find estimators of the parameters. In \pkg{CaliCo} another method called modularization by \citet{liu2009} is implemented. It consists in splitting the estimation step in two. First the maximum likelihood estimates (MLE) of the conditional likelihood is looked for to get the estimates of the parameters relative to the surrogate. Then, these estimators are plugged into the conditional likelihood (see appendix \ref{app:GP}) from which the posterior distribution is sampled with MCMC's. More insights on the writing of all the likelihoods for each models and details on parameter estimation are available in \citet{Carmassi2018}.

%% -- Illustrations ------------------------------------------------------------

%% - Virtually all JSS manuscripts list source code along with the generated
%%   output. The style files provide dedicated environments for this.
%% - In R, the environments {Sinput} and {Soutput} - as produced by Sweave() or
%%   or knitr using the render_sweave() hook - are used (without the need to
%%   load Sweave.sty).
%% - Equivalently, {CodeInput} and {CodeOutput} can be used.
%% - The code input should use "the usual" command prompt in the respective
%%   software system.
%% - For R code, the prompt "R> " should be used with "+  " as the
%%   continuation prompt.
%% - Comments within the code chunks should be avoided - these should be made
%%   within the regular LaTeX text.

\section{Guidelines for users}
\label{sec:guidelines}

\pkg{CaliCo} performs a Bayesian calibration through 3 different steps:
\begin{enumerate}
\item creation of the statistical model,
\item selection of the \prior distributions,
\item running calibration with some simulation options.
\end{enumerate}

\pkg{CaliCo} allows the user to easily take advantage of the calibration realized. Indeed, a prediction can be performed on a new data set using the calibrated code in the statistical model. The main functions of the package \pkg{CaliCo} are detailed Table \ref{tab:functions}. 

\begin{table}[h!]
\centering
\begin{tabular}{lllp{7.4cm}}
\hline
Function & Description \\ \hline
\code{model} & generates a statistical model\\
\code{prior} & creates one or a list of prior distributions\\
\code{calibrate} & realizes calibration for the \code{model} and \code{prior} specified\\
\code{forecast} & predicts the output over a new data set\\
\hline
\end{tabular}
\caption{\label{tab:functions} Main functions necessary for calibration in \pkg{CaliCo}.}
\end{table}

\pkg{CaliCo} is coded in \pkg{R6} \citep{Chang2017} which is an oriented object language. Each function generates an \pkg{R6} object that can be used by other functions (in this case methods) that are proper to the object. The \pkg{R6} layer is not visible to the user. The main classes implemented with the associated functions are detailed Table \ref{tab:classes}.

\begin{table}[h!]
\centering
\begin{tabular}{lllp{7.4cm}}
\hline
Function & \pkg{R6} class called \\ \hline
\code{model} & \code{model.class}\\
\code{prior} & \code{prior.class}\\
\code{calibrate} & \code{calibrate.class}\\
\code{forecast} & \code{forecast.class}\\
\hline
\end{tabular}
\caption{\label{tab:classes} \pkg{R6} classes called by the main functions in \pkg{CaliCo}.}
\end{table}

To define the statistical model, which is the first step in calibration, several elements are necessary. The code function must be defined in \proglang{R} and takes two arguments $\boldsymbol{X}$ and $\boldsymbol{\theta}$ respecting this order. 
For example:

\begin{knitrout}
\definecolor{shadecolor}{rgb}{0.969, 0.969, 0.969}\color{fgcolor}\begin{kframe}
\begin{alltt}
\hlstd{code} \hlkwb{<-} \hlkwa{function}\hlstd{(}\hlkwc{X}\hlstd{,}\hlkwc{theta}\hlstd{)}
\hlstd{\{}
   \hlkwd{return}\hlstd{((}\hlnum{6}\hlopt{*}\hlstd{X}\hlopt{-}\hlnum{2}\hlstd{)}\hlopt{^}\hlnum{2}\hlopt{*}\hlkwd{sin}\hlstd{(theta}\hlopt{*}\hlstd{X}\hlopt{-}\hlnum{4}\hlstd{))}
\hlstd{\}}
\end{alltt}
\end{kframe}
\end{knitrout}

If the numerical code is called from another language, one can implement a wrapper that calls from \proglang{R} the numerical code according to the above writting. It is also possible to build a design of experiements and run the code outside \proglang{R} to get the outputs. Then, the DOE and the relative outputs are used instead of the numerical code in the statistical model (more details below). The function \code{model} takes several other arguments (Table \ref{tab:model}) as for example the vector or the matrix of the input variables (described Section \ref{sec:statisticalBackground}), the vector of experimental data or the statistical model chosen for calibration. \newline

\begin{table}[h!]
\centering
\begin{tabular}{lllp{7.4cm}}
\hline
\code{model} description & Arguments to be specified \\ \hline
$f_c(\boldsymbol{x},\boldsymbol{\theta})$ the function to calibrate & \code{code} (defined as \code{code(x,theta)})\\
$\boldsymbol{X}$ the matrix of the input variables & \code{X}\\
$\boldsymbol{y}_{e}$ the vector of experimental data & \code{Yexp}\\
$\mathcal{M}$ the statistical model selected & (Default value \code{model1}) \code{model1}, \code{model2},\\
& \code{model3}, \code{model4}\\
Gaussian process options (optional only for  & (Optional) \code{opt.gp} (is a \code{list})\\
$\mathcal{M}_2$ and $\mathcal{M}_4$) & \\
Emulation options (optional only for  & (Optional) \code{opt.emul} (is a \code{list})\\
$\mathcal{M}_2$ and $\mathcal{M}_4$) & \\
Simulation options (optional only for  & (Optional) \code{opt.sim} (is a \code{list})\\
$\mathcal{M}_2$ and $\mathcal{M}_4$) & \\
Discrepancy options (necessary only for  & (Optional) \code{opt.disc} (is a \code{list})\\
$\mathcal{M}_3$ and $\mathcal{M}_4$) & \\
\hline
\end{tabular}
\caption{\label{tab:model} description of the arguments of the function \code{model}}
\end{table}

If the chosen model is $\mathcal{M}_2$ or $\mathcal{M}_4$, then a Gaussian process will be created as a surrogate of the function \code{code}. In each case the Gaussian process option (\code{opt.gp}, see Table \ref{tab:model}) is needed. It is a \code{list} which encompasses:
\begin{itemize}
\item \code{type}: type of covariance function chosen for the surrogate established by the package \pkg{DiceKriging} \citep{DiceKriging},
\item \code{DOE}: design of experiments for the surrogate (default value NULL).
\end{itemize}

Three cases can occur. First, the numerical code is available and the user does not possess any Design Of Experiments (DOE). In this case, only the Gaussian process option \code{opt.gp} and the emulation option \code{opt.emul} (Table \ref{tab:model}) are needed. The emulation option controls the establishment of the DOE. It is a \code{list} which contains:

\begin{itemize}
\item \code{p}: the number of parameters in the model,
\item \code{n.emul}: the number of points for constituting the DOE,
\item \code{binf}: the lower bound of the parameter vector,
\item \code{bsup}: the upper bound of the parameter vector.
\end{itemize}

The second possible case is when the user want to enforce a specific DOE. Note that in \code{opt.gp}, the \code{DOE} option is \code{NULL}. One can upload a specific DOE in this option. As the new DOE will be used, the emulation option \code{opt.emul} is not needed anymore. \newline

The third case is when no numerical code is available. The user is only in possession of a DOE and the corresponding code evaluations. Then, the simulation option \code{opt.sim} is added. This option encompasses:

\begin{itemize}
\item \code{Ysim}: the code evaluations of \code{DOEsim},
\item \code{DOEsim}: the specific DOE used to get simulated data.
\end{itemize}

When this option is added, the emulation option is not necessary anymore. The \code{code} argument in the function \code{model} can then be set to \code{code=NULL}. Table \ref{tab:optionsGP} presents a summary of these three cases and the options to add in the function \code{model}.

\begin{table}[h!]
\centering
\begin{tabular}{lllp{7.4cm}}
\hline
cases & options needed in the function \code{model} \\ \hline
numerical code without DOE & \code{opt.gp} and \code{opt.emul}  \\
numerical code with DOE & \code{opt.gp}\\
no numerical code & \code{opt.gp} and \code{opt.sim}\\
\hline
\end{tabular}
\caption{\label{tab:optionsGP} Summary of the options needed depending on the case}
\end{table}

If $\mathcal{M}_3$ or $\mathcal{M}_4$ is chosen, a discrepancy term is added in the statistical model. This discrepancy is created in \pkg{CaliCo} with the option \code{opt.disc} in the function \code{model}. It is a \code{list} composed of one component called \code{type.kernel} which corresponds to the correlation function of the discrepancy chosen. The list of the correlation functions implemented are detailed in Table \ref{tab:kernelD}.

\begin{table}[h!]
\begin{center}
\begin{tabular}{lp{7.4cm}l}
\hline
\code{kernel.type}  & covariance implemented \\ \hline
\code{gauss} & $g(d)=\sigma^2exp\Big(-\frac{1}{2}(\frac{d}{\psi})^2\Big)$ \\
\code{exp} & $g(d)=\sigma^2exp\Big(-\frac{1}{2}\frac{d}{\psi}\Big)$ \\
\code{matern3\_2} & $g(d)=\sigma^2\Big(1+\sqrt{3}\frac{d^2}{\psi}\Big)exp\Big(-\sqrt{3}\frac{d^2}{\psi}\Big)$ \\
\code{matern5\_2} & $g(d)=\sigma^2\Big(1+\sqrt{5}\frac{d^2}{\psi}+5\frac{d^2}{3\psi^2}\Big)exp\Big(-\sqrt{5}\frac{d^2}{\psi}\Big)$ \\
\hline
\end{tabular}
\caption{\label{tab:kernelD} Kernel implemented for the discrepancy covariance}
\end{center}
\end{table}

The \code{model} function creates an \pkg{R6} object in which two methods have been coded and are able to be used as regular functions. These function are \code{plot} and \code{print}. The function \code{print} gives an access to a short summary of the statistical model created. The function \code{plot} allows to get a visual representation. However, to get a visual representation, parameter values have to be specified in the model. A pipe \code{\%<\%} is available in \pkg{CaliCo} to parametrize a model. Let us consider a created random model called \code{myModel}. The code line \code{myModel \%<\% param} is the way to give the model parameter values. The \code{param} variable is a \code{list} containing values of $\boldsymbol{\theta}$ (named \code{theta} in the \code{list}), $\boldsymbol{\theta}_{\delta}$ for $\mathcal{M}_3$ and $\mathcal{M}_4$ (variance and correlation length of the discrepancy, named \code{thetaD} in the \code{list}) and $\sigma_e^2$ (named \code{var}). Section \ref{sec:exampleCaliCoMulti} gives an overview of how the pipe works for each models. The \code{plot} function takes two arguments that are the model generated by \code{model} and the x-axis to draw the results. An additional option \code{CI} (by default \code{CI="all"}) allows to select which credibility interval at $95\%$ one wants to display:
\begin{itemize}
\item \code{CI="err"} only the credibility interval of the measurement error with (or without) the discrepancy is given,
\item \code{CI="GP"}, only the credibility interval of the Gaussian process is plotted,
\item \code{CI="all"} all credibility intervals are displayed.
\end{itemize}

The second step for calibration is to define the \textit{prior} distributions of the parameters we seek to estimate. At least, two \prior distributions have to be set and it is in the case where the \code{code} function takes only one input parameter $\boldsymbol{\theta}$. That means, only the posterior distributions of this parameter and the variance $\sigma_e^2$ (for the model $\mathcal{M}_1$) are what we seek to sample in calibration.\newline

\begin{table}[h!]
\centering
\begin{tabular}{lllp{7.4cm}}
\hline
\code{prior} arguments & description \\ \hline
\code{type.prior}& vector or scalar of string among ("gaussian", "gamma" and "unif")  \\
\code{opt.prior} &  list of vector corresponding to the distribution parameters\\
\hline
\end{tabular}
\caption{\label{tab:prior} description of the arguments of the function \code{prior}}
\end{table}

Table \ref{tab:prior} describes the options needed into the function \code{prior}. Three choices of \code{type.prior} are available so far (\code{gaussian}, \code{gamma} and \code{unif} which respectively stands for Gaussian, Gamma and Uniform distributions, see Table \ref{tab:priorDensities} for details). For calibration with 2 parameters (which is the lower dimensional case), \code{type.prior} is a vector (\code{type.prior=c("gaussian","gamma")} for example). Then, \code{opt.prior} is a list containing characteristics of each distribution. For the Gaussian distribution, it will be a vector of the mean and the variance, for the Gamma distribution it will be the shape and the scale and for the Uniform distribution the lower bound and the upper bound (\code{opt.prior=list(c(1,0.1),c(0.01,1))} for example). \newline

\begin{table}[h!]
\centering
\begin{tabular}{llcp{7.4cm}}
\hline
\code{type.prior} & distribution & arguments in \code{opt.prior} vector \\ \hline
\code{gaussian} & $f(x)=\frac{1}{\sqrt{2\pi V}}exp\Big(-\frac{1}{2}(\frac{(x-m)^2}{V})\Big)$ & \code{c(m,V)}\\
\code{gamma} &  $f(x)=\frac{1}{(k^a*\Gamma(a))}x^{(a-1)} exp(-\frac{x}{k})$ & \code{c(a,k)}\\
\code{unif} & $f(x)=\frac{1}{b-a}$ & \code{c(a,b)}\\
\hline
\end{tabular}
\caption{\label{tab:priorDensities} description of the arguments of the function \code{prior}}
\end{table}

When the prior distributions and the model are defined, calibration can be run. The function \code{calibrate} implements a Markov Chain Monte Carlo (MCMC) according to specific conditions all controlled by the user. \newline

\begin{table}[h!]
\centering
\begin{tabular}{llp{7.4cm}}
\hline
\code{calibrate} arguments & Description \\ \hline
\code{md} & the model generated with the function \code{model}\\
\code{pr} & the list of prior generated by the function \code{prior}\\
\code{opt.estim} & estimation options for calibration\\
\code{opt.valid} & (optional) cross validation options (default value \code{NULL})\\
\hline
\end{tabular}
\caption{\label{tab:calibrate} description of the arguments of the function \code{calibrate}}
\end{table}

The MCMC implemented in \pkg{CaliCo} is composed of two algorithms as described in \citet{Carmassi2018}. The first algorithm is a Metropolis within Gibbs. As the access of conditional distribution is available, it is easy to propose a new "acceptable" point for each parameter. However, this operation is time consuming, especially if the code is long to run. That is why, this algorithm is run for a limited number of iterations and then the covariance matrix of all the simulated points generated is used for the second algorithm. This second algorithm is a Metropolis Hastings \citep{Metropolis1953,Hastings1970} which uses the previous covariance structure ($\Sigma$ in Algorithm appendix \ref{MCMC}) to be more efficient. These algorithms, described appendix \ref{MCMC}, are coded in \proglang{C++} thanks to \pkg{Rcpp} package \citep{Rcpp} in order to limit the time consuming aspect of these non-parallelizable loops. Note that there is an adaptability present to regulate the parameter $k$ according to the acceptation rate. The user is free to set that regulation at the wanted percentage. \newline

Then, the \code{opt.estim} option is a \code{list} composed of:
\begin{itemize}
\item \code{Ngibbs}: the number of iterations of the Metropolis within Gibbs algorithm,
\item \code{Nmh}: the number of iteration of the Metropolis Hastings algorithm,
\item \code{thetaInit}: the starting point,
\item \code{r}: the vector of regulation of the covariance in the Metropolis within Gibbs algorithm (in the proposition distribution the variance is $k\Sigma$),
\item \code{sig}: the variance of the proposition distribution $\Sigma$,
\item \code{Nchains}: (default value 1) the number of MCMC chains to run,
\item \code{burnIn}: the number of iteration to withdraw from the Metropolis Hastings algorithm.
\end{itemize}

In the function \code{calibrate}, one optional argument is available to run a cross validation. This option called \code{opt.valid}, is a list composed of two options which have to be filled:
\begin{itemize}
\item \code{type.CV}: the type of cross validation wanted (leave one out is the only cross validation implemented so far \code{type.CV="loo"}),
\item \code{nCV}: the number of iteration to run in the cross validation.
\end{itemize}

After calibration is complete, an \pkg{R6} object is created and two methods (\code{print} and \code{plot}) are available and are also able to be used as regular functions. The \code{print} function is a summary that recalls the selected model, the code used for calibration, the acceptation rate of the Metropolis within Gibbs algorithm, the acceptation rate of the Metropolis Hastings algorithm, the maximum \textit{a posteriori} and the mean \textit{a posteriori}. It allows to quickly check the acceptation ratios and see if the chains have properly mixed. The \code{plot} function generates automatically, a series of graphs that displays, notably, the ouput of the calibrated code. Two arguments are necessary to run this function: the calibrated model and the x-axis to draw the results. An additional option \code{graph} (by default \code{graph="all"}) allows to control which graphic layout one wants to plot:
\begin{itemize}
\item if \code{graph="chains"} a layout containing the autocorrelation graphs, the MCMC chains and the \textit{prior} and \textit{posterior} distributions for each parameter is given,
\item if \code{graph="corr"} a layout containing in the diagonal the \textit{prior} and \textit{posterior} distributions for the parameter vector $\boldsymbol{\theta}$ and the scatterplot between each pair of parameters is plotted,
\item if \code{graph="results"} the result of calibration is displayed,
\item if \code{graph="all} all of them are printed.
\end{itemize}

Note that all these graphs (made in \pkg{ggplot2}) are proposed in a particular layout but one can easily load all of them into a variable and extract the particular graph one wants. Indeed, if a variable \code{p} is used to store all the graphs, then \code{p} is a \code{list} containing "ACF" (the autocorrelation graphs), "MCMC" (the MCMC chains), "corr" (the scatterplot between each pair of parameters), "dens" (\textit{prior} and \textit{posterior} distributions) and "out" (the calibration result graph) variables.\newline

Two external functions can be run on an object generated by the function \code{calibrate}:
\begin{itemize}
\item \code{chain}: function that allows to extract the chains sampled in the \textit{posterior} distribution. If the variable \code{Nchains}, in \code{opt.estim} option, is higher than $1$ then the function \code{chain} return a \pkg{coda} \citep{coda} object with the sampled chains,
\item \code{estimators}: function that accesses the maximum \textit{a posteriori}(MAP) and the mean \textit{a posteriori}.
\end{itemize}

Sequential design introduced in \citet{damblin2018} allows to improve the Gaussian process estimation for $\mathcal{M}_2$ and $\mathcal{M}_4$. Based on the expected improvement (EI), introduced in \citet{jones1998efficient}, new points are added to an initial DOE in order to improve the quality of calibration. 
The arguments of the function are given Table \ref{tab:seqDesign}. \newline

\begin{table}[h!]
\centering
\begin{tabular}{llp{7.4cm}}
\hline
\code{sequentialDesign} arguments & Description \\ \hline
\code{md} & the model generated with the function \code{model}\\
&(for $\mathcal{M}_2$ or $\mathcal{M}_4$)\\
\code{pr} & the list of prior generated by the function \code{prior}\\
\code{opt.estim} & estimation options for calibration\\
\code{k} & number of points to add in the design\\
\hline
\end{tabular}
\caption{\label{tab:seqDesign} description of the arguments of the function \code{sequentialDesign}}
\end{table}

The last main function in \pkg{CaliCo} is \code{forecast} which produces a prediction of a selected model on a new data and based on previous calibration. \newline

\begin{table}[h!]
\centering
\begin{tabular}{llp{7.4cm}}
\hline
\code{forecast} arguments & Description \\ \hline
\code{modelfit} & calibrated model (run by \code{calibrate} function)\\
\code{x.new} & new data for prediction\\
\hline
\end{tabular}
\caption{\label{tab:calibrate} description of the arguments of the function \code{calibrate}}
\end{table}

The object generated by \code{forecast.class} possesses the two similar methods \code{print} and \code{plot}. The \code{print} function gives a summary identical to the one in \code{model.class} exept that it adds the MAP estimator. The \code{plot} function displays the calibration results and also adds the predicted results. The arguments of \code{plot} are the forecasted model and the x-axis which is the axis corresponding to calibration extended with the axis corresponding to the forecast.

\section{Multidimensional example with CaliCo}
\label{sec:exampleCaliCoMulti}

\begin{figure}[h!]
\begin{center}
\begin{knitrout}
\definecolor{shadecolor}{rgb}{0.969, 0.969, 0.969}\color{fgcolor}
\includegraphics[width=0.45\linewidth,height=0.45\linewidth]{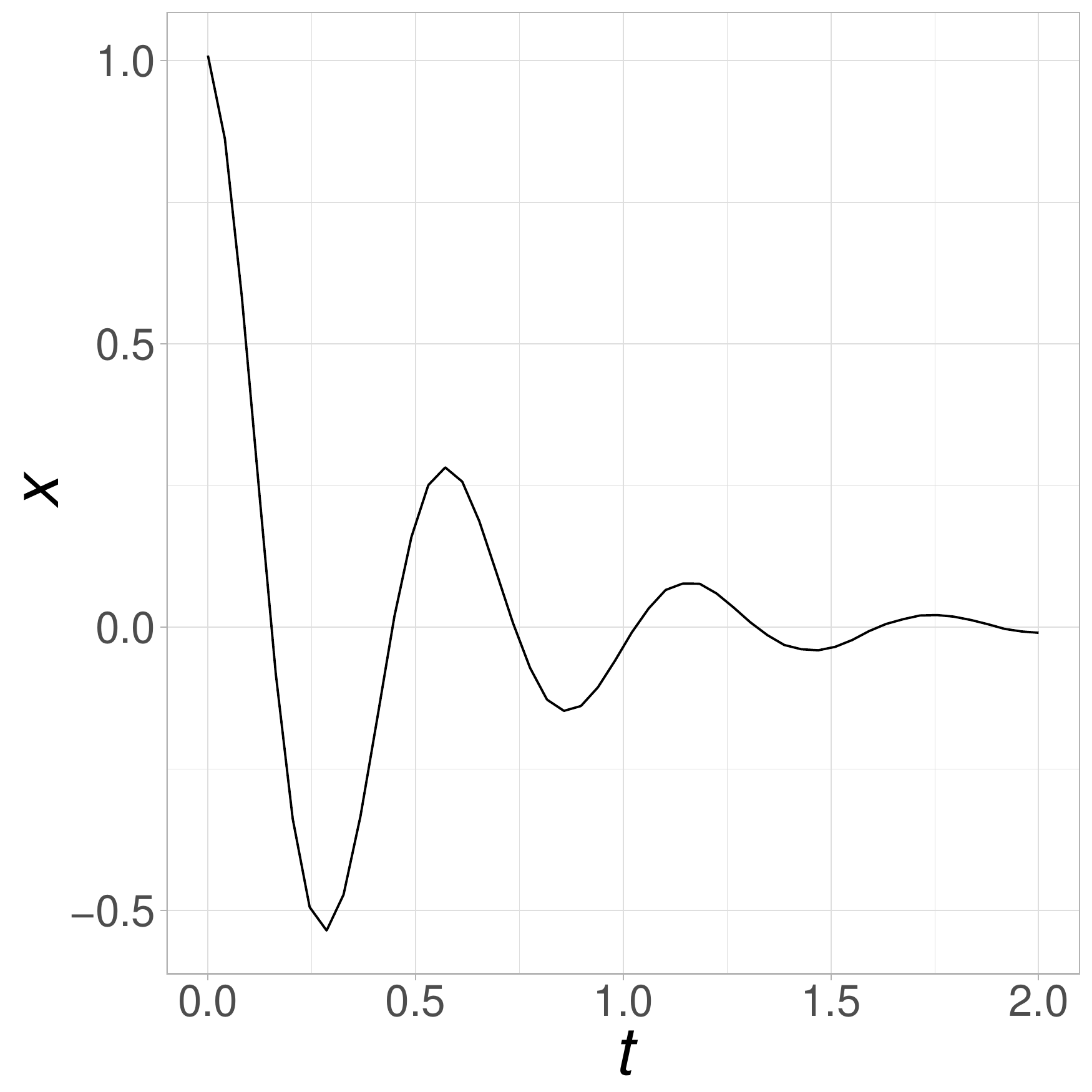} 

\end{knitrout}
\caption{Displacement of the oscillator simulated}
\label{fig:OscillatorDisplacement}\end{center}
\end{figure}

An illustration is provided in this section to help the user to easily handle the functionalities of \pkg{CaliCo}. This example, represents a damped harmonic oscillator and experimental data are simulated for specific values of the parameter vector $\boldsymbol{\theta}$. These parameters to calibrate are $A$ the constant amplitude, the damping ratio $\xi$, the spring constant $k$, the mass of the spring $m$ and $\phi$ the phase. The recorded displacement of the damped oscillator is represented Figure \ref{fig:OscillatorDisplacement} and the equation of the displacement of a damped harmonic oscillator is:

\begin{align}
x(t):
\mathbb{R}^6 & \mapsto \mathbb{R} \\
(t,\theta=(A,\xi,k,m,\phi)^T) &\rightarrow A e ^{-\xi\sqrt{\frac{k}{m}}t}sin(\sqrt{1-\xi^2}\sqrt{\frac{k}{m}}t + \phi)
\end{align}

There is five parameters to calibrate. Let us consider that experiments are available for the 2 first seconds of the movement (these experiments have been simulated for the interval time $[0,2]$ with the time step of $40ms$ and for specific parameter values). Visually, at time $t=0$ the position of the mass seems to be at the position $x=1$. So the values \textit{a priori} of $A$ and $\phi$ are $A=1$ and $\phi=\frac{\pi}{2}$. The company states that the spring has a constant of $k=6 N/m$ and the mass is weighing at $m=50g$. The major uncertainty lies in the knowledge of $\xi$. It is indeed a difficult parameter to estimate. However, the value of the damping ratio $\xi$ determines the behavior of the system. A damped harmonic oscillator can be:
\begin{itemize}
\item over-damped ($\xi > 1$): the system exponentially decays to steady state without oscillating,
\item critically damped ($\xi = 1$): the system returns to steady state as quickly as possible without oscillating,
\item under-damped ($\xi < 1$): The system oscillates with the amplitude gradually decreasing to zero.
\end{itemize}

Physical experts provide us a value of $\xi=0.3$ but says that the parameter can oscillate between the value $[0.15,0.45]$ at $95\%$.

\subsection{The models}

To define the first statistical model, the function \code{code} has to be defined such as:

\begin{knitrout}
\definecolor{shadecolor}{rgb}{0.969, 0.969, 0.969}\color{fgcolor}\begin{kframe}
\begin{alltt}
\hlstd{n} \hlkwb{<-} \hlnum{50}
\hlstd{t} \hlkwb{<-} \hlkwd{seq}\hlstd{(}\hlnum{0}\hlstd{,}\hlnum{2}\hlstd{,}\hlkwc{length.out}\hlstd{=n)}
\hlstd{code} \hlkwb{<-} \hlkwa{function}\hlstd{(}\hlkwc{t}\hlstd{,}\hlkwc{theta}\hlstd{)}
\hlstd{\{}
  \hlstd{w0} \hlkwb{<-} \hlkwd{sqrt}\hlstd{(theta[}\hlnum{3}\hlstd{]}\hlopt{/}\hlstd{theta[}\hlnum{4}\hlstd{])}
  \hlkwd{return}\hlstd{(theta[}\hlnum{1}\hlstd{]}\hlopt{*}\hlkwd{exp}\hlstd{(}\hlopt{-}\hlstd{theta[}\hlnum{2}\hlstd{]}\hlopt{*}\hlstd{w0}\hlopt{*}\hlstd{t)}\hlopt{*}\hlkwd{sin}\hlstd{(}\hlkwd{sqrt}\hlstd{(}\hlnum{1}\hlopt{-}\hlstd{theta[}\hlnum{2}\hlstd{]}\hlopt{^}\hlnum{2}\hlstd{)}\hlopt{*}\hlstd{w0}\hlopt{*}\hlstd{t}\hlopt{+}\hlstd{theta[}\hlnum{5}\hlstd{]))}
\hlstd{\}}
\end{alltt}
\end{kframe}
\end{knitrout}

In \pkg{CaliCo}, one function allows to define the statistical model. This function \code{model} takes as inputs the \code{code} function, the input variables \code{X}, experimental data and the model choice. If a numerical code has no input variables, it is just enough to put \code{X=0}.

\begin{knitrout}
\definecolor{shadecolor}{rgb}{0.969, 0.969, 0.969}\color{fgcolor}\begin{kframe}
\begin{alltt}
\hlstd{model1} \hlkwb{<-} \hlkwd{model}\hlstd{(code,}\hlkwc{X}\hlstd{=t,Yexp,}\hlstr{"model1"}\hlstd{)}
\end{alltt}
\end{kframe}
\end{knitrout}

In this particular case where, the input variables are unidimensional, it is easy to choose a graphical representation of the model. As mentioned Section \ref{sec:guidelines}, when the function \code{model} is called, a \code{model.class} object is generated. This object owns several methods as \code{plot} or \code{print} that behave as regular functions.

\begin{knitrout}
\definecolor{shadecolor}{rgb}{0.969, 0.969, 0.969}\color{fgcolor}\begin{kframe}
\begin{alltt}
\hlkwd{print}\hlstd{(model1)}
\end{alltt}
\begin{verbatim}
## Call:
## [1] "model1"
## 
## With the function:
## function(t,theta)
## {
##   w0 <- sqrt(theta[3]/theta[4])
##   return(theta[1]*exp(-theta[2]*w0*t)*sin(sqrt(1-theta[2]^2)*w0*t+theta[5]))
## }
## 
## No surrogate is selected
## 
## No discrepancy is added
\end{verbatim}
\end{kframe}
\end{knitrout}

To get a visual representation, parameter values need to be added to the model. To realize such an operation in \pkg{CaliCo}, one can use the defined pipe \code{\%<\%}. Following the pipe, a list containing all parameter values allows to select these values for the visual representation. The parameter vector (\code{theta}) and the value of the variance of the measurement error (\code{var}), here, are needed in the list to set a proper parametrization of the model: \newline

\begin{knitrout}
\definecolor{shadecolor}{rgb}{0.969, 0.969, 0.969}\color{fgcolor}\begin{kframe}
\begin{alltt}
\hlstd{model1} \hlopt{%<%} \hlkwd{list}\hlstd{(}\hlkwc{theta}\hlstd{=}\hlkwd{c}\hlstd{(}\hlnum{1}\hlstd{,}\hlnum{0.3}\hlstd{,}\hlnum{6}\hlstd{,}\hlnum{50e-3}\hlstd{,pi}\hlopt{/}\hlnum{2}\hlstd{),}\hlkwc{var}\hlstd{=}\hlnum{1e-4}\hlstd{)}
\end{alltt}

{\ttfamily\noindent\color{warningcolor}{\#\# Warning: Please be carefull to the size of the parameter vector}}\end{kframe}
\end{knitrout}

The \code{Warning} is present at each use of the pipe. It appears as a reminder for the user to be careful with the size of the parameter vector. When the model is defined nothing indicates the number of parameter within. To get a visual representation of the model with such parameters values, the \code{plot} function can be straightforwardly applied on the model object created by \code{model} and completed by the pipe \code{\%<\%}. The x-axis needs to be filled in \code{plot} to get an x-axis for display. The left panel of Figure \ref{fig:PlotModel1&2} is the result of:

\begin{knitrout}
\definecolor{shadecolor}{rgb}{0.969, 0.969, 0.969}\color{fgcolor}\begin{kframe}
\begin{alltt}
\hlkwd{plot}\hlstd{(model1,t)}
\end{alltt}
\end{kframe}
\end{knitrout}

If no parameter value is added to the model and the visual representation is required, a \code{Warning} appears and remind the user that no parameter value has been defined and only experiments are plotted (Figure \ref{FakeModel1}):
\begin{figure}[h!]
\begin{knitrout}
\definecolor{shadecolor}{rgb}{0.969, 0.969, 0.969}\color{fgcolor}\begin{kframe}
\begin{alltt}
\hlstd{model1bis} \hlkwb{<-} \hlkwd{model}\hlstd{(code,}\hlkwc{X}\hlstd{=t,Yexp,}\hlstr{"model1"}\hlstd{)}
\hlkwd{plot}\hlstd{(model1bis,t)}
\end{alltt}
\end{kframe}
\end{knitrout}
\centering
\begin{knitrout}
\definecolor{shadecolor}{rgb}{0.969, 0.969, 0.969}\color{fgcolor}\begin{kframe}

{\ttfamily\noindent\color{warningcolor}{\#\# Warning: no theta and var has been given to the model, experiments only are plotted}}\end{kframe}
\includegraphics[width=0.4\linewidth,height=0.35\linewidth]{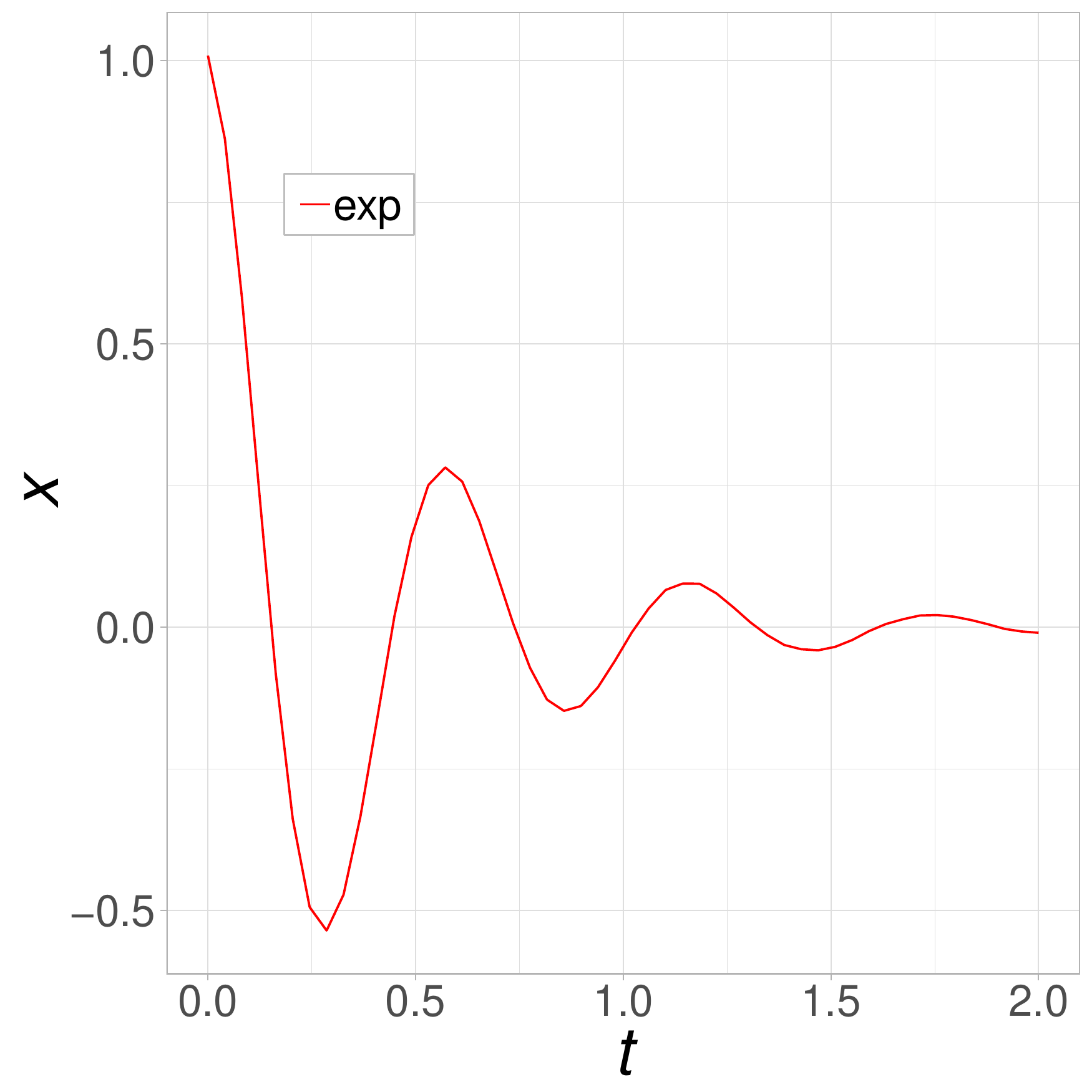} 

\end{knitrout}
\caption{Experimental data displayed when no parameter values are set in the model}
\label{FakeModel1}
\end{figure}

If no x-axis is defined, then no visual representation is possible and the function \code{plot} breaks:

\begin{knitrout}
\definecolor{shadecolor}{rgb}{0.969, 0.969, 0.969}\color{fgcolor}\begin{kframe}
\begin{alltt}
\hlstd{model1bis} \hlkwb{<-} \hlkwd{model}\hlstd{(code,}\hlkwc{X}\hlstd{=t,Yexp,}\hlstr{"model1"}\hlstd{)}
\hlkwd{plot}\hlstd{(model1bis)}
\end{alltt}

{\ttfamily\noindent\bfseries\color{errorcolor}{\#\# Error: No x-axis selected, no graph is displayed}}\end{kframe}
\end{knitrout}

For $\mathcal{M}_2$ several cases may occur (see Table \ref{tab:optionsGP} for more details). First the user only has the time consuming code without any Design Of Experiments (DOE). Then, the definition of the model is done by delimiting the boundaries of the parameters. The option \code{opt.gp} allows the user to set the kernel type of the Gaussian process and to specify if the user has a particular DOE. In this first case the DOE is not available, so \code{DOE=NULL} in the \code{opt.gp} option. To parametrize the DOE created in the function \code{model}, the option called \code{opt.emul} needs to be filled by \code{p}, \code{n.emul}, \code{binf}, \code{bsup}. Where \code{p} stands for the number of parameter to calibrate, \code{n.emul} for the number of experiments in the DOE, \code{binf} and \code{bsup} for the lower and upper bounds of the parameter vector.

\begin{knitrout}
\definecolor{shadecolor}{rgb}{0.969, 0.969, 0.969}\color{fgcolor}\begin{kframe}
\begin{alltt}
\hlstd{binf} \hlkwb{<-} \hlkwd{c}\hlstd{(}\hlnum{0.9}\hlstd{,}\hlnum{0.15}\hlstd{,}\hlnum{5.8}\hlstd{,}\hlnum{48e-3}\hlstd{,}\hlnum{1.49}\hlstd{)}
\hlstd{bsup} \hlkwb{<-} \hlkwd{c}\hlstd{(}\hlnum{1.1}\hlstd{,}\hlnum{0.45}\hlstd{,}\hlnum{6.2}\hlstd{,}\hlnum{52e-3}\hlstd{,}\hlnum{1.6}\hlstd{)}

\hlstd{model2} \hlkwb{<-} \hlkwd{model}\hlstd{(code,t,Yexp,}\hlstr{"model2"}\hlstd{,}
                \hlkwc{opt.gp} \hlstd{=} \hlkwd{list}\hlstd{(}\hlkwc{type}\hlstd{=}\hlstr{"matern5_2"}\hlstd{,}\hlkwc{DOE}\hlstd{=}\hlkwa{NULL}\hlstd{),}
                \hlkwc{opt.emul} \hlstd{=} \hlkwd{list}\hlstd{(}\hlkwc{p}\hlstd{=}\hlnum{5}\hlstd{,}\hlkwc{n.emul}\hlstd{=}\hlnum{60}\hlstd{,}\hlkwc{binf}\hlstd{=binf,}\hlkwc{bsup}\hlstd{=bsup))}
\end{alltt}
\end{kframe}
\end{knitrout}

The second case is when the users has a numerical code and a specific DOE. In \pkg{CaliCo}, the option \code{DOE} in \code{opt.gp} allows to consider a particular DOE wanted by the user. As no DOE is build with the function \code{model}, the option \code{opt.emul} is not necessary anymore:

\begin{knitrout}
\definecolor{shadecolor}{rgb}{0.969, 0.969, 0.969}\color{fgcolor}\begin{kframe}
\begin{alltt}
\hlkwd{library}\hlstd{(DiceDesign)}
\hlstd{DOE} \hlkwb{<-} \hlkwd{maximinSA_LHS}\hlstd{(}\hlkwd{lhsDesign}\hlstd{(}\hlnum{60}\hlstd{,}\hlnum{6}\hlstd{)}\hlopt{$}\hlstd{design)}\hlopt{$}\hlstd{design}
\hlstd{DOE} \hlkwb{<-} \hlkwd{unscale}\hlstd{(DOE,}\hlkwd{c}\hlstd{(}\hlnum{0}\hlstd{,binf),}\hlkwd{c}\hlstd{(}\hlnum{2}\hlstd{,bsup))}

\hlstd{model2doe} \hlkwb{<-} \hlkwd{model}\hlstd{(code,t,Yexp,}\hlstr{"model2"}\hlstd{,}
                   \hlkwc{opt.gp}\hlstd{=}\hlkwd{list}\hlstd{(}\hlkwc{type}\hlstd{=}\hlstr{"matern5_2"}\hlstd{,}\hlkwc{DOE}\hlstd{=DOE))}
\end{alltt}
\end{kframe}
\end{knitrout}

When one does not possess any numerical code, but only the DOE and the corresponding output, another option, called \code{opt.sim}, needs to be filled. The \code{opt.gp} option is still needed to specify the chosen kernel but the \code{opt.emul} option is no longer necessary (for the same reasons as in the second case). The \code{opt.sim} option is the list containing the DOE and the output of the code. As the user does not possess the numerical code, the \code{code} option in the function \code{model} can be set to \code{code=NULL}.

\begin{knitrout}
\definecolor{shadecolor}{rgb}{0.969, 0.969, 0.969}\color{fgcolor}\begin{kframe}
\begin{alltt}
\hlstd{Ysim} \hlkwb{<-} \hlkwd{code}\hlstd{(DOE[}\hlnum{1}\hlstd{,}\hlnum{1}\hlstd{],DOE[}\hlnum{1}\hlstd{,}\hlnum{2}\hlopt{:}\hlnum{6}\hlstd{])}
\hlkwa{for} \hlstd{(i} \hlkwa{in} \hlnum{2}\hlopt{:}\hlnum{60}\hlstd{)\{Ysim} \hlkwb{<-} \hlkwd{c}\hlstd{(Ysim,}\hlkwd{code}\hlstd{(DOE[i,}\hlnum{1}\hlstd{],DOE[i,}\hlnum{2}\hlopt{:}\hlnum{6}\hlstd{]))\}}

\hlstd{model2code} \hlkwb{<-} \hlkwd{model}\hlstd{(}\hlkwc{code}\hlstd{=}\hlkwa{NULL}\hlstd{,t,Yexp,}\hlstr{"model2"}\hlstd{,}
                    \hlkwc{opt.gp} \hlstd{=} \hlkwd{list}\hlstd{(}\hlkwc{type}\hlstd{=}\hlstr{"matern5_2"}\hlstd{,} \hlkwc{DOE}\hlstd{=}\hlkwa{NULL}\hlstd{),}
                    \hlkwc{opt.sim} \hlstd{=} \hlkwd{list}\hlstd{(}\hlkwc{Ysim}\hlstd{=Ysim,}\hlkwc{DOEsim}\hlstd{=DOE))}
\end{alltt}
\end{kframe}
\end{knitrout}

The package \pkg{CaliCo} is comfortable with these three situations and bring flexibility according to the different problems of the users. Similarly as before, parameter values need to be added to each models and the function \code{print} and \code{plot} can be directly used:

\begin{knitrout}
\definecolor{shadecolor}{rgb}{0.969, 0.969, 0.969}\color{fgcolor}\begin{kframe}
\begin{alltt}
\hlstd{ParamList} \hlkwb{<-} \hlkwd{list}\hlstd{(}\hlkwc{theta}\hlstd{=}\hlkwd{c}\hlstd{(}\hlnum{1}\hlstd{,}\hlnum{0.3}\hlstd{,}\hlnum{6}\hlstd{,}\hlnum{50e-3}\hlstd{,pi}\hlopt{/}\hlnum{2}\hlstd{),}\hlkwc{var}\hlstd{=}\hlnum{1e-4}\hlstd{)}
\hlstd{model2} \hlopt{%<%} \hlstd{ParamList}
\hlstd{model2doe} \hlopt{%<%} \hlstd{ParamList}
\hlstd{model2code} \hlopt{%<%} \hlstd{ParamList}
\end{alltt}
\end{kframe}
\end{knitrout}

\begin{knitrout}
\definecolor{shadecolor}{rgb}{0.969, 0.969, 0.969}\color{fgcolor}\begin{kframe}
\begin{alltt}
\hlkwd{plot}\hlstd{(model2,t)}
\hlkwd{plot}\hlstd{(model2doe,t)}
\hlkwd{plot}\hlstd{(model2code,t)}
\end{alltt}
\end{kframe}
\end{knitrout}

These three lines of code produce the same graphs because \pkg{CaliCo} uses a maximin Latin Hypercube Sample (LHS) to establish the DOE. Several credibility interval are displayed. For the first model only the $95\%$ credibility interval of the measurement error is available. For the second model, the $95\%$ credibility interval of the Gaussian process can also be shown. Figure \ref{fig:PlotModel1&2} illustrates $\mathcal{M}_1$ and $\mathcal{M}_2$. \newline

\begin{figure}[h!]
\begin{center}
    \begin{tabular}{cc}
    $\mathcal{M}_1$ & $\mathcal{M}_2$ \\
\begin{knitrout}
\definecolor{shadecolor}{rgb}{0.969, 0.969, 0.969}\color{fgcolor}
\includegraphics[width=0.4\linewidth,height=0.35\linewidth]{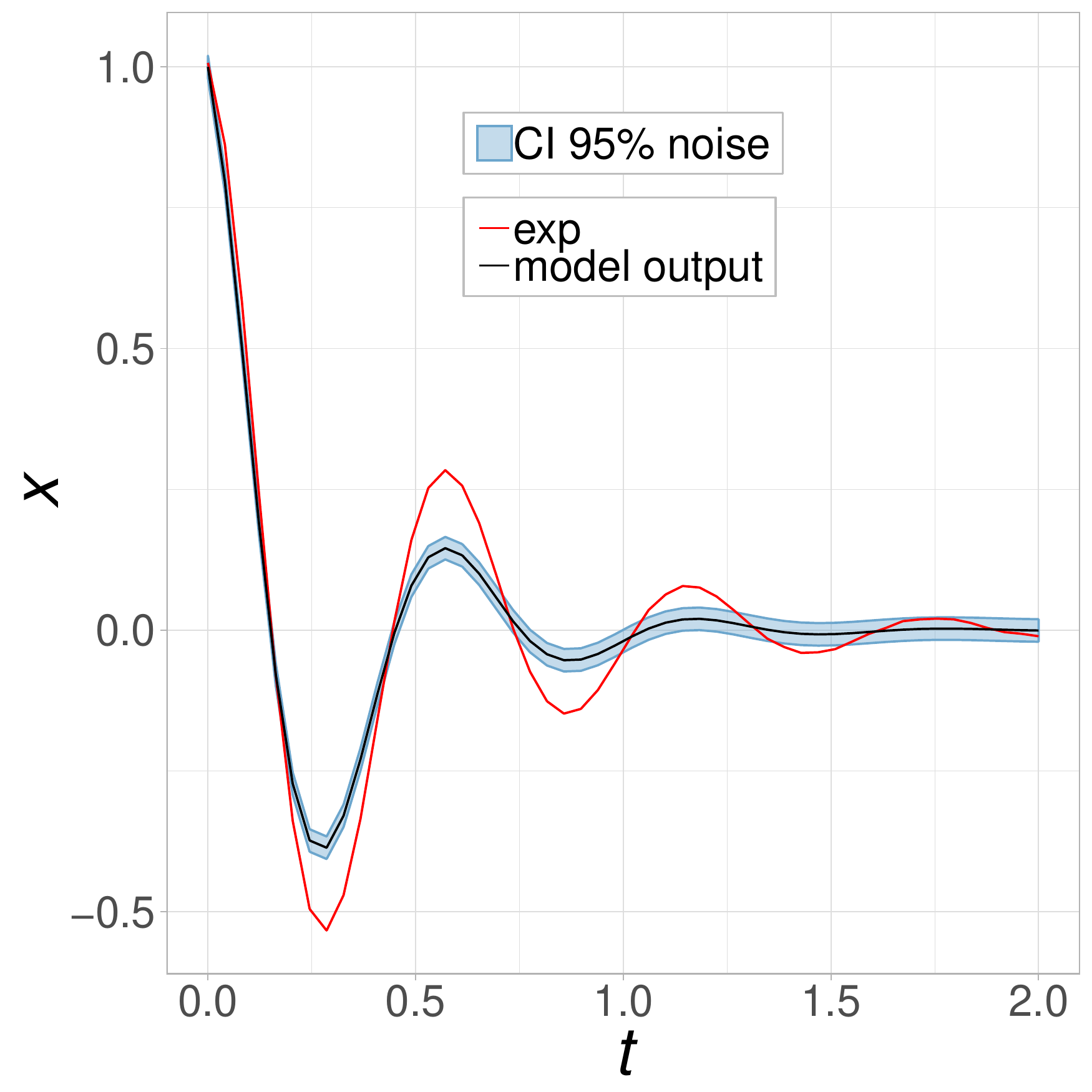} 

\end{knitrout}
    &
\begin{knitrout}
\definecolor{shadecolor}{rgb}{0.969, 0.969, 0.969}\color{fgcolor}
\includegraphics[width=0.4\linewidth,height=0.35\linewidth]{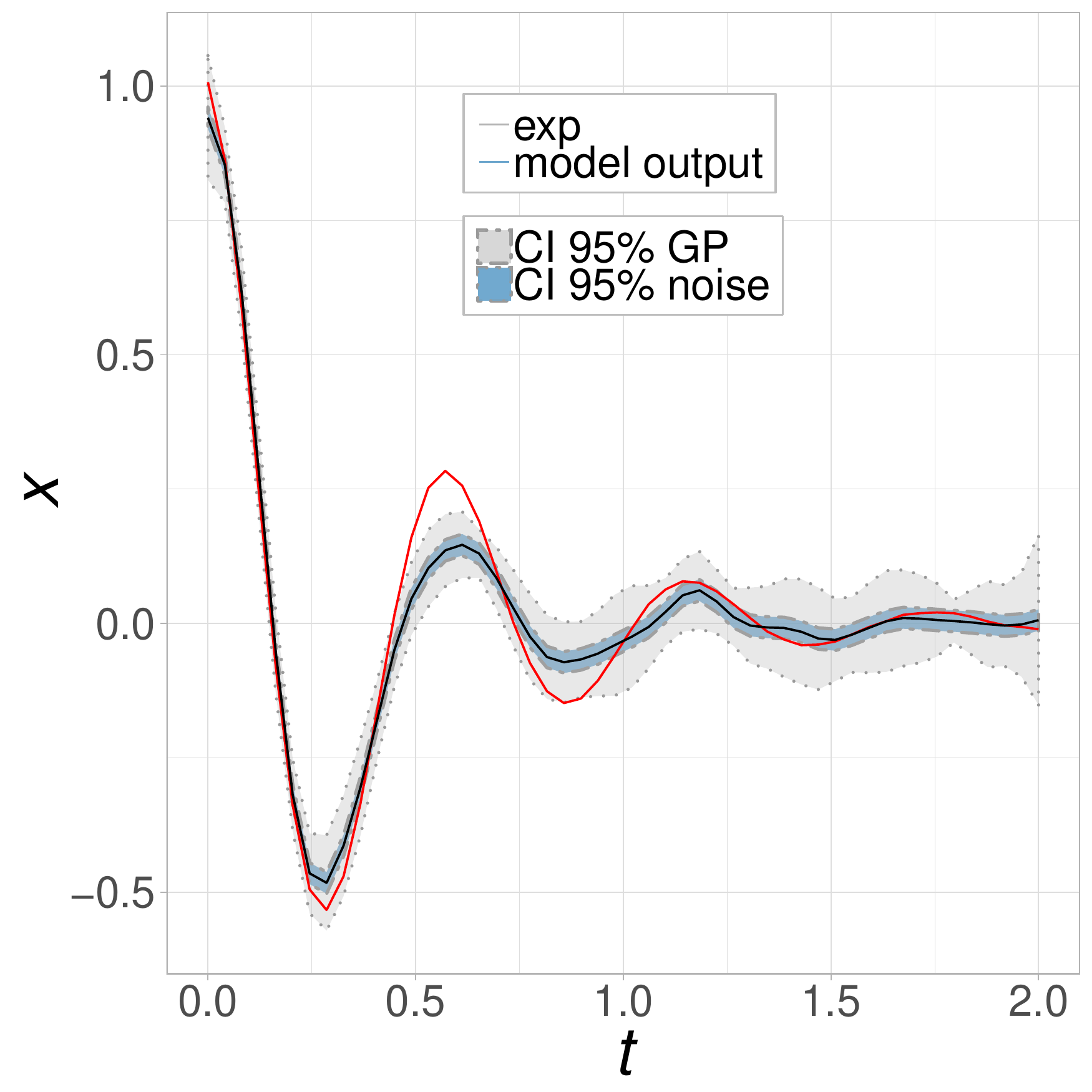} 

\end{knitrout}
  \end{tabular}
\caption{First and second model output for prior belief on parameter values. The left panel illustrates the first model and the right panel the second model with the Gaussian process estimated.}
\label{fig:PlotModel1&2}
\end{center}
\end{figure}

One can be interested in modeling a code discrepancy. When the code is not time consuming, the model to choose is $\mathcal{M}_3$. The \code{opt.disc} option allows to specify the kernel type of the discrepancy. Note that the visual representation requires initial values for the discrepancy and are needed in the pipe \code{\%<\%}. This vector \code{thetaD} is composed of $\sigma^2$ and $\psi$ according to Table \ref{tab:kernelD}.

\begin{knitrout}
\definecolor{shadecolor}{rgb}{0.969, 0.969, 0.969}\color{fgcolor}\begin{kframe}
\begin{alltt}
\hlstd{model3} \hlkwb{<-} \hlkwd{model}\hlstd{(code,t,Yexp,}\hlstr{"model3"}\hlstd{,}
                \hlkwc{opt.disc} \hlstd{=} \hlkwd{list}\hlstd{(}\hlkwc{kernel.type}\hlstd{=}\hlstr{"gauss"}\hlstd{))}
\hlstd{model3} \hlopt{%<%} \hlkwd{list}\hlstd{(}\hlkwc{theta}\hlstd{=}\hlkwd{c}\hlstd{(}\hlnum{1}\hlstd{,}\hlnum{0.3}\hlstd{,}\hlnum{6}\hlstd{,}\hlnum{50e-3}\hlstd{,pi}\hlopt{/}\hlnum{2}\hlstd{),}\hlkwc{thetaD}\hlstd{=}\hlkwd{c}\hlstd{(}\hlnum{1e-4}\hlstd{,}\hlnum{0.2}\hlstd{),}\hlkwc{var}\hlstd{=}\hlnum{1e-4}\hlstd{)}
\end{alltt}
\end{kframe}
\end{knitrout}

When the code is time consuming, then $\mathcal{M}_4$ is selected. The same cases can occur as for $\mathcal{M}_2$ but only the case where the code is not available will be considered, here, for $\mathcal{M}_4$.

\begin{knitrout}
\definecolor{shadecolor}{rgb}{0.969, 0.969, 0.969}\color{fgcolor}\begin{kframe}
\begin{alltt}
\hlstd{model4} \hlkwb{<-} \hlkwd{model}\hlstd{(}\hlkwc{code}\hlstd{=}\hlkwa{NULL}\hlstd{,t,Yexp,}\hlstr{"model4"}\hlstd{,}
                \hlkwc{opt.gp} \hlstd{=} \hlkwd{list}\hlstd{(}\hlkwc{type}\hlstd{=}\hlstr{"matern5_2"}\hlstd{,} \hlkwc{DOE}\hlstd{=}\hlkwa{NULL}\hlstd{),}
                \hlkwc{opt.sim} \hlstd{=} \hlkwd{list}\hlstd{(}\hlkwc{Ysim}\hlstd{=Ysim,}\hlkwc{DOEsim}\hlstd{=DOE),}
                \hlkwc{opt.disc} \hlstd{=} \hlkwd{list}\hlstd{(}\hlkwc{kernel.type}\hlstd{=}\hlstr{"gauss"}\hlstd{))}
\hlstd{model4} \hlopt{%<%} \hlkwd{list}\hlstd{(}\hlkwc{theta}\hlstd{=}\hlkwd{c}\hlstd{(}\hlnum{1}\hlstd{,}\hlnum{0.3}\hlstd{,}\hlnum{6}\hlstd{,}\hlnum{50e-3}\hlstd{,pi}\hlopt{/}\hlnum{2}\hlstd{),}\hlkwc{thetaD}\hlstd{=}\hlkwd{c}\hlstd{(}\hlnum{1e-4}\hlstd{,}\hlnum{0.2}\hlstd{),}\hlkwc{var}\hlstd{=}\hlnum{1e-4}\hlstd{)}
\end{alltt}
\end{kframe}
\end{knitrout}

To get a visual representation of $\mathcal{M}_3$ and $\mathcal{M}_4$, the function \code{plot} is defined identically as before. The results are displayed Figure \ref{fig:PlotModel3&4}.

\begin{knitrout}
\definecolor{shadecolor}{rgb}{0.969, 0.969, 0.969}\color{fgcolor}\begin{kframe}
\begin{alltt}
\hlkwd{plot}\hlstd{(model3,t)}
\hlkwd{plot}\hlstd{(model4,t)}
\end{alltt}
\end{kframe}
\end{knitrout}

\begin{figure}[h!]
\begin{center}
    \begin{tabular}{cc}
    $\mathcal{M}_3$ & $\mathcal{M}_4$ \\
\begin{knitrout}
\definecolor{shadecolor}{rgb}{0.969, 0.969, 0.969}\color{fgcolor}
\includegraphics[width=0.4\linewidth,height=0.35\linewidth]{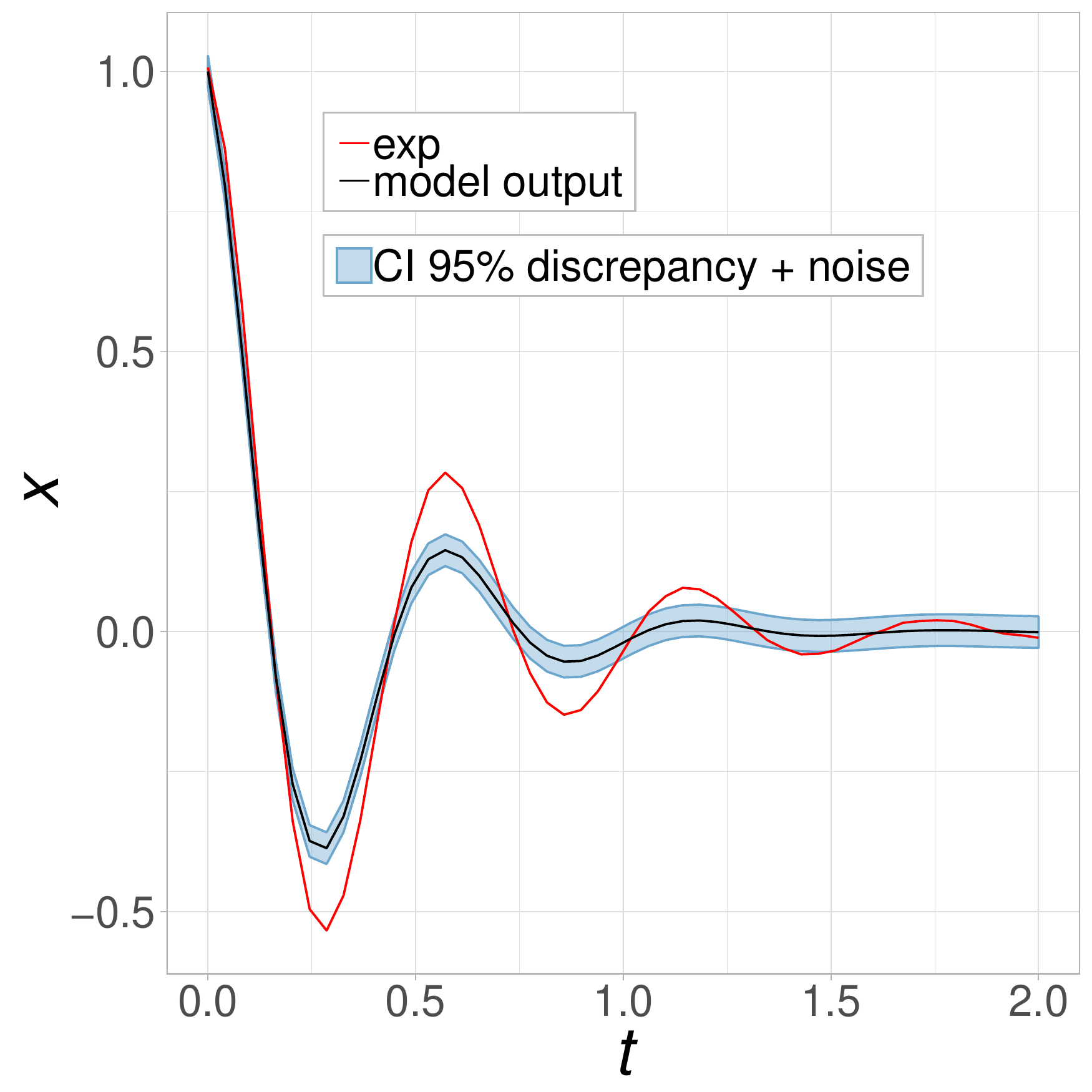} 

\end{knitrout}
    &
\begin{knitrout}
\definecolor{shadecolor}{rgb}{0.969, 0.969, 0.969}\color{fgcolor}
\includegraphics[width=0.4\linewidth,height=0.35\linewidth]{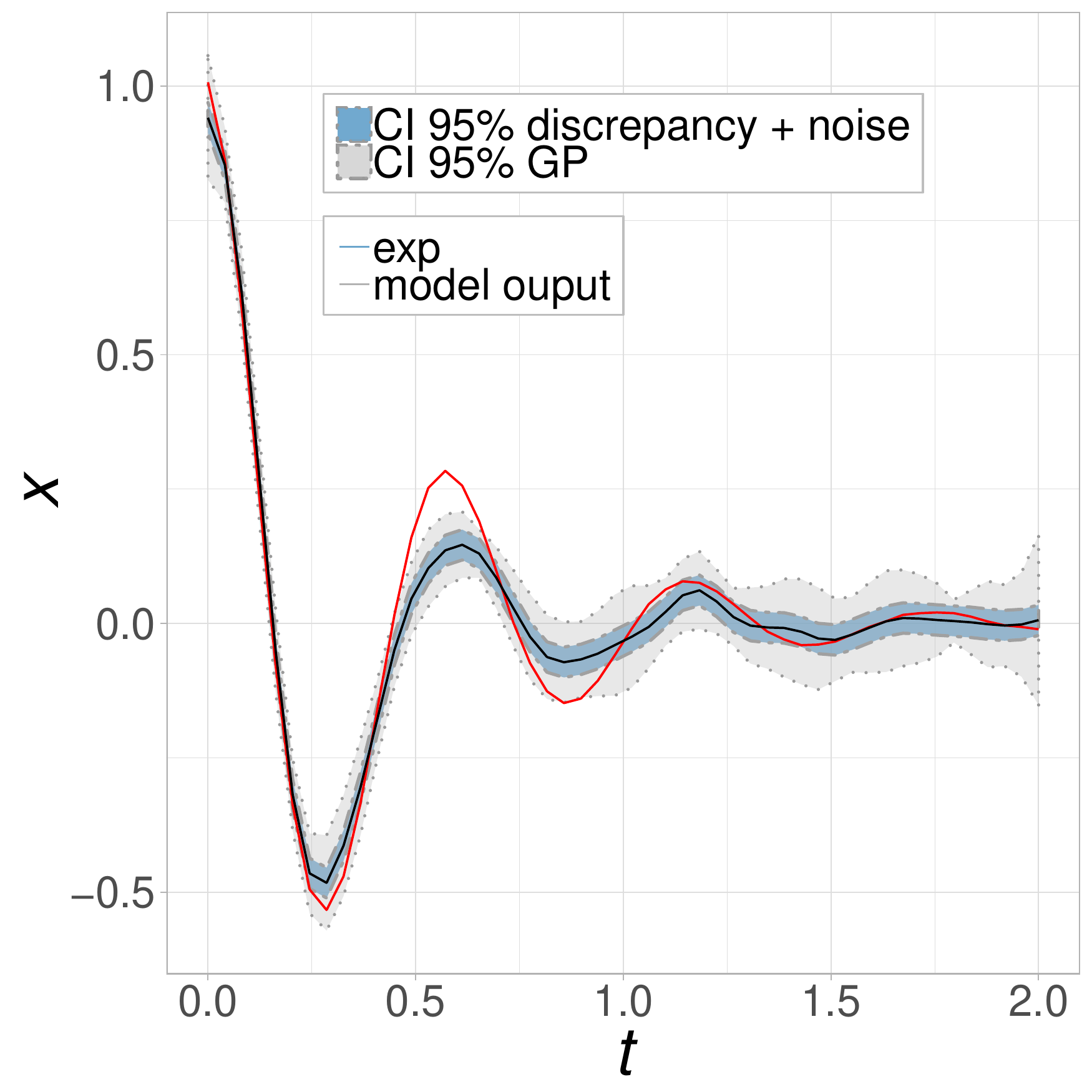} 

\end{knitrout}
\\
  \end{tabular}
\caption{Third and fourth model output for prior belief on parameter values. The left panel illustrates the third model and the right one, the fourth model with the Gaussian process estimated. Both are encompassing the discrepancy.}\label{fig:PlotModel3&4}
\end{center}
\end{figure}

Note that several credibility intervals can be displayed. By default all of them are shown (for example see right panels of Figures \ref{fig:PlotModel1&2} and \ref{fig:PlotModel3&4}). For $\mathcal{M}_1$ and $\mathcal{M}_3$ only one credibility interval is given. It represents the $95\%$ credibility interval of the measurement error, in the case of $\mathcal{M}_1$, and the $95\%$ credibility interval of the measurement error plus the discrepancy, in the case of $\mathcal{M}_3$. For $\mathcal{M}_2$ and $\mathcal{M}_4$, two credibility intervals are available. Compared to $\mathcal{M}_1$ and $\mathcal{M}_3$ the credibility interval at $95\%$ of the Gaussian process, that emulates the code, is added. It allows to quickly visualize from where the variability of the model comes before calibration.\newline

With the option \code{CI}, one can deactivate or select which credibility interval (CI) one wants to display. By default \code{CI="all"}, but if \code{CI="err"} only the $95\%$ CI of the measurement error with, or without, the discrepancy is given. Similarly, for $\mathcal{M}_2$ and $\mathcal{M}_4$, if \code{CI="GP"}, only the $95\%$ CI of the Gaussian process is shown. For example, for $\mathcal{M}_4$ the three possibilities are obtained with the following code and are displayed Figure \ref{fig:Model4CI}.

\begin{knitrout}
\definecolor{shadecolor}{rgb}{0.969, 0.969, 0.969}\color{fgcolor}\begin{kframe}
\begin{alltt}
\hlkwd{plot}\hlstd{(model4,t,}\hlkwc{CI}\hlstd{=}\hlstr{"err"}\hlstd{)}
\hlkwd{plot}\hlstd{(model4,t,}\hlkwc{CI}\hlstd{=}\hlstr{"GP"}\hlstd{)}
\hlkwd{plot}\hlstd{(model4,t,}\hlkwc{CI}\hlstd{=}\hlstr{"all"}\hlstd{)}
\end{alltt}
\end{kframe}
\end{knitrout}

\begin{figure}[h!]
\begin{center}
    \begin{tabular}{ccc}
\begin{knitrout}
\definecolor{shadecolor}{rgb}{0.969, 0.969, 0.969}\color{fgcolor}
\includegraphics[width=0.29\linewidth,height=0.25\linewidth]{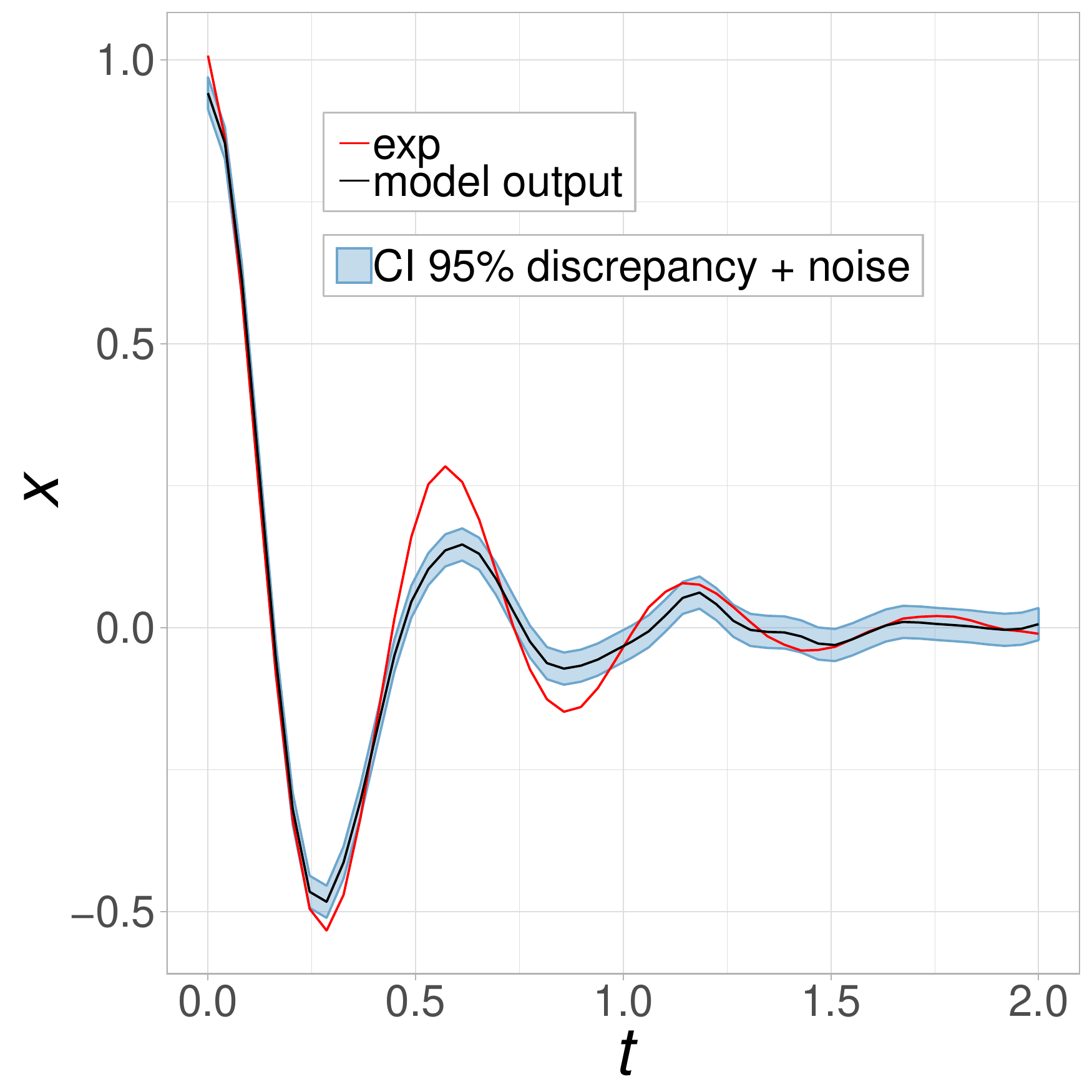} 

\end{knitrout}
    &
\begin{knitrout}
\definecolor{shadecolor}{rgb}{0.969, 0.969, 0.969}\color{fgcolor}
\includegraphics[width=0.29\linewidth,height=0.25\linewidth]{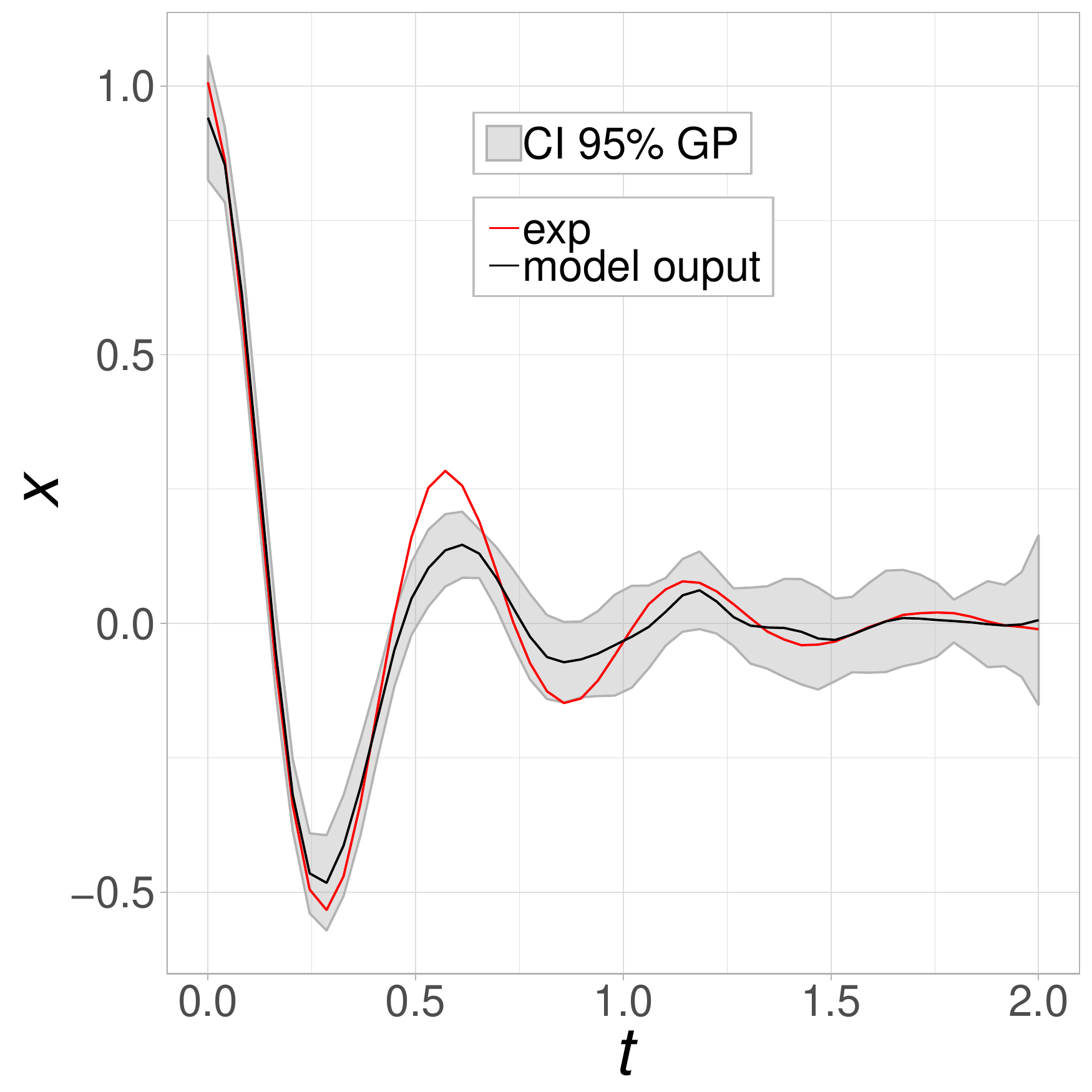} 

\end{knitrout}
&
\begin{knitrout}
\definecolor{shadecolor}{rgb}{0.969, 0.969, 0.969}\color{fgcolor}
\includegraphics[width=0.29\linewidth,height=0.25\linewidth]{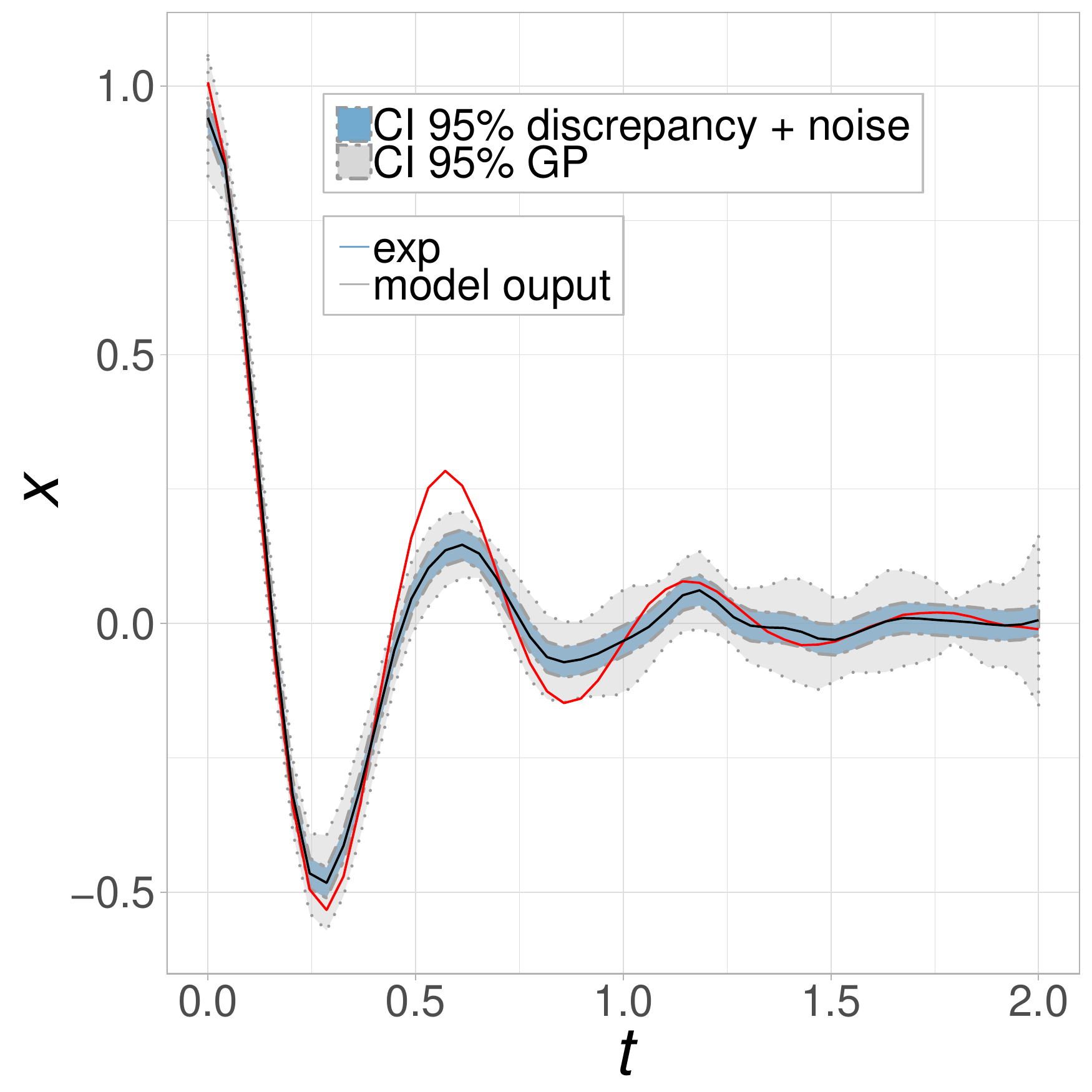} 

\end{knitrout}
  \end{tabular}
\caption{$\mathcal{M}_4$ displayed for some guessed values with the CI relative to the measurement error on the left panel, with the CI relative to the Gaussian process only on the middle panel and both credibility intervals on the right panel.}
\label{fig:Model4CI}
\end{center}
\end{figure}

\subsection{Priors}
To a proper Bayesian calibration, prior distributions have to be defined on every parameters we want to estimate (parameters of interest as $\boldsymbol{\theta}$ or nuisance parameter such as $\boldsymbol{\theta}_{\delta}$ or $\sigma_{err}^2$). It means that the number of parameters to estimate differs according to the model. In \pkg{CaliCo}, the possible distributions are detailed
in Table \ref{tab:priorDensities}.\newline

To define a proper prior distribution in \pkg{CaliCo}, a \code{prior.class} object with the function \code{prior} is generated. One or several prior distributions can be produced with this function. Two arguments have to be completed: \code{type.prior} and \code{opt.prior}. The argument \code{type.prior} can be a string (if only one prior distribution is looked for) or a vector of strings (if several prior distributions are wanted). Similarly, the argument \code{opt.prior} can be a vector of the distribution parameters or a list of vectors.\newline

In this example, $5$ parameters have to be calibrated. For $\mathcal{M}_3$ and $\mathcal{M}_4$, the discrepancy is added and the variance $\sigma_{\delta}^2$ with the correlation length $\psi$ have to be estimated as much as the other parameters. It means that for these models, two more prior distributions have to be added compared at $\mathcal{M}_1$ and $\mathcal{M}_2$. The order to define them are, first the parameters $\boldsymbol{\theta}$, then $\boldsymbol{\theta}_{\delta}$ and $\sigma_{err}^2$. In the following code lines, \code{pr1} stands for the prior distributions for $\mathcal{M}_1$ and $\mathcal{M}_2$ where \code{pr2} for $\mathcal{M}_3$ and $\mathcal{M}_4$. In the first prior definition, only the $5$ parameters and $\sigma_{err}^2$ prior distributions are defined. In the second definition, the $\boldsymbol{\theta}_{\delta}$ prior distributions are added between $\boldsymbol{\theta}$ and $\sigma_{err}^2$ ones.\newline

\begin{knitrout}
\definecolor{shadecolor}{rgb}{0.969, 0.969, 0.969}\color{fgcolor}\begin{kframe}
\begin{alltt}
\hlstd{type.prior} \hlkwb{<-} \hlkwd{c}\hlstd{(}\hlkwd{rep}\hlstd{(}\hlstr{"gaussian"}\hlstd{,}\hlnum{5}\hlstd{),}\hlstr{"gamma"}\hlstd{)}
\hlstd{opt.prior} \hlkwb{<-} \hlkwd{list}\hlstd{(}\hlkwd{c}\hlstd{(}\hlnum{1}\hlstd{,}\hlnum{1e-3}\hlstd{),}\hlkwd{c}\hlstd{(}\hlnum{0.3}\hlstd{,}\hlnum{1e-3}\hlstd{),}\hlkwd{c}\hlstd{(}\hlnum{6}\hlstd{,}\hlnum{1e-3}\hlstd{),}\hlkwd{c}\hlstd{(}\hlnum{50e-3}\hlstd{,}\hlnum{1e-5}\hlstd{),}
                  \hlkwd{c}\hlstd{(pi}\hlopt{/}\hlnum{2}\hlstd{,}\hlnum{1e-2}\hlstd{),}\hlkwd{c}\hlstd{(}\hlnum{1}\hlstd{,}\hlnum{1e-3}\hlstd{))}
\hlstd{pr1} \hlkwb{<-} \hlkwd{prior}\hlstd{(type.prior,opt.prior)}
\end{alltt}
\end{kframe}
\end{knitrout}

\begin{knitrout}
\definecolor{shadecolor}{rgb}{0.969, 0.969, 0.969}\color{fgcolor}\begin{kframe}
\begin{alltt}
\hlstd{type.prior} \hlkwb{<-} \hlkwd{c}\hlstd{(}\hlkwd{rep}\hlstd{(}\hlstr{"gaussian"}\hlstd{,}\hlnum{5}\hlstd{),}\hlstr{"gamma"}\hlstd{,}\hlstr{"unif"}\hlstd{,}\hlstr{"gamma"}\hlstd{)}
\hlstd{opt.prior} \hlkwb{<-} \hlkwd{list}\hlstd{(}\hlkwd{c}\hlstd{(}\hlnum{1}\hlstd{,}\hlnum{1e-3}\hlstd{),}\hlkwd{c}\hlstd{(}\hlnum{0.3}\hlstd{,}\hlnum{1e-3}\hlstd{),}\hlkwd{c}\hlstd{(}\hlnum{6}\hlstd{,}\hlnum{1e-3}\hlstd{),}\hlkwd{c}\hlstd{(}\hlnum{50e-3}\hlstd{,}\hlnum{1e-5}\hlstd{),}
                  \hlkwd{c}\hlstd{(pi}\hlopt{/}\hlnum{2}\hlstd{,}\hlnum{1e-2}\hlstd{),}\hlkwd{c}\hlstd{(}\hlnum{1}\hlstd{,}\hlnum{1e-3}\hlstd{),}\hlkwd{c}\hlstd{(}\hlnum{0}\hlstd{,}\hlnum{1}\hlstd{),}\hlkwd{c}\hlstd{(}\hlnum{1}\hlstd{,}\hlnum{1e-3}\hlstd{))}
\hlstd{pr2} \hlkwb{<-} \hlkwd{prior}\hlstd{(type.prior,opt.prior)}
\end{alltt}
\end{kframe}
\end{knitrout}

\begin{figure}[h!]
\begin{center}
    \begin{tabular}{cccc}
\begin{knitrout}
\definecolor{shadecolor}{rgb}{0.969, 0.969, 0.969}\color{fgcolor}
\includegraphics[width=0.21\linewidth,height=0.15\linewidth]{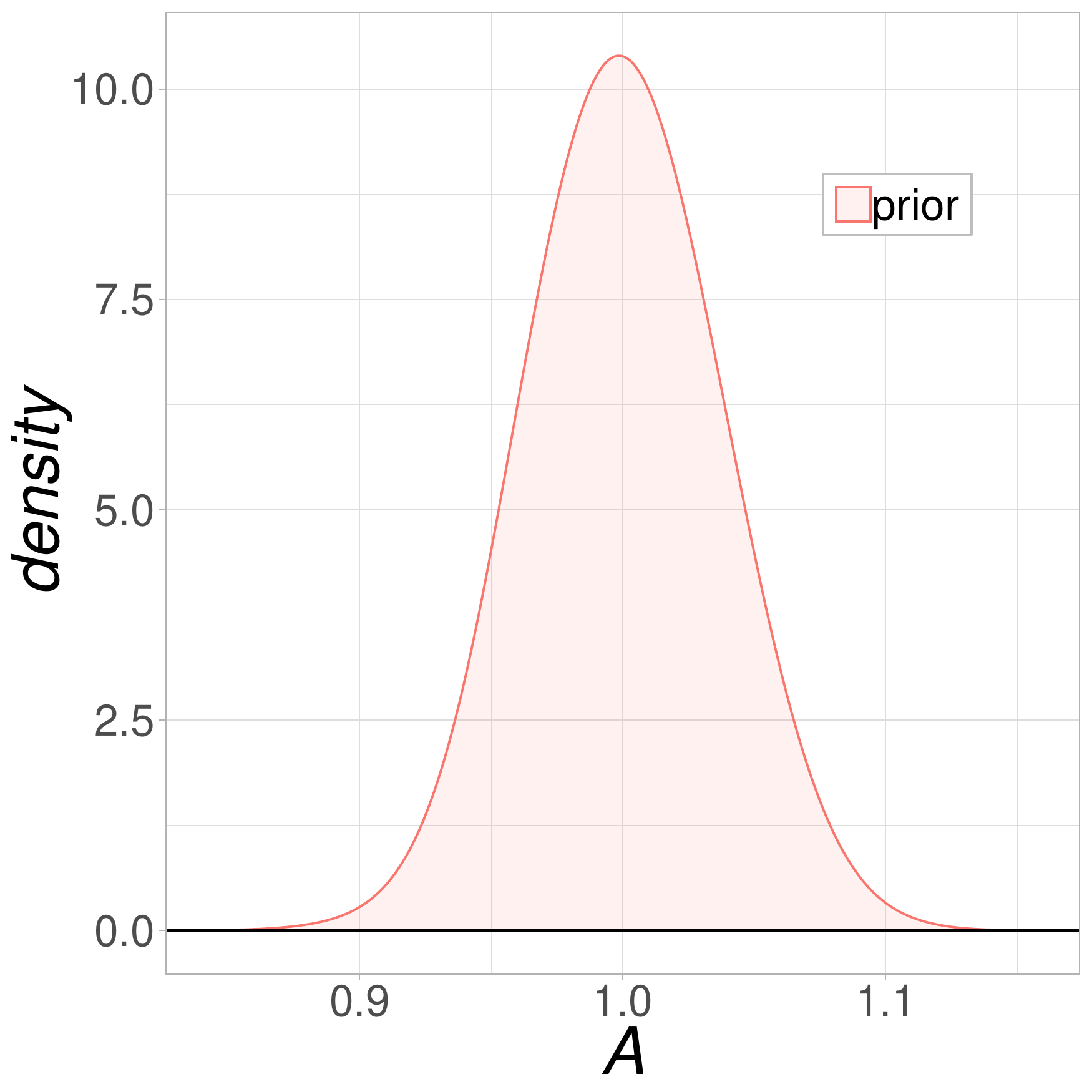} 

\end{knitrout}
    &
\begin{knitrout}
\definecolor{shadecolor}{rgb}{0.969, 0.969, 0.969}\color{fgcolor}
\includegraphics[width=0.21\linewidth,height=0.15\linewidth]{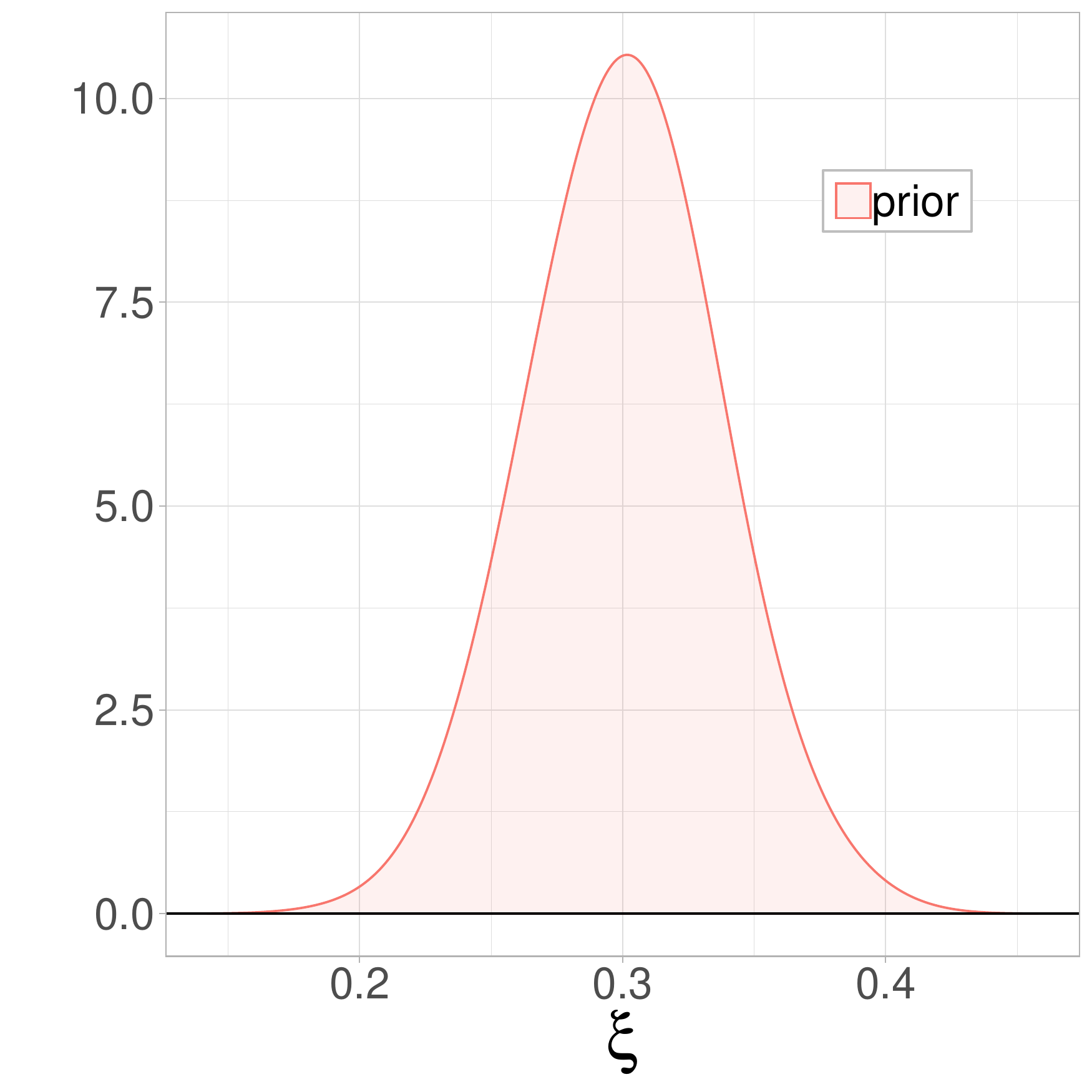} 

\end{knitrout}
&
\begin{knitrout}
\definecolor{shadecolor}{rgb}{0.969, 0.969, 0.969}\color{fgcolor}
\includegraphics[width=0.21\linewidth,height=0.15\linewidth]{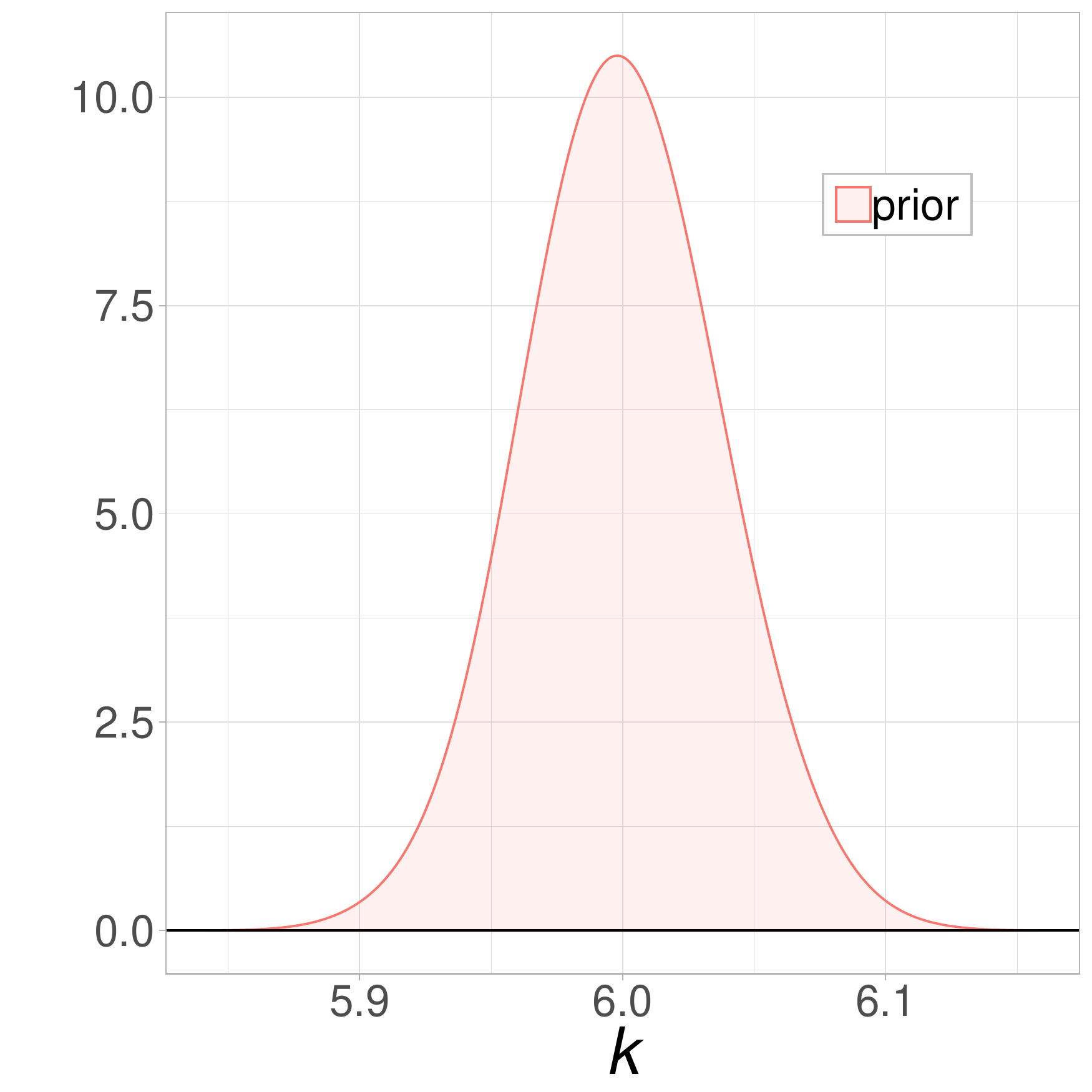} 

\end{knitrout}
    &
\begin{knitrout}
\definecolor{shadecolor}{rgb}{0.969, 0.969, 0.969}\color{fgcolor}
\includegraphics[width=0.21\linewidth,height=0.15\linewidth]{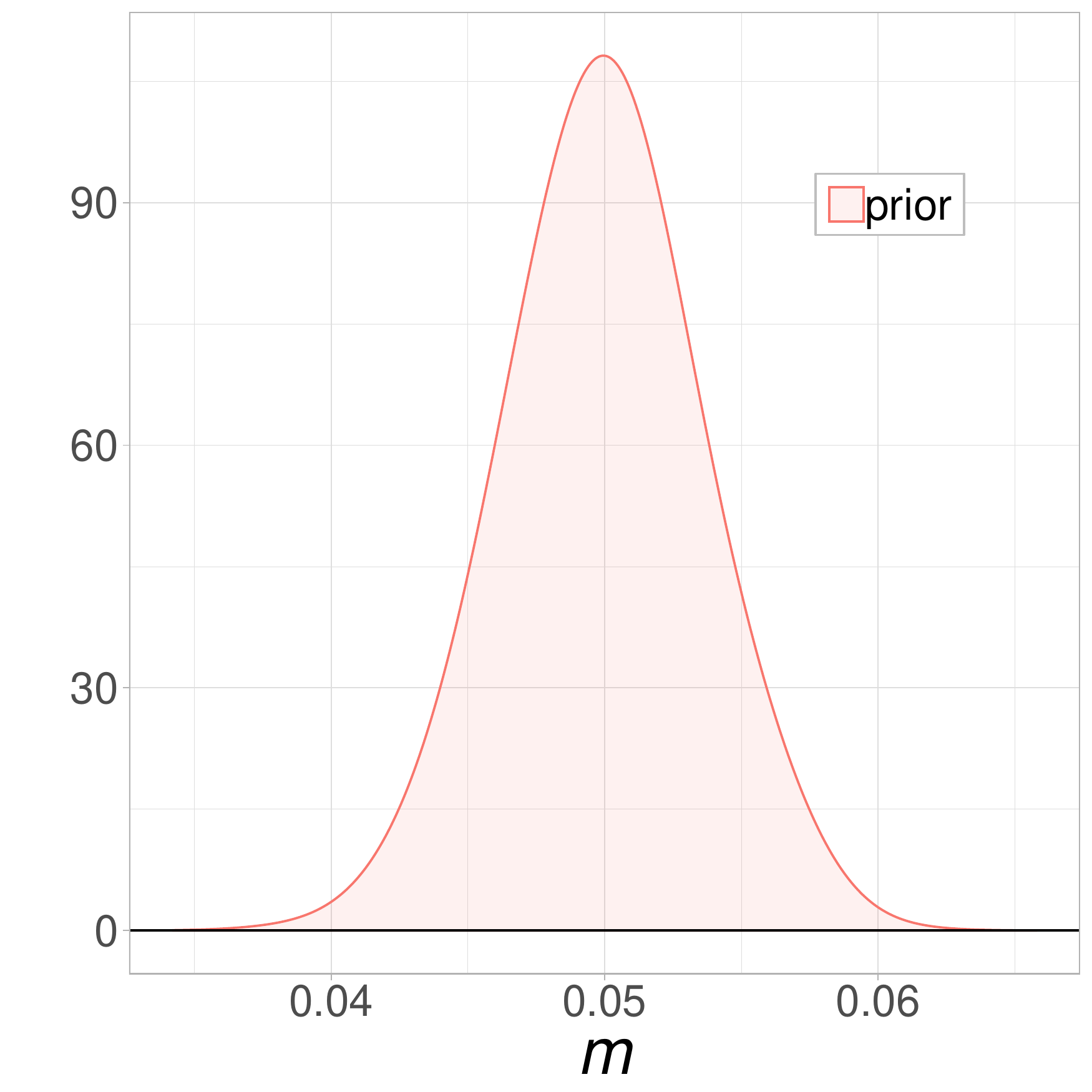} 

\end{knitrout}
\\
\begin{knitrout}
\definecolor{shadecolor}{rgb}{0.969, 0.969, 0.969}\color{fgcolor}
\includegraphics[width=0.21\linewidth,height=0.15\linewidth]{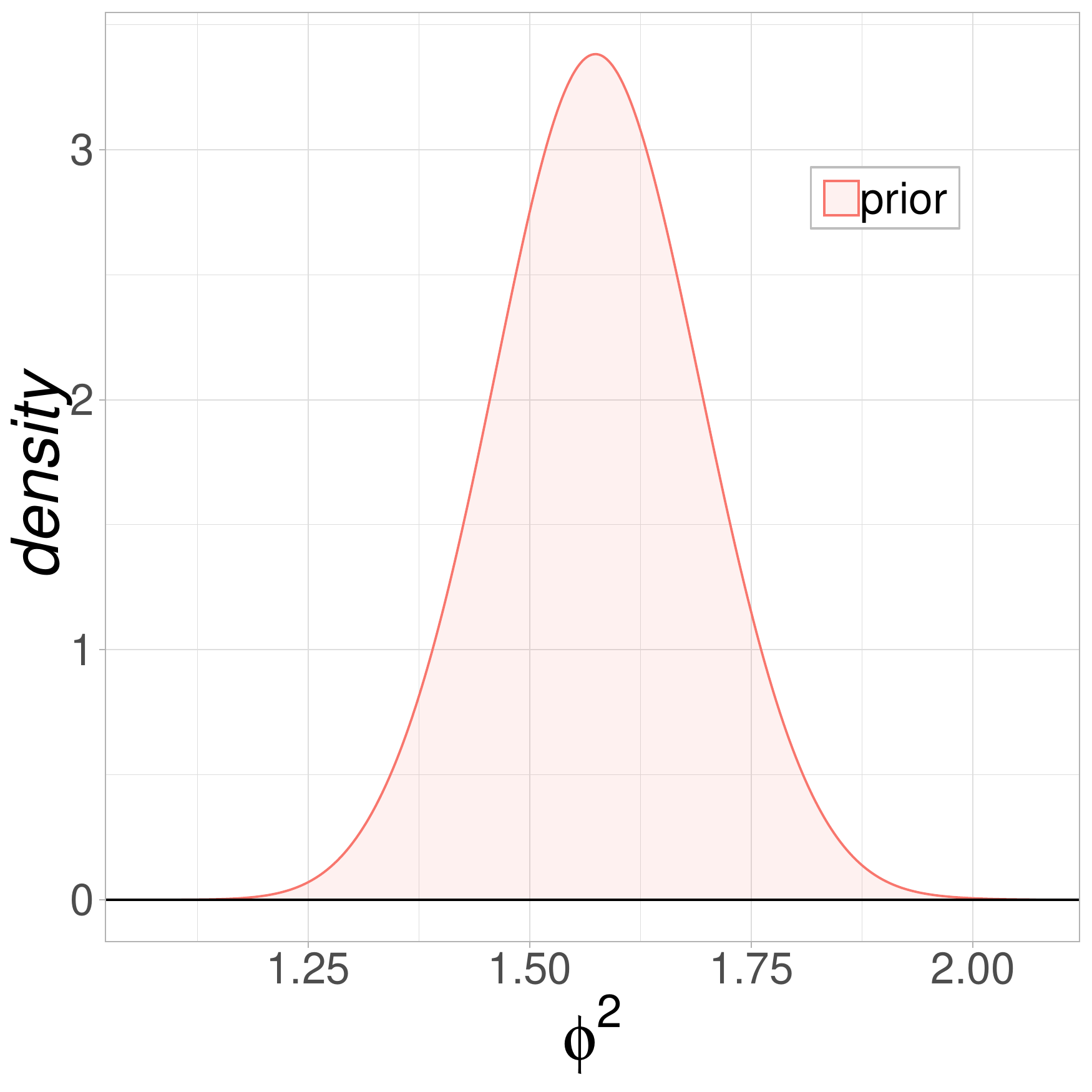} 

\end{knitrout}
    &
\begin{knitrout}
\definecolor{shadecolor}{rgb}{0.969, 0.969, 0.969}\color{fgcolor}
\includegraphics[width=0.21\linewidth,height=0.15\linewidth]{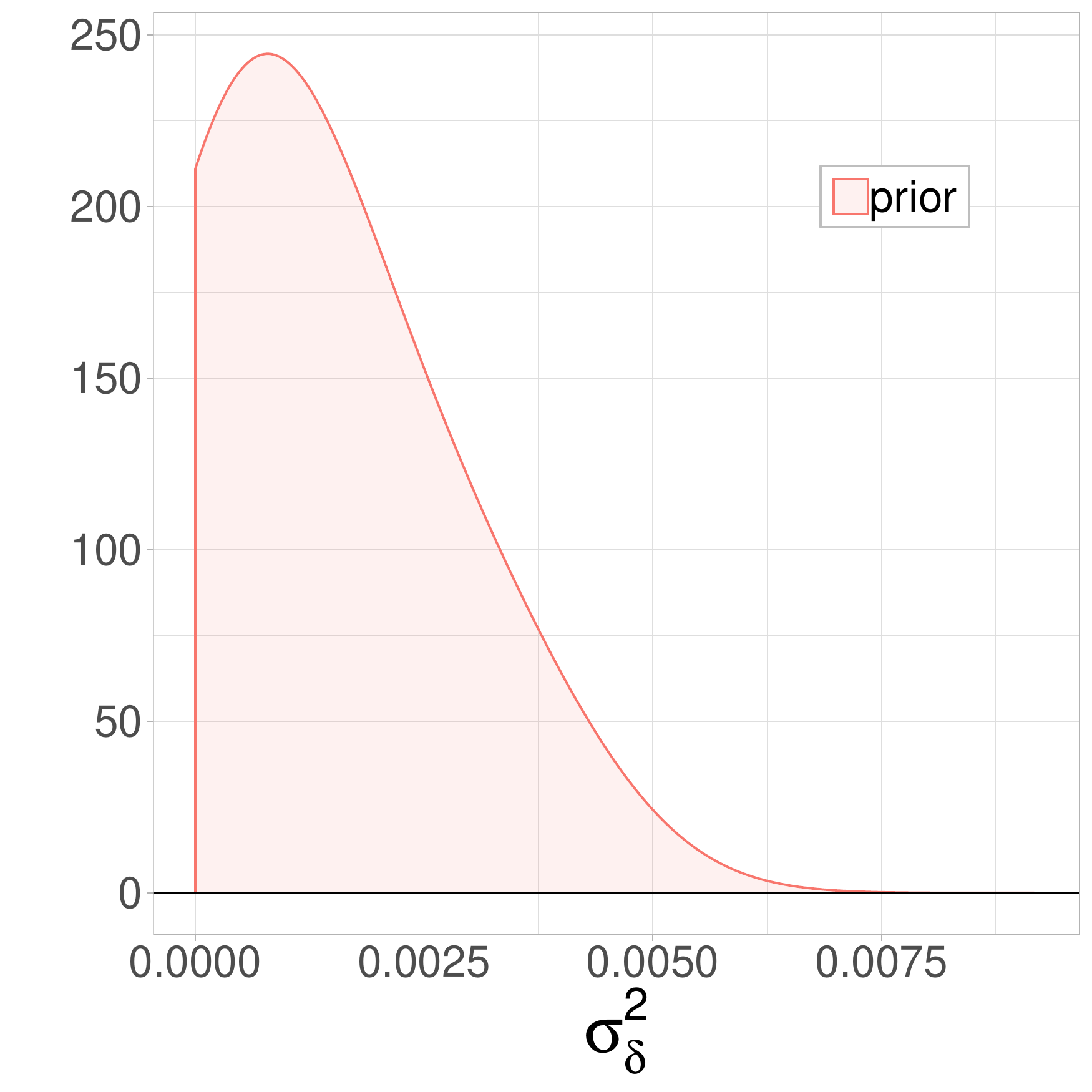} 

\end{knitrout}
&
\begin{knitrout}
\definecolor{shadecolor}{rgb}{0.969, 0.969, 0.969}\color{fgcolor}
\includegraphics[width=0.21\linewidth,height=0.15\linewidth]{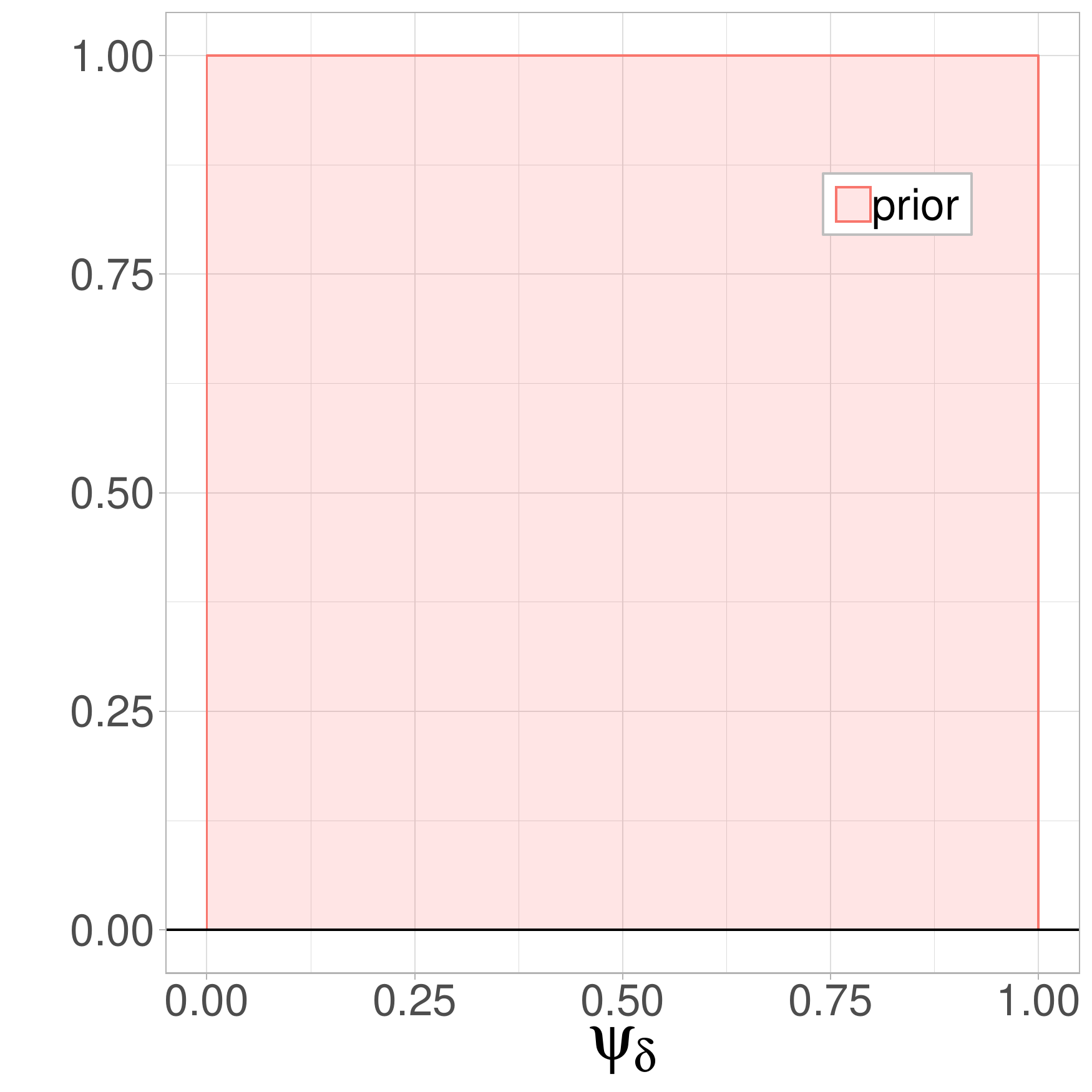} 

\end{knitrout}
    &
\begin{knitrout}
\definecolor{shadecolor}{rgb}{0.969, 0.969, 0.969}\color{fgcolor}
\includegraphics[width=0.21\linewidth,height=0.15\linewidth]{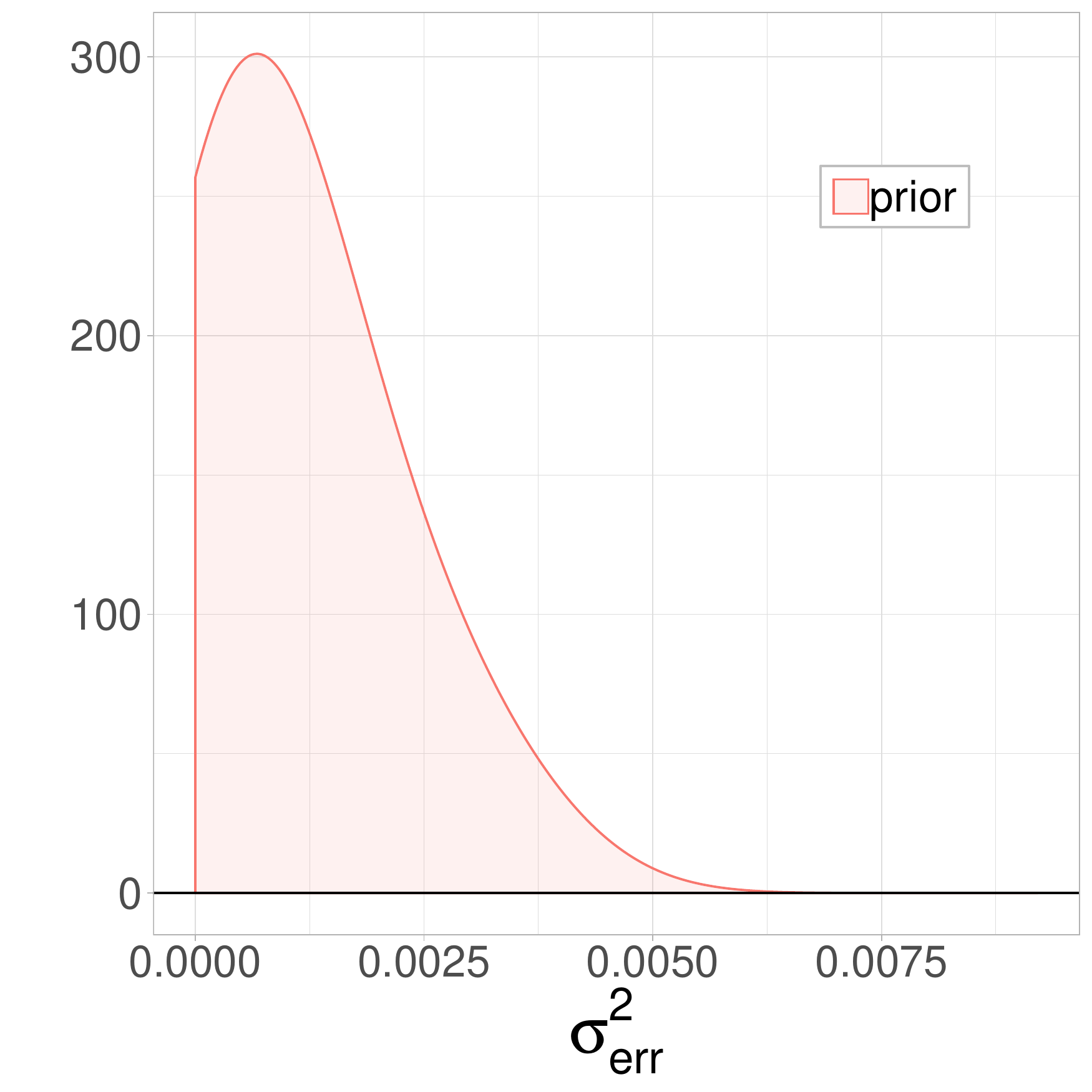} 

\end{knitrout}
    \\
  \end{tabular}
\caption{Prior distributions for each parameter to calibrate in the application case.}
\label{fig:PriorApp}
\end{center}
\end{figure}

Figure \ref{fig:PriorApp} illustrates the prior distributions considered for the parameters of the damped harmonic oscillator code. For $\mathcal{M}_1$ and $\mathcal{M}_2$, only $A$, $\xi$, $k$, $m$, $\phi^2$ and $\sigma_{err}^2$ distributions are useful. Calibration with $\mathcal{M}_3$ or $\mathcal{M}_4$ the two last distributions ($\sigma_{\delta}^2$ and $\psi_{\delta}$) are then used.

\subsection{Calibration}

Calibration is run thanks to the function \code{calibrate} in \pkg{CaliCo}. Estimation option (\code{opt.estim}) has to be filled to run the algorithm properly. As it is described in Section \ref{sec:guidelines}, two MCMC algorithms are run by the function \code{calibrate}.

\begin{knitrout}
\definecolor{shadecolor}{rgb}{0.969, 0.969, 0.969}\color{fgcolor}\begin{kframe}
\begin{alltt}
\hlstd{opt.estim} \hlkwb{=} \hlkwd{list}\hlstd{(}\hlkwc{Ngibbs}\hlstd{=}\hlnum{1000}\hlstd{,}\hlkwc{Nmh}\hlstd{=}\hlnum{5000}\hlstd{,}\hlkwc{thetaInit}\hlstd{=}\hlkwd{c}\hlstd{(}\hlnum{1}\hlstd{,}\hlnum{0.25}\hlstd{,}\hlnum{6}\hlstd{,}\hlnum{50e-3}\hlstd{,pi}\hlopt{/}\hlnum{2}\hlstd{,}\hlnum{1e-3}\hlstd{),}
                 \hlkwc{r}\hlstd{=}\hlkwd{c}\hlstd{(}\hlnum{0.05}\hlstd{,}\hlnum{0.05}\hlstd{),}\hlkwc{sig}\hlstd{=}\hlkwd{diag}\hlstd{(}\hlnum{6}\hlstd{),}\hlkwc{Nchains}\hlstd{=}\hlnum{1}\hlstd{,}\hlkwc{burnIn}\hlstd{=}\hlnum{2000}\hlstd{)}
\hlstd{mdfit1} \hlkwb{<-} \hlkwd{calibrate}\hlstd{(model1,pr1,opt.estim)}
\end{alltt}
\end{kframe}
\end{knitrout}

In the terminal, a loading bar represents the execution time of the inference algorithm.
Then, the method \code{print} can be used to quickly access some information (see Section \ref{sec:guidelines}).

\begin{knitrout}
\definecolor{shadecolor}{rgb}{0.969, 0.969, 0.969}\color{fgcolor}\begin{kframe}
\begin{alltt}
\hlkwd{print}\hlstd{(mdfit1)}
\end{alltt}
\begin{verbatim}
## Call:
## 
## With the function:
## function(t,theta)
## {
##   w0 <- sqrt(theta[3]/theta[4])
##   return(theta[1]*exp(-theta[2]*w0*t)*sin(sqrt(1-theta[2]^2)*w0*t+theta[5]))
## }
## <bytecode: 0x561a96e4f068>
## 
## Selected model : model1 
## 
## Acceptation rate of the Metropolis within Gibbs algorithm:
## [1] "46.6%" "27%"   "27.9%" "8.9%"  "5.1%"  "4.4%" 
## 
## Acceptation rate of the Metropolis Hastings algorithm:
## [1] "44.52%"
## 
## Maximum a posteriori:
## [1] 1.0138252830 0.2032794864 5.9924680899 0.0505938887 1.5199179525
## [6] 0.0002565248
## 
## Mean a posteriori:
## [1] 1.0107256606 0.2010363776 5.9755232004 0.0497000370 1.5042868046
## [6] 0.0002493378
\end{verbatim}
\end{kframe}
\end{knitrout}

To visualize the results, the \code{plot} method allows to generate \pkg{ggplot2} objects and if the option \code{graph} is not deactivated, the function \code{plot} will create a layout of graphs displayed in Figure \ref{fig:PlotCalib1} which contains:
\begin{enumerate}
\item the auto-correlation graphs,
\item the chains trajectories,
\item the \prior and \posterior distributions,
\item the correlation between parameters,
\item the results on the quantity of interest.
\end{enumerate}

To generate all the graphs of Figure \ref{fig:PlotCalib1}, the \code{plot} function is used similarly as for a \code{model.class} object with the following code line.

\begin{knitrout}
\definecolor{shadecolor}{rgb}{0.969, 0.969, 0.969}\color{fgcolor}\begin{kframe}
\begin{alltt}
\hlkwd{plot}\hlstd{(mdfit1,t)}
\end{alltt}
\end{kframe}
\end{knitrout}

\begin{figure}[h!]
\begin{center}
    \begin{tabular}{cccccc}
\begin{knitrout}
\definecolor{shadecolor}{rgb}{0.969, 0.969, 0.969}\color{fgcolor}
\includegraphics[width=0.12\linewidth]{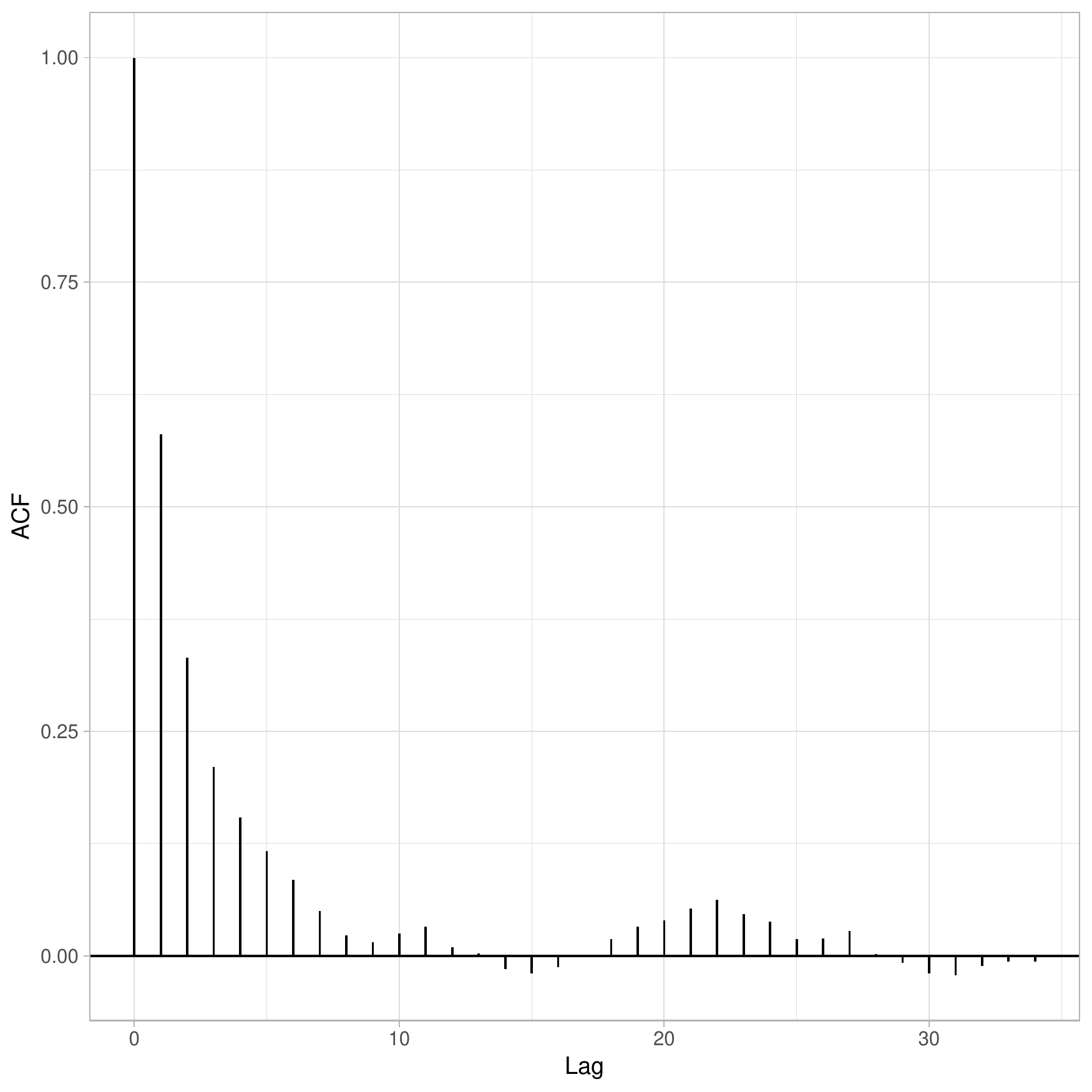} 

\end{knitrout}
&
\begin{knitrout}
\definecolor{shadecolor}{rgb}{0.969, 0.969, 0.969}\color{fgcolor}
\includegraphics[width=0.12\linewidth]{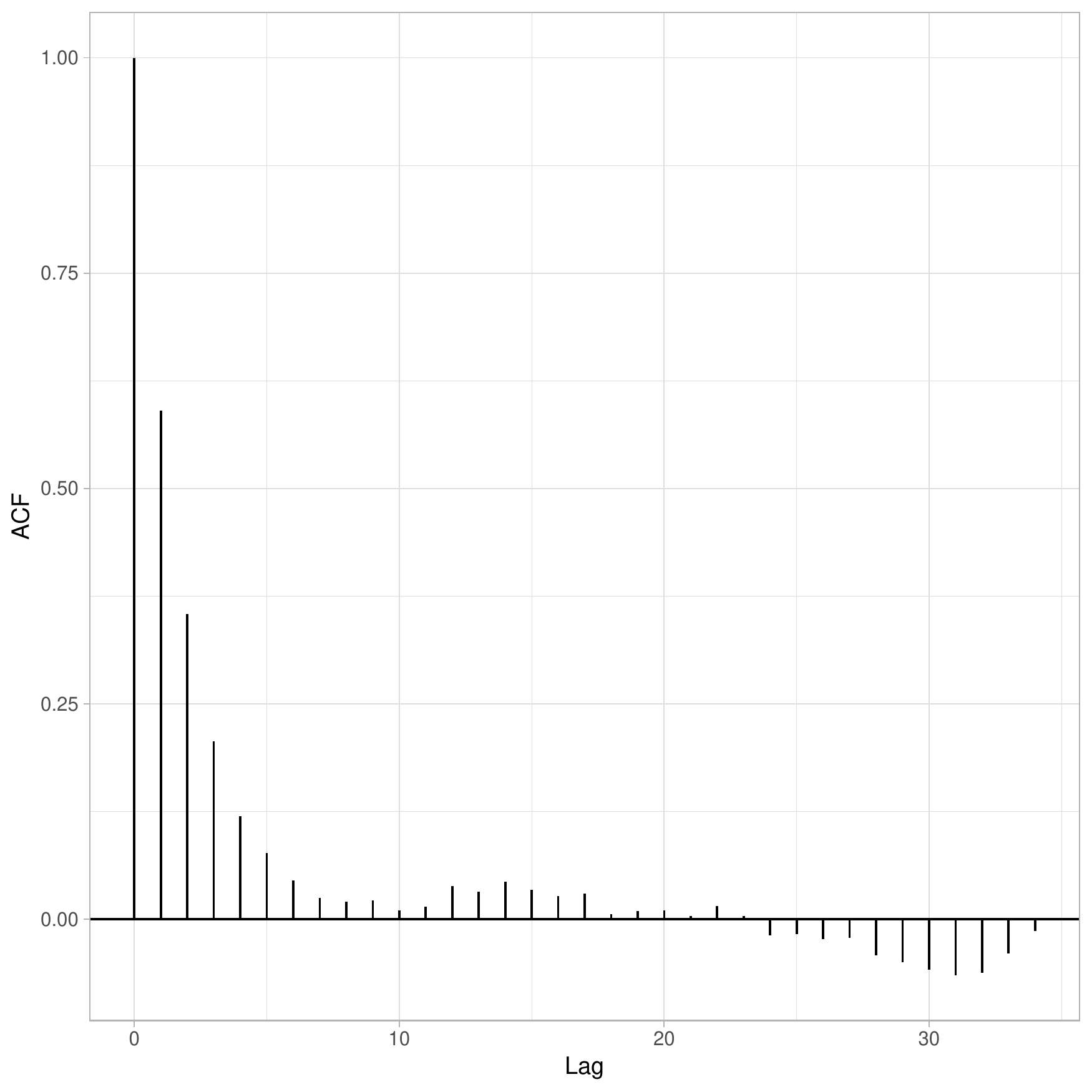} 

\end{knitrout}
&
\begin{knitrout}
\definecolor{shadecolor}{rgb}{0.969, 0.969, 0.969}\color{fgcolor}
\includegraphics[width=0.12\linewidth]{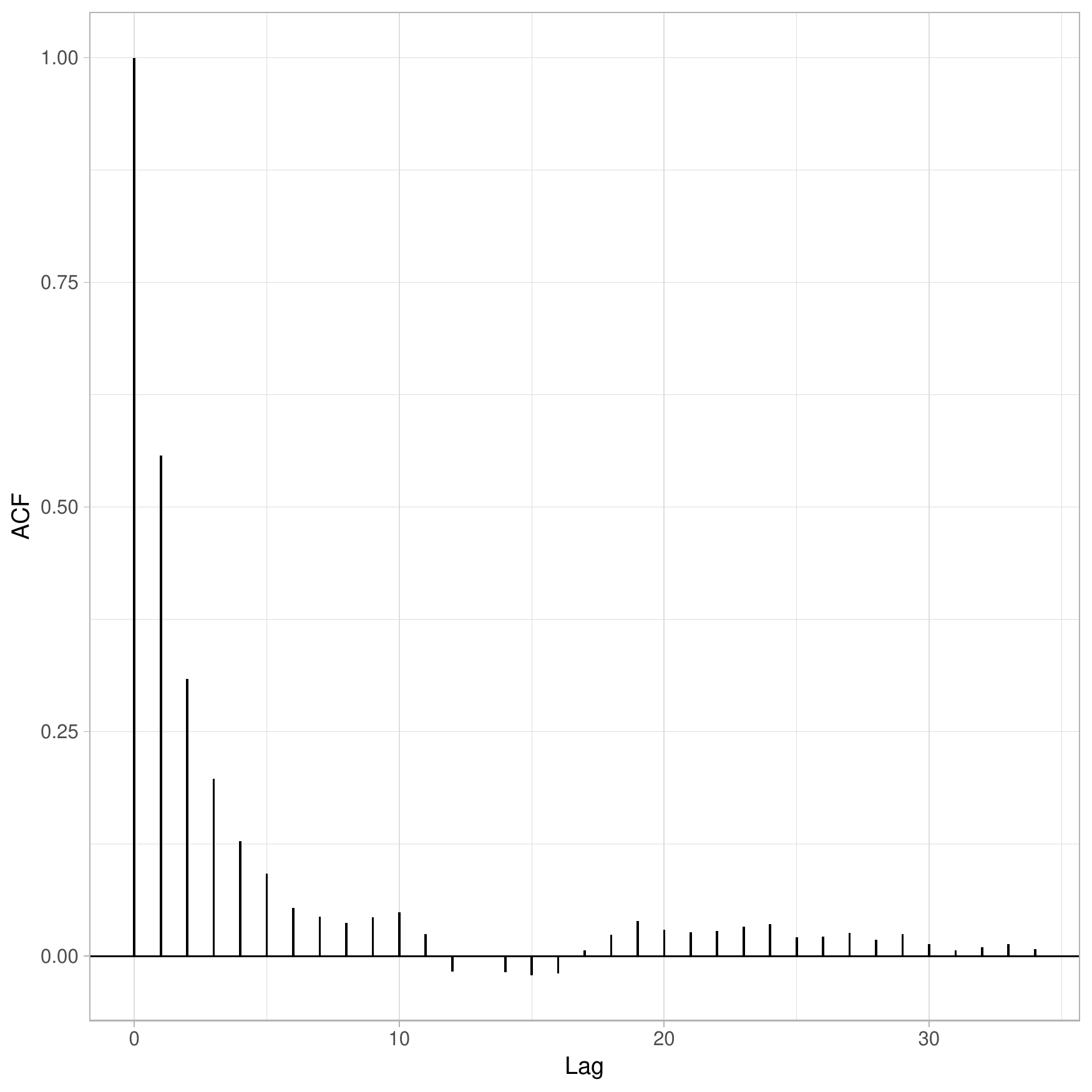} 

\end{knitrout}
&
\begin{knitrout}
\definecolor{shadecolor}{rgb}{0.969, 0.969, 0.969}\color{fgcolor}
\includegraphics[width=0.12\linewidth]{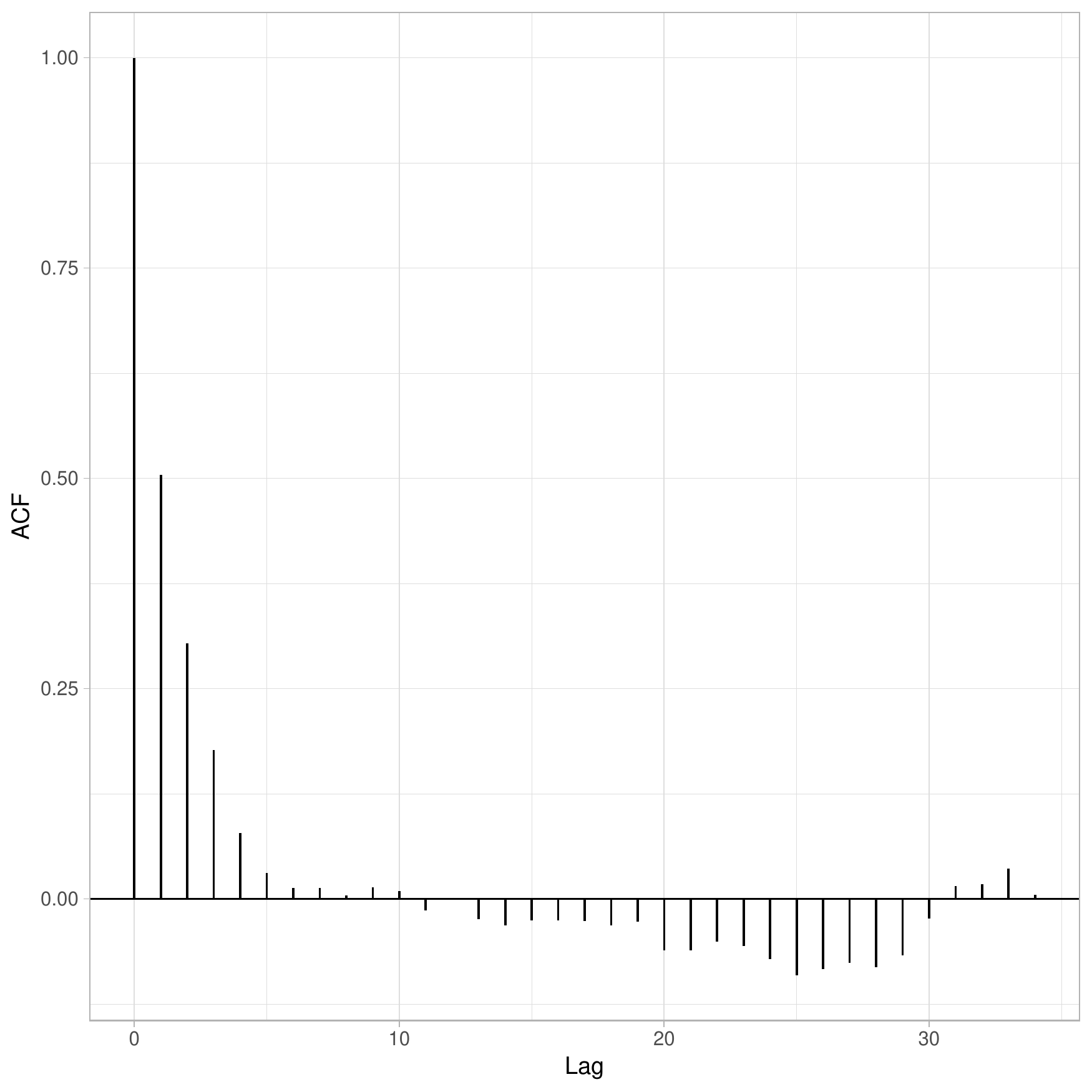} 

\end{knitrout}
&
\begin{knitrout}
\definecolor{shadecolor}{rgb}{0.969, 0.969, 0.969}\color{fgcolor}
\includegraphics[width=0.12\linewidth]{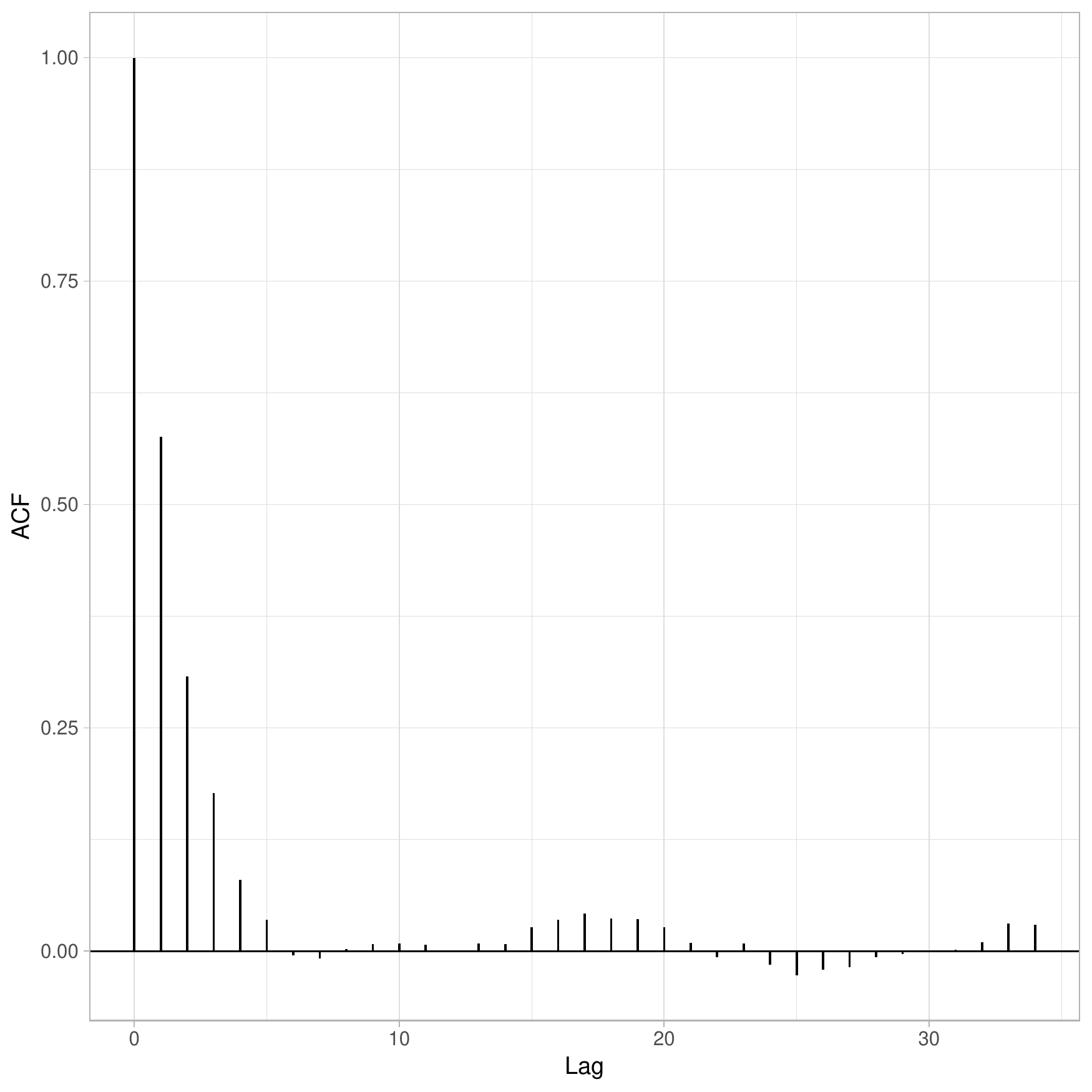} 

\end{knitrout}
&
\begin{knitrout}
\definecolor{shadecolor}{rgb}{0.969, 0.969, 0.969}\color{fgcolor}
\includegraphics[width=0.12\linewidth]{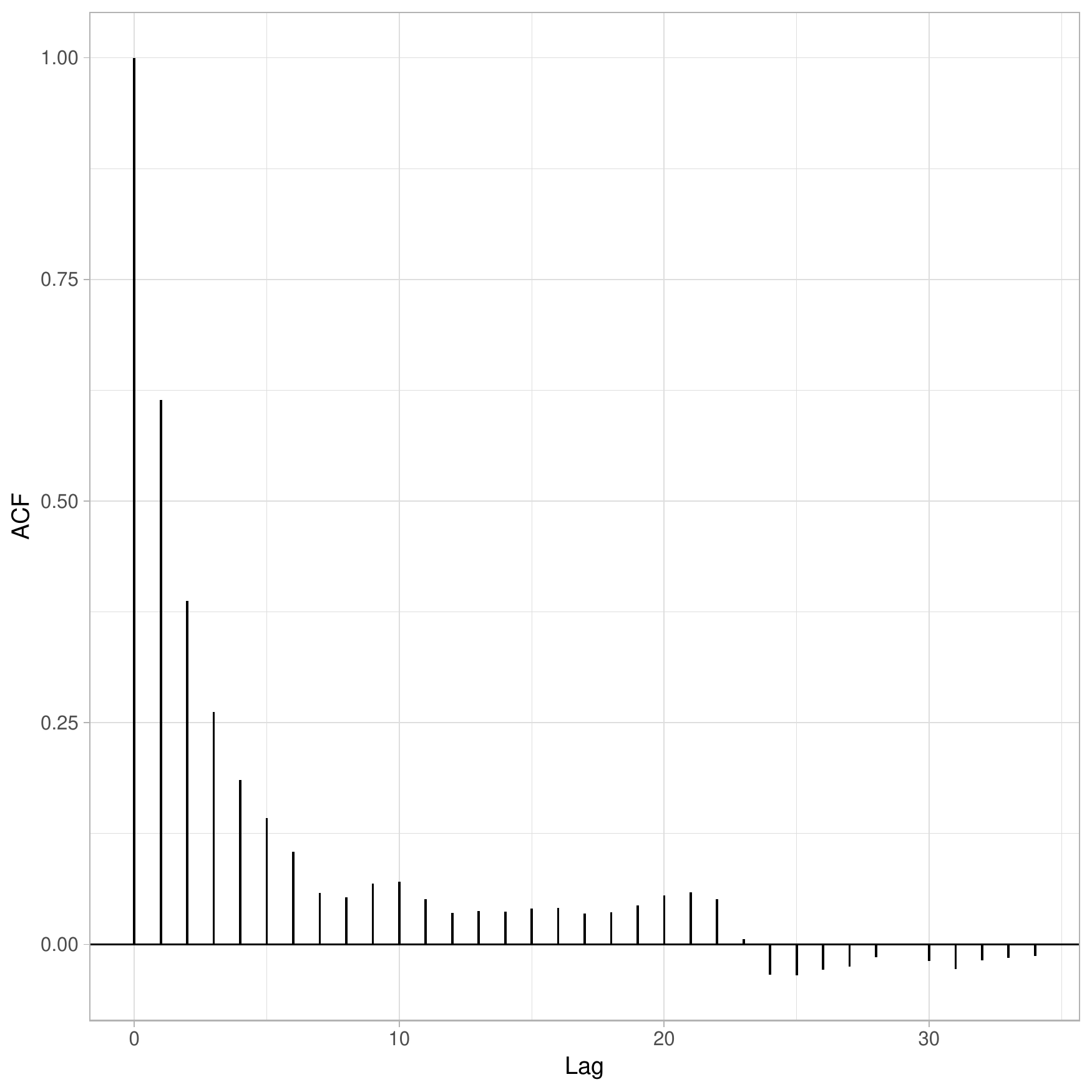} 

\end{knitrout}
\\
\begin{knitrout}
\definecolor{shadecolor}{rgb}{0.969, 0.969, 0.969}\color{fgcolor}
\includegraphics[width=0.12\linewidth]{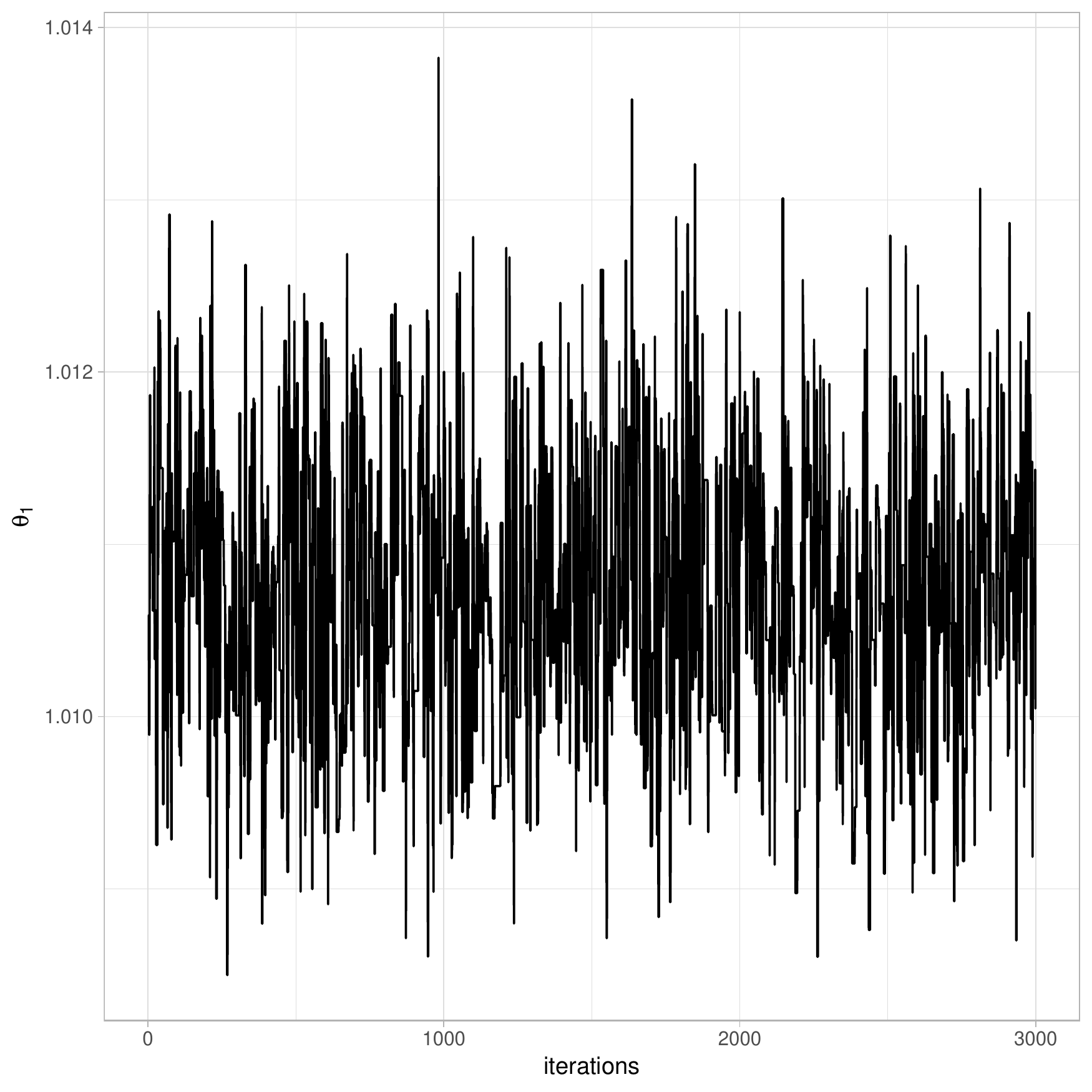} 

\end{knitrout}
&
\begin{knitrout}
\definecolor{shadecolor}{rgb}{0.969, 0.969, 0.969}\color{fgcolor}
\includegraphics[width=0.12\linewidth]{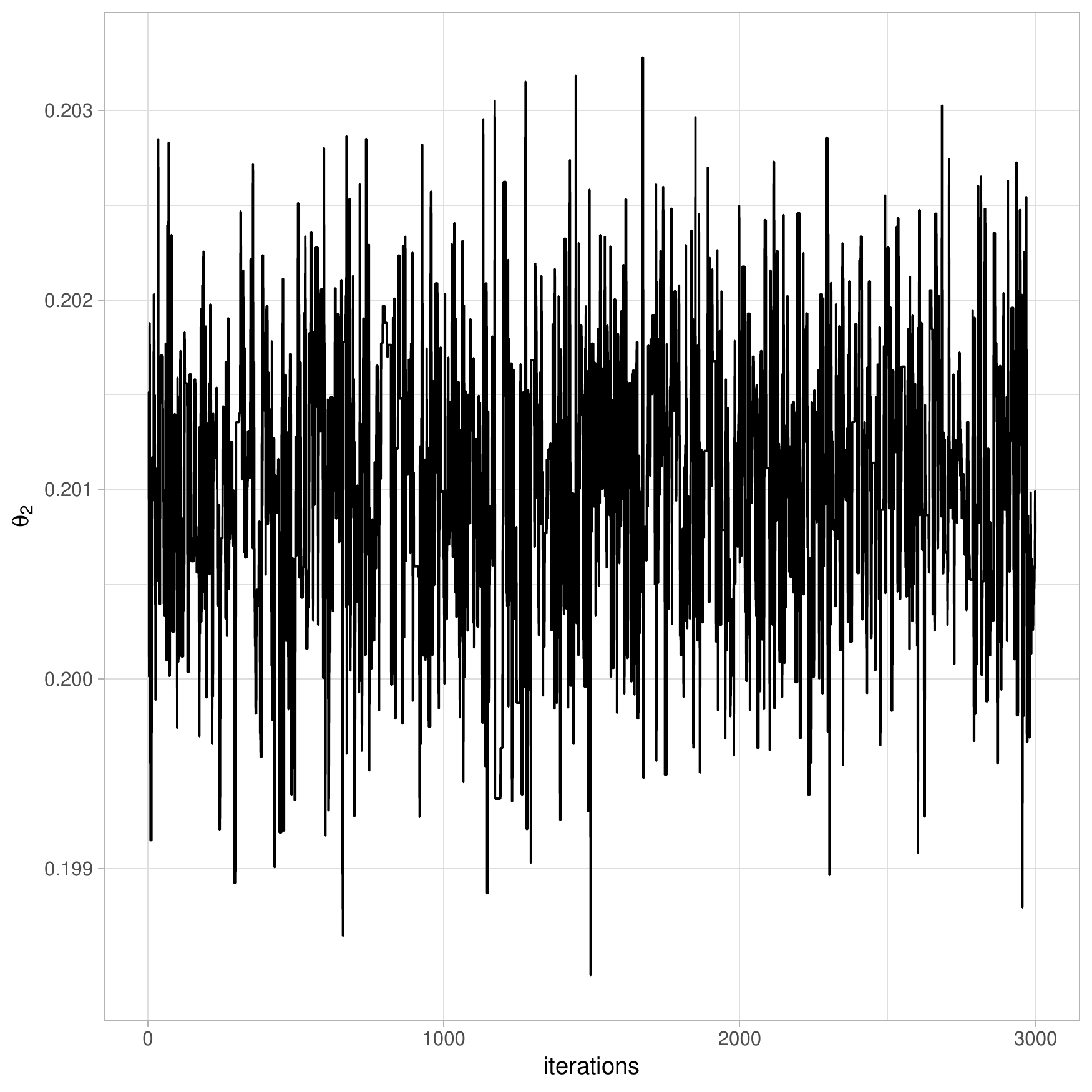} 

\end{knitrout}
&
\begin{knitrout}
\definecolor{shadecolor}{rgb}{0.969, 0.969, 0.969}\color{fgcolor}
\includegraphics[width=0.12\linewidth]{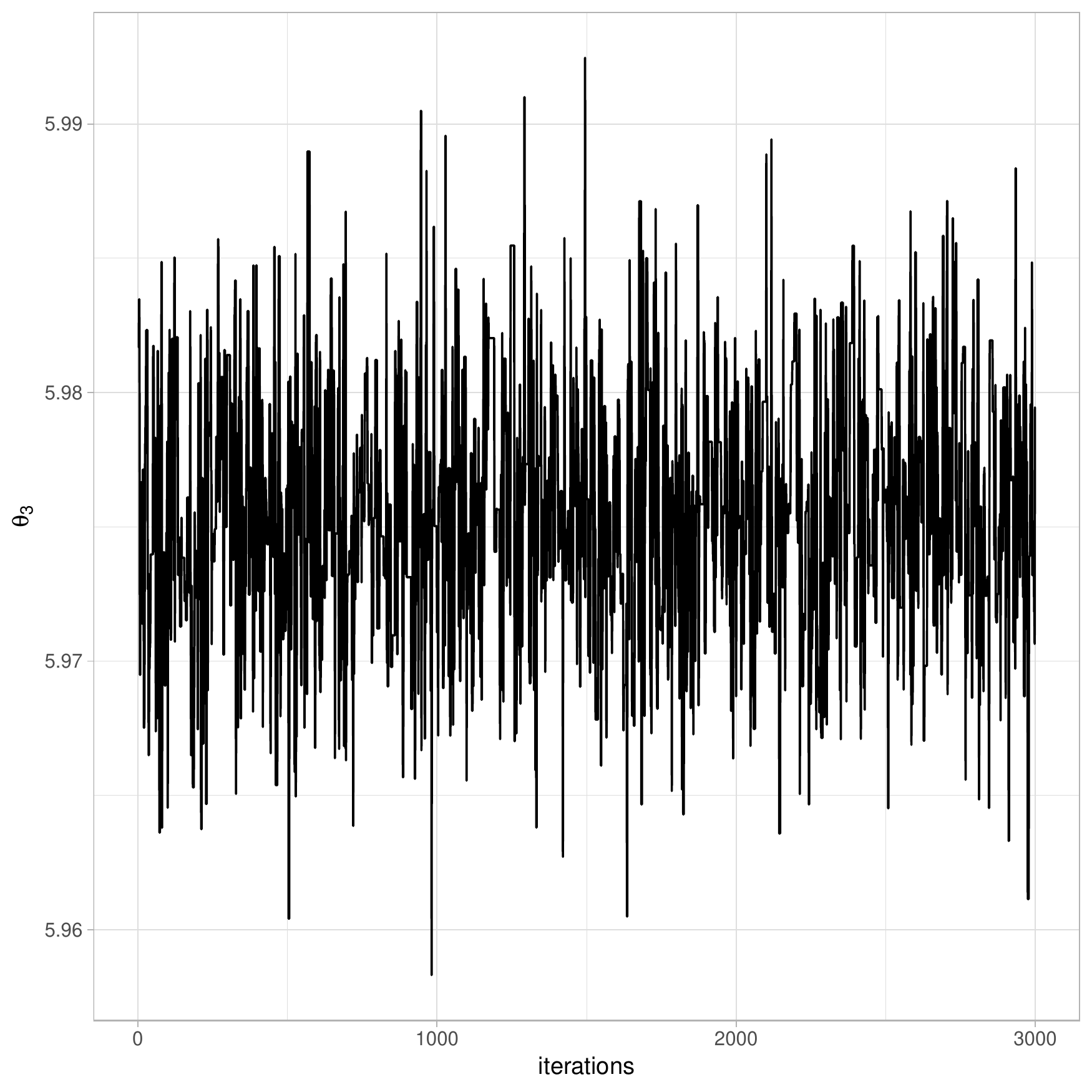} 

\end{knitrout}
&
\begin{knitrout}
\definecolor{shadecolor}{rgb}{0.969, 0.969, 0.969}\color{fgcolor}
\includegraphics[width=0.12\linewidth]{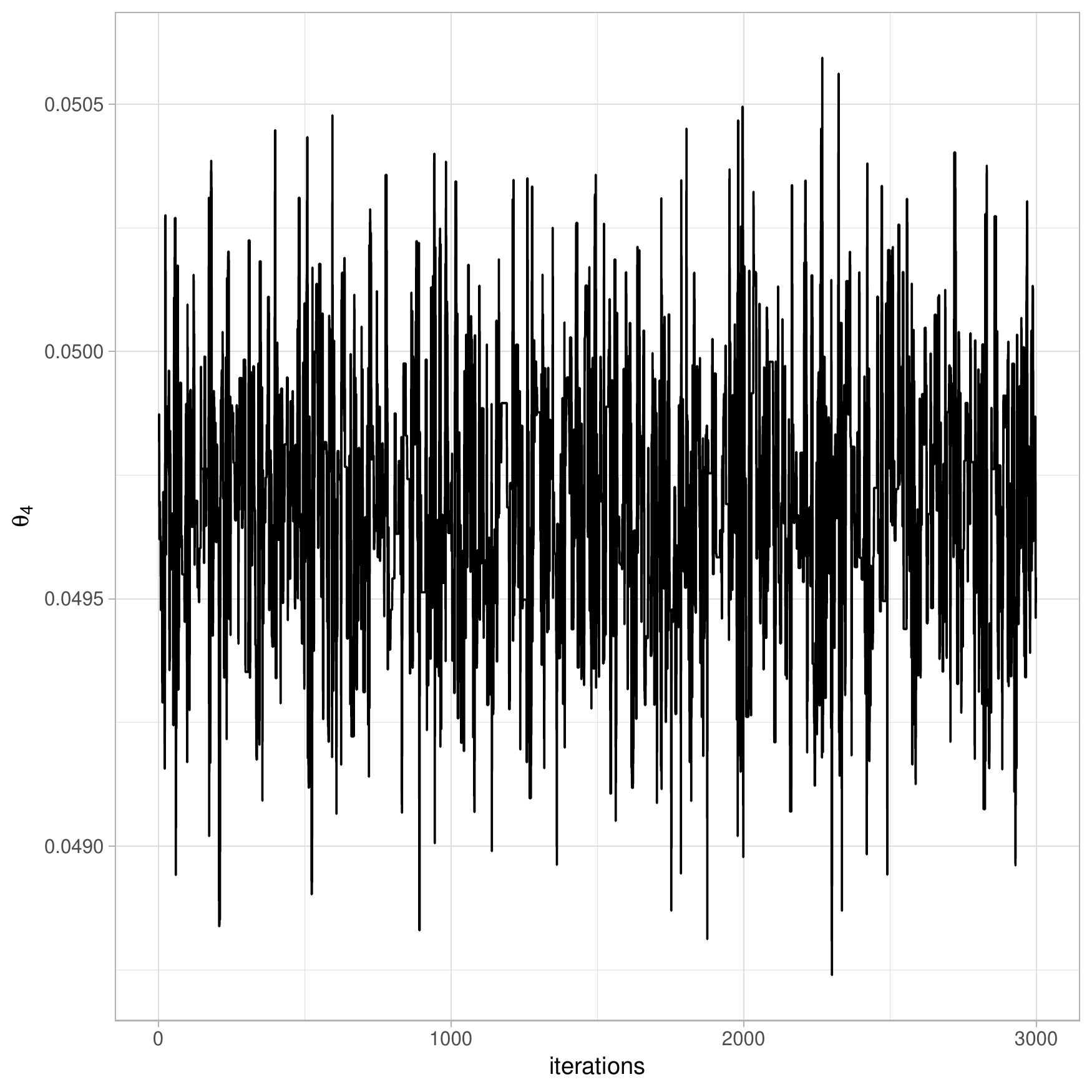} 

\end{knitrout}
&
\begin{knitrout}
\definecolor{shadecolor}{rgb}{0.969, 0.969, 0.969}\color{fgcolor}
\includegraphics[width=0.12\linewidth]{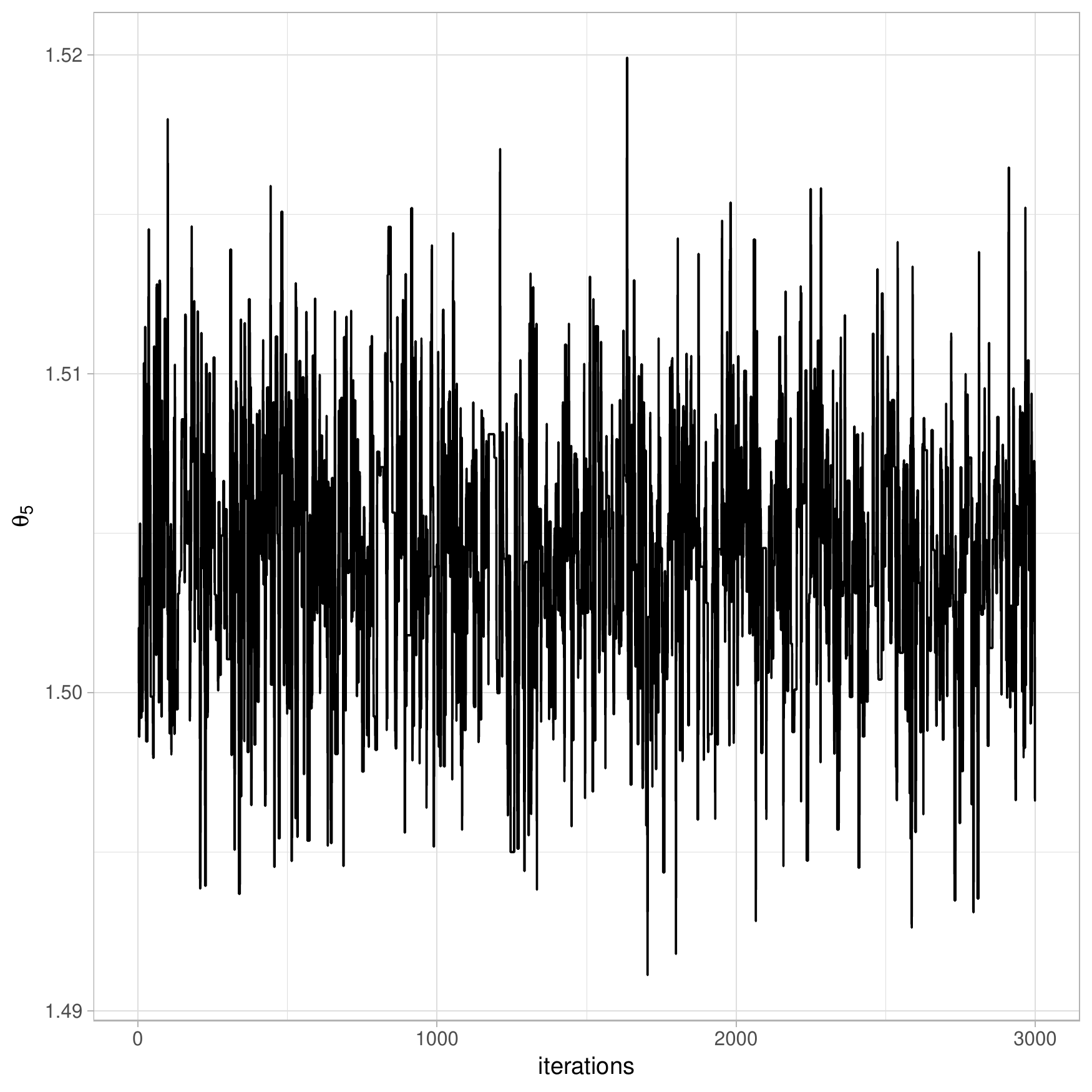} 

\end{knitrout}
&
\begin{knitrout}
\definecolor{shadecolor}{rgb}{0.969, 0.969, 0.969}\color{fgcolor}
\includegraphics[width=0.12\linewidth]{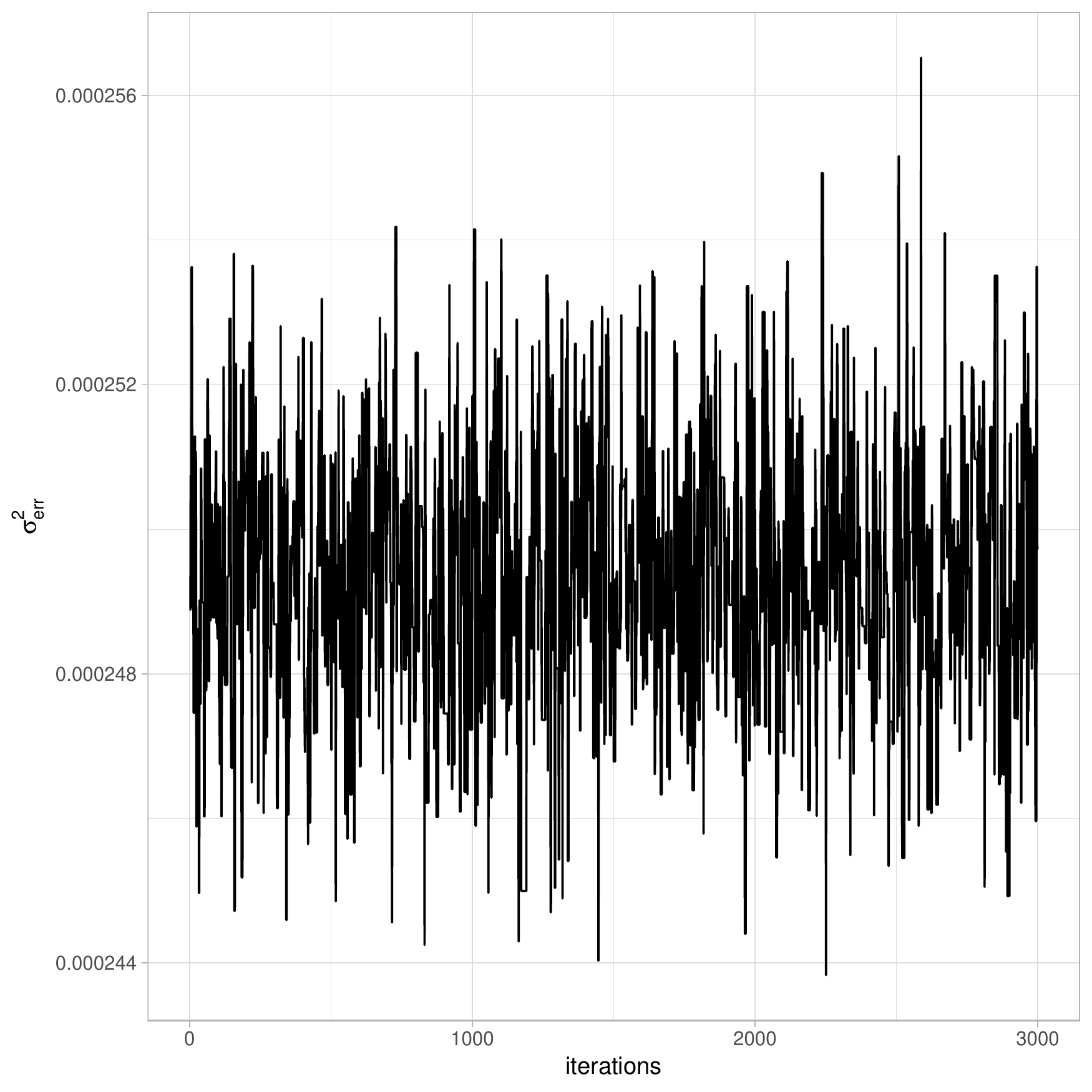} 

\end{knitrout}
\\
\begin{knitrout}
\definecolor{shadecolor}{rgb}{0.969, 0.969, 0.969}\color{fgcolor}
\includegraphics[width=0.12\linewidth]{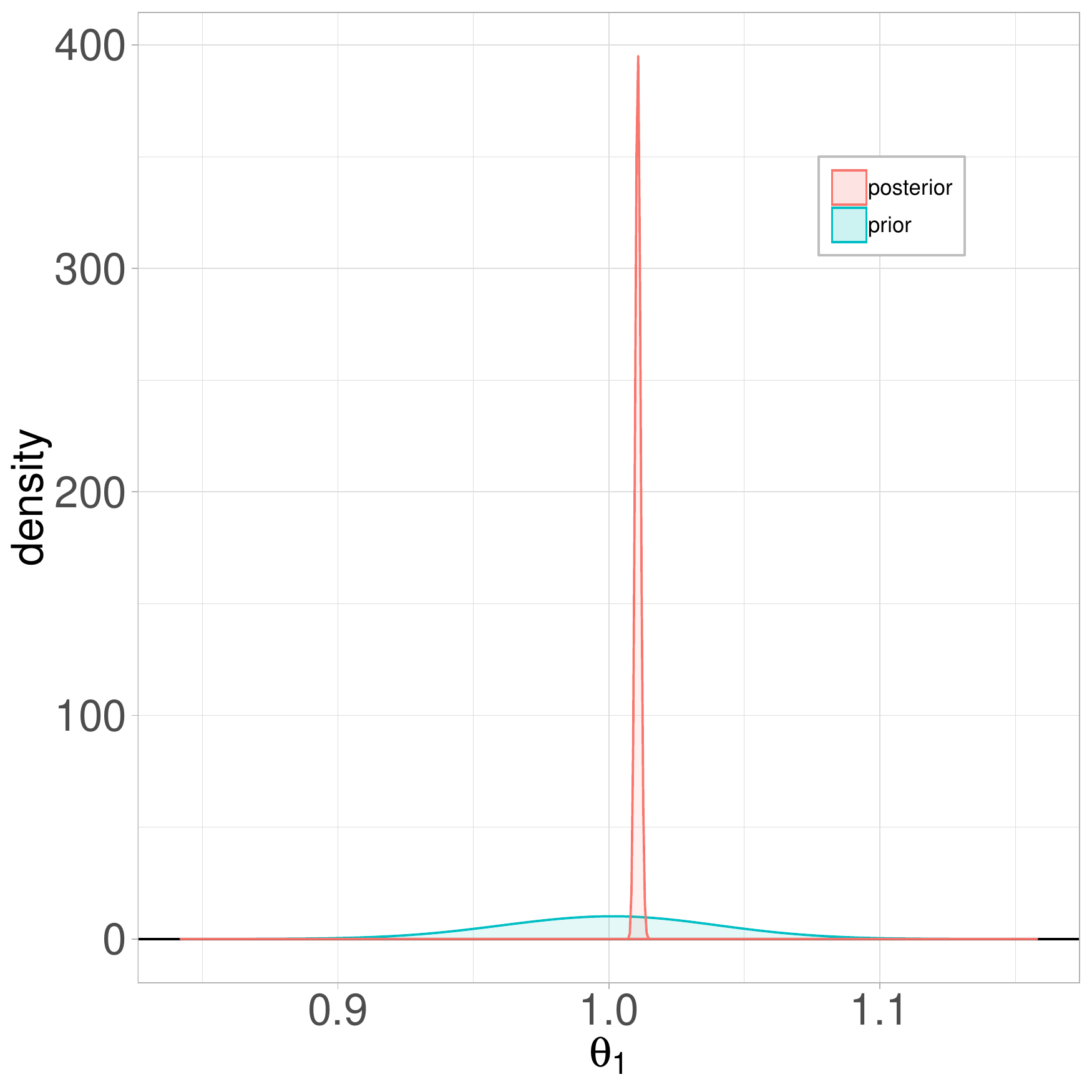} 

\end{knitrout}
&
\begin{knitrout}
\definecolor{shadecolor}{rgb}{0.969, 0.969, 0.969}\color{fgcolor}
\includegraphics[width=0.12\linewidth]{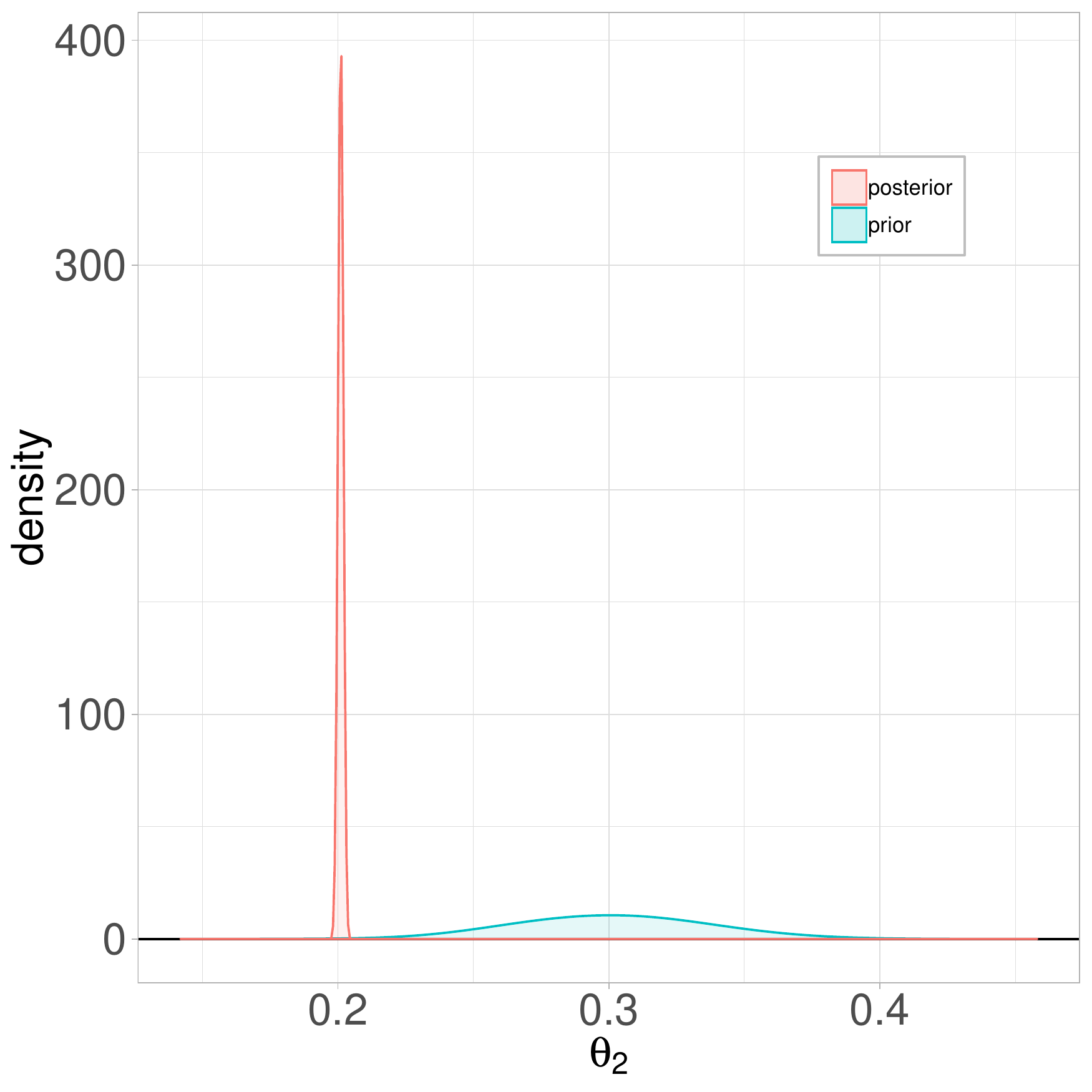} 

\end{knitrout}
&
\begin{knitrout}
\definecolor{shadecolor}{rgb}{0.969, 0.969, 0.969}\color{fgcolor}
\includegraphics[width=0.12\linewidth]{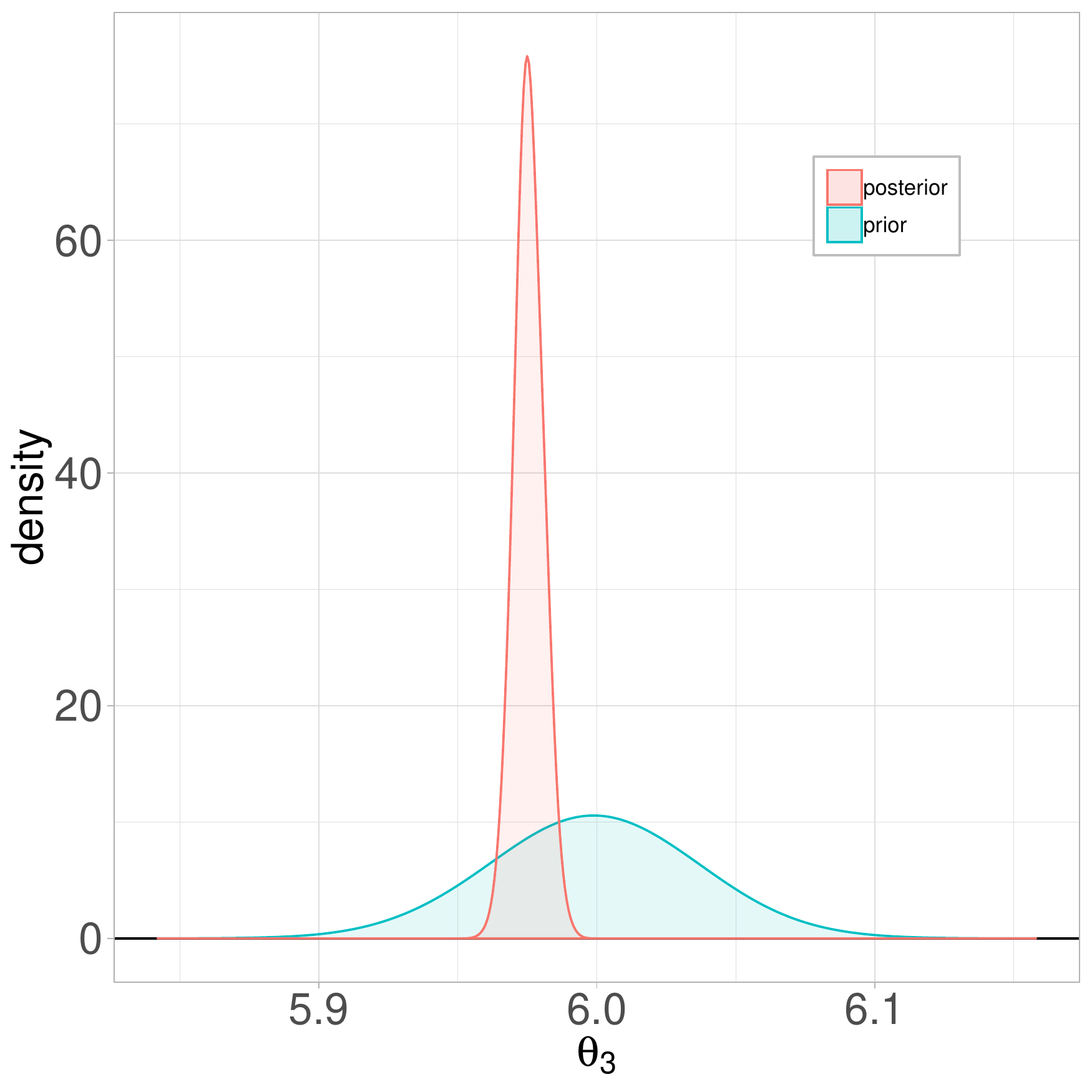} 

\end{knitrout}
&
\begin{knitrout}
\definecolor{shadecolor}{rgb}{0.969, 0.969, 0.969}\color{fgcolor}
\includegraphics[width=0.12\linewidth]{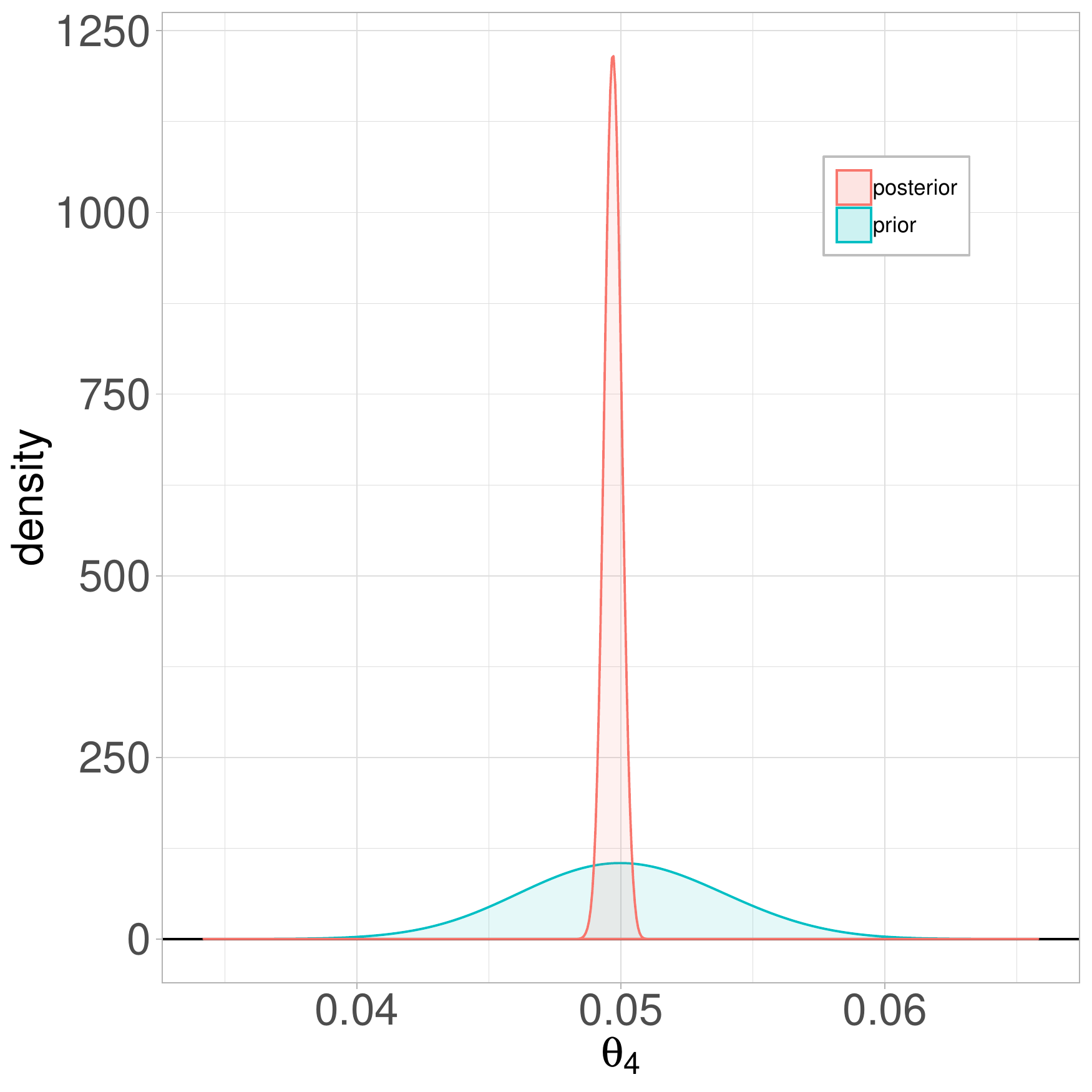} 

\end{knitrout}
&
\begin{knitrout}
\definecolor{shadecolor}{rgb}{0.969, 0.969, 0.969}\color{fgcolor}
\includegraphics[width=0.12\linewidth]{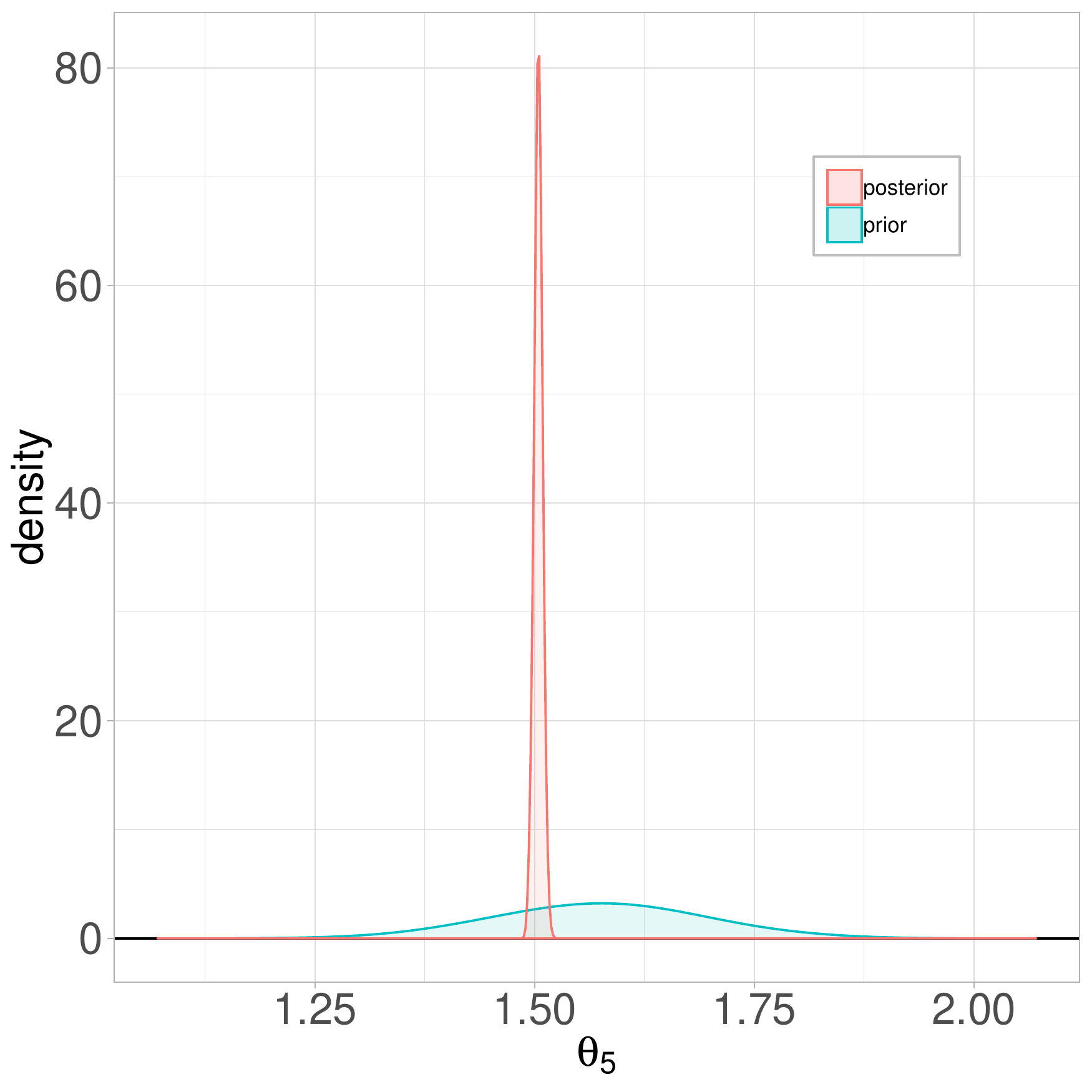} 

\end{knitrout}
&
\begin{knitrout}
\definecolor{shadecolor}{rgb}{0.969, 0.969, 0.969}\color{fgcolor}
\includegraphics[width=0.12\linewidth]{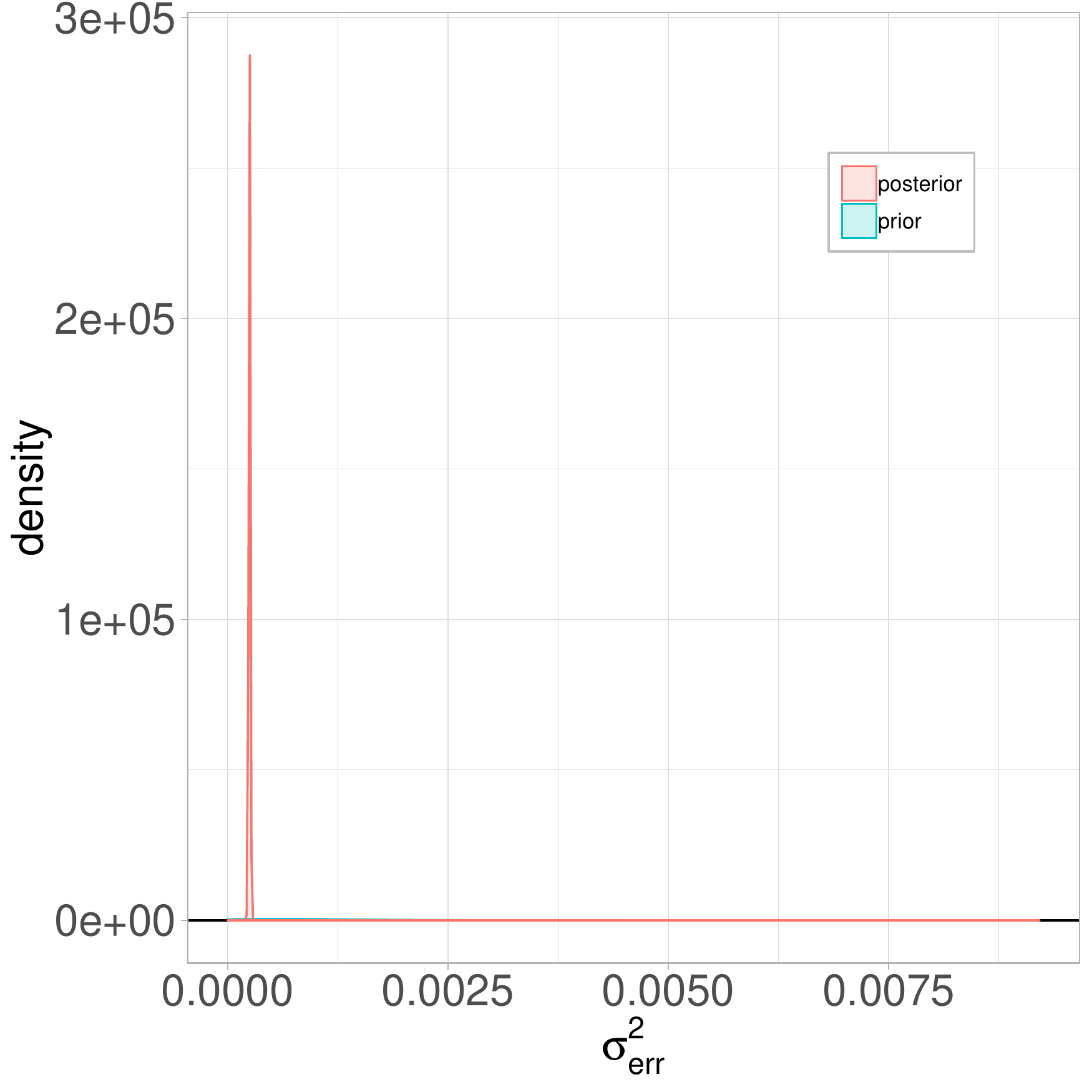} 

\end{knitrout}
  \end{tabular}
\begin{tikzpicture}
		\tikzstyle{m1}=[]
		
		\node[m1] (N1) at (0,0) {
		
  \begin{tabular}{ccccc}
\begin{knitrout}
\definecolor{shadecolor}{rgb}{0.969, 0.969, 0.969}\color{fgcolor}
\includegraphics[width=0.08\linewidth]{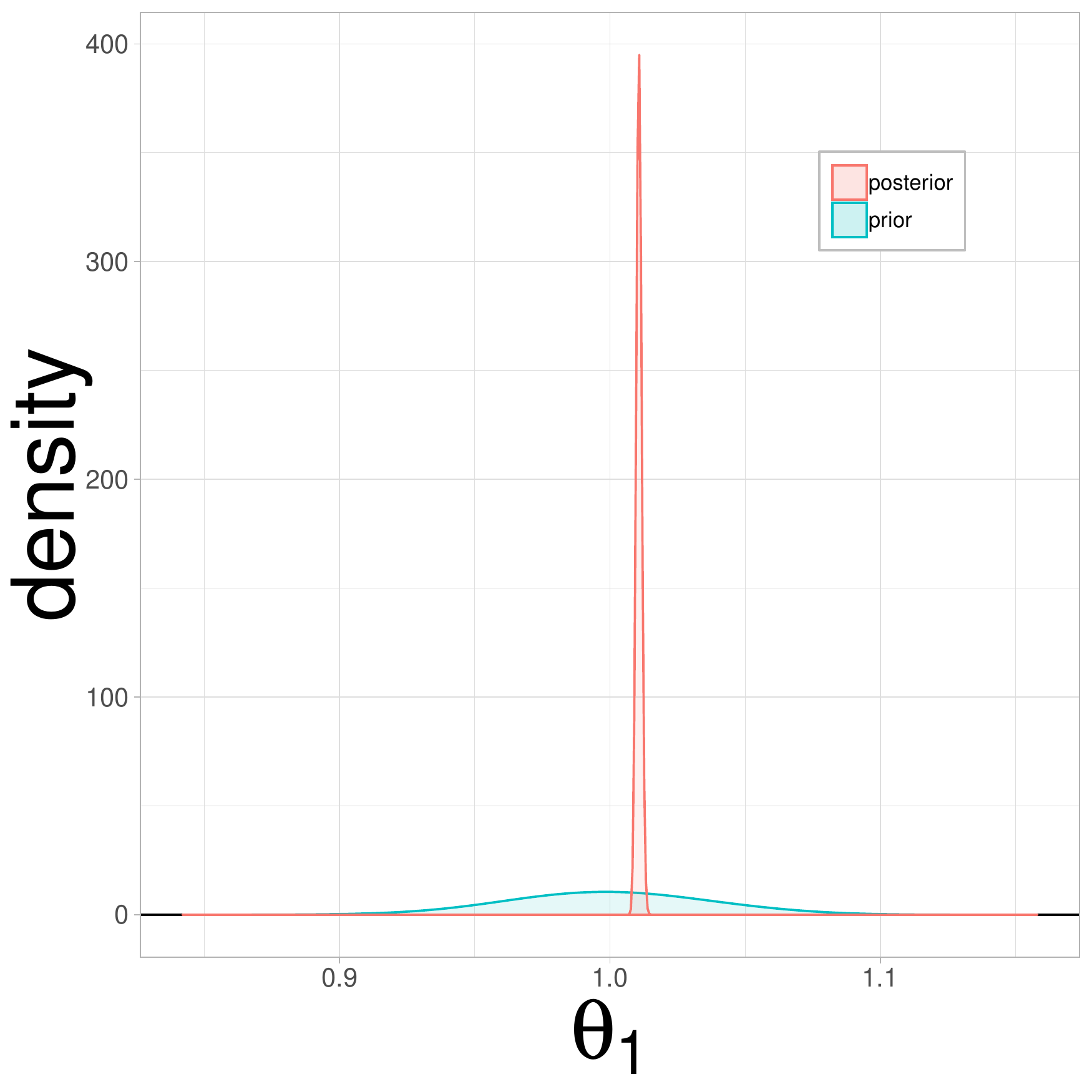} 

\end{knitrout}
&
\begin{knitrout}
\definecolor{shadecolor}{rgb}{0.969, 0.969, 0.969}\color{fgcolor}
\includegraphics[width=0.08\linewidth]{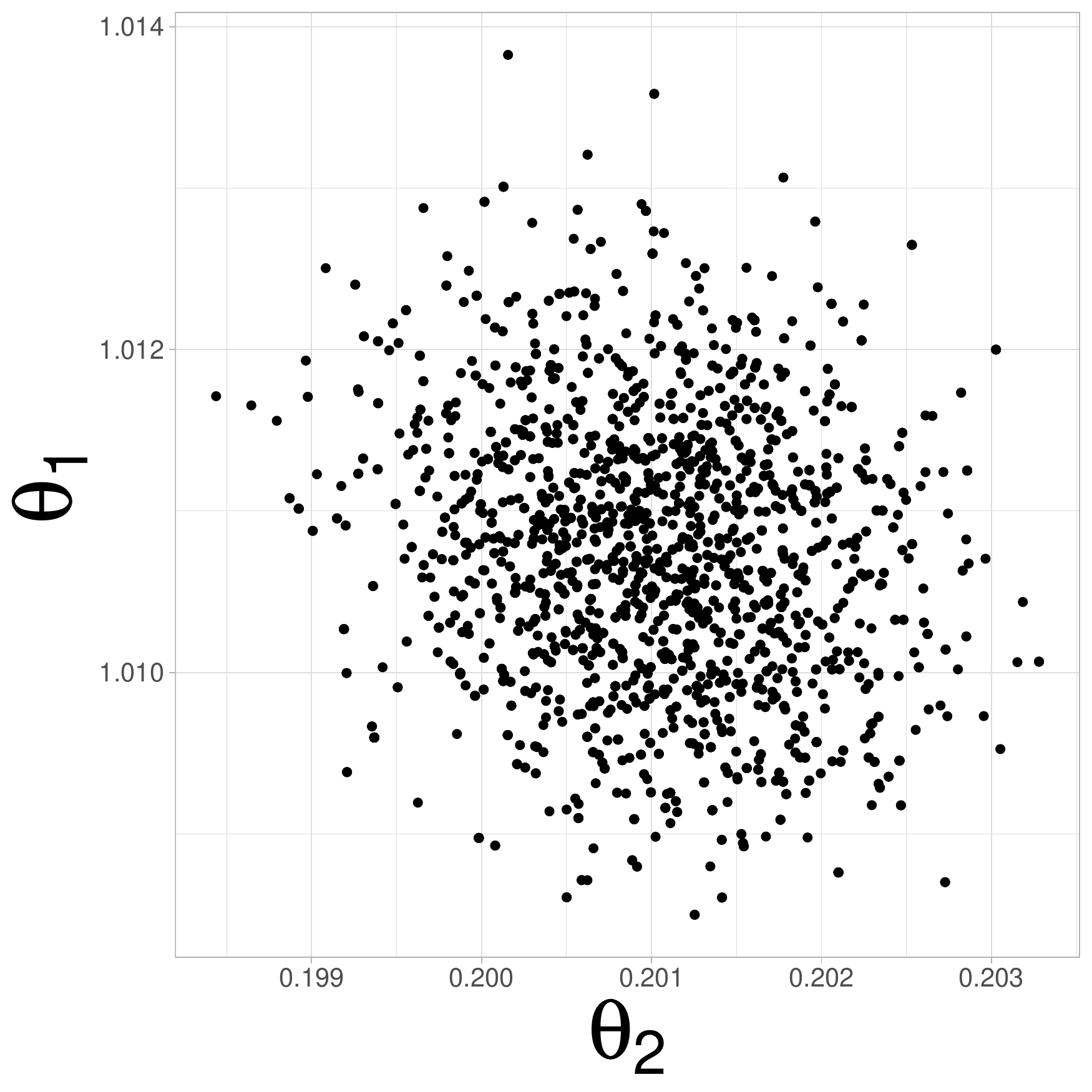} 

\end{knitrout}
&
\begin{knitrout}
\definecolor{shadecolor}{rgb}{0.969, 0.969, 0.969}\color{fgcolor}
\includegraphics[width=0.08\linewidth]{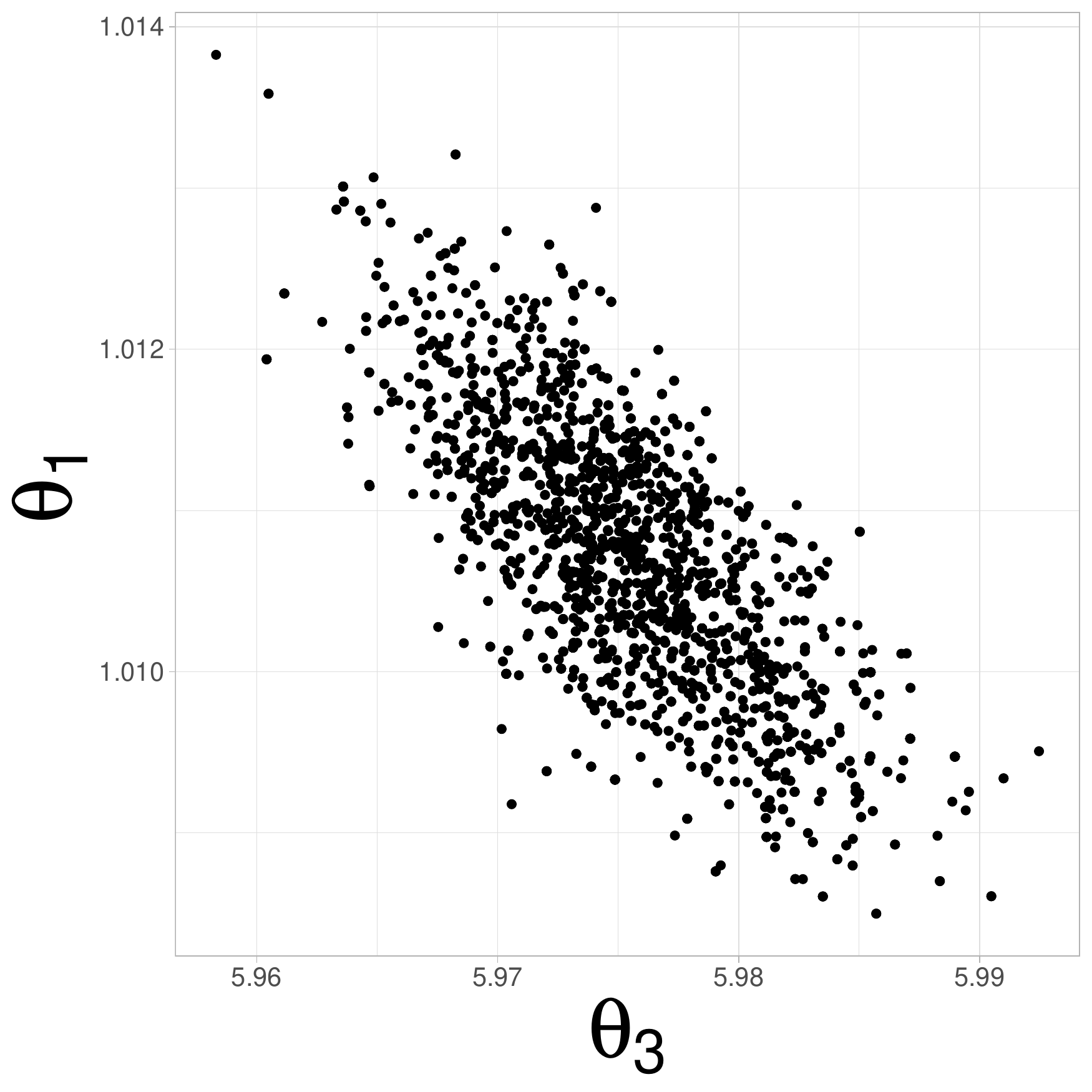} 

\end{knitrout}
&
\begin{knitrout}
\definecolor{shadecolor}{rgb}{0.969, 0.969, 0.969}\color{fgcolor}
\includegraphics[width=0.08\linewidth]{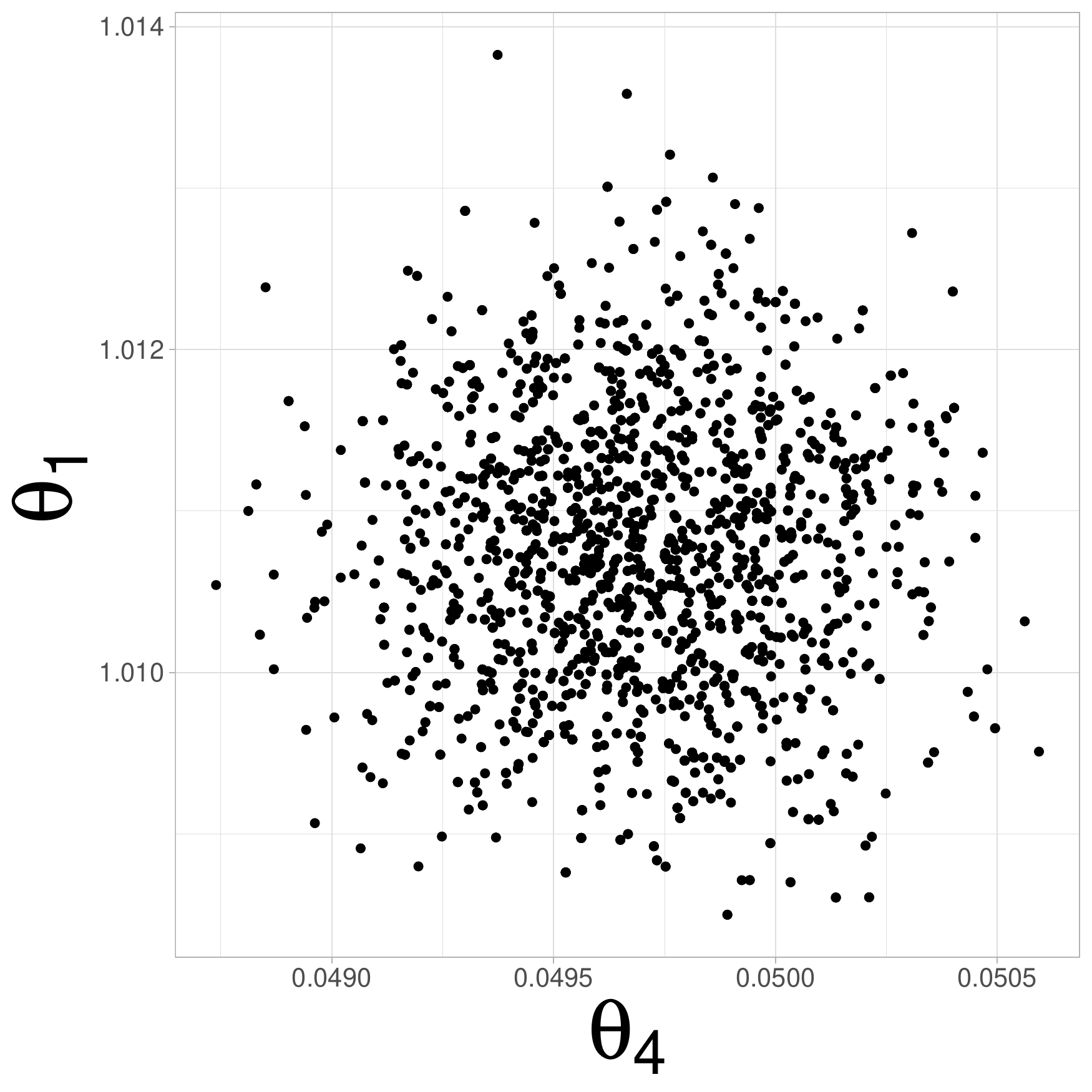} 

\end{knitrout}
&
\begin{knitrout}
\definecolor{shadecolor}{rgb}{0.969, 0.969, 0.969}\color{fgcolor}
\includegraphics[width=0.08\linewidth]{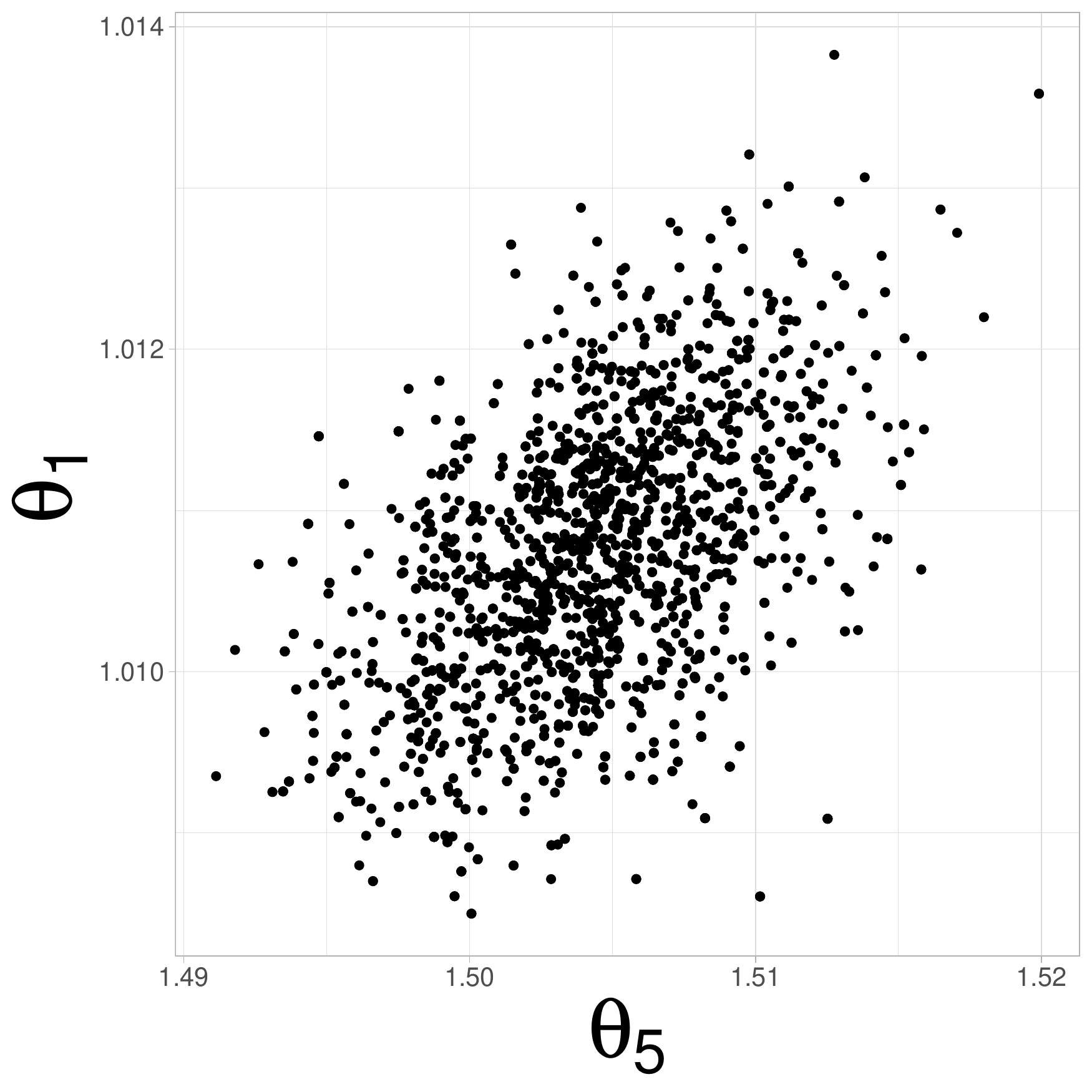} 

\end{knitrout}

\\
\begin{knitrout}
\definecolor{shadecolor}{rgb}{0.969, 0.969, 0.969}\color{fgcolor}
\includegraphics[width=0.08\linewidth]{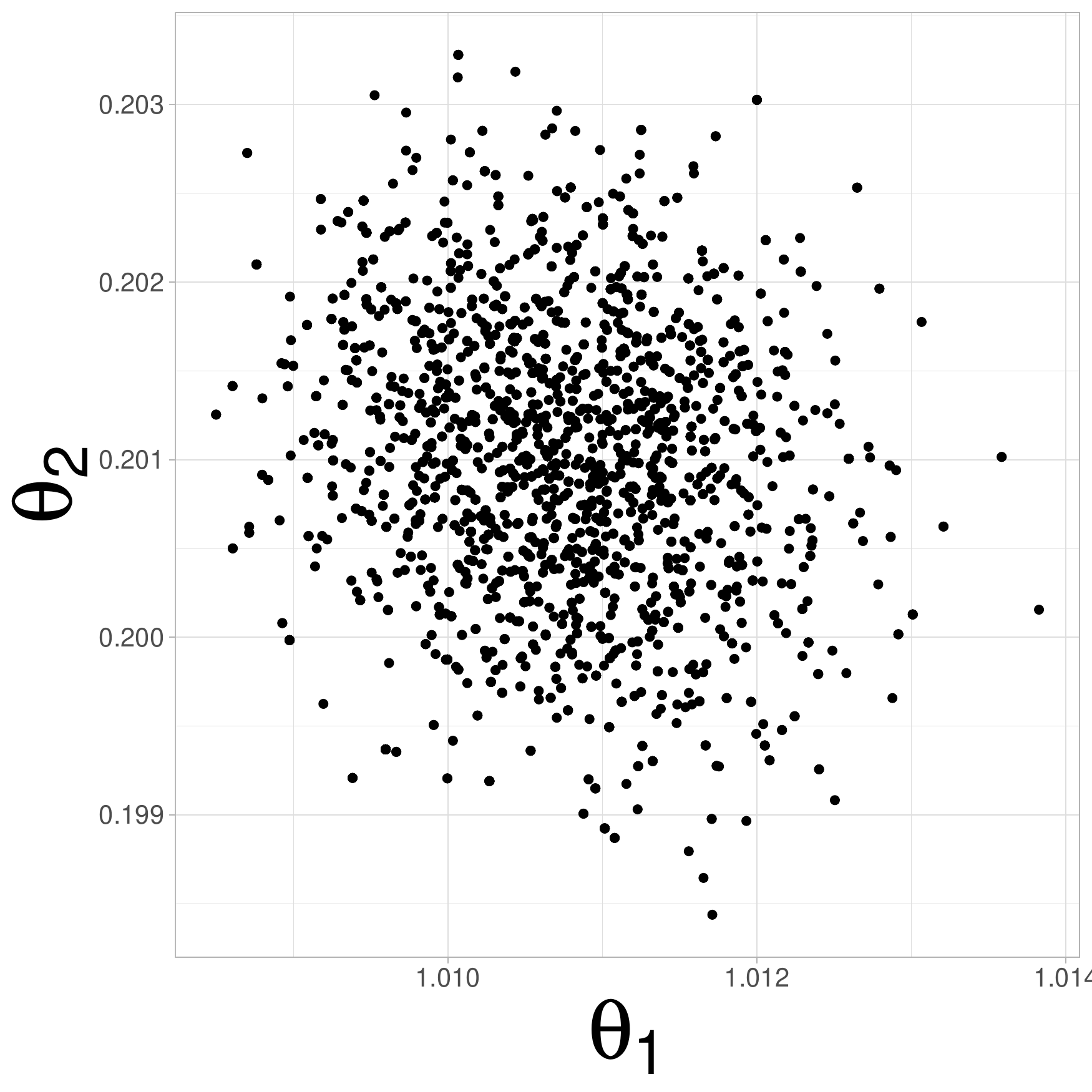} 

\end{knitrout}
&
\begin{knitrout}
\definecolor{shadecolor}{rgb}{0.969, 0.969, 0.969}\color{fgcolor}
\includegraphics[width=0.08\linewidth]{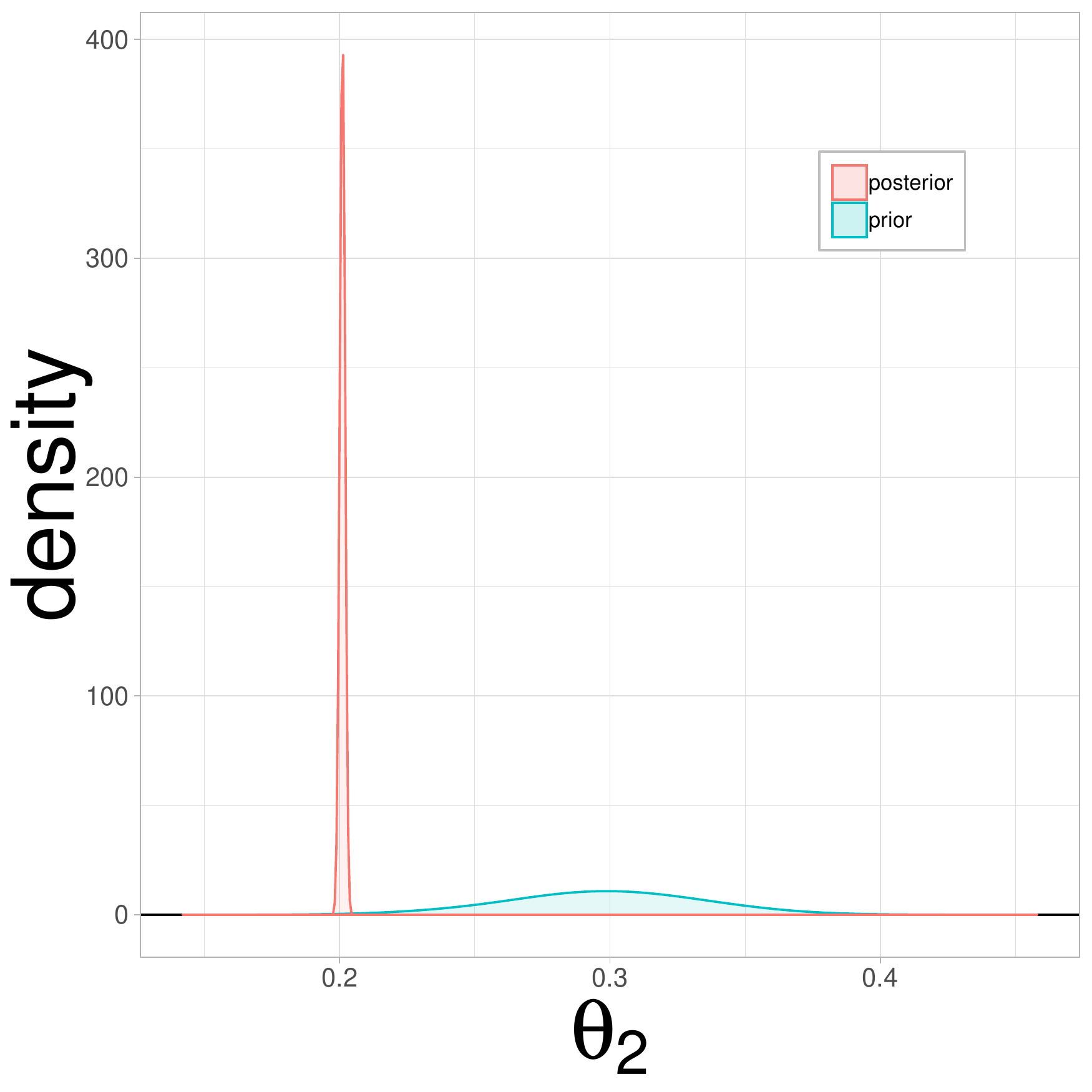} 

\end{knitrout}
&
\begin{knitrout}
\definecolor{shadecolor}{rgb}{0.969, 0.969, 0.969}\color{fgcolor}
\includegraphics[width=0.08\linewidth]{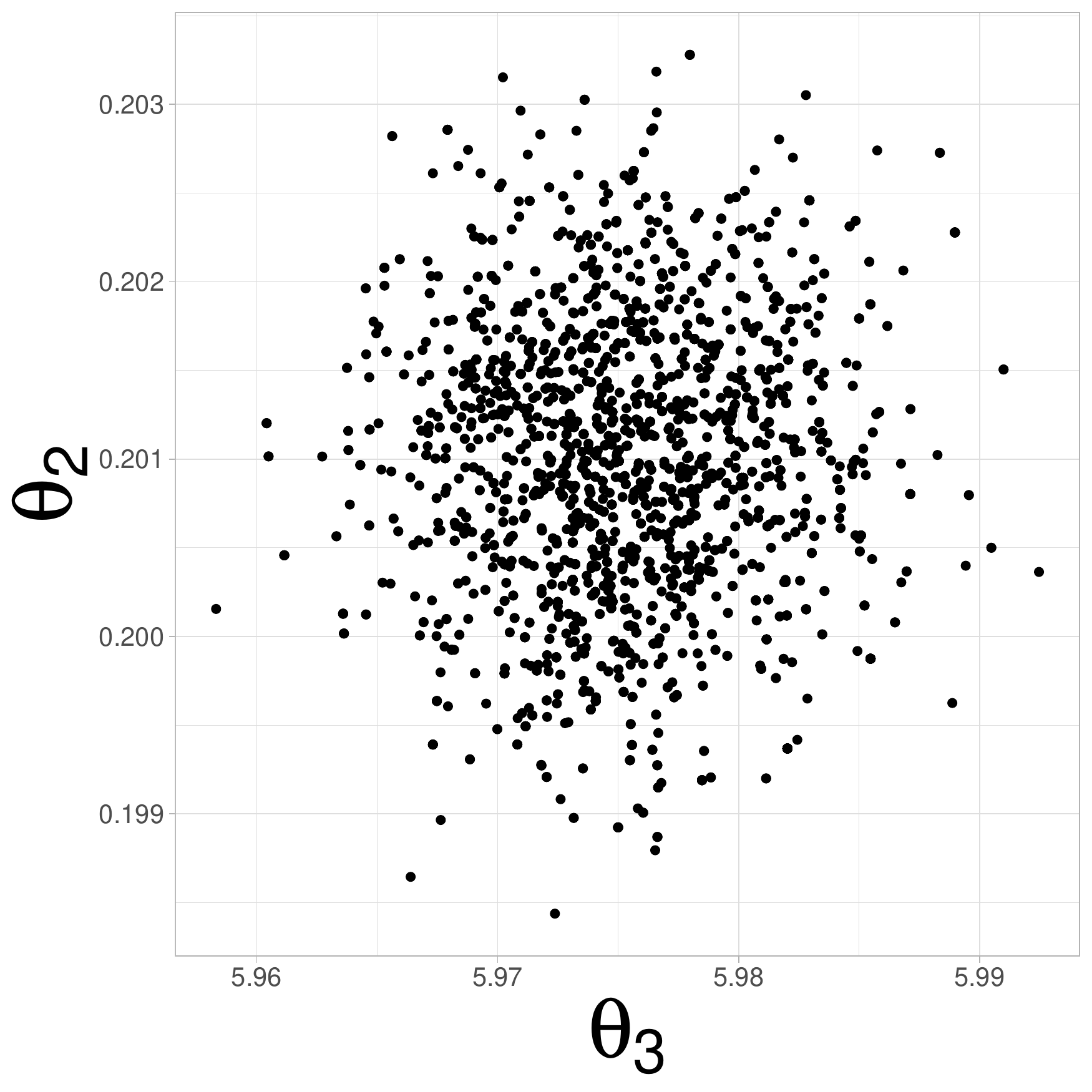} 

\end{knitrout}
&
\begin{knitrout}
\definecolor{shadecolor}{rgb}{0.969, 0.969, 0.969}\color{fgcolor}
\includegraphics[width=0.08\linewidth]{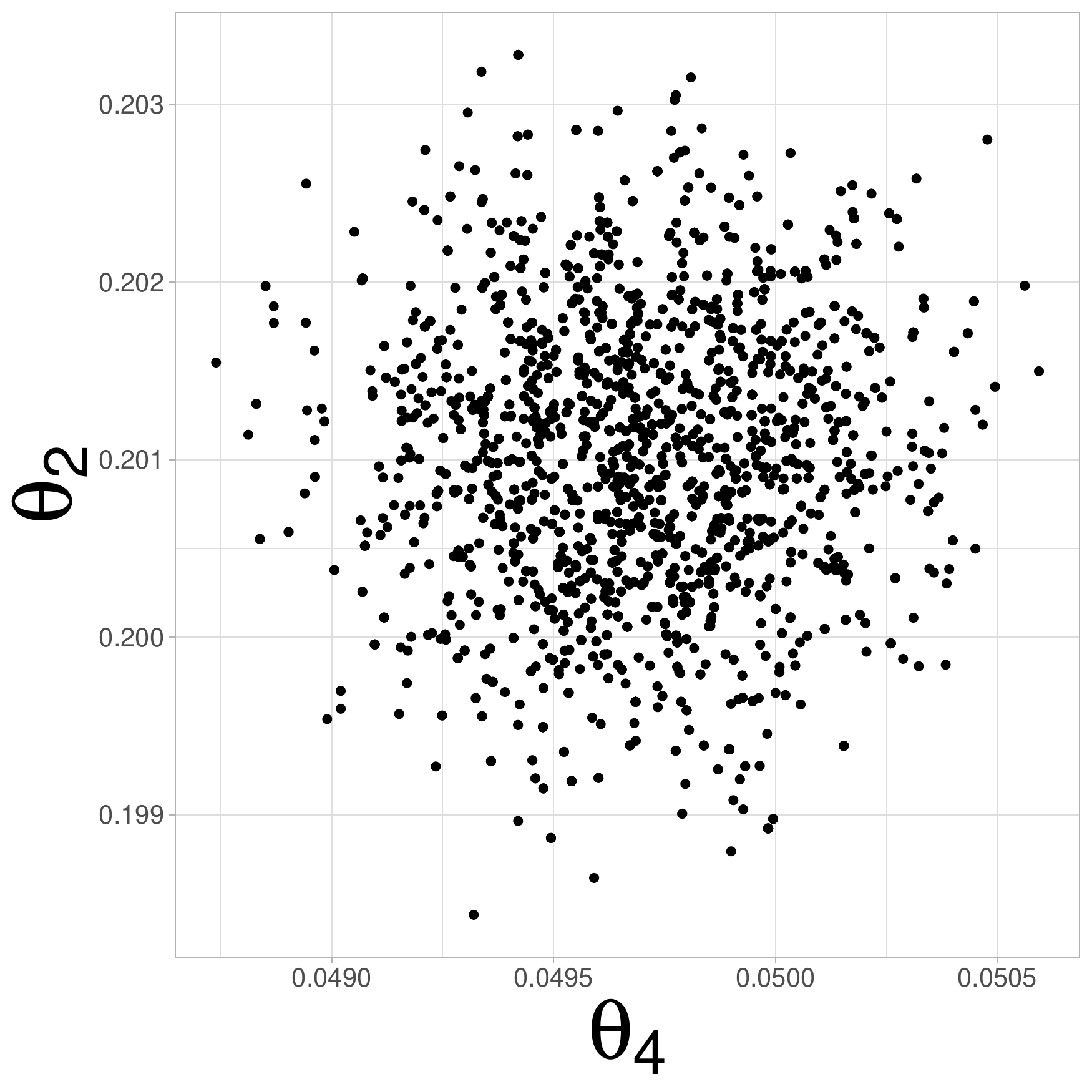} 

\end{knitrout}
&
\begin{knitrout}
\definecolor{shadecolor}{rgb}{0.969, 0.969, 0.969}\color{fgcolor}
\includegraphics[width=0.08\linewidth]{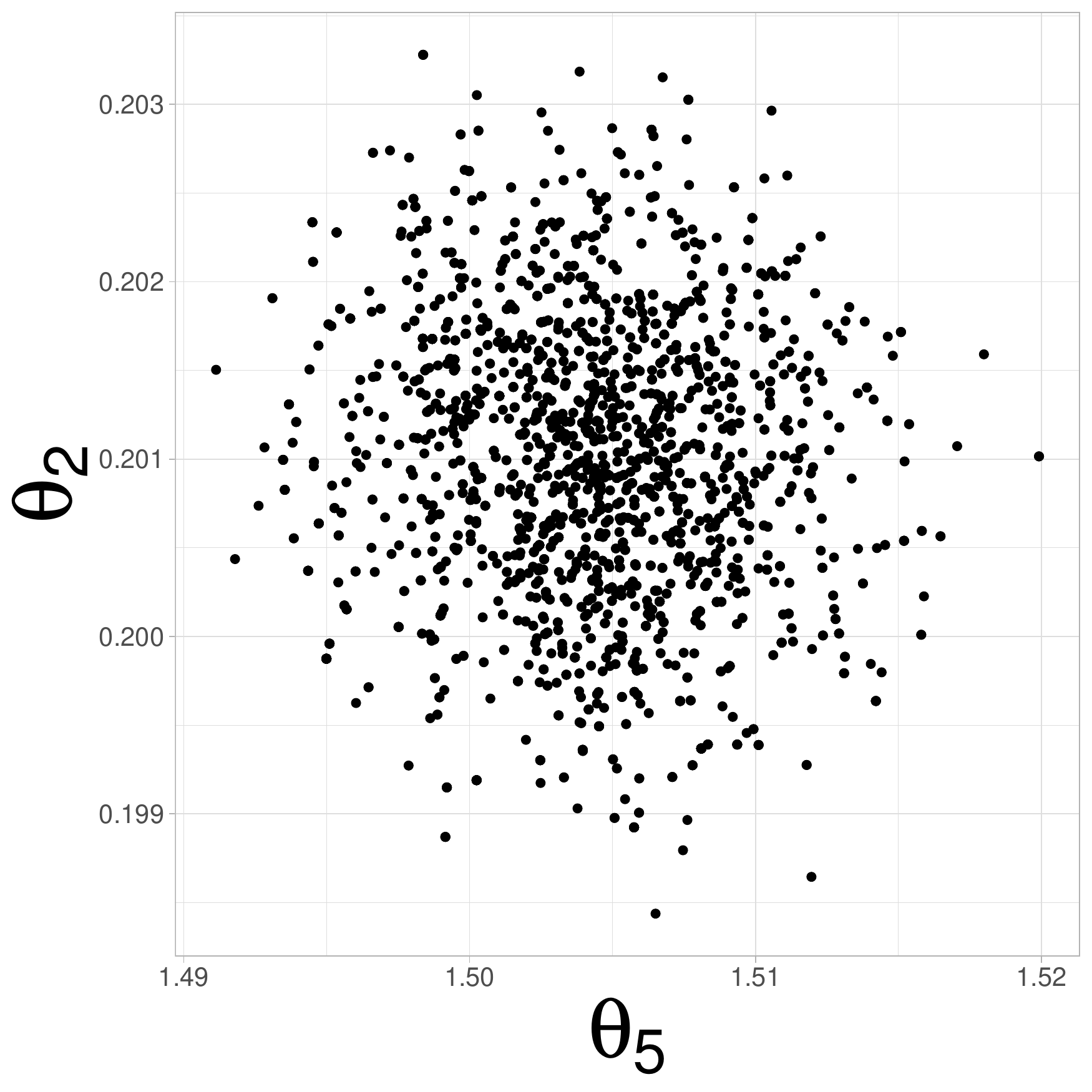} 

\end{knitrout}

\\
\begin{knitrout}
\definecolor{shadecolor}{rgb}{0.969, 0.969, 0.969}\color{fgcolor}
\includegraphics[width=0.08\linewidth]{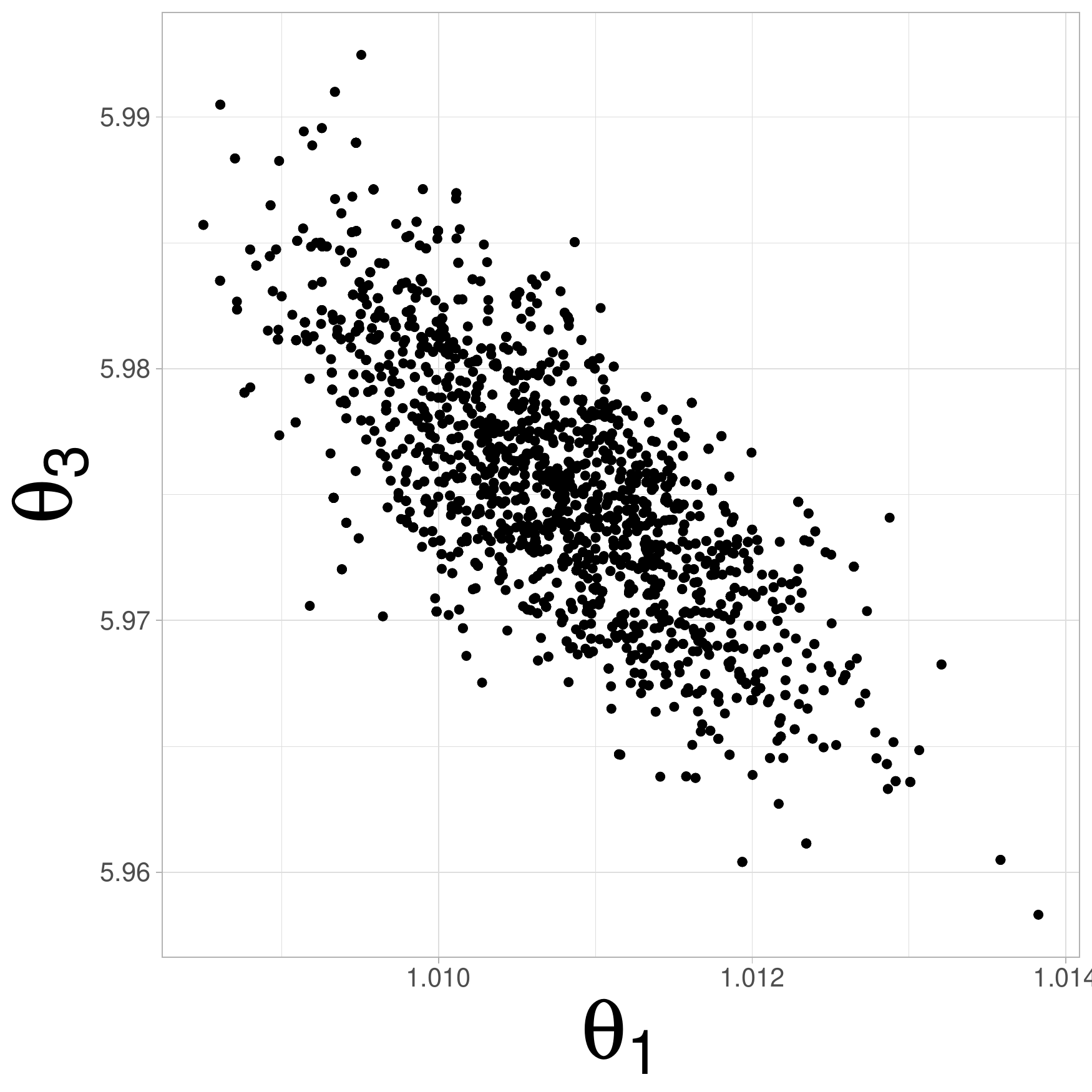} 

\end{knitrout}
&
\begin{knitrout}
\definecolor{shadecolor}{rgb}{0.969, 0.969, 0.969}\color{fgcolor}
\includegraphics[width=0.08\linewidth]{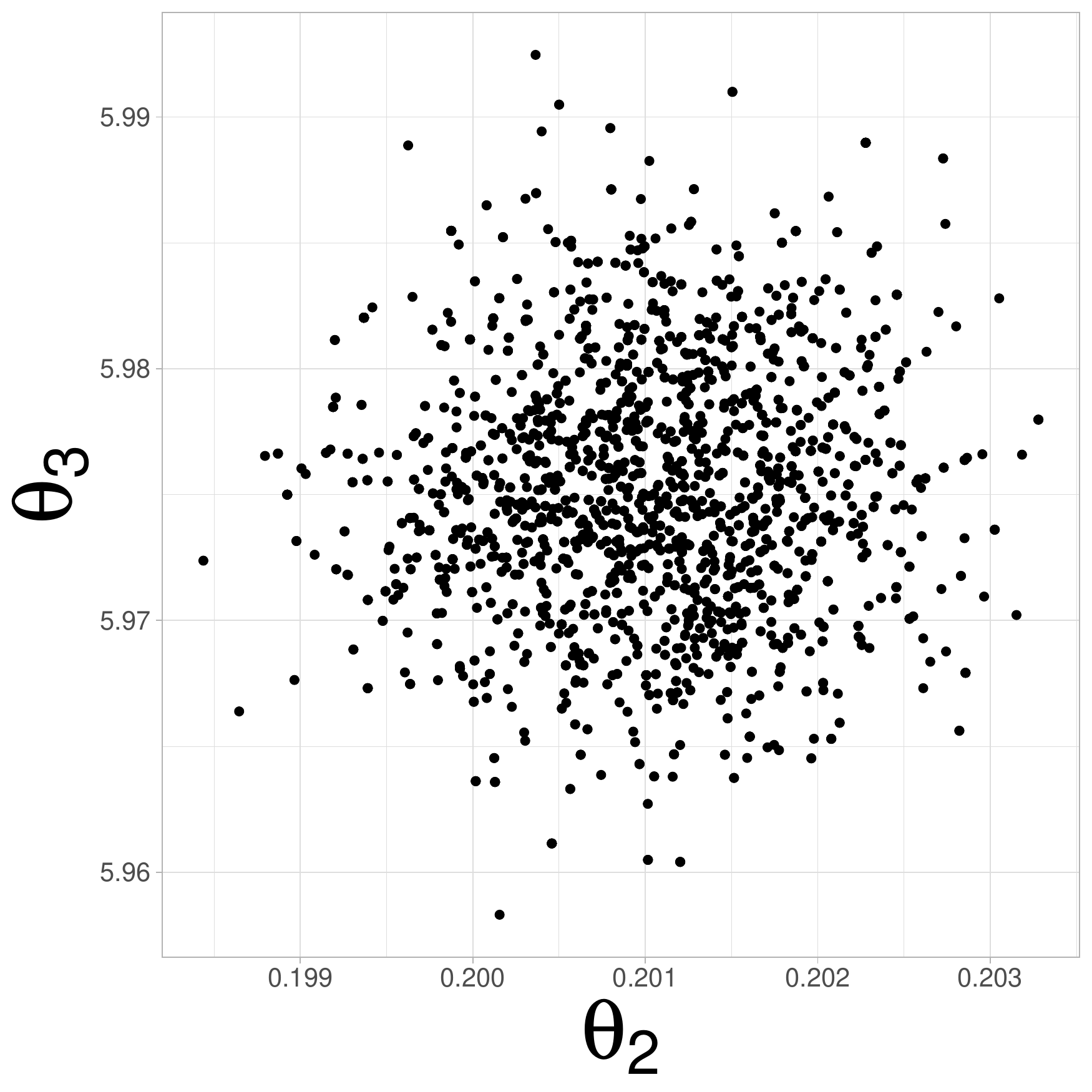} 

\end{knitrout}
&
\begin{knitrout}
\definecolor{shadecolor}{rgb}{0.969, 0.969, 0.969}\color{fgcolor}
\includegraphics[width=0.08\linewidth]{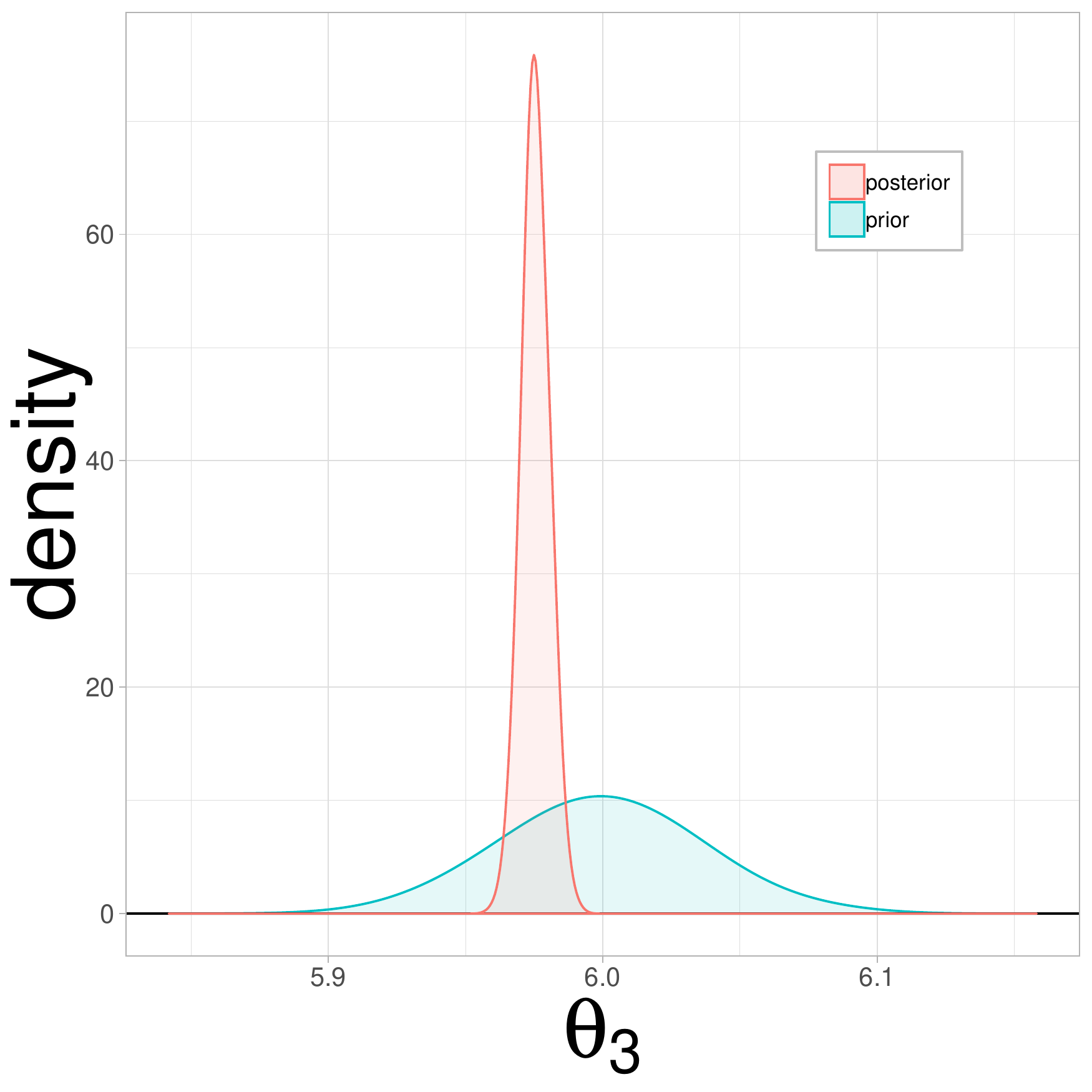} 

\end{knitrout}
&
\begin{knitrout}
\definecolor{shadecolor}{rgb}{0.969, 0.969, 0.969}\color{fgcolor}
\includegraphics[width=0.08\linewidth]{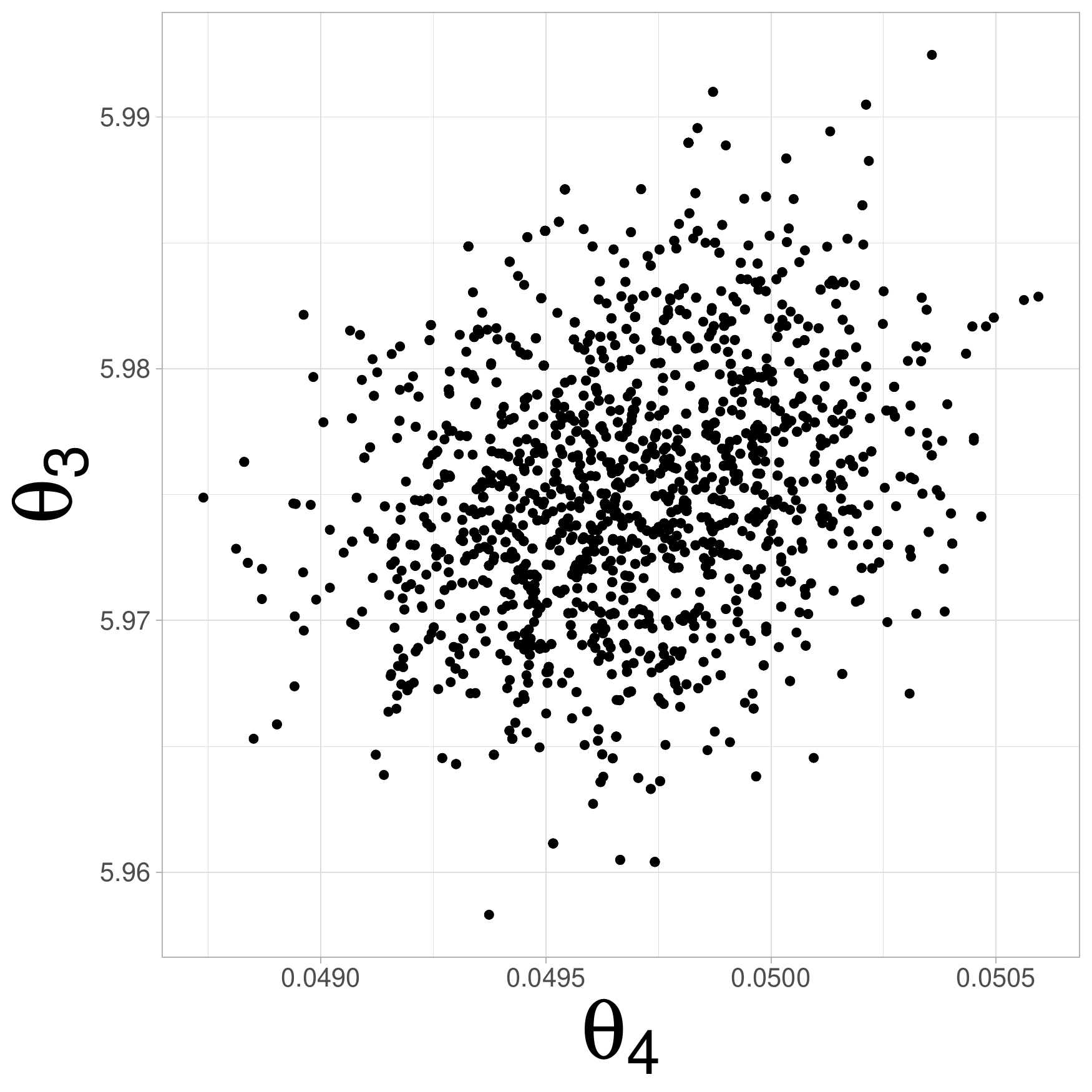} 

\end{knitrout}
&
\begin{knitrout}
\definecolor{shadecolor}{rgb}{0.969, 0.969, 0.969}\color{fgcolor}
\includegraphics[width=0.08\linewidth]{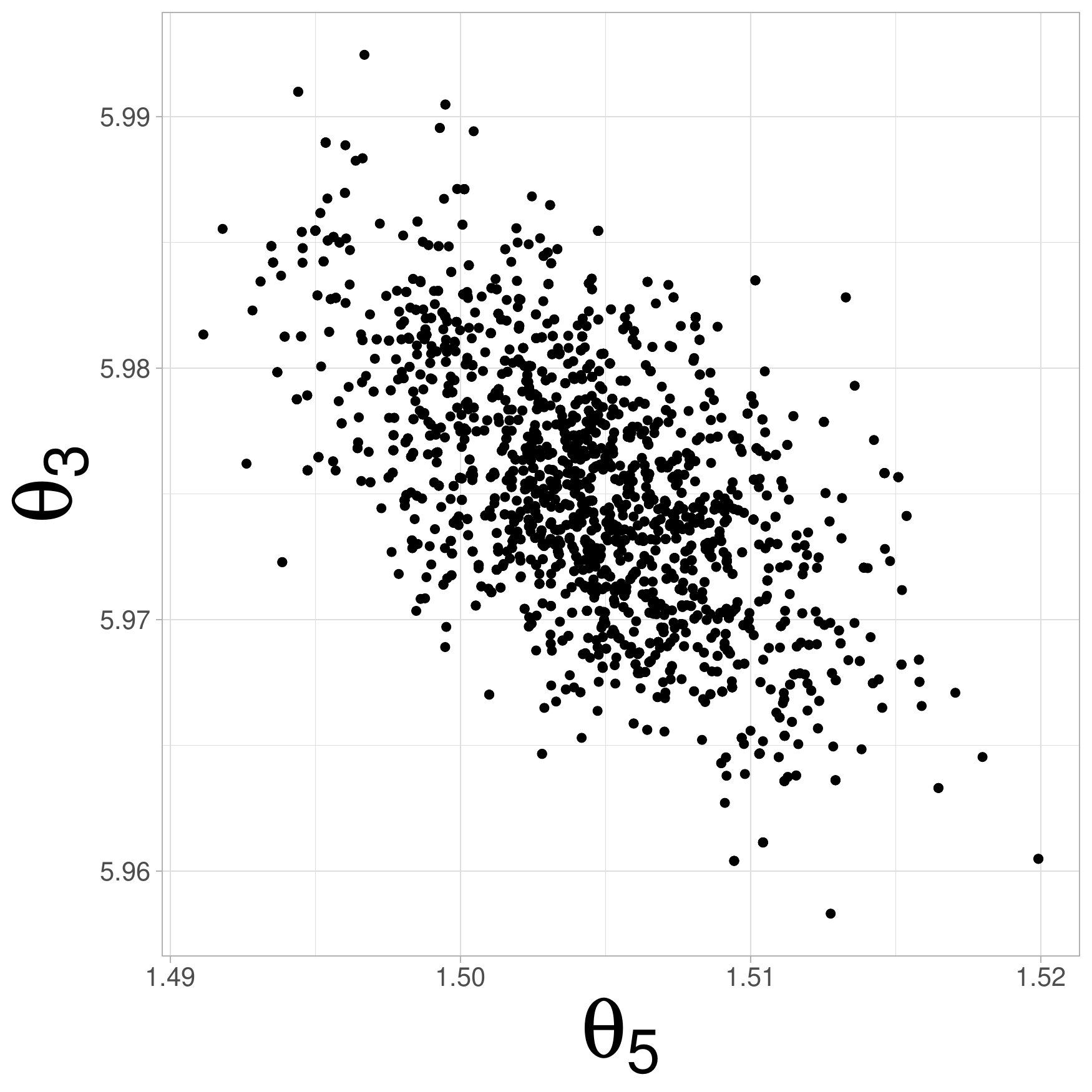} 

\end{knitrout}

\\
\begin{knitrout}
\definecolor{shadecolor}{rgb}{0.969, 0.969, 0.969}\color{fgcolor}
\includegraphics[width=0.08\linewidth]{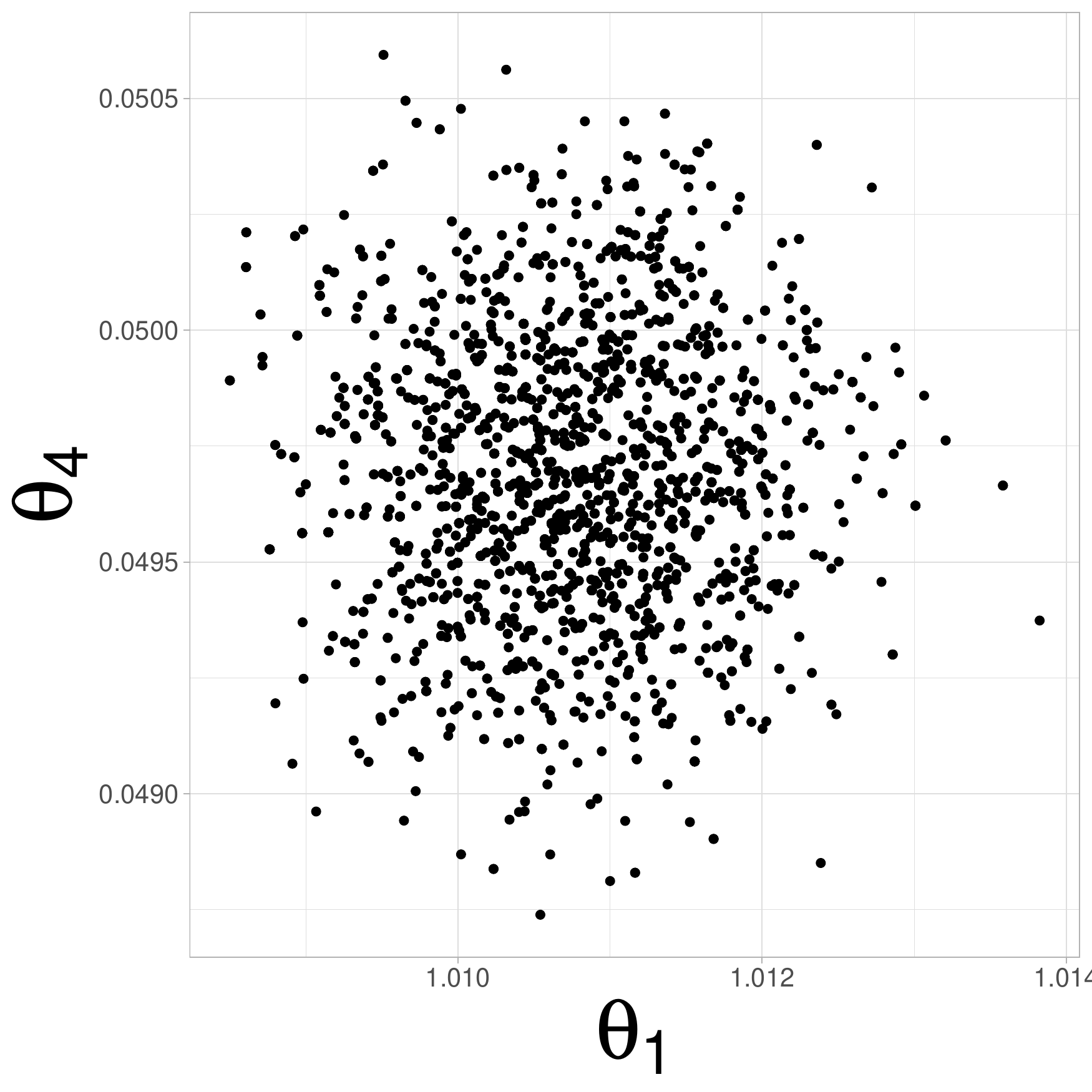} 

\end{knitrout}
&
\begin{knitrout}
\definecolor{shadecolor}{rgb}{0.969, 0.969, 0.969}\color{fgcolor}
\includegraphics[width=0.08\linewidth]{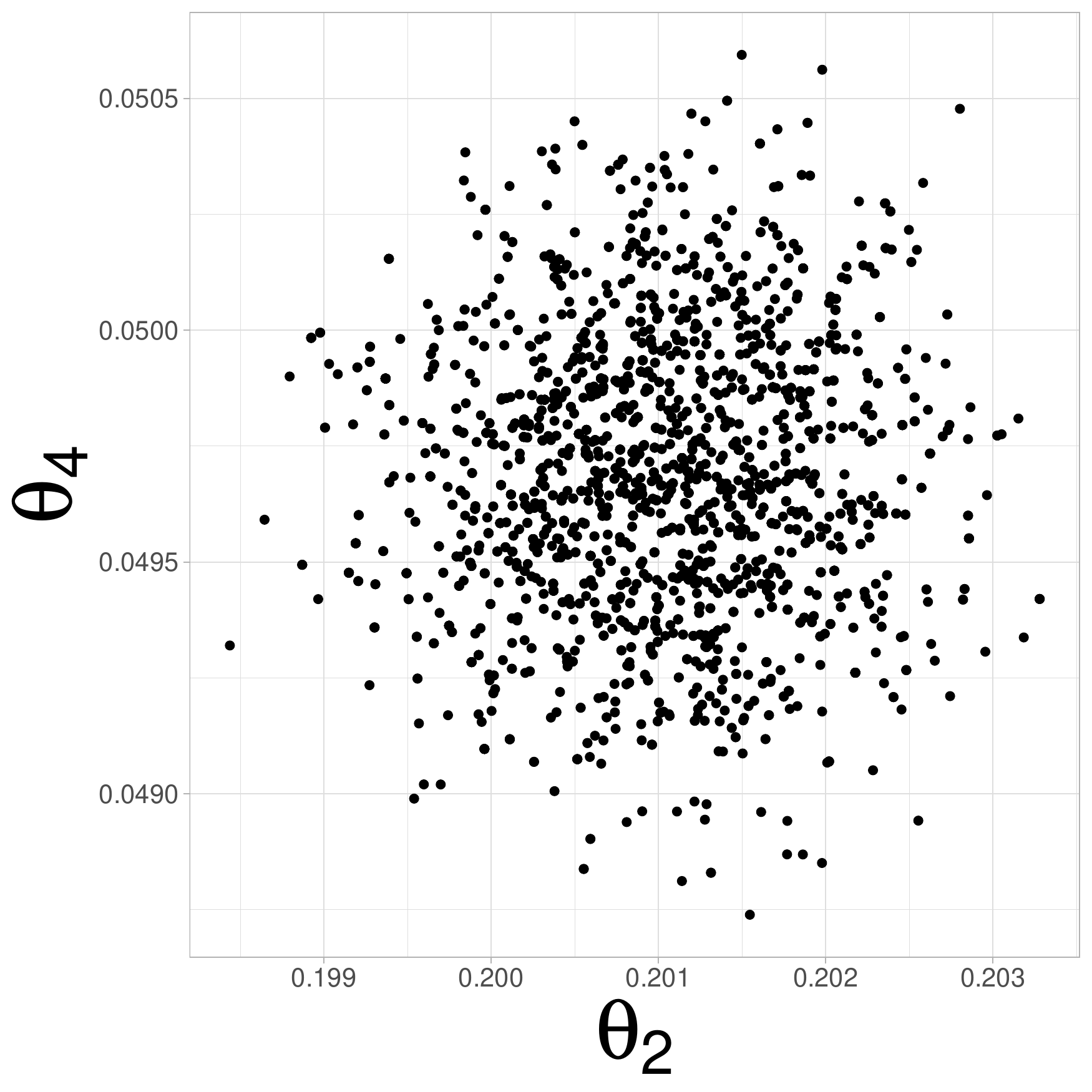} 

\end{knitrout}
&
\begin{knitrout}
\definecolor{shadecolor}{rgb}{0.969, 0.969, 0.969}\color{fgcolor}
\includegraphics[width=0.08\linewidth]{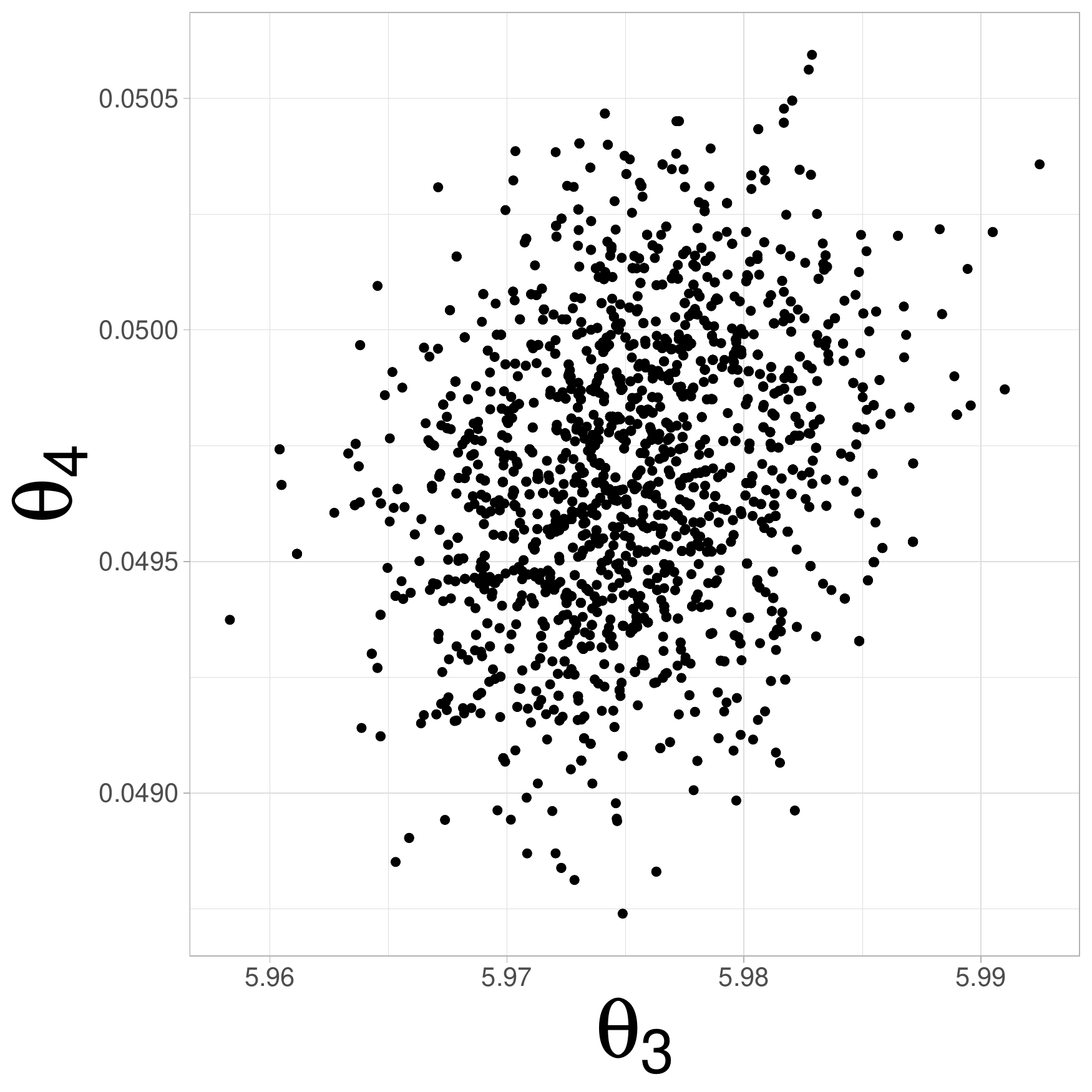} 

\end{knitrout}
&
\begin{knitrout}
\definecolor{shadecolor}{rgb}{0.969, 0.969, 0.969}\color{fgcolor}
\includegraphics[width=0.08\linewidth]{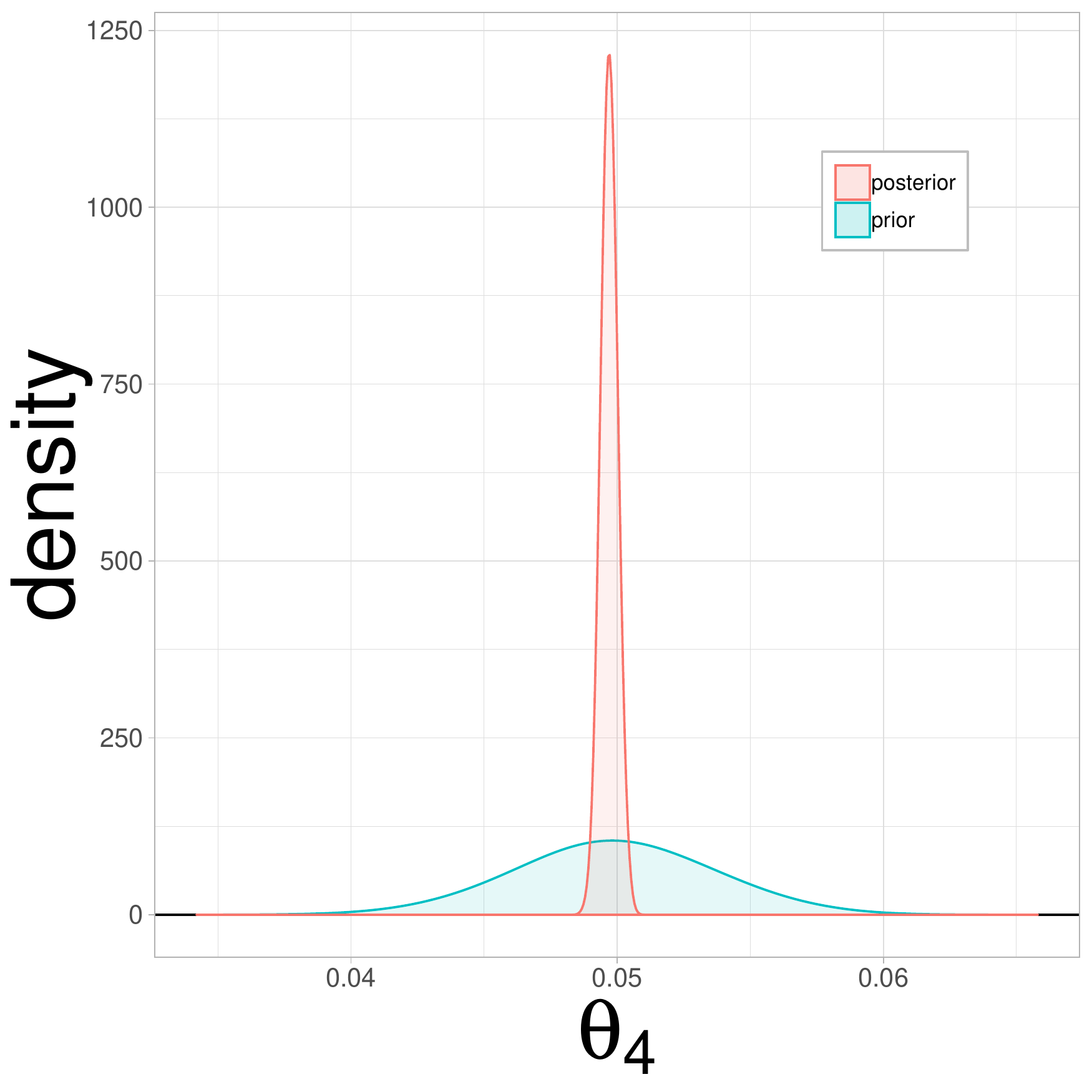} 

\end{knitrout}
&
\begin{knitrout}
\definecolor{shadecolor}{rgb}{0.969, 0.969, 0.969}\color{fgcolor}
\includegraphics[width=0.08\linewidth]{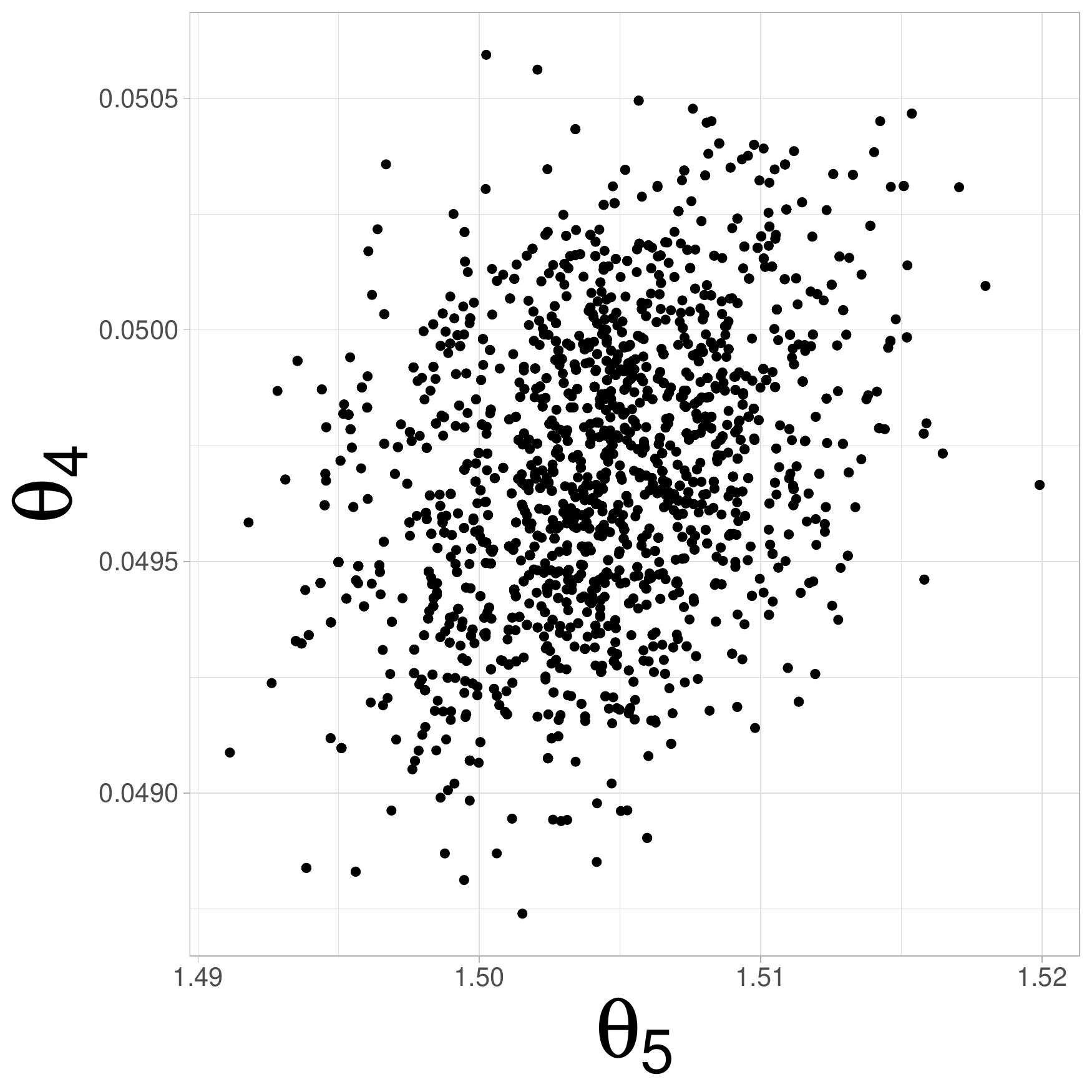} 

\end{knitrout}
\\
\begin{knitrout}
\definecolor{shadecolor}{rgb}{0.969, 0.969, 0.969}\color{fgcolor}
\includegraphics[width=0.08\linewidth]{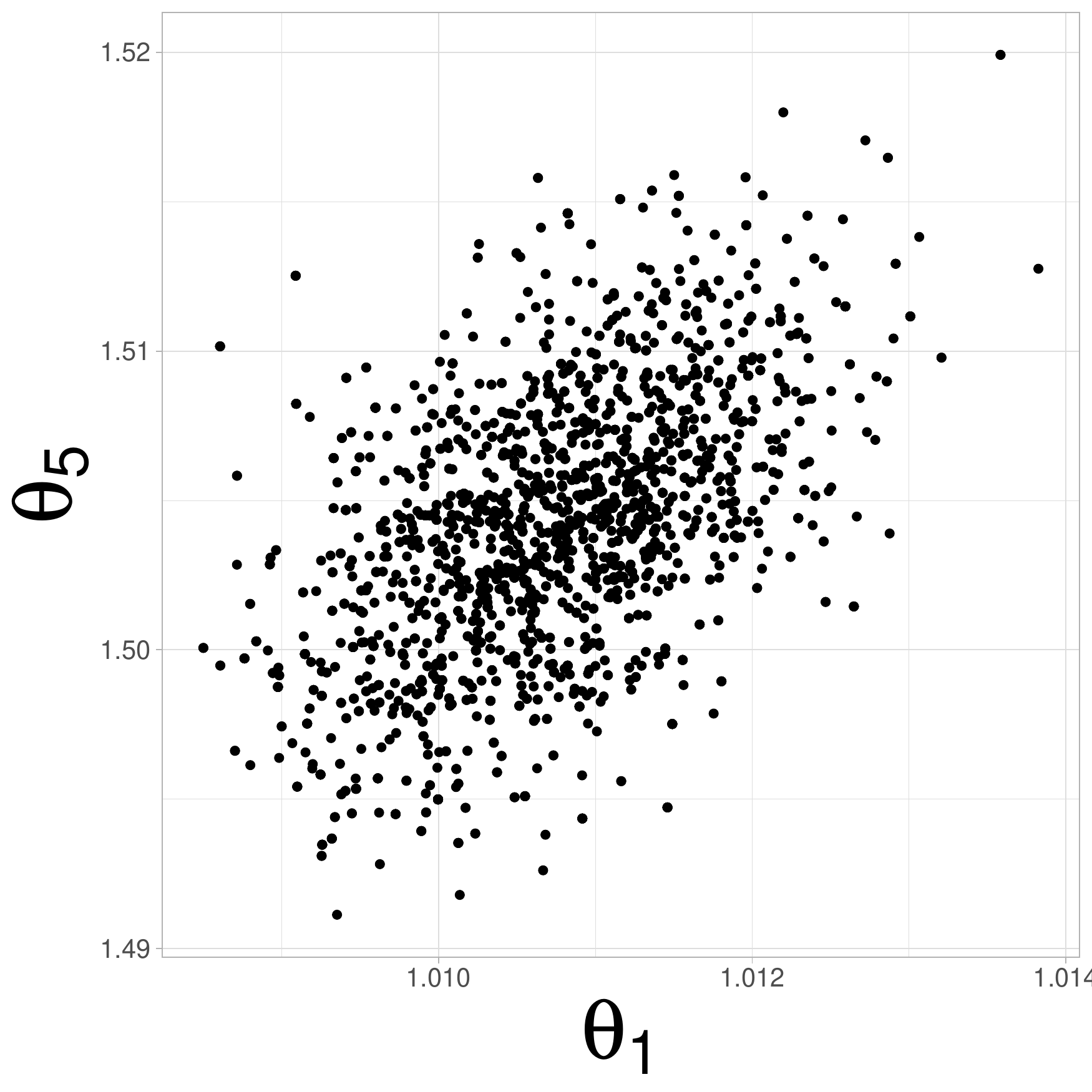} 

\end{knitrout}
&
\begin{knitrout}
\definecolor{shadecolor}{rgb}{0.969, 0.969, 0.969}\color{fgcolor}
\includegraphics[width=0.08\linewidth]{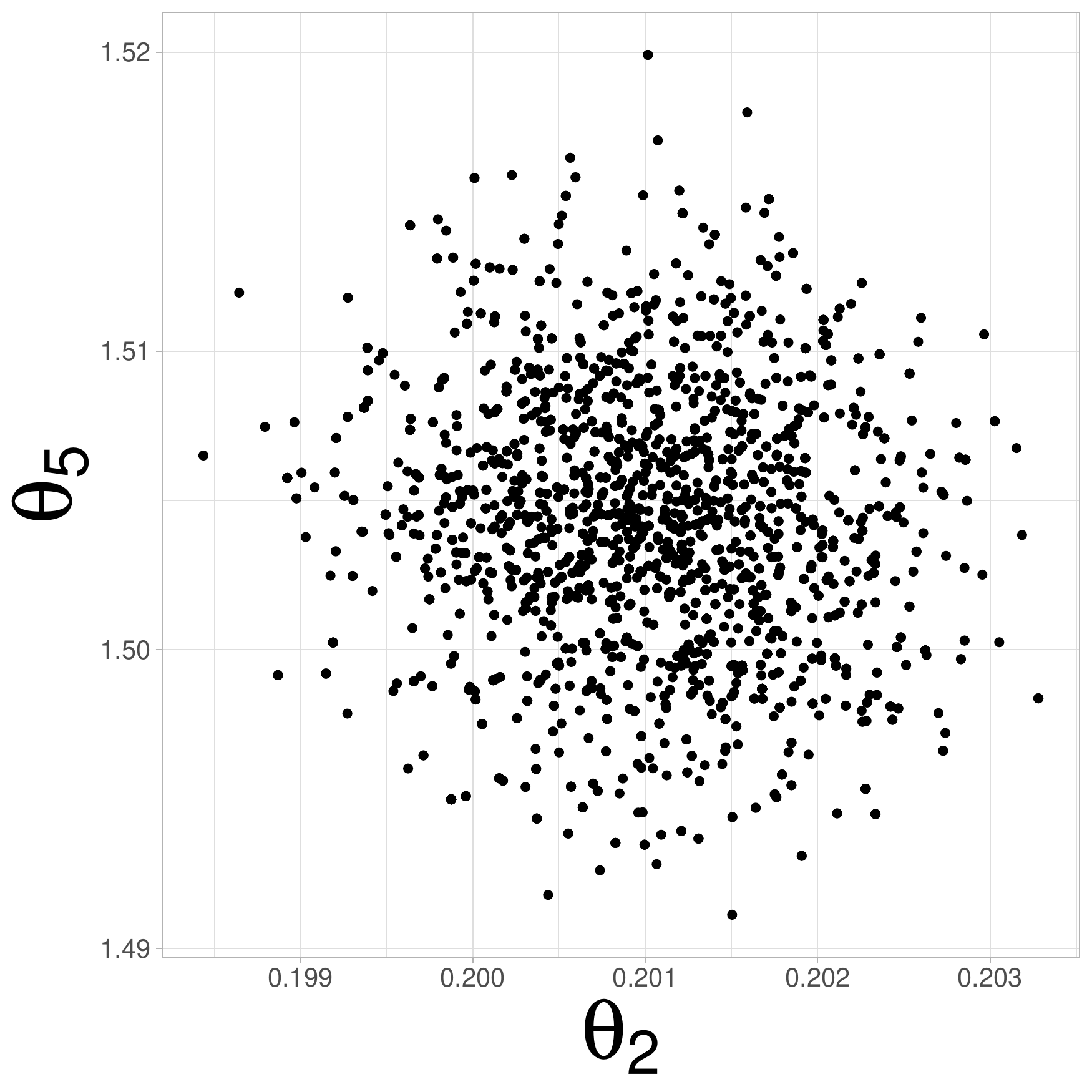} 

\end{knitrout}
&
\begin{knitrout}
\definecolor{shadecolor}{rgb}{0.969, 0.969, 0.969}\color{fgcolor}
\includegraphics[width=0.08\linewidth]{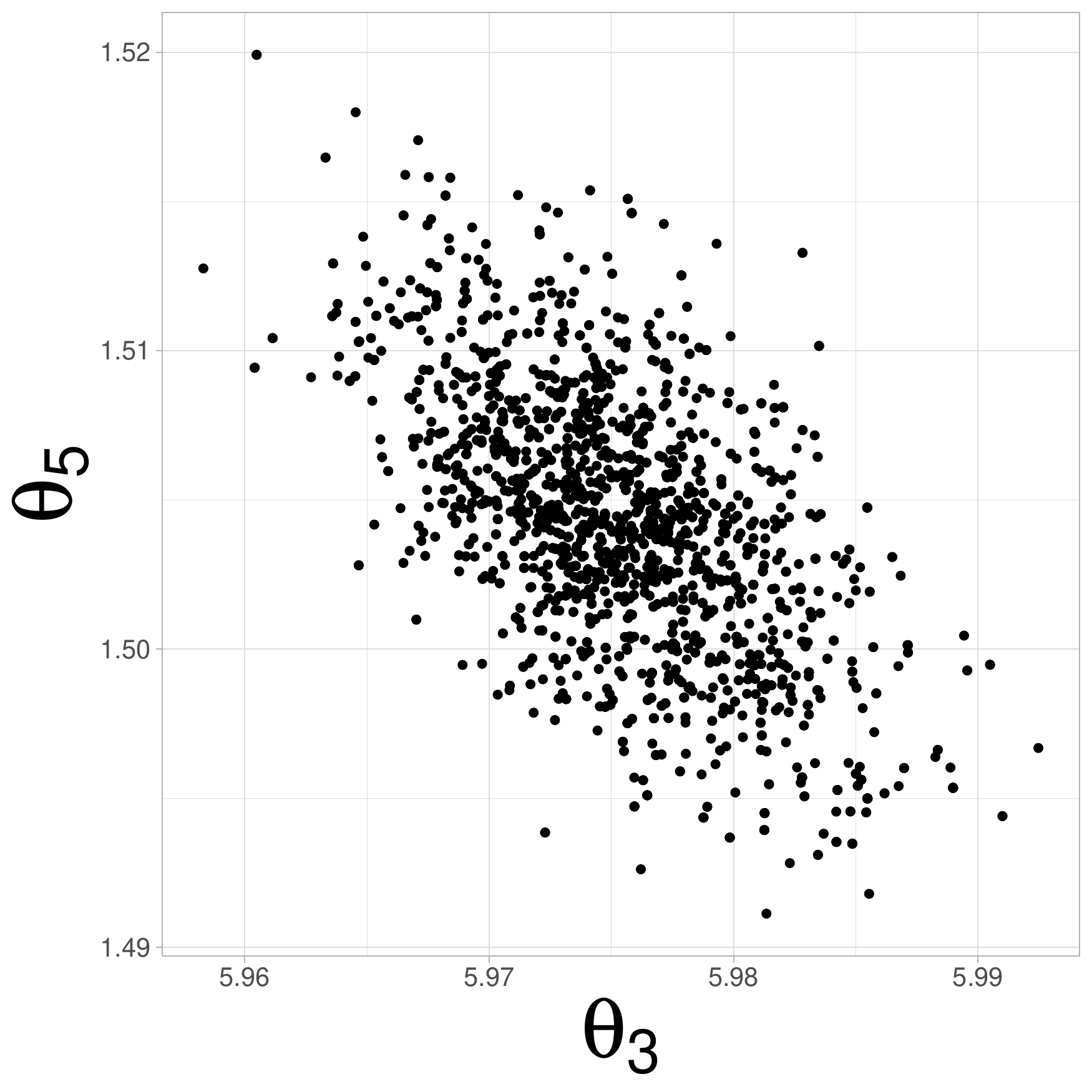} 

\end{knitrout}
&
\begin{knitrout}
\definecolor{shadecolor}{rgb}{0.969, 0.969, 0.969}\color{fgcolor}
\includegraphics[width=0.08\linewidth]{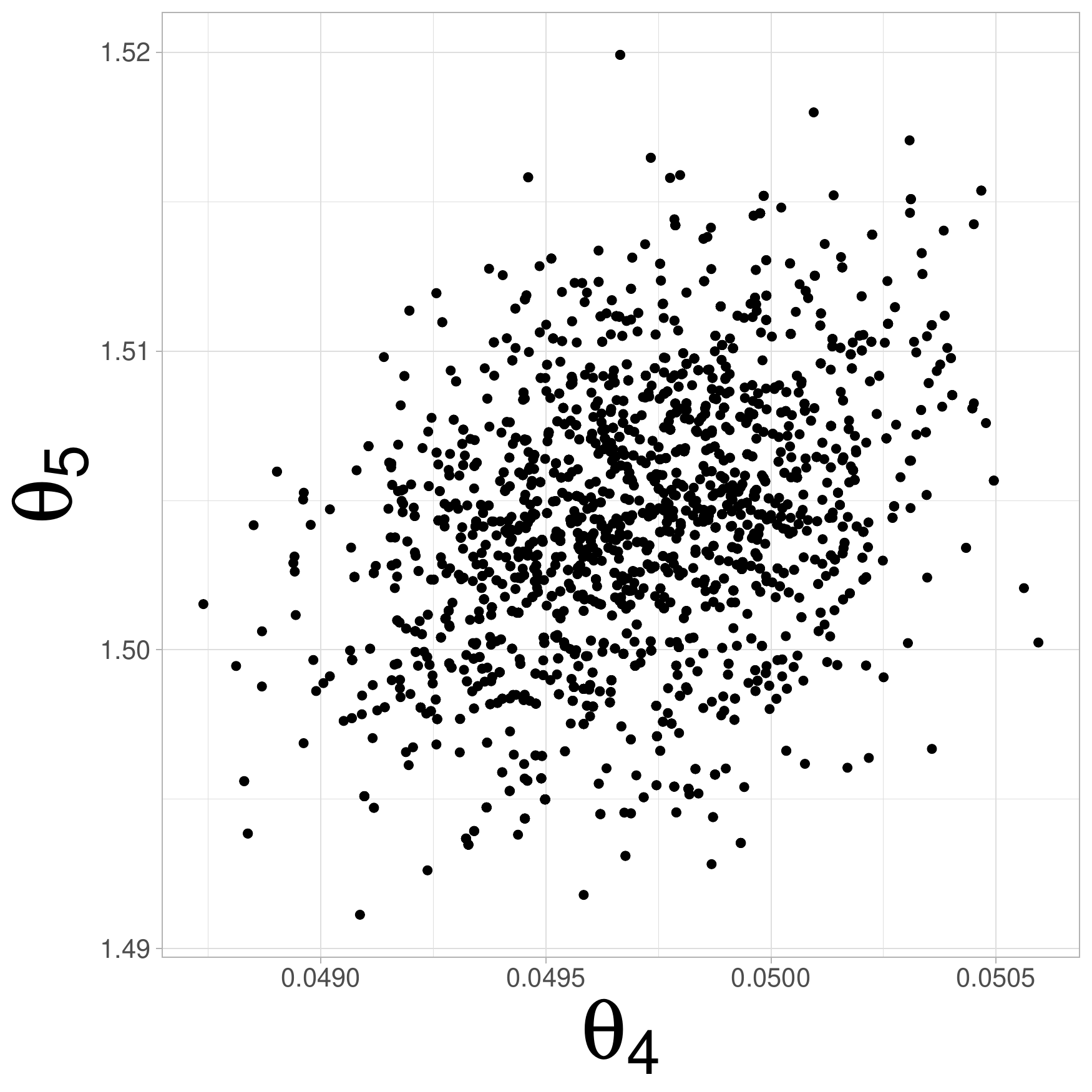} 

\end{knitrout}
&
\begin{knitrout}
\definecolor{shadecolor}{rgb}{0.969, 0.969, 0.969}\color{fgcolor}
\includegraphics[width=0.08\linewidth]{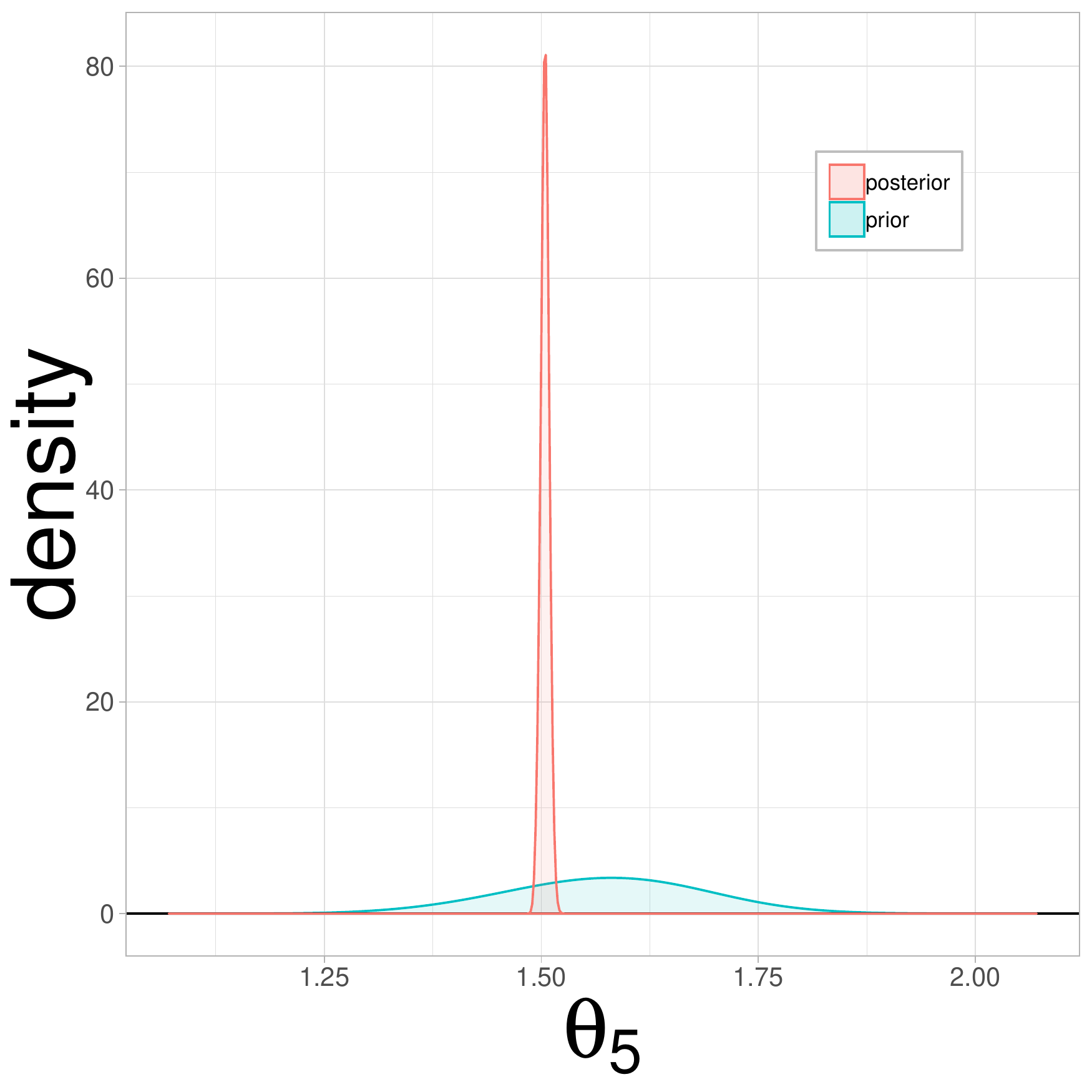} 

\end{knitrout}
\\
\end{tabular}};
\node[m1] (N2) at (7.5,0) {
\begin{knitrout}
\definecolor{shadecolor}{rgb}{0.969, 0.969, 0.969}\color{fgcolor}
\includegraphics[width=0.35\linewidth]{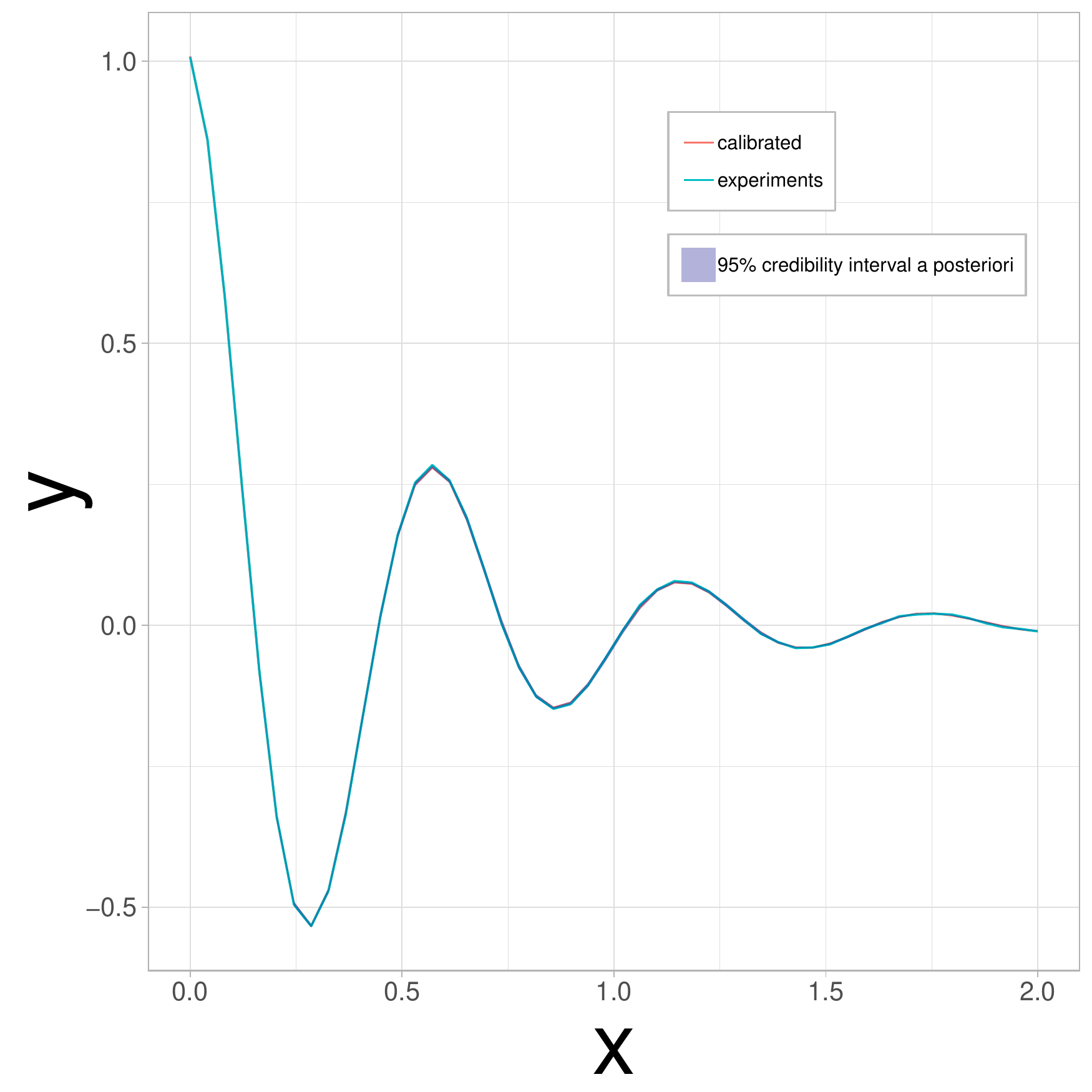} 

\end{knitrout}
};
\end{tikzpicture}
\end{center}
\caption{Series of plot generated by the function \code{plot} for calibration on $\mathcal{M}_1$}
\label{fig:PlotCalib1}
\end{figure}

Same procedures are available for $\mathcal{M}_2$.

\begin{knitrout}
\definecolor{shadecolor}{rgb}{0.969, 0.969, 0.969}\color{fgcolor}\begin{kframe}
\begin{alltt}
\hlstd{mdfit2} \hlkwb{<-} \hlkwd{calibrate}\hlstd{(model2,pr1,opt.estim)}
\end{alltt}
\end{kframe}
\end{knitrout}

For $\mathcal{M}_3$ and $\mathcal{M}_4$, the estimation options are slightly different because the number of parameter to estimate has increased. The \prior object also has changed.

\begin{knitrout}
\definecolor{shadecolor}{rgb}{0.969, 0.969, 0.969}\color{fgcolor}\begin{kframe}
\begin{alltt}
\hlstd{opt.estim2}\hlkwb{=}\hlkwd{list}\hlstd{(}\hlkwc{Ngibbs}\hlstd{=}\hlnum{1000}\hlstd{,}\hlkwc{Nmh}\hlstd{=}\hlnum{5000}\hlstd{,}\hlkwc{thetaInit}\hlstd{=}\hlkwd{c}\hlstd{(}\hlnum{1}\hlstd{,}\hlnum{0.3}\hlstd{,}\hlnum{6}\hlstd{,}\hlnum{50e-3}\hlstd{,pi}\hlopt{/}\hlnum{2}\hlstd{,}\hlnum{1e-3}\hlstd{,}\hlnum{0.5}\hlstd{,}\hlnum{1e-3}\hlstd{),}
                \hlkwc{r}\hlstd{=}\hlkwd{c}\hlstd{(}\hlnum{0.05}\hlstd{,}\hlnum{0.05}\hlstd{),}\hlkwc{sig}\hlstd{=}\hlkwd{diag}\hlstd{(}\hlnum{8}\hlstd{),}\hlkwc{Nchains}\hlstd{=}\hlnum{1}\hlstd{,}\hlkwc{burnIn}\hlstd{=}\hlnum{2000}\hlstd{)}
\end{alltt}
\end{kframe}
\end{knitrout}

\begin{knitrout}
\definecolor{shadecolor}{rgb}{0.969, 0.969, 0.969}\color{fgcolor}\begin{kframe}
\begin{alltt}
\hlstd{mdfit3} \hlkwb{<-} \hlkwd{calibrate}\hlstd{(model3,pr2,opt.estim2)}
\end{alltt}
\end{kframe}
\end{knitrout}

\begin{knitrout}
\definecolor{shadecolor}{rgb}{0.969, 0.969, 0.969}\color{fgcolor}\begin{kframe}
\begin{alltt}
\hlstd{mdfit4} \hlkwb{<-} \hlkwd{calibrate}\hlstd{(model4,pr2,opt.estim2)}
\end{alltt}
\end{kframe}
\end{knitrout}

\begin{knitrout}
\definecolor{shadecolor}{rgb}{0.969, 0.969, 0.969}\color{fgcolor}\begin{kframe}
\begin{alltt}
\hlkwd{print}\hlstd{(mdfit4)}
\end{alltt}
\begin{verbatim}
## Call:
## 
## With the function:
## NULL
## 
## Selected model : model4 
## 
## Acceptation rate of the Metropolis within Gibbs algorithm:
## [1] "97.8%" "94.7%" "96.8%" "90.3%" "87.5%" "94.3%" "97.5%" "94.1%"
## 
## Acceptation rate of the Metropolis Hastings algorithm:
## [1] "59.2%"
## 
## Maximum a posteriori:
## [1] 1.0145661817 0.3052534056 6.0274228120 0.0521952278 1.6229728079
## [6] 0.0009970529 0.4769160005 0.0007652975
## 
## Mean a posteriori:
## [1] 0.9968290203 0.2835127409 6.0032790767 0.0506295380 1.5845041957
## [6] 0.0009225639 0.4235883680 0.0006717480
\end{verbatim}
\end{kframe}
\end{knitrout}

Figure \ref{fig:PlotCalib1} illustrates the several graphs layout one can obtain with the use of the function \code{plot}. To select which specific graph one wants to display, the option \code{graph} can be added to the function \code{plot}:
\begin{itemize}
\item \code{graph="chains"}: only the table of the autocorrelation, chains points and distributions \textit{a priori} and \textit{a posteriori} is produced . It represents only the top part of the Figure \ref{fig:PlotCalib1},
\item \code{graph="corr"}: only the table of the correlation graph between each parameter is displayed. It represents only the bottom left part of the Figure \ref{fig:PlotCalib1},
\item \code{graph="result"}: only the result on the quantity of interest is given. It represents only the bottom right part of the Figure \ref{fig:PlotCalib1},
\item \code{graph=NULL}: no graphs are produced automatically.
\end{itemize}

If one does not want to produce these graphs automatically, one can set the \code{graph} option to \code{NULL}. As the \code{plot} function generates \pkg{ggplot2} objects, it is possible to load all the generated graphs apart.

\begin{knitrout}
\definecolor{shadecolor}{rgb}{0.969, 0.969, 0.969}\color{fgcolor}\begin{kframe}
\begin{alltt}
\hlstd{p} \hlkwb{<-} \hlkwd{plot}\hlstd{(mdfit4,t,}\hlkwc{graph}\hlstd{=}\hlkwa{NULL}\hlstd{)}
\end{alltt}
\end{kframe}
\end{knitrout}

The variable $p$ is a \code{list} of all the graphs displayed Figure \ref{fig:PlotCalib1}. The elements in $p$ are:
\begin{itemize}
\item \code{ACF} a \code{list} of all autocorrelation graphs in the chains for each variable,
\item \code{MCMC} a \code{list} of all the MCMC chains for each variable,
\item \code{corrplot} a \code{list} of all correlation graphs between each parameter,
\item \code{dens} a \code{list} of all distribution \textit{a priori} and \textit{a posteriori} graphs for each variable,
\item \code{out} the \pkg{ggplot2} object of the result on the quantity of interest.
\end{itemize}

\begin{figure}[h!]
\begin{center}
    \begin{tabular}{cccc}
\begin{knitrout}
\definecolor{shadecolor}{rgb}{0.969, 0.969, 0.969}\color{fgcolor}
\includegraphics[width=0.21\linewidth,height=0.21\linewidth]{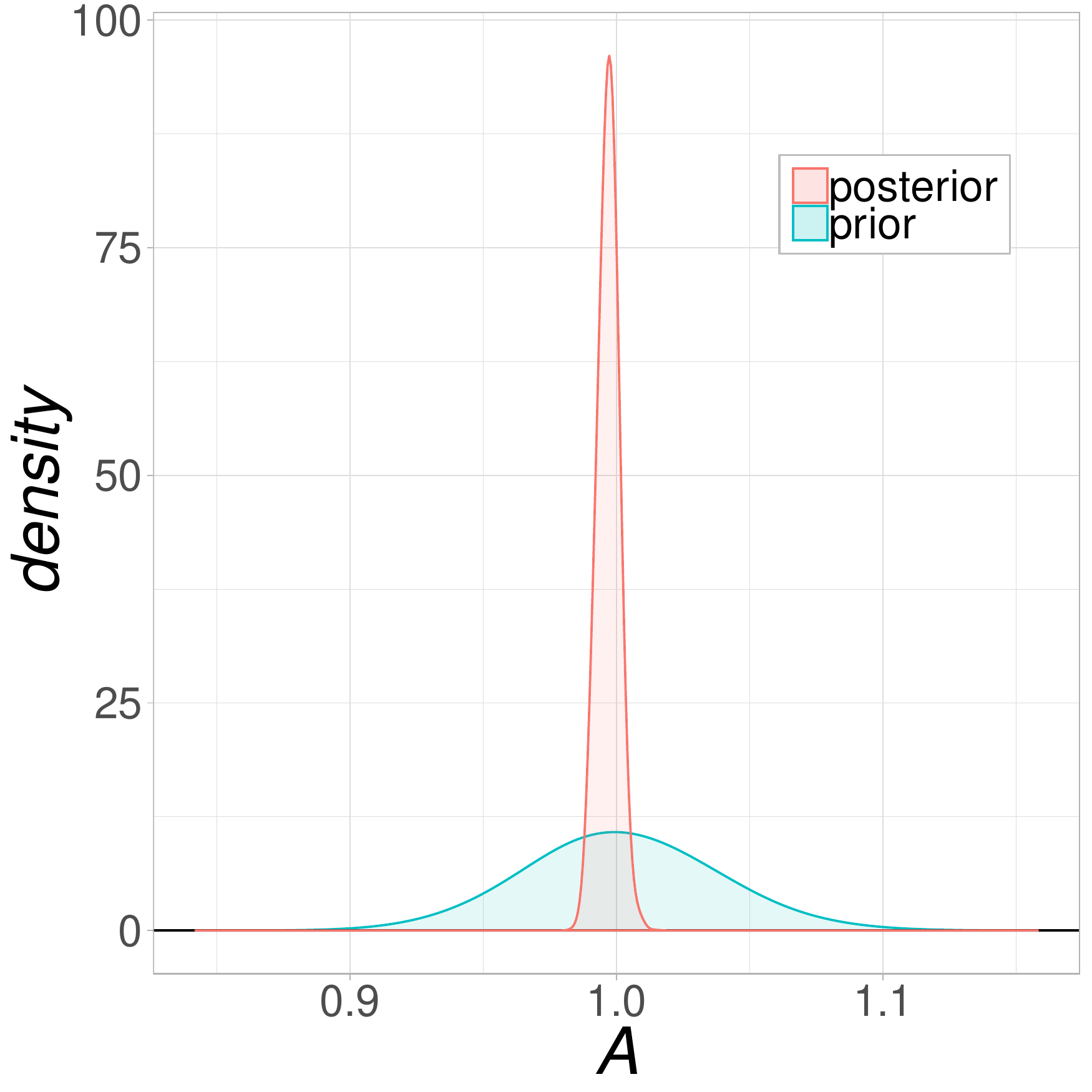} 

\end{knitrout}
    &
\begin{knitrout}
\definecolor{shadecolor}{rgb}{0.969, 0.969, 0.969}\color{fgcolor}
\includegraphics[width=0.21\linewidth,height=0.21\linewidth]{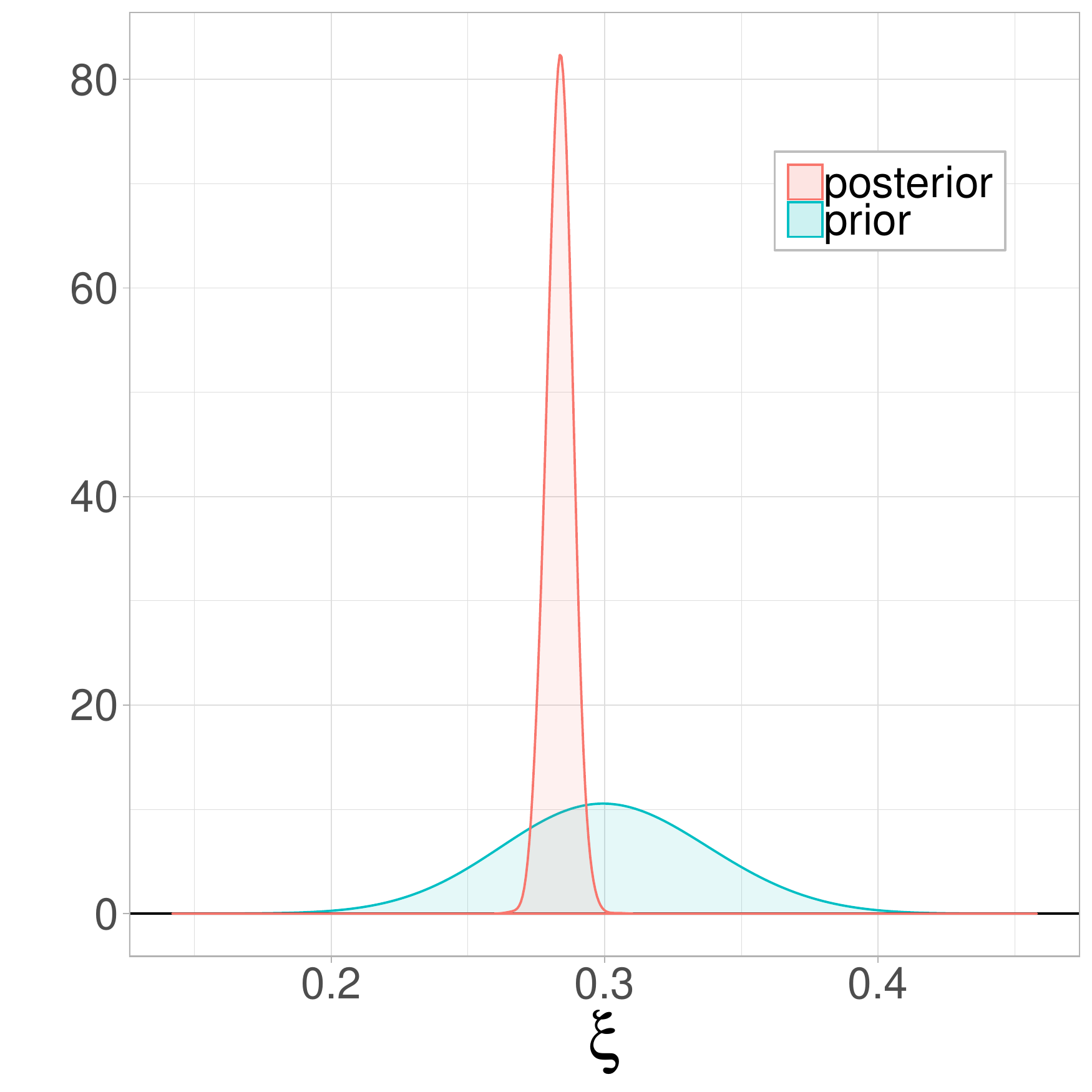} 

\end{knitrout}
&
\begin{knitrout}
\definecolor{shadecolor}{rgb}{0.969, 0.969, 0.969}\color{fgcolor}
\includegraphics[width=0.21\linewidth,height=0.21\linewidth]{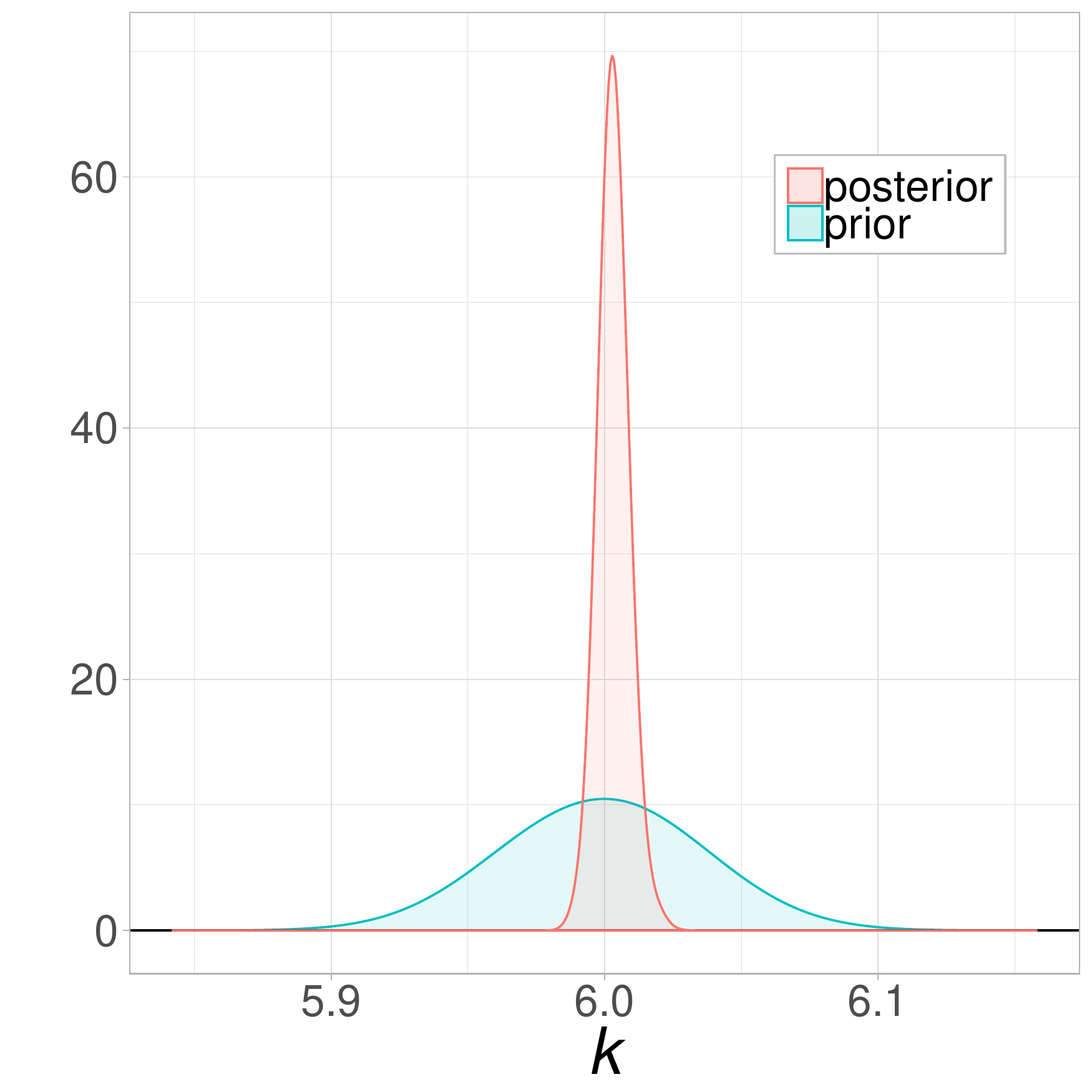} 

\end{knitrout}
    &
\begin{knitrout}
\definecolor{shadecolor}{rgb}{0.969, 0.969, 0.969}\color{fgcolor}
\includegraphics[width=0.21\linewidth,height=0.21\linewidth]{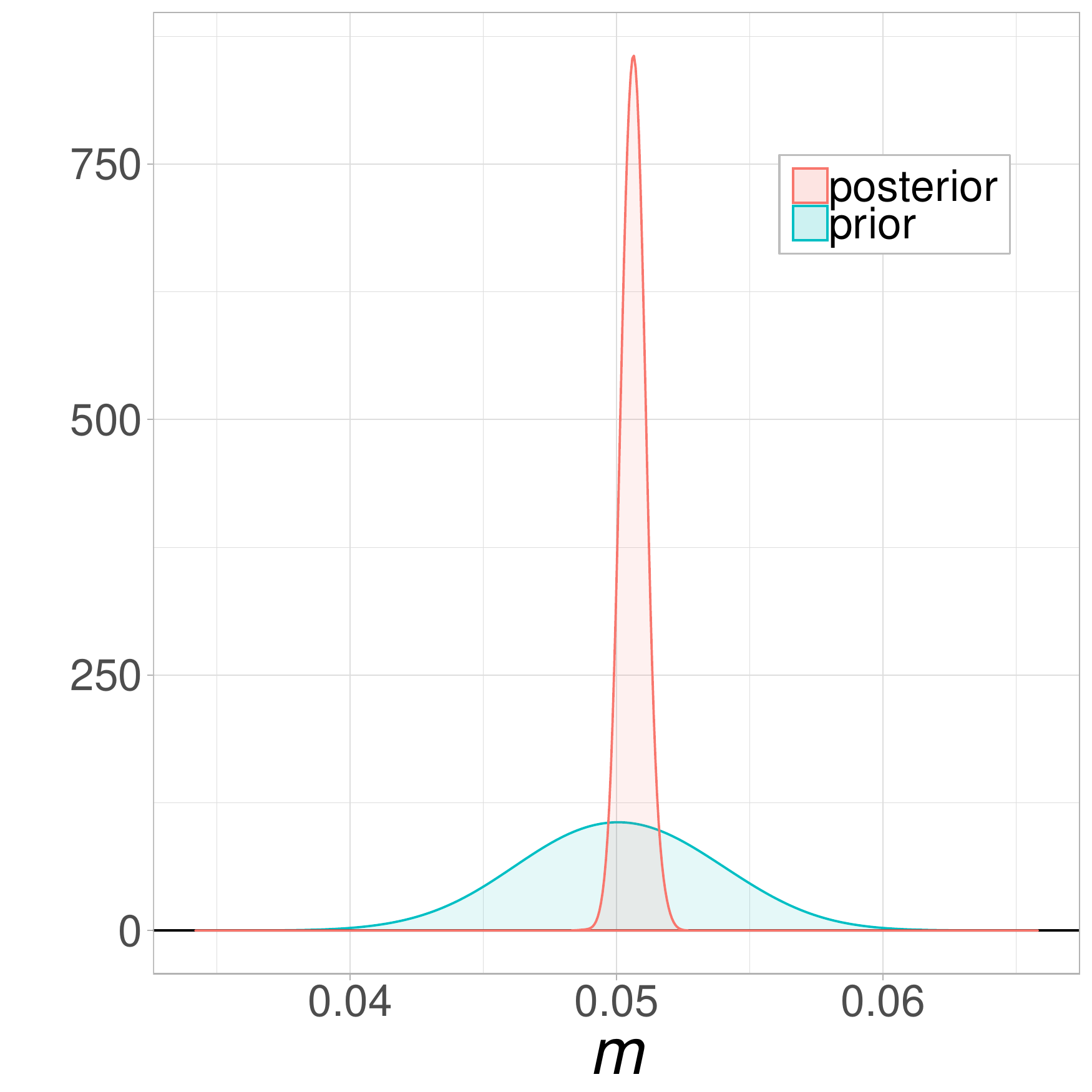} 

\end{knitrout}
\\
\begin{knitrout}
\definecolor{shadecolor}{rgb}{0.969, 0.969, 0.969}\color{fgcolor}
\includegraphics[width=0.21\linewidth,height=0.21\linewidth]{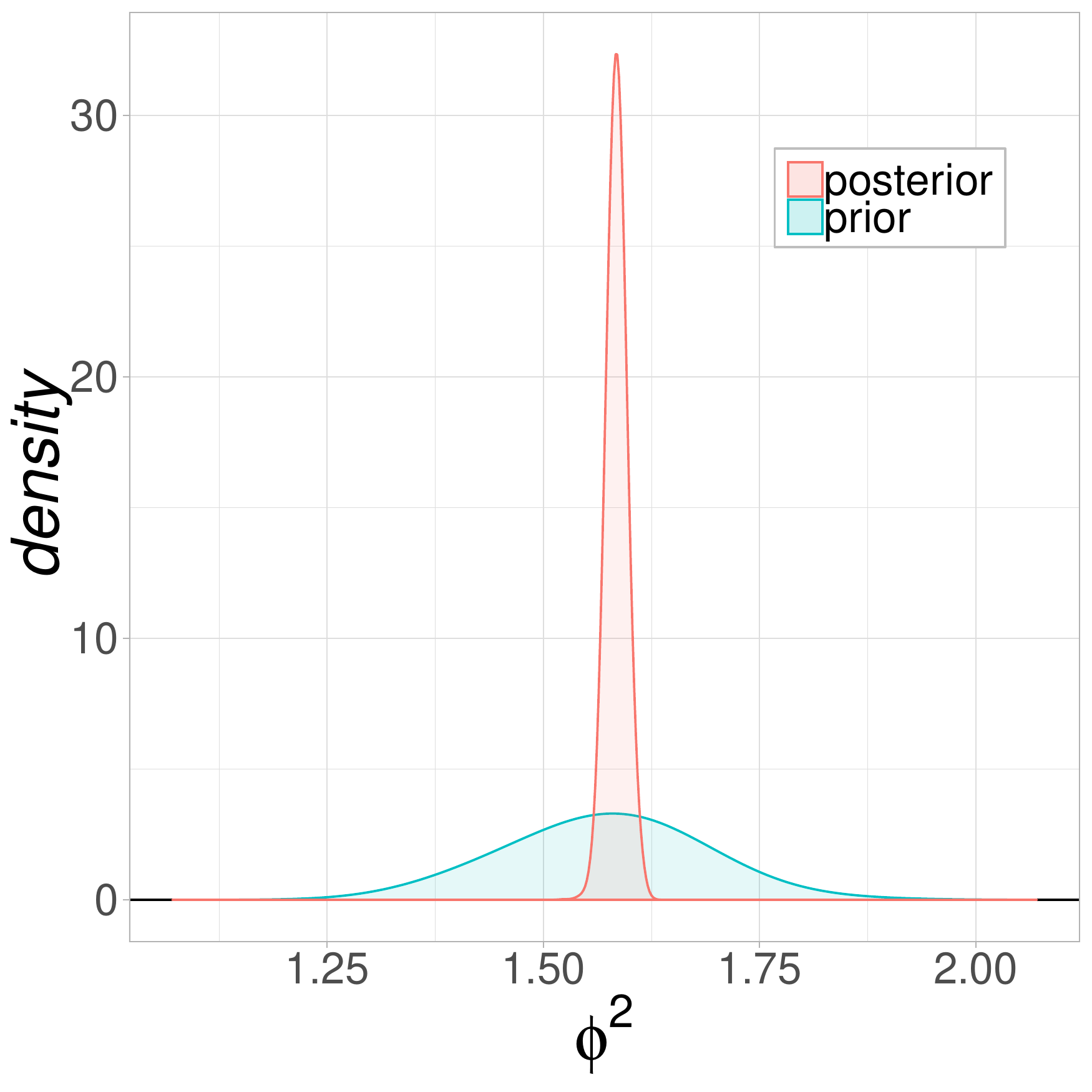} 

\end{knitrout}
    &
\begin{knitrout}
\definecolor{shadecolor}{rgb}{0.969, 0.969, 0.969}\color{fgcolor}
\includegraphics[width=0.21\linewidth,height=0.21\linewidth]{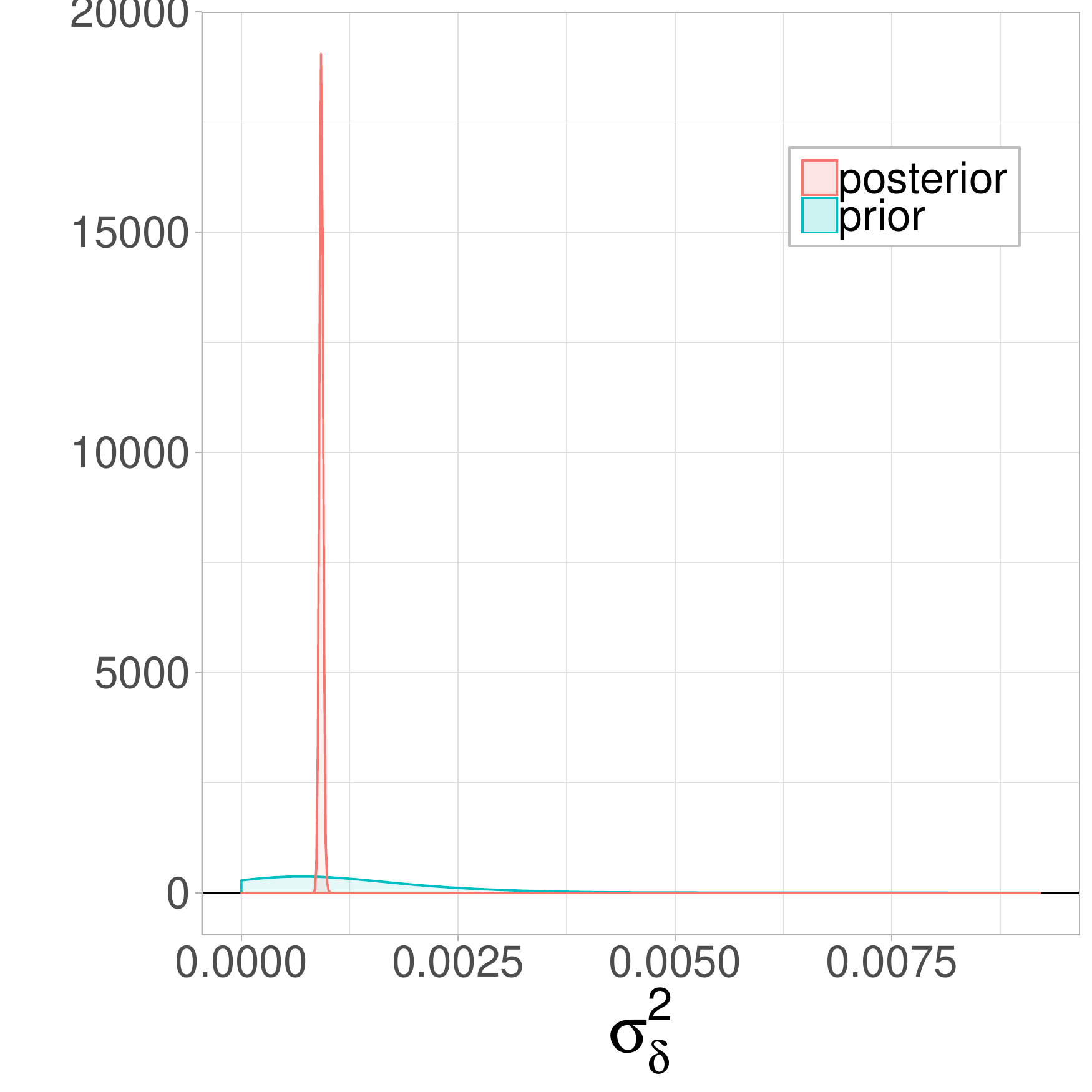} 

\end{knitrout}
&
\begin{knitrout}
\definecolor{shadecolor}{rgb}{0.969, 0.969, 0.969}\color{fgcolor}
\includegraphics[width=0.21\linewidth,height=0.21\linewidth]{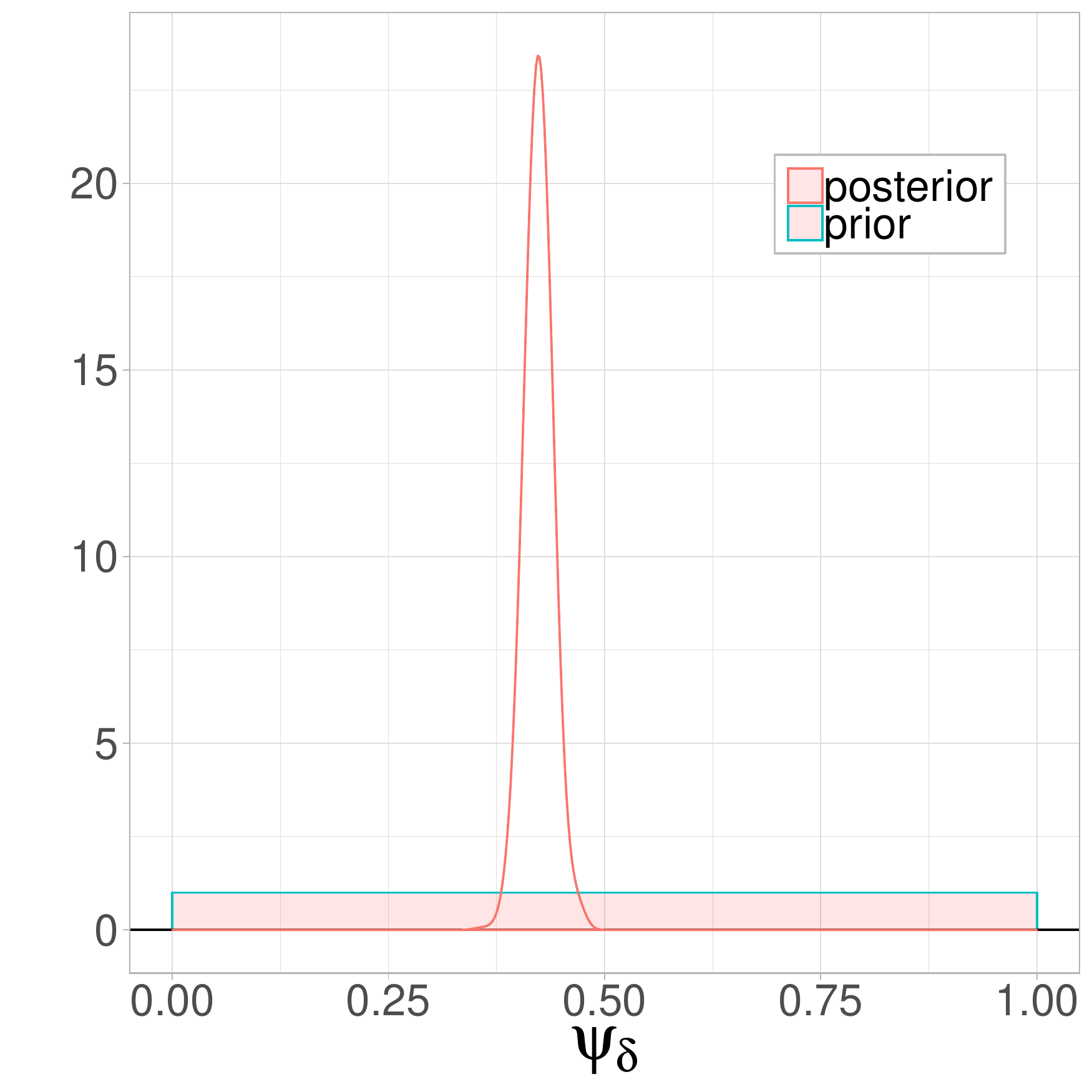} 

\end{knitrout}
    &
\begin{knitrout}
\definecolor{shadecolor}{rgb}{0.969, 0.969, 0.969}\color{fgcolor}
\includegraphics[width=0.21\linewidth,height=0.21\linewidth]{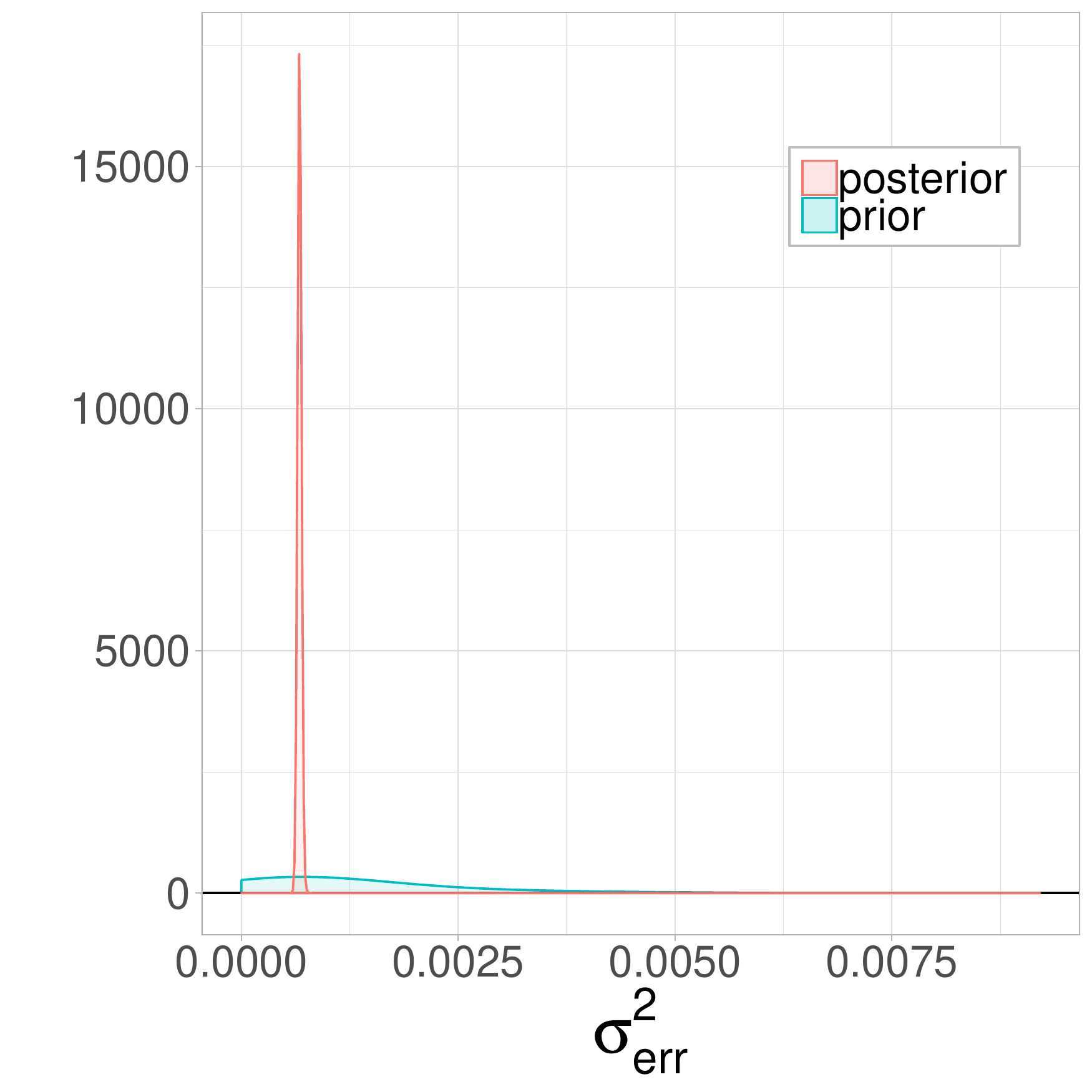} 

\end{knitrout}
    \\
  \end{tabular}
\caption{\textit{prior} and \textit{posterior} distributions for each parameter for calibration on $\mathcal{M}_4$}
\label{fig:PlotCalib4}
\end{center}
\end{figure}

Figure \ref{fig:PlotCalib4} illustrates the \textit{prior} and \textit{posterior} distributions resulted from calibration on $\mathcal{M}_4$. \newline

Similarly, if one desires to access the graph of the result on the quantity of interest one only needs to run \code{p\$out}.
\begin{figure}[h!]
\begin{center}
\begin{knitrout}
\definecolor{shadecolor}{rgb}{0.969, 0.969, 0.969}\color{fgcolor}\begin{kframe}
\begin{alltt}
\hlstd{p}\hlopt{$}\hlstd{out}
\end{alltt}
\end{kframe}
\end{knitrout}

\begin{knitrout}
\definecolor{shadecolor}{rgb}{0.969, 0.969, 0.969}\color{fgcolor}
\includegraphics[width=0.45\linewidth,height=0.45\linewidth]{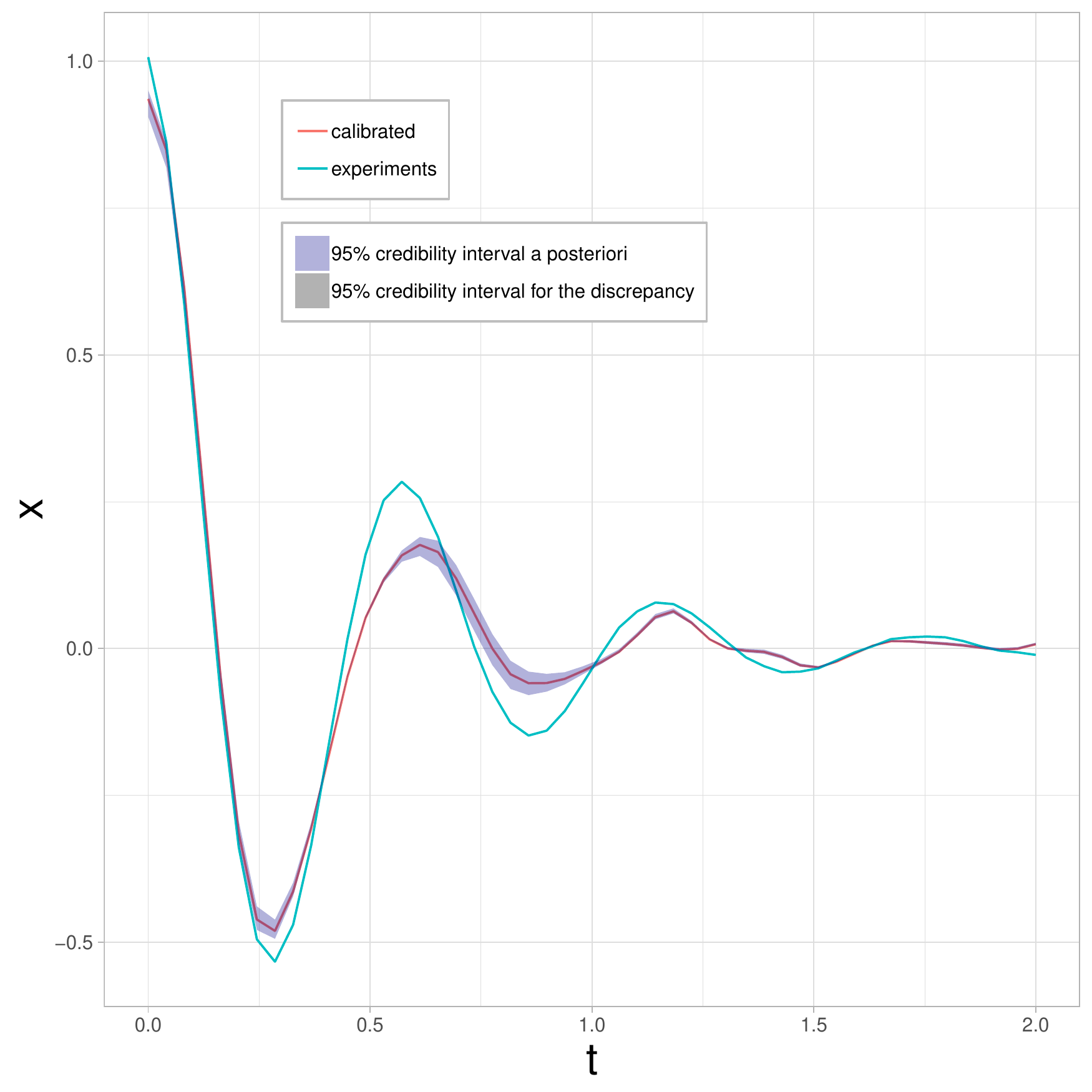} 

\end{knitrout}
\caption{Result of calibration on $\mathcal{M}_4$ for the quantity of interest with the credibility interval at $95\%$ \textit{a posteriori}}
\label{fig:PlotCalib42}
\end{center}
\end{figure}

\newpage

\subsection{Additionnal tools}

A function in \pkg{CaliCo} called \code{estimators} allows to access estimators as the MAP and the mean \textit{a posteriori}.
\begin{knitrout}
\definecolor{shadecolor}{rgb}{0.969, 0.969, 0.969}\color{fgcolor}\begin{kframe}
\begin{alltt}
\hlkwd{estimators}\hlstd{(mdfit1)}
\end{alltt}
\begin{verbatim}
## $MAP
## [1] 1.0105967765 0.2011318973 5.9741254796 0.0496559638 1.5042583733
## [6] 0.0002488385
## 
## $MEAN
## [1] 1.0107256606 0.2010363776 5.9755232004 0.0497000370 1.5042868046
## [6] 0.0002493378
\end{verbatim}
\end{kframe}
\end{knitrout}

If one is interested in running convergence diagnostics on the MCMC chains run by the function \code{calibrate}, one is free to increase the number of chains in the \code{opt.estim} options. This operation is realized in parallel with an automatically detected number of cores.

\begin{knitrout}
\definecolor{shadecolor}{rgb}{0.969, 0.969, 0.969}\color{fgcolor}\begin{kframe}
\begin{alltt}
\hlstd{opt.estim}\hlkwb{=}\hlkwd{list}\hlstd{(}\hlkwc{Ngibbs}\hlstd{=}\hlnum{1000}\hlstd{,}\hlkwc{Nmh}\hlstd{=}\hlnum{5000}\hlstd{,}\hlkwc{thetaInit}\hlstd{=}\hlkwd{c}\hlstd{(}\hlnum{1}\hlstd{,}\hlnum{0.25}\hlstd{,}\hlnum{6}\hlstd{,}\hlnum{50e-3}\hlstd{,pi}\hlopt{/}\hlnum{2}\hlstd{,}\hlnum{1e-3}\hlstd{),}
                \hlkwc{r}\hlstd{=}\hlkwd{c}\hlstd{(}\hlnum{0.05}\hlstd{,}\hlnum{0.05}\hlstd{),}\hlkwc{sig}\hlstd{=}\hlkwd{diag}\hlstd{(}\hlnum{6}\hlstd{),}\hlkwc{Nchains}\hlstd{=}\hlnum{3}\hlstd{,}\hlkwc{burnIn}\hlstd{=}\hlnum{2000}\hlstd{)}

\hlstd{mdfitMCMC} \hlkwb{<-} \hlkwd{calibrate}\hlstd{(model1,pr1,opt.estim)}
\end{alltt}
\end{kframe}
\end{knitrout}

By setting \code{Nchains=3}, calibration is run 3 times. The function \code{chain} allows to load the \pkg{coda} object generated and then to use \pkg{coda} tools as Gelman-Rubin diagnostics \citep{gelman1992} for example.

\begin{knitrout}
\definecolor{shadecolor}{rgb}{0.969, 0.969, 0.969}\color{fgcolor}\begin{kframe}
\begin{alltt}
\hlstd{mcmc} \hlkwb{<-} \hlkwd{chain}\hlstd{(mdfitMCMC)}
\hlkwd{library}\hlstd{(coda)}
\hlkwd{gelman.diag}\hlstd{(mcmc)}
\end{alltt}
\begin{verbatim}
## Potential scale reduction factors:
## 
##      Point est. Upper C.I.
## [1,]       4.00       7.81
## [2,]       1.55       2.39
## [3,]       8.58      16.65
## [4,]       1.25       1.68
## [5,]       3.52       7.23
## [6,]      38.66      75.48
## 
## Multivariate psrf
## 
## 31.3
\end{verbatim}
\end{kframe}
\end{knitrout}

The user can also run very easily a cross validation (a leave one out) to estimate how accurately the model prediction will perform in practice. An additional option, called \code{opt.valid}, is then necessary to run this cross validation. This option is a list containing the number of iteration (\code{nCV}) and the type cross validation method (\code{type.valid}).

\begin{knitrout}
\definecolor{shadecolor}{rgb}{0.969, 0.969, 0.969}\color{fgcolor}\begin{kframe}
\begin{alltt}
\hlstd{mdfitCV} \hlkwb{<-} \hlkwd{calibrate}\hlstd{(model1,pr1,}
                     \hlkwc{opt.estim} \hlstd{=} \hlkwd{list}\hlstd{(}\hlkwc{Ngibbs}\hlstd{=}\hlnum{1000}\hlstd{,}
                                      \hlkwc{Nmh}\hlstd{=}\hlnum{5000}\hlstd{,}
                                      \hlkwc{thetaInit}\hlstd{=}\hlkwd{c}\hlstd{(}\hlnum{1}\hlstd{,}\hlnum{0.25}\hlstd{,}\hlnum{6}\hlstd{,}\hlnum{50e-3}\hlstd{,pi}\hlopt{/}\hlnum{2}\hlstd{,}\hlnum{1e-3}\hlstd{),}
                                      \hlkwc{r}\hlstd{=}\hlkwd{c}\hlstd{(}\hlnum{0.05}\hlstd{,}\hlnum{0.05}\hlstd{),}
                                      \hlkwc{sig}\hlstd{=}\hlkwd{diag}\hlstd{(}\hlnum{6}\hlstd{),}
                                      \hlkwc{Nchains}\hlstd{=}\hlnum{1}\hlstd{,}
                                      \hlkwc{burnIn}\hlstd{=}\hlnum{2000}\hlstd{),}
                     \hlkwc{opt.valid} \hlstd{=} \hlkwd{list}\hlstd{(}\hlkwc{type.valid}\hlstd{=}\hlstr{"loo"}\hlstd{,}
                                      \hlkwc{nCV}\hlstd{=}\hlnum{50}\hlstd{))}
\end{alltt}
\end{kframe}
\end{knitrout}

The activation of the cross validation will run the regular calibration and then the \code{nCV} iterations requested by the user. To decrease the computational burden of such operation, a parallel operation is realized by to the package \pkg{parallel} present in \code{R core}.

\begin{knitrout}
\definecolor{shadecolor}{rgb}{0.969, 0.969, 0.969}\color{fgcolor}\begin{kframe}
\begin{alltt}
\hlkwd{print}\hlstd{(mdfitCV)}
\end{alltt}
\begin{verbatim}
## Call:
## 
## With the function:
## function(t,theta)
## {
##   w0 <- sqrt(theta[3]/theta[4])
##   return(theta[1]*exp(-theta[2]*w0*t)*sin(sqrt(1-theta[2]^2)*w0*t+theta[5]))
## }
## <bytecode: 0x561a93b33348>
## 
## Selected model : model1 
## 
## Acceptation rate of the Metropolis within Gibbs algorithm:
## [1] "43.8%" "27.5%" "27.3%" "9.1%"  "4.3%"  "4.4%" 
## 
## Acceptation rate of the Metropolis Hastings algorithm:
## [1] "48.56%"
## 
## Maximum a posteriori:
## [1] 1.0144067620 0.2032408379 6.0077799738 0.0508023686 1.5169972084
## [6] 0.0002536493
## 
## Mean a posteriori:
## [1] 1.0121094549 0.2016018588 5.9975259667 0.0499179223 1.5060238122
## [6] 0.0002488799
## 
## 
## Cross validation:
##  Method: loo 
##     Predicted       Real        Error
## 1 0.581834516 0.58394961 0.0021150930
## 2 0.251110861 0.25245428 0.0013434211
## 3 0.860725971 0.86208615 0.0013601740
## 4 0.860627965 0.86208615 0.0014581796
## 5 0.009987004 0.00981318 0.0001738244
## 6 0.251052820 0.25245428 0.0014014613
## 
## RMSE: [1] 0.03737526
## 
## Cover rate: 
## [1] "94%"
\end{verbatim}
\end{kframe}
\end{knitrout}

The \code{print} method displays the head of the first iterations of the cross validation and the root mean square error (RMSE) associated. The coverage rate is also printed to check the accuracy of the \posterior credibility interval. \newline

The implemented function \code{sequentialDesign} is available only for $\mathcal{M}_2$ and $\mathcal{M}_4$. This function allows to run a sequential design as described in \citet{damblin2018}. Based on the expected improvement \citep{jones1998efficient}, it improves the estimation of the Gaussian process that emulates the code by adding new points in the design. Calibration quality is, as expected, increased.

\begin{knitrout}
\definecolor{shadecolor}{rgb}{0.969, 0.969, 0.969}\color{fgcolor}\begin{kframe}
\begin{alltt}
\hlstd{binf} \hlkwb{<-} \hlkwd{c}\hlstd{(}\hlnum{0.9}\hlstd{,}\hlnum{0.05}\hlstd{,}\hlnum{5.8}\hlstd{,}\hlnum{40e-3}\hlstd{,}\hlnum{1.49}\hlstd{)}
\hlstd{bsup} \hlkwb{<-} \hlkwd{c}\hlstd{(}\hlnum{1.1}\hlstd{,}\hlnum{0.55}\hlstd{,}\hlnum{6.2}\hlstd{,}\hlnum{60e-3}\hlstd{,}\hlnum{1.6}\hlstd{)}

\hlstd{model2} \hlkwb{<-} \hlkwd{model}\hlstd{(code,t,Yexp,}\hlstr{"model2"}\hlstd{,}
                \hlkwc{opt.gp} \hlstd{=} \hlkwd{list}\hlstd{(}\hlkwc{type}\hlstd{=}\hlstr{"matern5_2"}\hlstd{,}\hlkwc{DOE}\hlstd{=}\hlkwa{NULL}\hlstd{),}
                \hlkwc{opt.emul} \hlstd{=} \hlkwd{list}\hlstd{(}\hlkwc{p}\hlstd{=}\hlnum{5}\hlstd{,}\hlkwc{n.emul}\hlstd{=}\hlnum{200}\hlstd{,}\hlkwc{binf}\hlstd{=binf,}\hlkwc{bsup}\hlstd{=bsup))}

\hlstd{type.prior} \hlkwb{<-} \hlkwd{c}\hlstd{(}\hlkwd{rep}\hlstd{(}\hlstr{"gaussian"}\hlstd{,}\hlnum{5}\hlstd{),}\hlstr{"gamma"}\hlstd{)}
\hlstd{opt.prior} \hlkwb{<-} \hlkwd{list}\hlstd{(}\hlkwd{c}\hlstd{(}\hlnum{1}\hlstd{,}\hlnum{1e-3}\hlstd{),}\hlkwd{c}\hlstd{(}\hlnum{0.3}\hlstd{,}\hlnum{1e-3}\hlstd{),}\hlkwd{c}\hlstd{(}\hlnum{6}\hlstd{,}\hlnum{1e-3}\hlstd{),}\hlkwd{c}\hlstd{(}\hlnum{50e-3}\hlstd{,}\hlnum{1e-5}\hlstd{),}
                  \hlkwd{c}\hlstd{(pi}\hlopt{/}\hlnum{2}\hlstd{,}\hlnum{1e-2}\hlstd{),}\hlkwd{c}\hlstd{(}\hlnum{1}\hlstd{,}\hlnum{1e-3}\hlstd{))}
\hlstd{pr1} \hlkwb{<-} \hlkwd{prior}\hlstd{(type.prior,opt.prior)}

\hlstd{newModel2} \hlkwb{<-} \hlkwd{sequentialDesign}\hlstd{(model2,pr1,}
                    \hlkwc{opt.estim} \hlstd{=} \hlkwd{list}\hlstd{(}\hlkwc{Ngibbs}\hlstd{=}\hlnum{100}\hlstd{,}
                                     \hlkwc{Nmh}\hlstd{=}\hlnum{600}\hlstd{,}
                                     \hlkwc{thetaInit}\hlstd{=}\hlkwd{c}\hlstd{(}\hlnum{1}\hlstd{,}\hlnum{0.25}\hlstd{,}\hlnum{6}\hlstd{,}\hlnum{50e-3}\hlstd{,pi}\hlopt{/}\hlnum{2}\hlstd{,}\hlnum{1e-3}\hlstd{),}
                                     \hlkwc{r}\hlstd{=}\hlkwd{c}\hlstd{(}\hlnum{0.05}\hlstd{,}\hlnum{0.05}\hlstd{),}
                                     \hlkwc{sig}\hlstd{=}\hlkwd{diag}\hlstd{(}\hlnum{6}\hlstd{),}
                                     \hlkwc{Nchains}\hlstd{=}\hlnum{1}\hlstd{,}
                                     \hlkwc{burnIn}\hlstd{=}\hlnum{200}\hlstd{),}
                    \hlkwc{k}\hlstd{=}\hlnum{20}\hlstd{)}
\end{alltt}
\end{kframe}
\end{knitrout}

\begin{figure}[h!]
\begin{center}
   \begin{tabular}{ccccc}
\begin{knitrout}
\definecolor{shadecolor}{rgb}{0.969, 0.969, 0.969}\color{fgcolor}
\includegraphics[width=0.15\linewidth,height=0.15\linewidth]{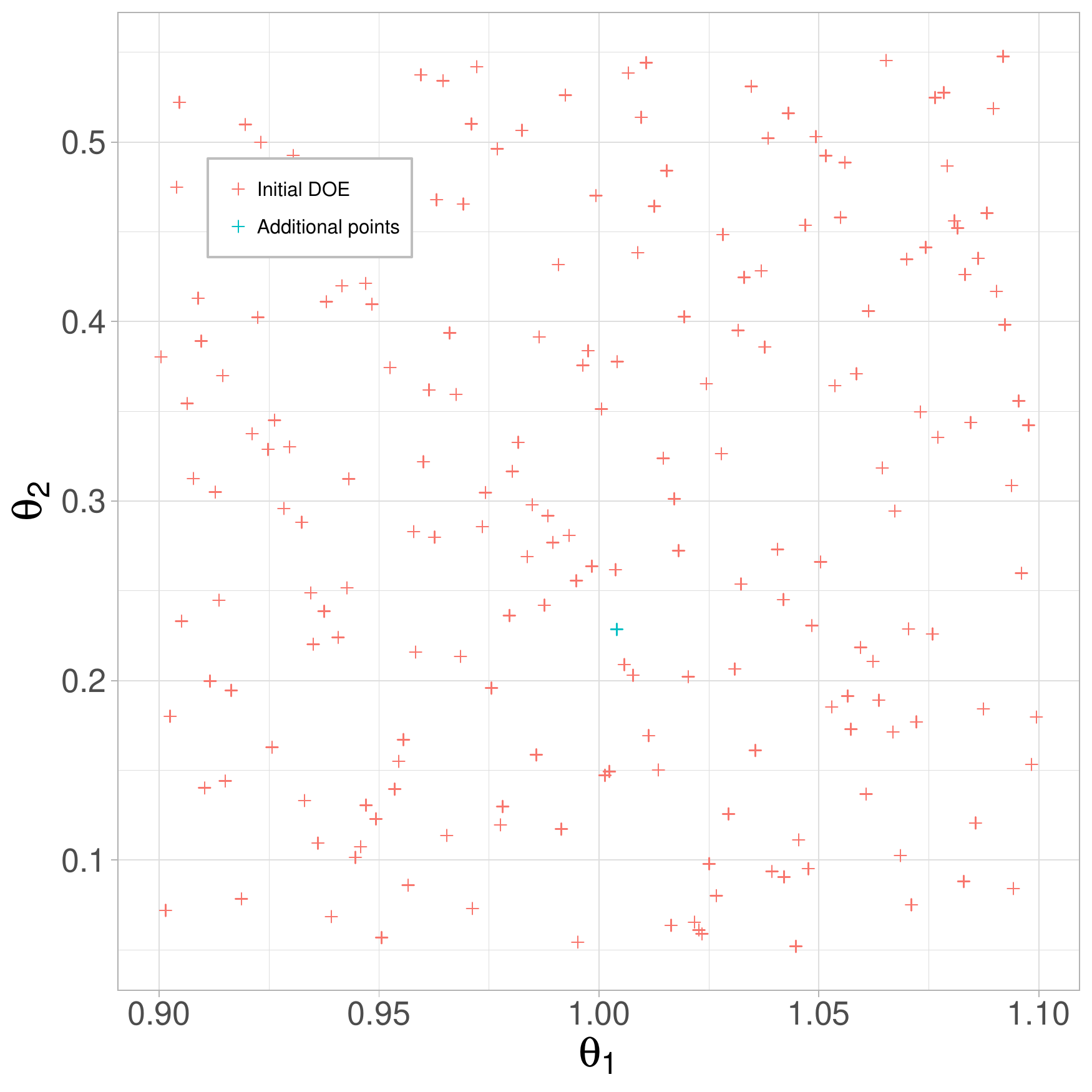} 

\end{knitrout}
&
\begin{knitrout}
\definecolor{shadecolor}{rgb}{0.969, 0.969, 0.969}\color{fgcolor}
\includegraphics[width=0.15\linewidth,height=0.15\linewidth]{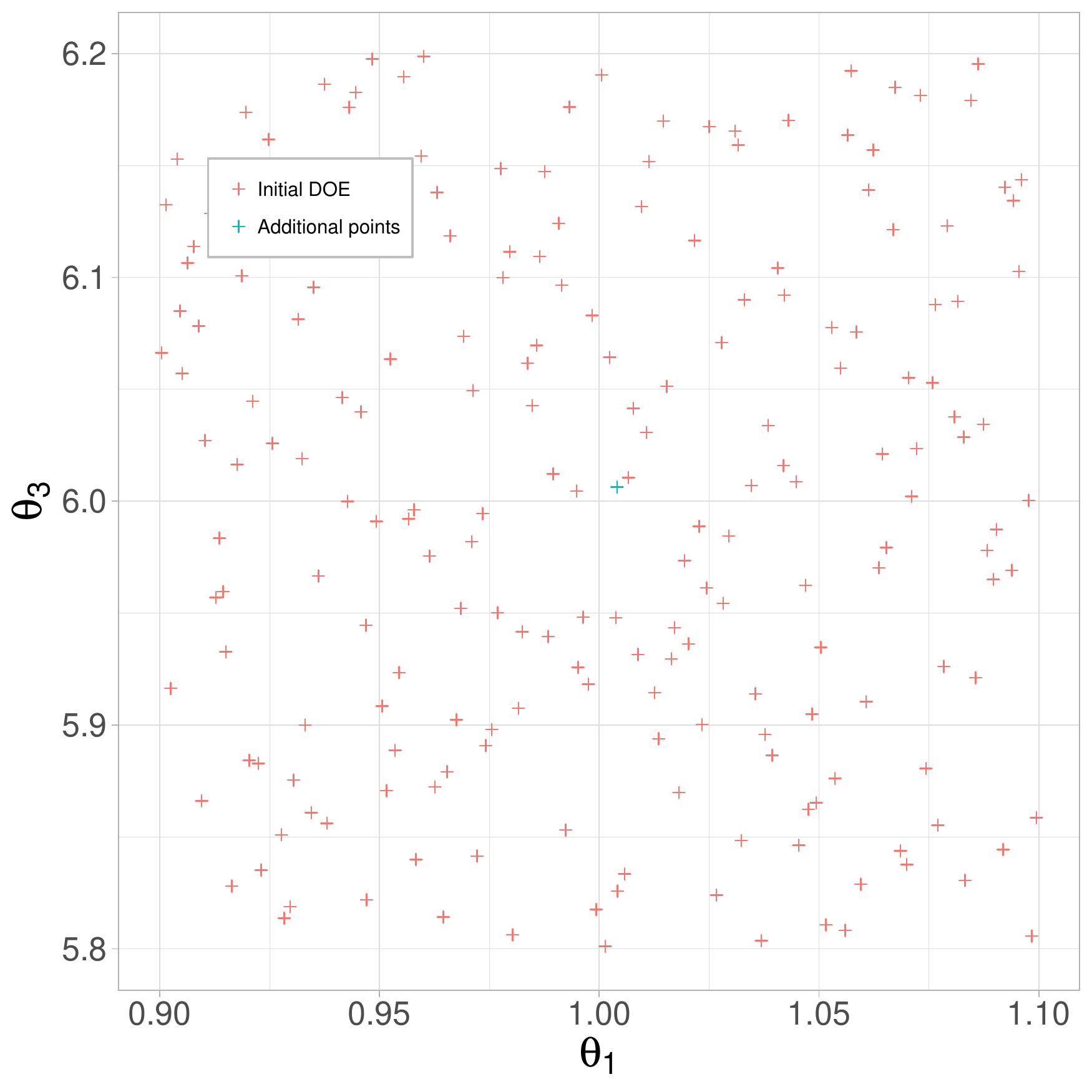} 

\end{knitrout}
&
\begin{knitrout}
\definecolor{shadecolor}{rgb}{0.969, 0.969, 0.969}\color{fgcolor}
\includegraphics[width=0.15\linewidth,height=0.15\linewidth]{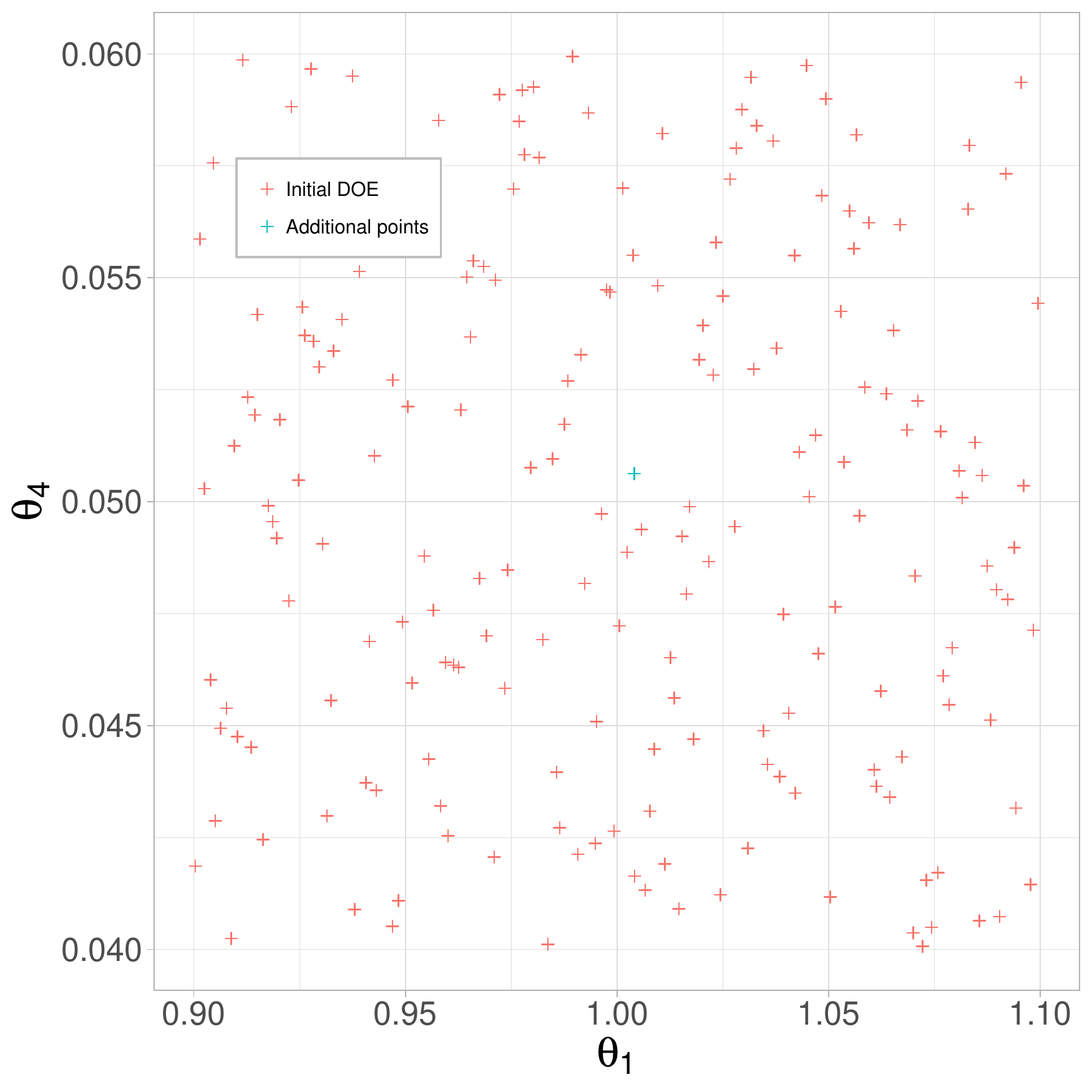} 

\end{knitrout}
&
\begin{knitrout}
\definecolor{shadecolor}{rgb}{0.969, 0.969, 0.969}\color{fgcolor}
\includegraphics[width=0.15\linewidth,height=0.15\linewidth]{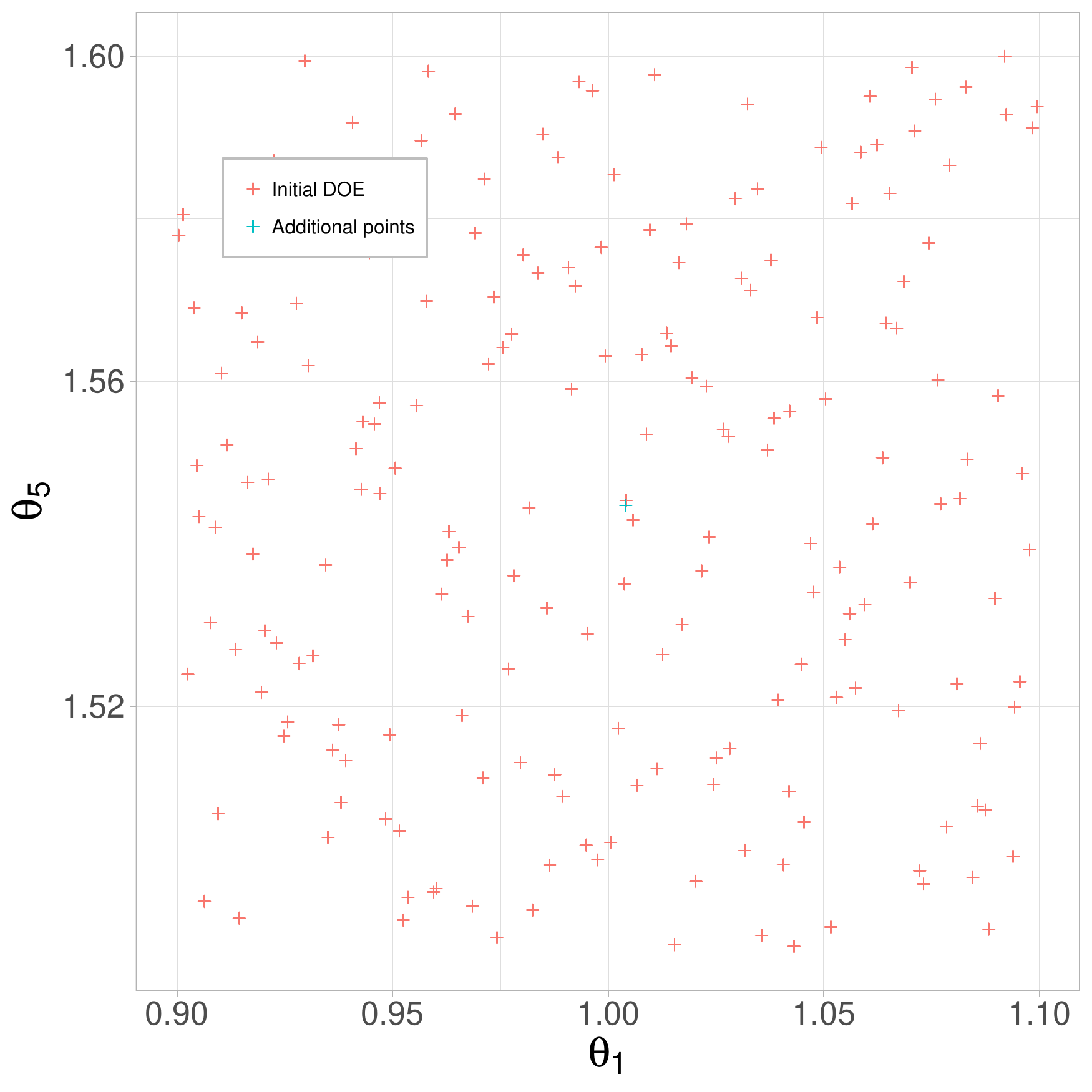} 

\end{knitrout}
&
\begin{knitrout}
\definecolor{shadecolor}{rgb}{0.969, 0.969, 0.969}\color{fgcolor}
\includegraphics[width=0.15\linewidth,height=0.15\linewidth]{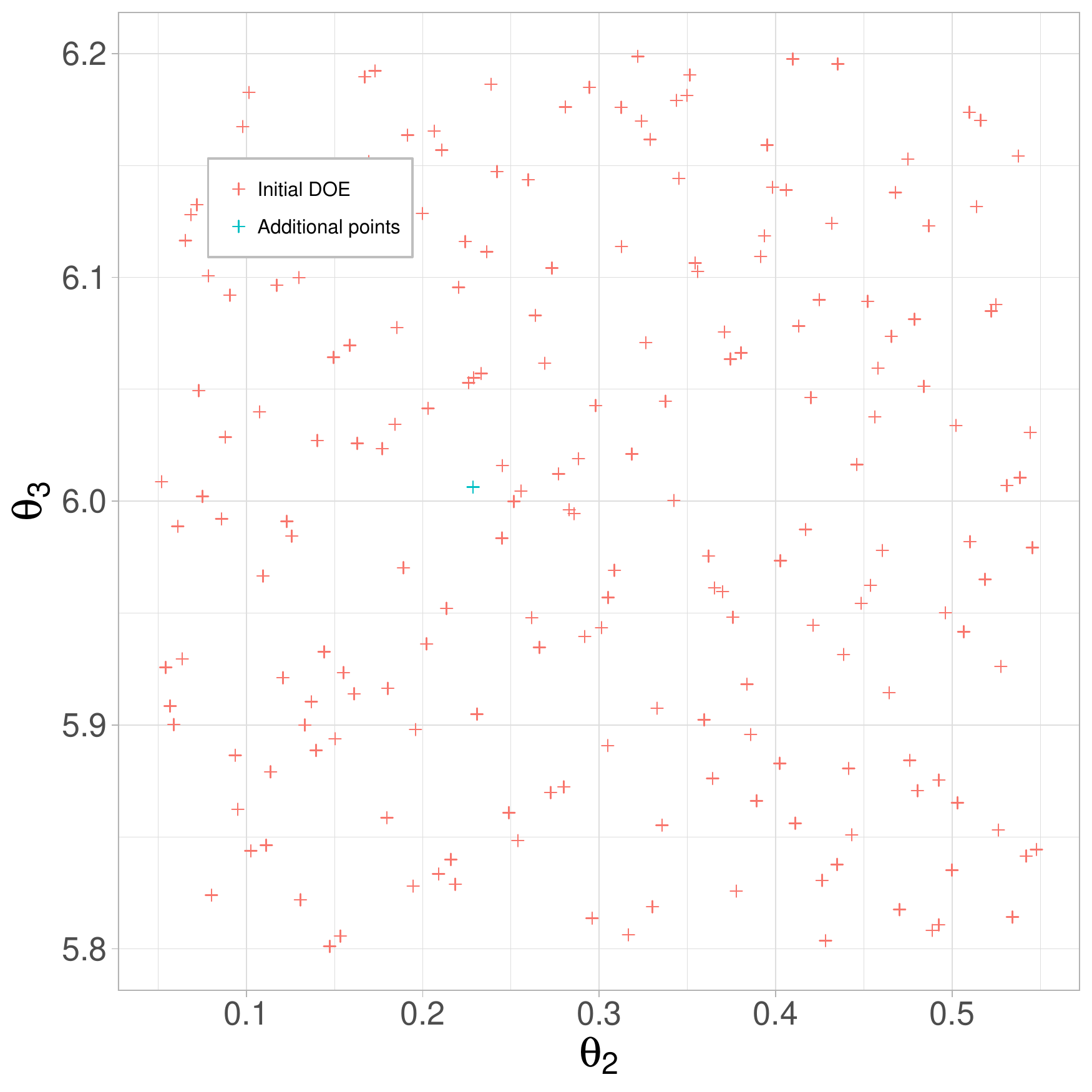} 

\end{knitrout}
\\
\begin{knitrout}
\definecolor{shadecolor}{rgb}{0.969, 0.969, 0.969}\color{fgcolor}
\includegraphics[width=0.15\linewidth,height=0.15\linewidth]{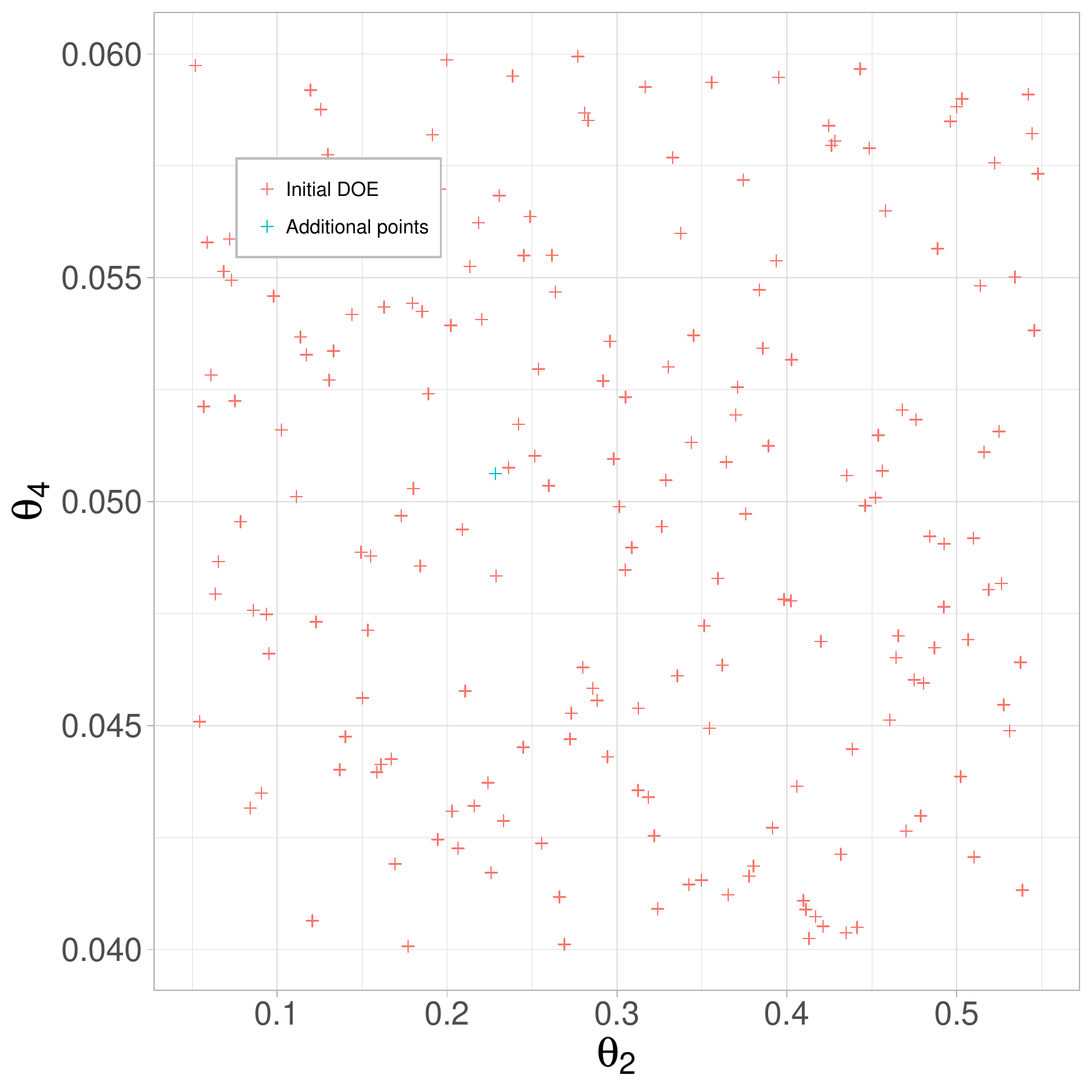} 

\end{knitrout}
&
\begin{knitrout}
\definecolor{shadecolor}{rgb}{0.969, 0.969, 0.969}\color{fgcolor}
\includegraphics[width=0.15\linewidth,height=0.15\linewidth]{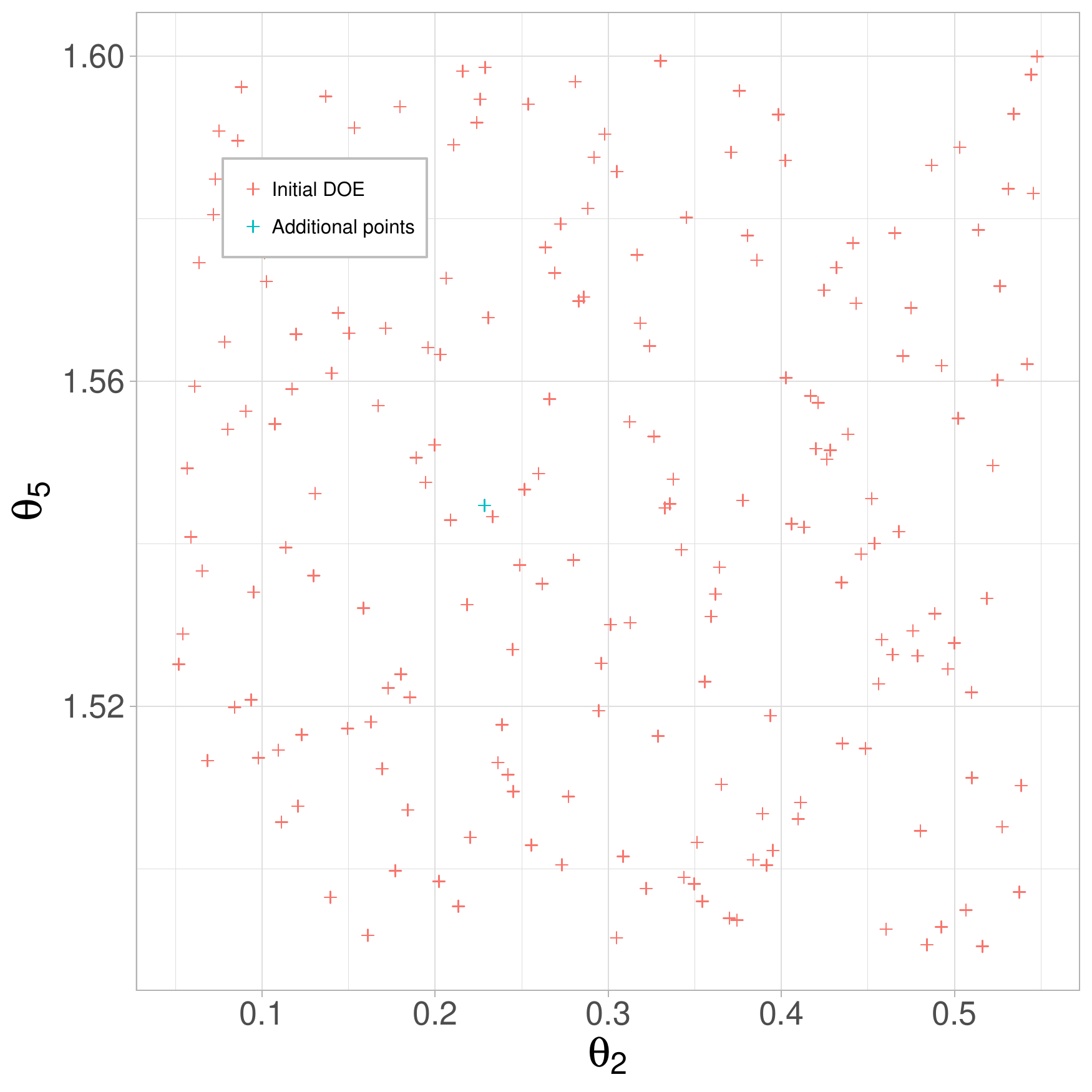} 

\end{knitrout}
&
\begin{knitrout}
\definecolor{shadecolor}{rgb}{0.969, 0.969, 0.969}\color{fgcolor}
\includegraphics[width=0.15\linewidth,height=0.15\linewidth]{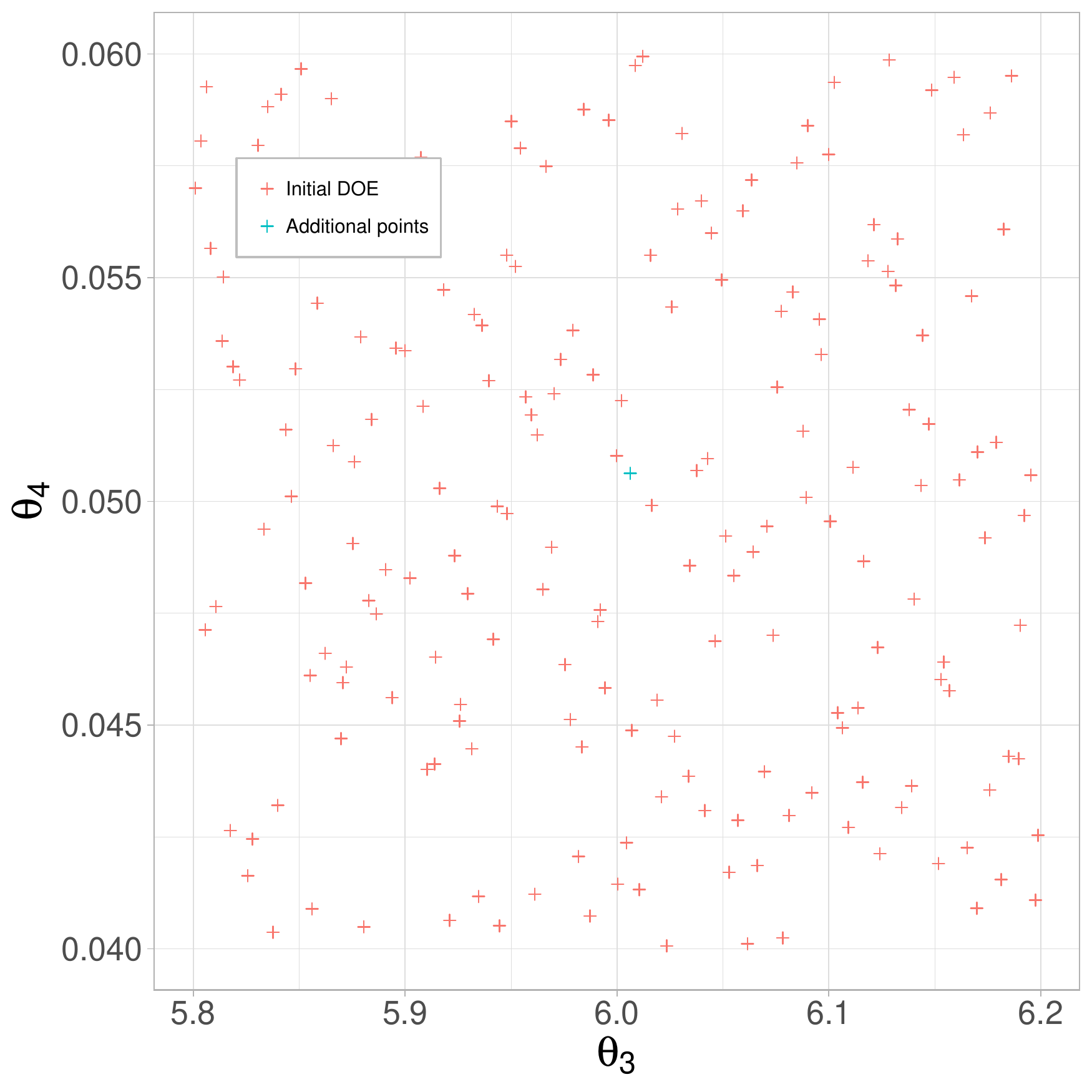} 

\end{knitrout}
&
\begin{knitrout}
\definecolor{shadecolor}{rgb}{0.969, 0.969, 0.969}\color{fgcolor}
\includegraphics[width=0.15\linewidth,height=0.15\linewidth]{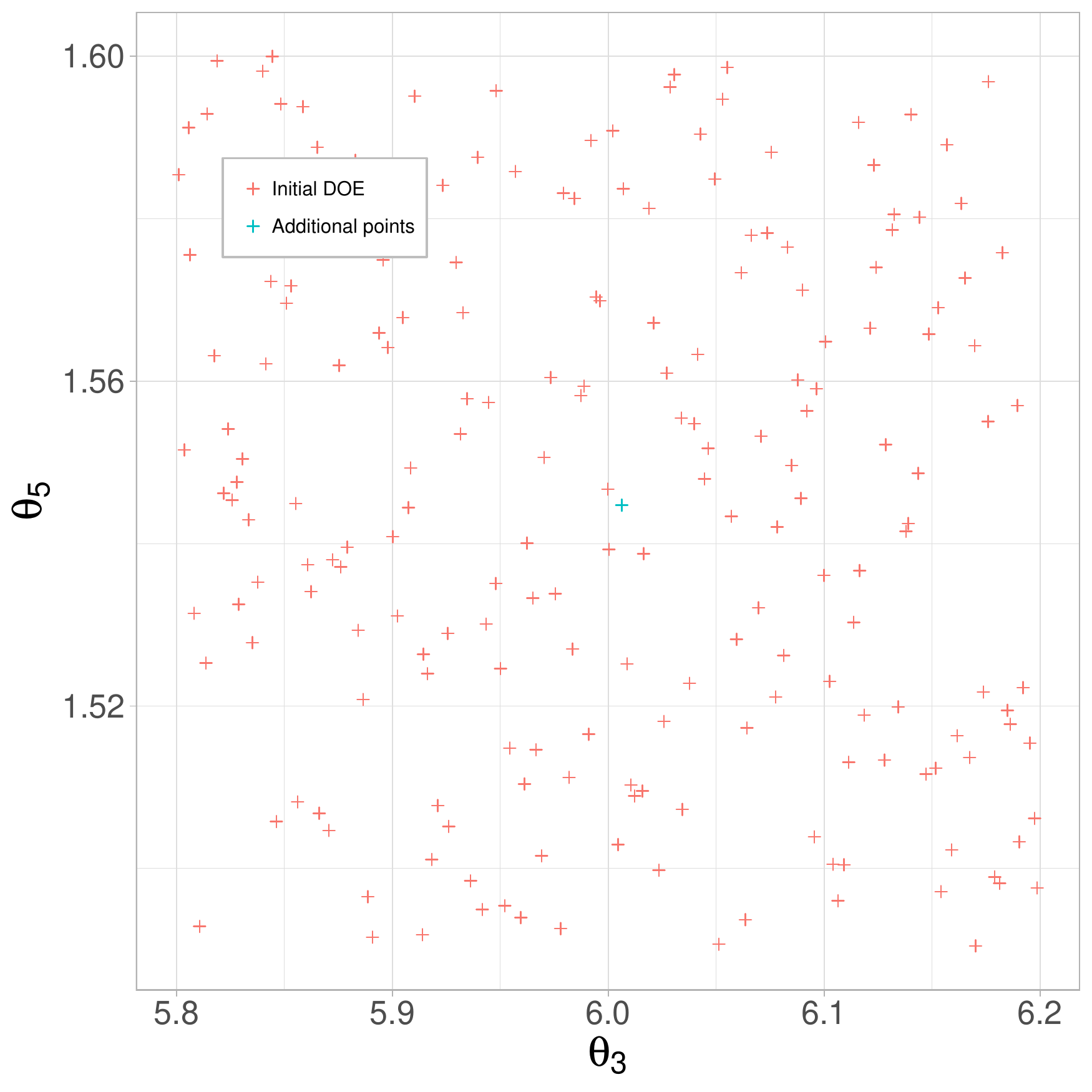} 

\end{knitrout}
&
\begin{knitrout}
\definecolor{shadecolor}{rgb}{0.969, 0.969, 0.969}\color{fgcolor}
\includegraphics[width=0.15\linewidth,height=0.15\linewidth]{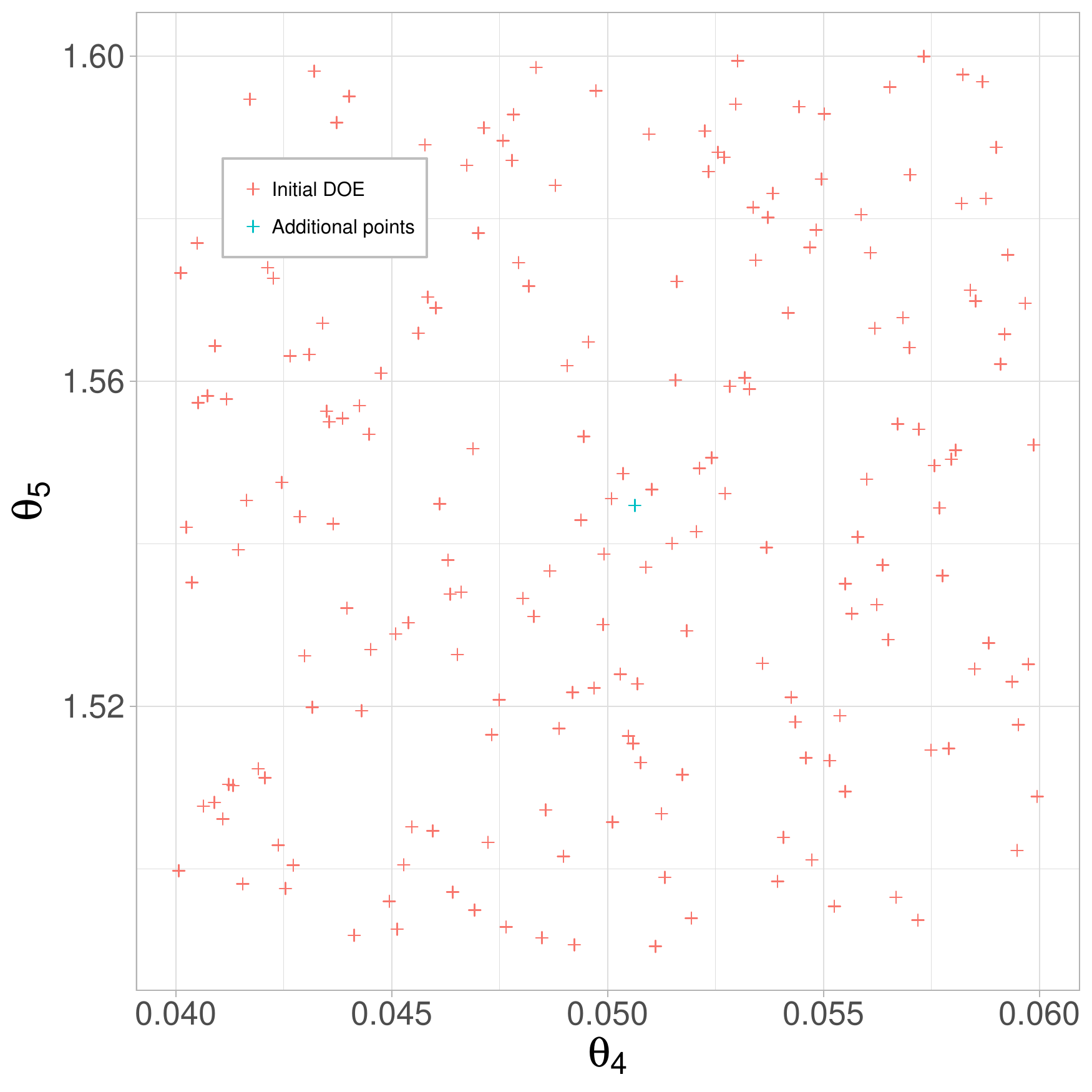} 

\end{knitrout}

% <<sequential23, out.width='0.15\\linewidth', out.height='0.15\\linewidth',echo=FALSE,cache=FALSE>>=
% p$doe[[11]]+tt
% @
% &
% <<sequential24, out.width='0.15\\linewidth', out.height='0.15\\linewidth',echo=FALSE,cache=FALSE>>=
% p$doe[[12]]+tt
% @
% &
% <<sequential25, out.width='0.15\\linewidth', out.height='0.15\\linewidth',echo=FALSE,cache=FALSE>>=
% p$doe[[13]]+tt
% @
% &
% <<sequential26, out.width='0.15\\linewidth', out.height='0.15\\linewidth',echo=FALSE,cache=FALSE>>=
% p$doe[[14]]+tt
% @
% &
% <<sequential27, out.width='0.15\\linewidth', out.height='0.15\\linewidth',echo=FALSE,cache=FALSE>>=
% p$doe[[15]]+tt
% @
% \\
\end{tabular}
\\
\begin{tabular}{cccccc}
    \multicolumn{6}{c}{Before sequential design}\\
\begin{knitrout}
\definecolor{shadecolor}{rgb}{0.969, 0.969, 0.969}\color{fgcolor}
\includegraphics[width=0.128\linewidth,height=0.128\linewidth]{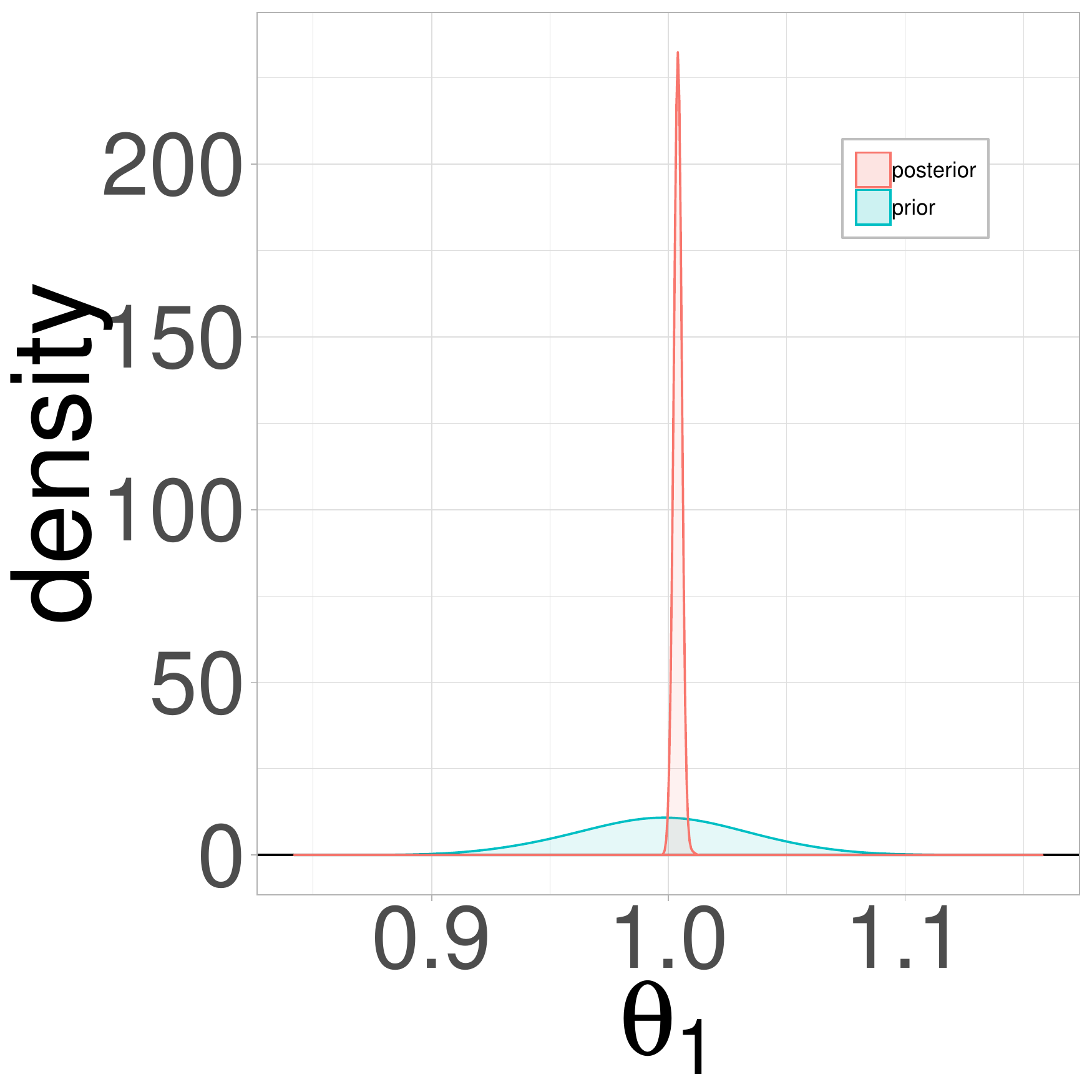} 

\end{knitrout}
&
\begin{knitrout}
\definecolor{shadecolor}{rgb}{0.969, 0.969, 0.969}\color{fgcolor}
\includegraphics[width=0.128\linewidth,height=0.128\linewidth]{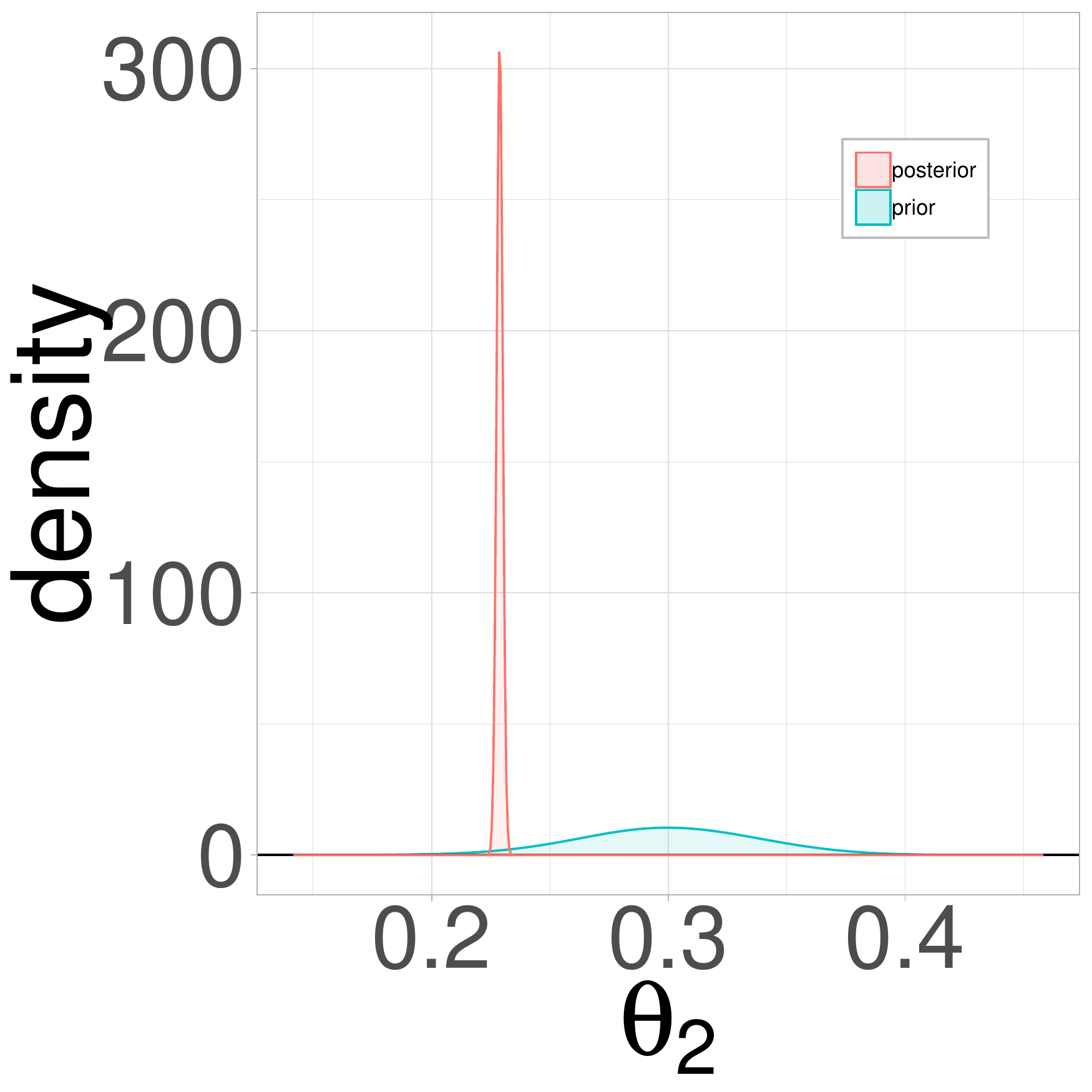} 

\end{knitrout}
&
\begin{knitrout}
\definecolor{shadecolor}{rgb}{0.969, 0.969, 0.969}\color{fgcolor}
\includegraphics[width=0.128\linewidth,height=0.128\linewidth]{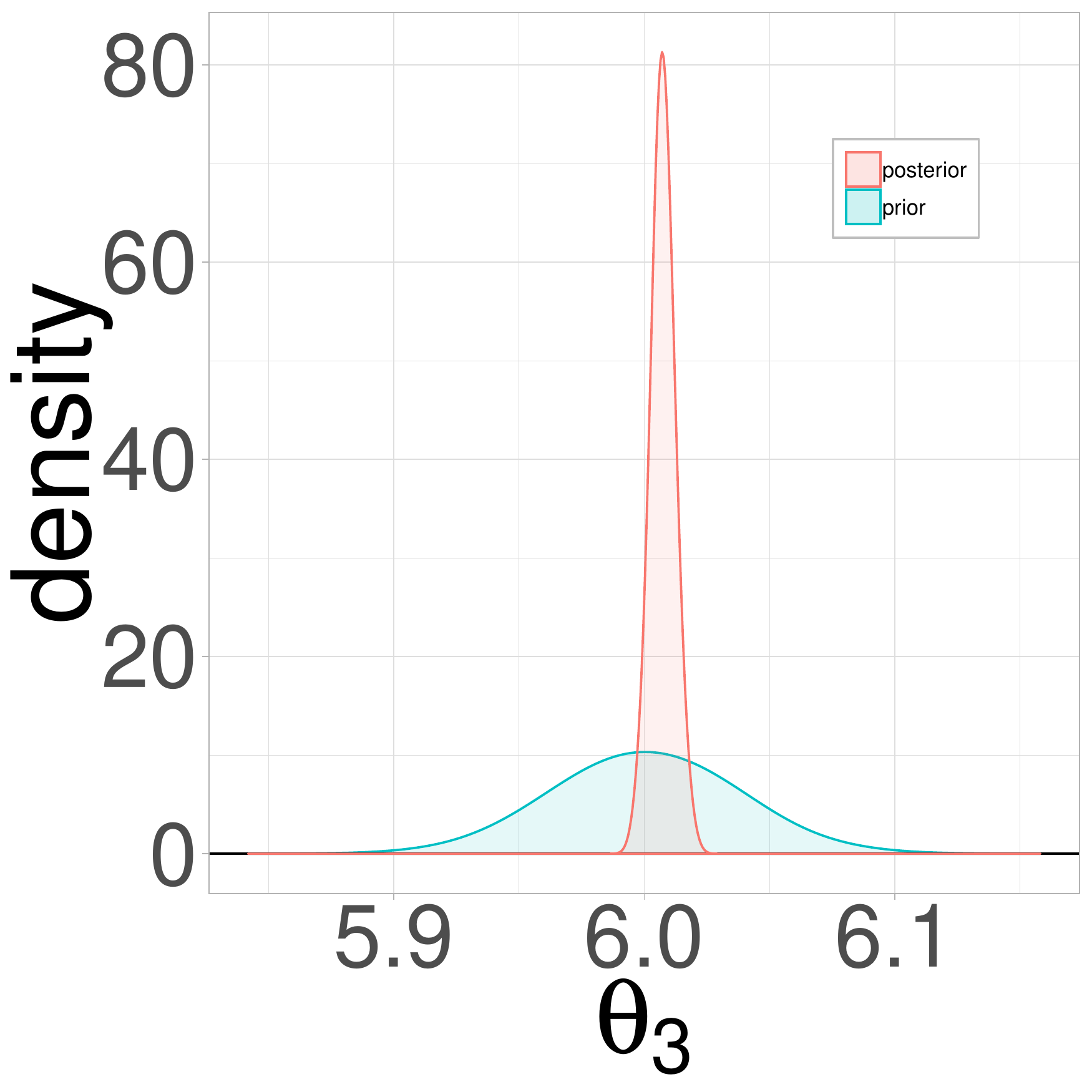} 

\end{knitrout}
&
\begin{knitrout}
\definecolor{shadecolor}{rgb}{0.969, 0.969, 0.969}\color{fgcolor}
\includegraphics[width=0.128\linewidth,height=0.128\linewidth]{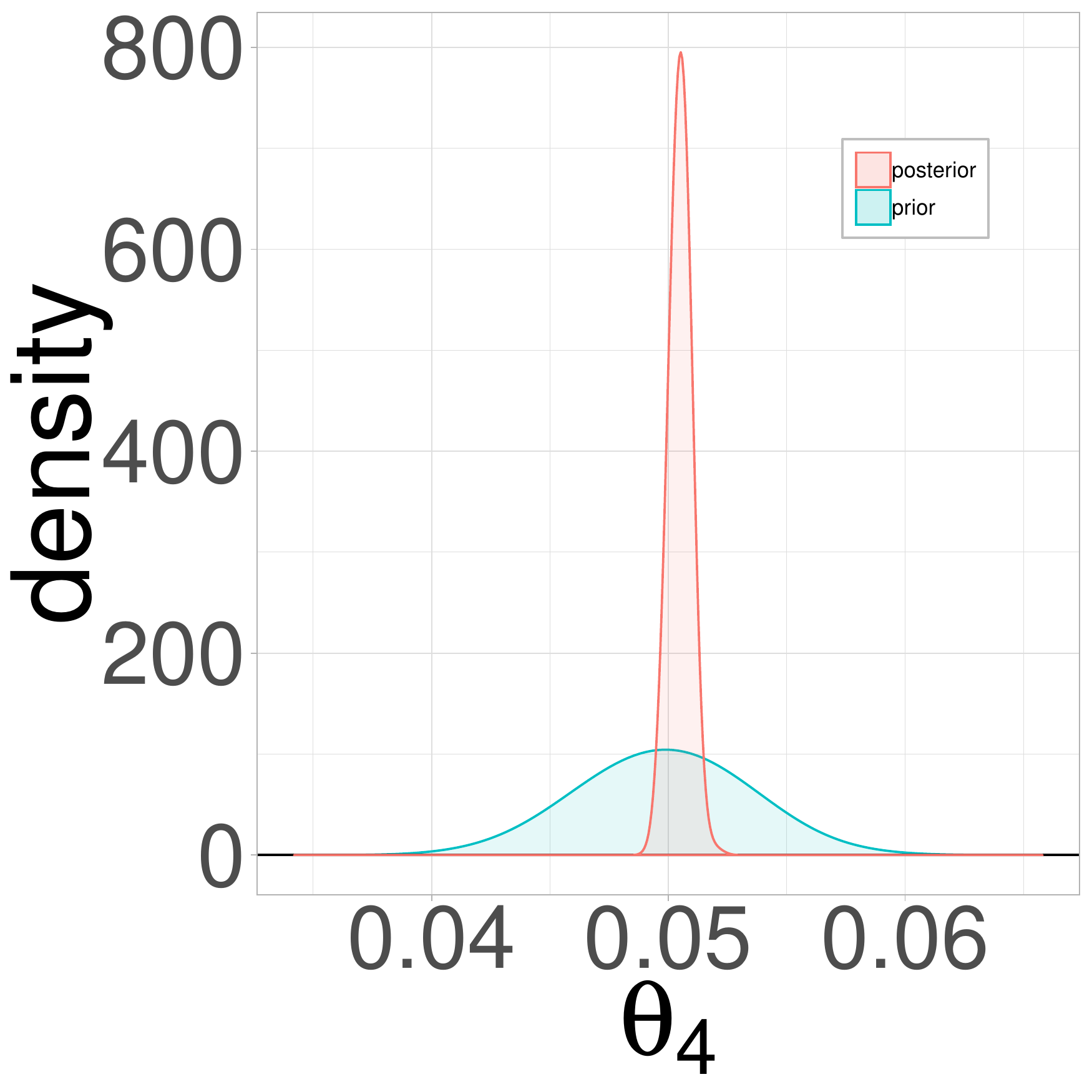} 

\end{knitrout}
&
\begin{knitrout}
\definecolor{shadecolor}{rgb}{0.969, 0.969, 0.969}\color{fgcolor}
\includegraphics[width=0.128\linewidth,height=0.128\linewidth]{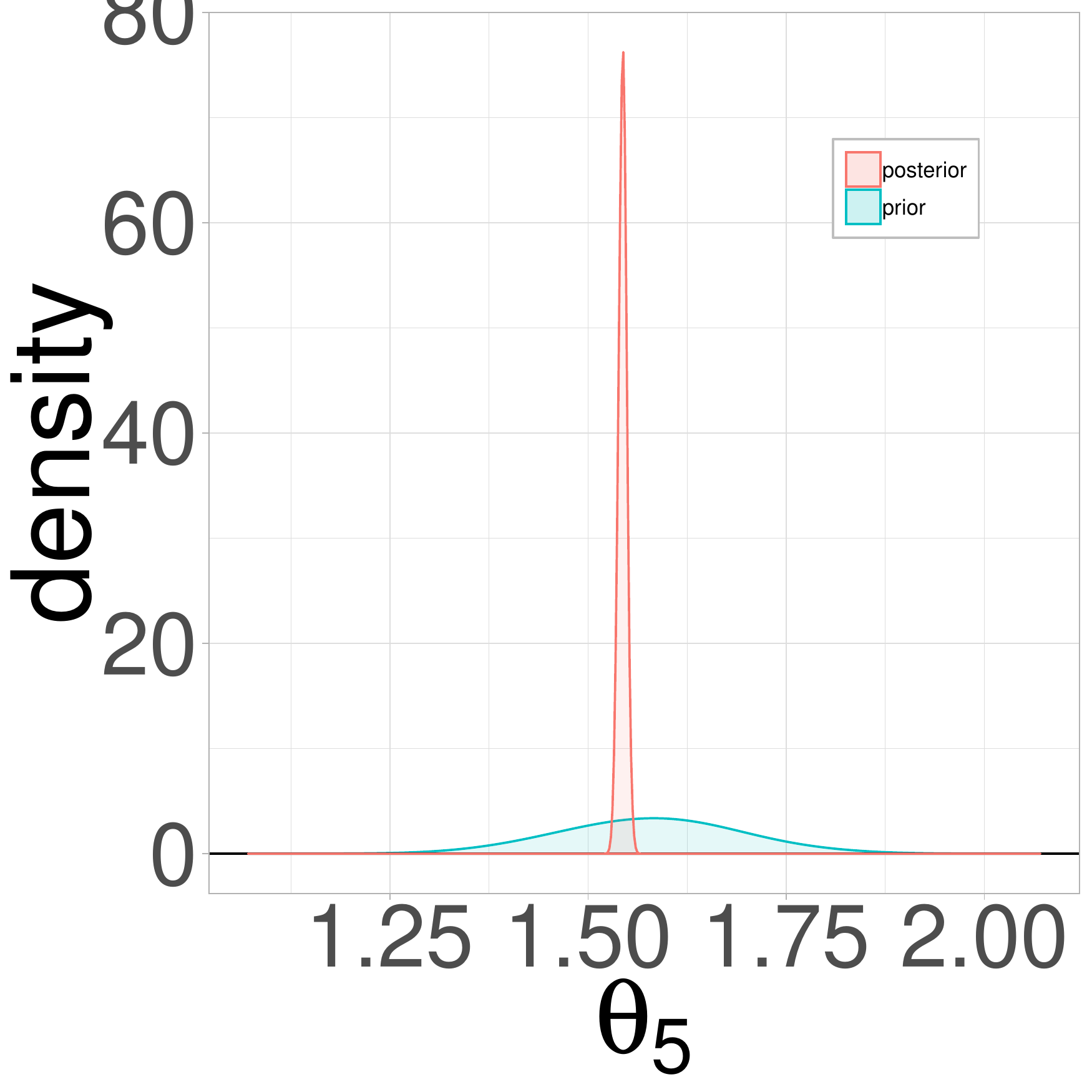} 

\end{knitrout}
&
\begin{knitrout}
\definecolor{shadecolor}{rgb}{0.969, 0.969, 0.969}\color{fgcolor}
\includegraphics[width=0.128\linewidth,height=0.128\linewidth]{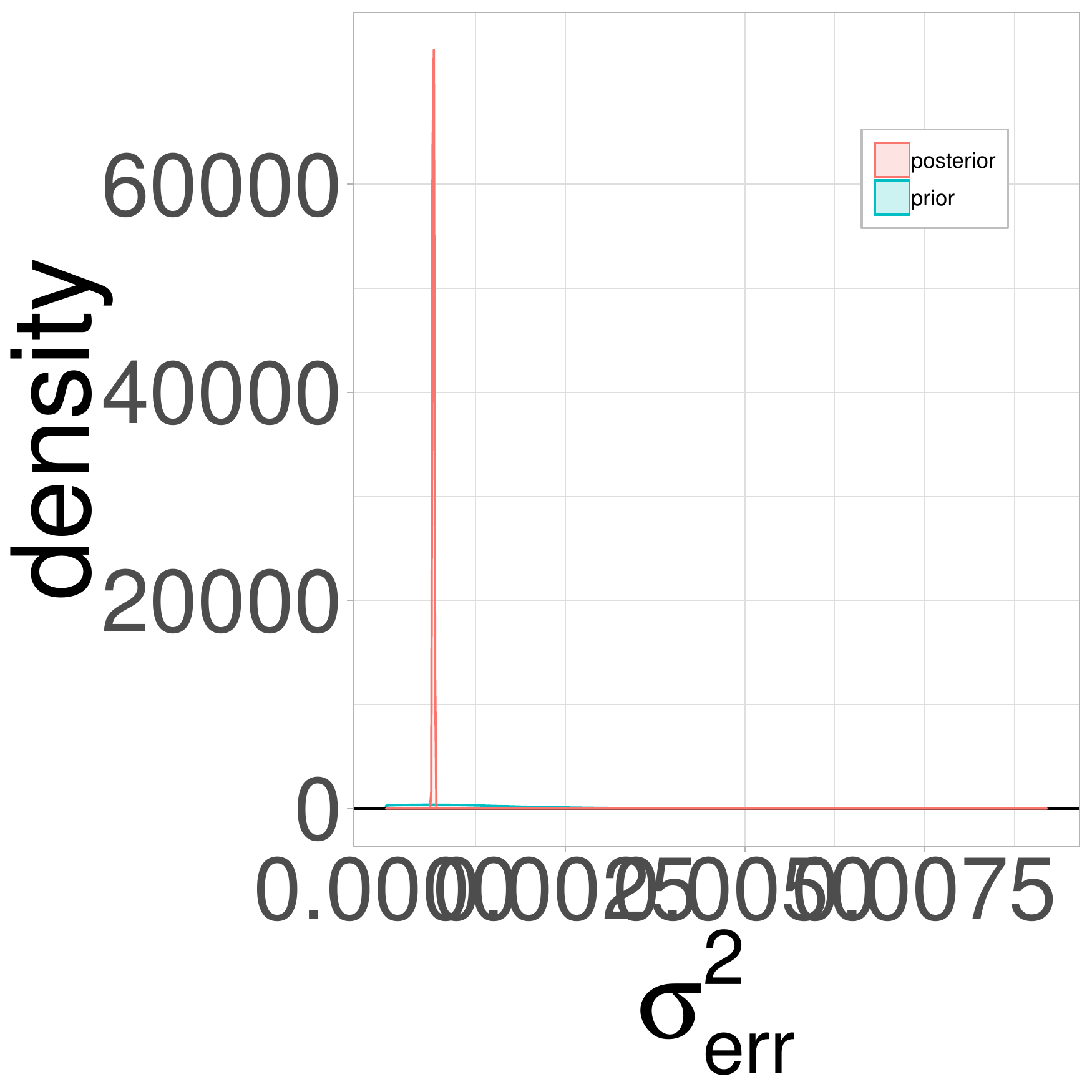} 

\end{knitrout}
\\
\multicolumn{6}{c}{After sequential design}\\
\begin{knitrout}
\definecolor{shadecolor}{rgb}{0.969, 0.969, 0.969}\color{fgcolor}
\includegraphics[width=0.128\linewidth,height=0.128\linewidth]{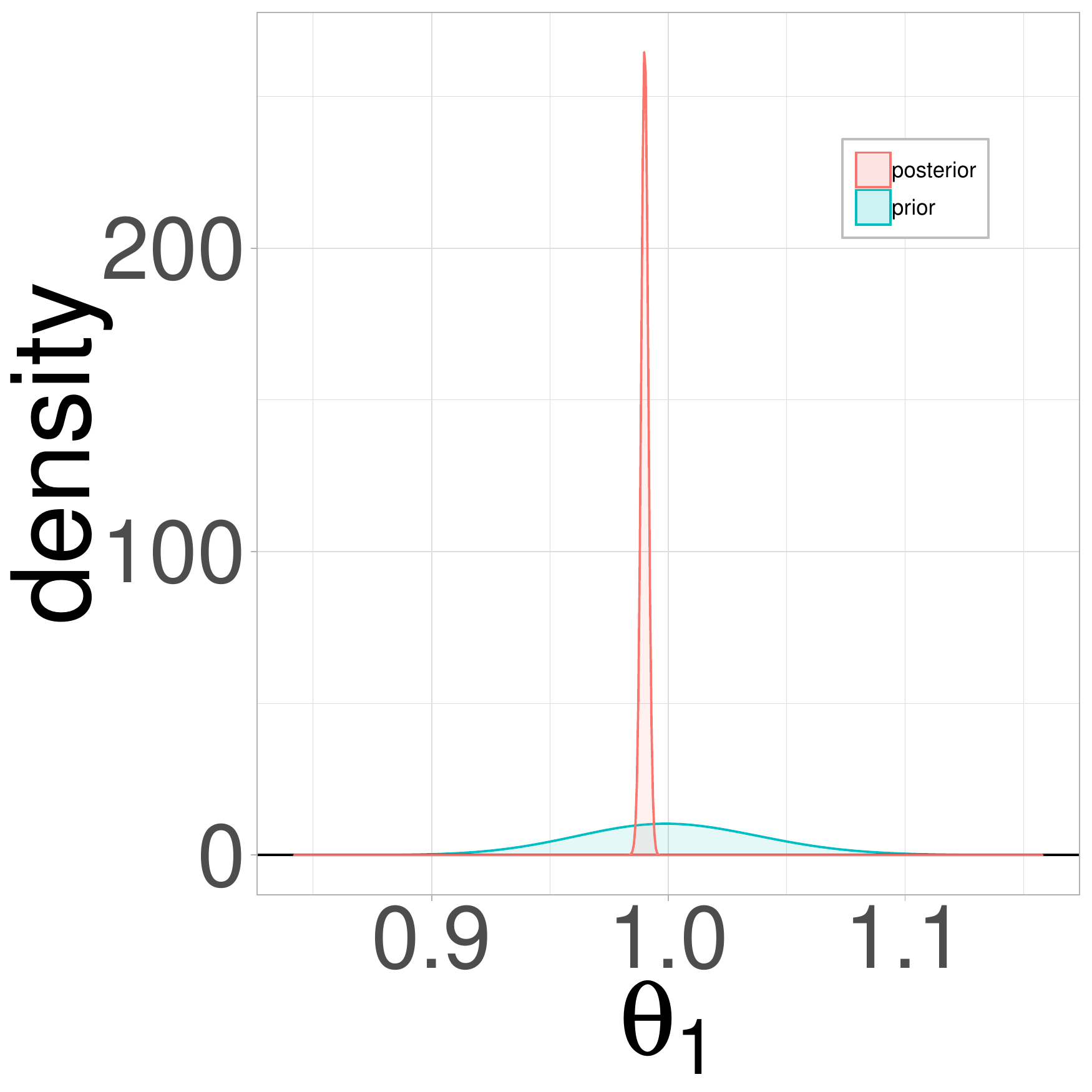} 

\end{knitrout}
&
\begin{knitrout}
\definecolor{shadecolor}{rgb}{0.969, 0.969, 0.969}\color{fgcolor}
\includegraphics[width=0.128\linewidth,height=0.128\linewidth]{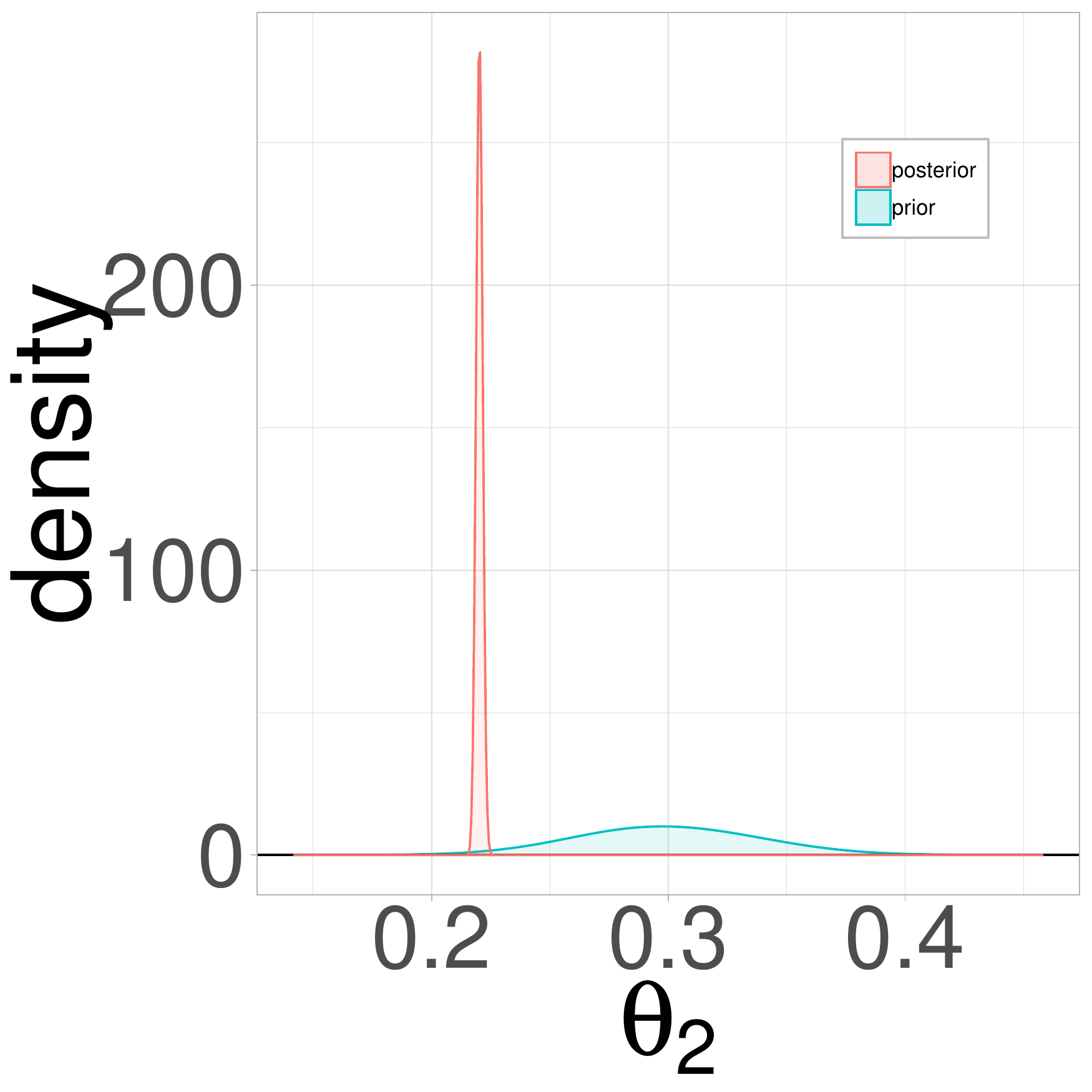} 

\end{knitrout}
&
\begin{knitrout}
\definecolor{shadecolor}{rgb}{0.969, 0.969, 0.969}\color{fgcolor}
\includegraphics[width=0.128\linewidth,height=0.128\linewidth]{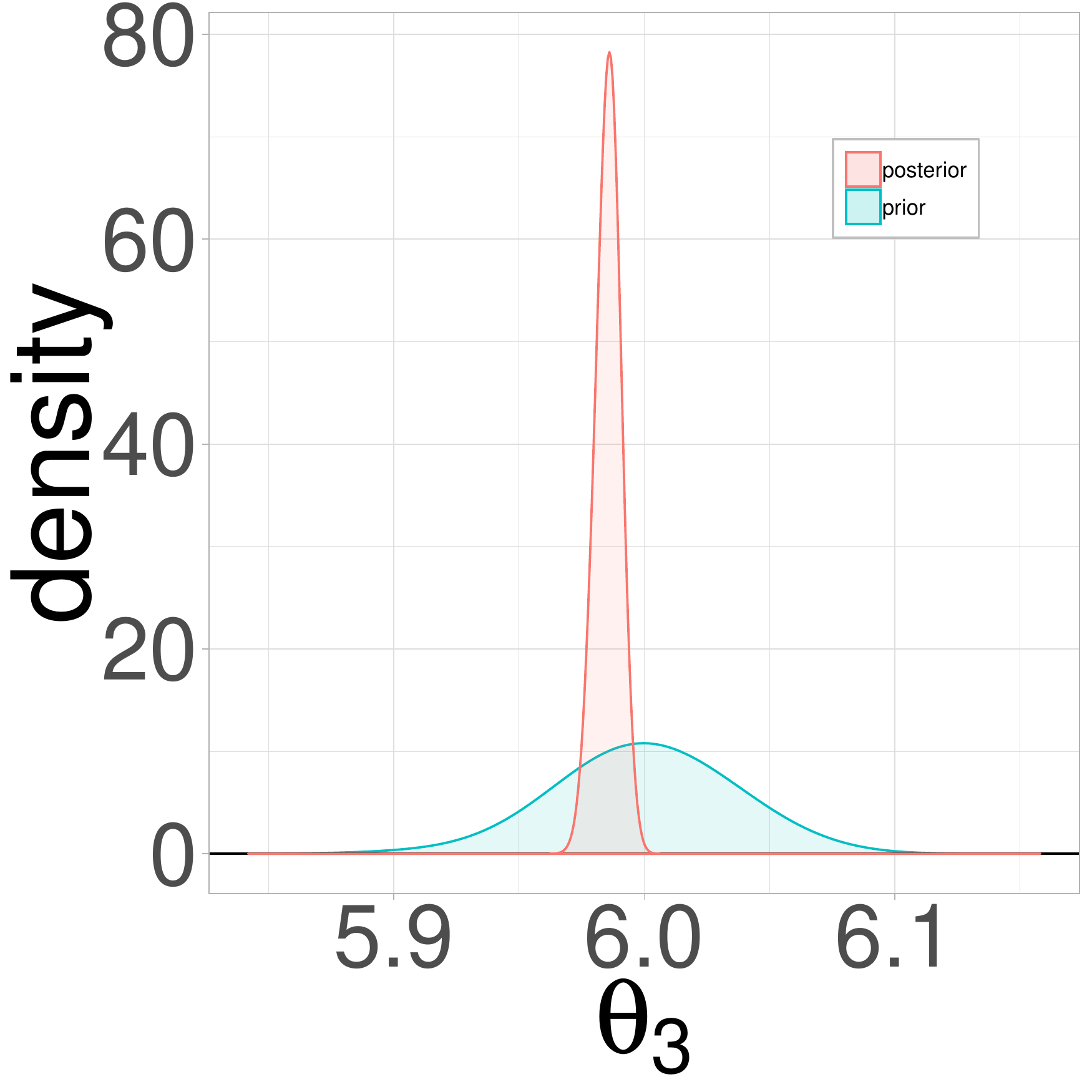} 

\end{knitrout}
&
\begin{knitrout}
\definecolor{shadecolor}{rgb}{0.969, 0.969, 0.969}\color{fgcolor}
\includegraphics[width=0.128\linewidth,height=0.128\linewidth]{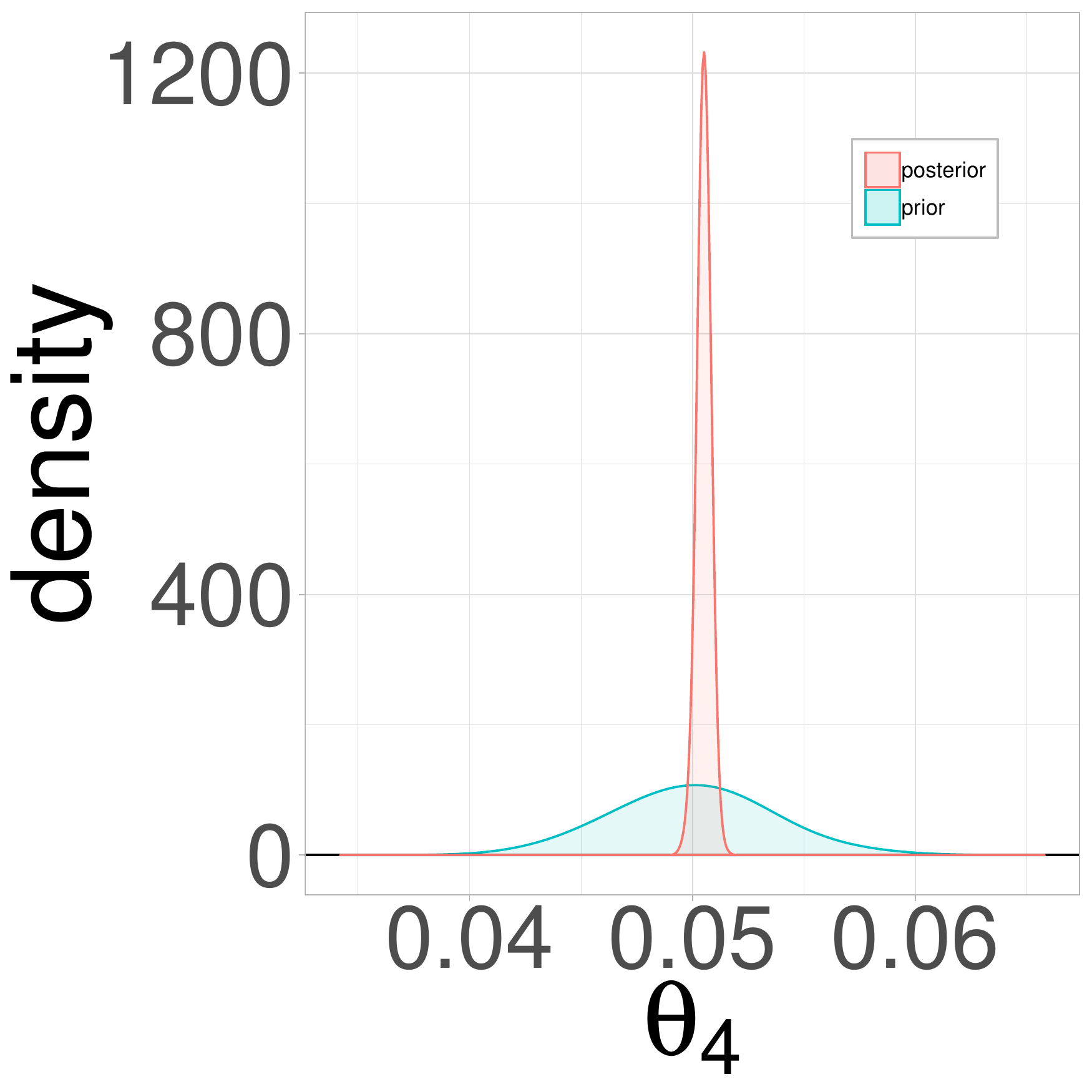} 

\end{knitrout}
&
\begin{knitrout}
\definecolor{shadecolor}{rgb}{0.969, 0.969, 0.969}\color{fgcolor}
\includegraphics[width=0.128\linewidth,height=0.128\linewidth]{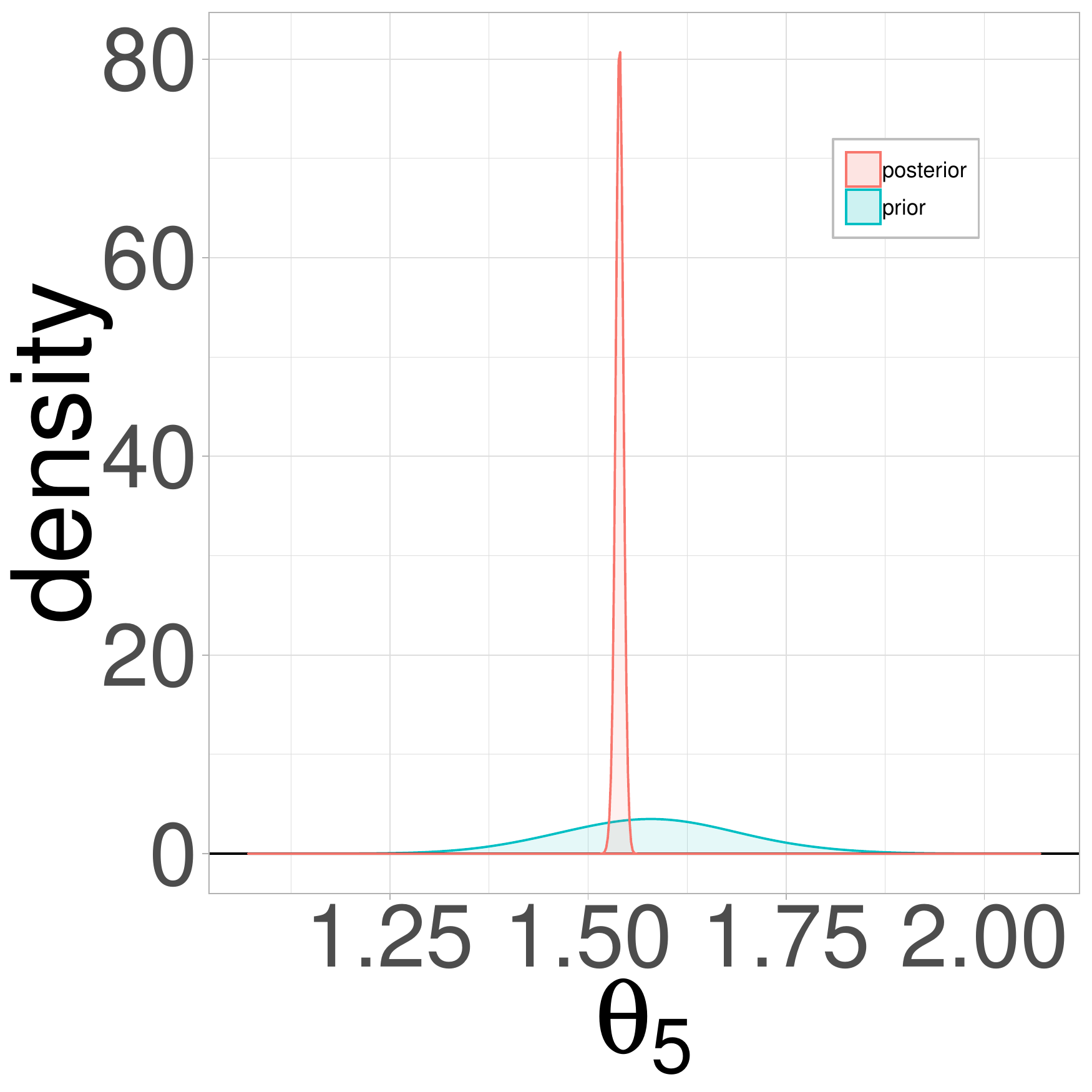} 

\end{knitrout}
&
\begin{knitrout}
\definecolor{shadecolor}{rgb}{0.969, 0.969, 0.969}\color{fgcolor}
\includegraphics[width=0.128\linewidth,height=0.128\linewidth]{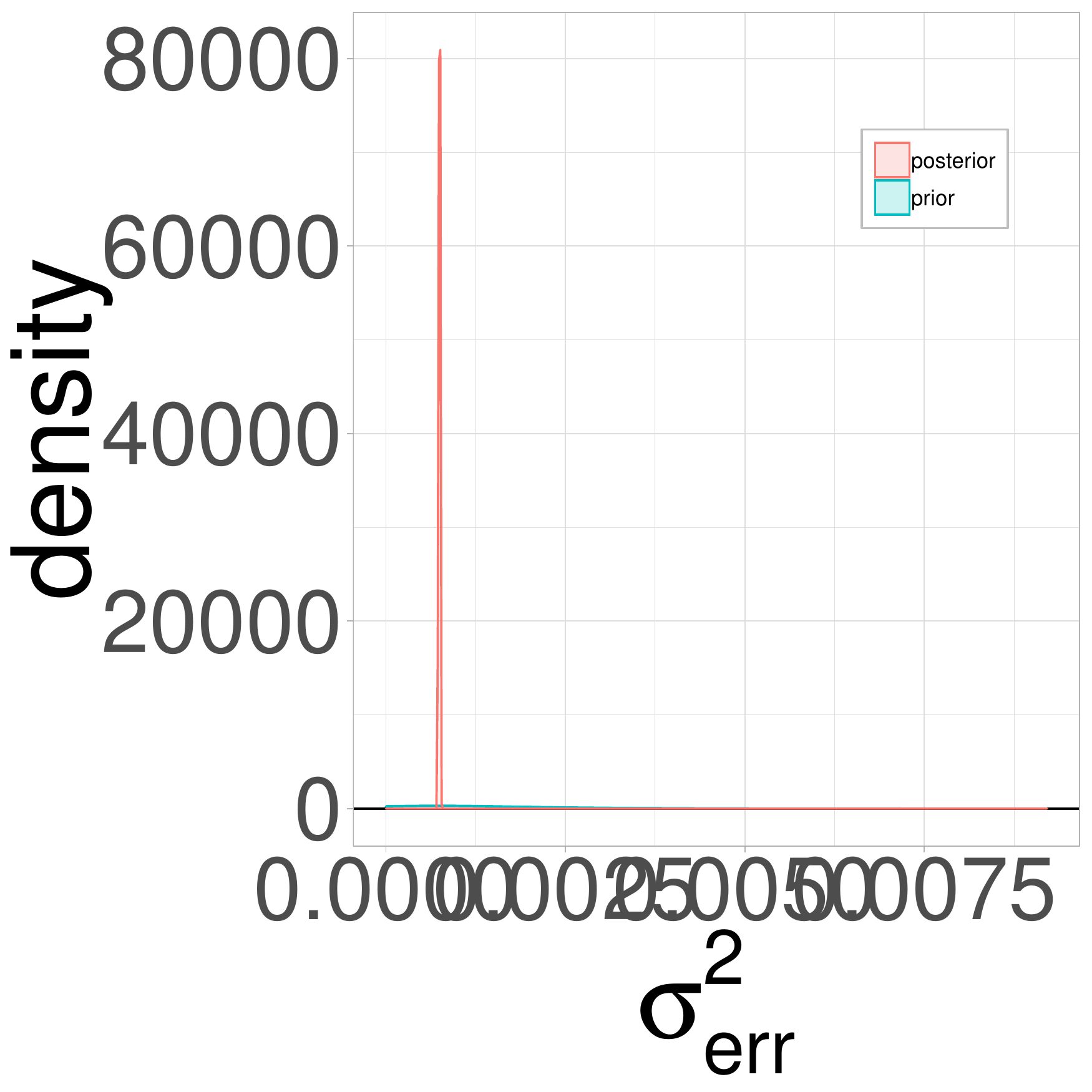} 

\end{knitrout}
\\
\end{tabular}
\caption{Series of plot generated by the function \code{plot} for the sequential design on $\mathcal{M}_2$}
\label{fig:PlotSequential}
\end{center}
\end{figure}

\newpage

\section{Conclusion}
\label{sec:conclusion}
In conclusion, \pkg{CaliCo} is a package that deals with Bayesian calibration through four main functions. For an industrial numerical code, every specific cases is covered by \pkg{CaliCo} (if the user has a DOE or not, with a numerical code or not). The \pkg{R6} classes used in the implementation makes the package more robust. Even if the class layer is not visible to the user, the standardized formulation allows a rigorous treatment. The multiple \pkg{ggplot2} graphs available for each class allow the user to take advantage of the graphical display without any knowledge of \pkg{ggplot2}. The flexibility of \pkg{ggplot2} enables also the user to modify the frame, scale, title, labels of the graphs really quickly. All the MCMC calls are implemented in
\proglang{C++}, which reduces the time of these time-consuming algorithms. The Metropolis within Gibbs algorithm provides a better learning of the covariance matrix that the Metropolis Hastings will use in its proposition distribution. That improves the performance of the algorithms. \newline

Many developments can be brought to the package. For example, statistical validation can be added to the package and permit the user to elect the best model according to the data. Based on \citet{damblin2016} a validation using the Bayes factor or mixture models can be implemented. The dependences on \pkg{DiceKriging} or \pkg{DiceDesign} can also be a weakness of the package. When too many dependencies are implemented, the chances to have bad configuration also increase.

%% -- Summary/conclusions/discussion -------------------------------------------
\section*{Acknowledgments}

\begin{leftbar}
This work was supported by the research contract CIFRE n◦2015/0974 between Élec- tricité de France and AgroParisTech.
\end{leftbar}

\newpage

\bibliographystyle{apalike}
\bibliography{biblio}

\newpage

\begin{appendix}

\section{Gaussian processes}
\label{app:GP}

Let us consider a probability space $(\Omega,\mathcal{F},\mathbb{P})$ where $\Omega$ stands for a sample space, $\mathcal{F}$ a $\sigma$-algebra on $\Omega$ and $\mathbb{P}$ a probability on $\mathcal{F}$. A stochastic process $X$ is a family as $\{ X_t\ ;\ t\in\mathcal{T}\}$ where $\mathcal{T}\subset\mathbb{R}$. It is said that the aleatory process is indexed by the indexes of $\mathcal{T}$. At $t$ fixed, the application $X_t \ : \ \Omega \rightarrow \mathbb{R}$ is an random variable. However at $\omega \in \Omega$ fixed, the application $ t \rightarrow X_t(\omega)$ is a trajectory of the stochastic process.\newline

For $t_1 \in \mathcal{T},\dots, t_n\in\mathcal{T}$, the probability distribution of the random vector $(X_{t_1},\dots,X_{t_n})$ is called finite-dimensional distributions of the stochastic process $\{X_t\}_{t\in\mathcal{T}}$. Hence, the probability distribution of an aleatory process is determined by its finite-dimensional distributions. Kolmogorov's theorem guaranties the existence of such a stochastic process if a suitably collection of coherent finite-dimensional distributions is provided.\newline

An random vector $\boldsymbol{Z}$ such as $\boldsymbol{Z}=(Z_1,\dots,Z_n)$ is Gaussian if $\forall \lambda_1,\dots,\lambda_n \in \mathbb{R}$ the random variable $\sum_{i=1}^n\lambda_iZ_i$ is Gaussian. The distribution of $Z$ is straightforwardly determined by its two first moments : the mean $\boldsymbol{\mu}=(\mathbb{E}[Z_1],\dots,\mathbb{E}[Z_n])$ and the variance covariance matrix $\Sigma = cov(Z_i,Z_j)_{1\leq i,\ j\leq n}$. When $\Sigma$ is positive definite, $Z$ has a probability distribution defined by equation (\ref{eq:densityGaussian}). \newline

\begin{equation}
f(\bm{z})=\frac{|\Sigma|^{-1/2}}{(2\pi)^{n/2}}\exp{-\frac{1}{2}(\boldsymbol{z}-\boldsymbol{\mu})^T\Sigma^{-1}(\boldsymbol{z}-\boldsymbol{\mu})}
\label{eq:densityGaussian}
\end{equation}

Let us consider two Gaussian vectors called $\bm{U_1}$ and $\boldsymbol{U_2}$ such as: \newline

\begin{equation*}
\begin{pmatrix}
\bm{U_1}\\
\bm{U_2}
\end{pmatrix} \sim \mathcal{N} \Big( \begin{pmatrix}
\bm{\mu_1}\\
\bm{\mu_2}
\end{pmatrix}, \begin{pmatrix}
\Sigma_{1,1} & \Sigma_{1,2}\\
\Sigma_{2,1} & \Sigma_{2,2}
\end{pmatrix} \Big)
\end{equation*}

The conditional distribution $\bm{U_2}|\bm{U_1}$ is also Gaussian (equation (\ref{eq:conditionalGaussian})). This property is especially useful when a surrogate model is created from a code. \newline

\begin{equation}
\bm{U_2}|\bm{U_1} \sim \mathcal{N}(\bm{\mu_2}+\Sigma_{2,1}\Sigma_{1,1}^{-1}(\bm{U_1}-\bm{\mu_1}), \Sigma_{2,2}-\Sigma_{2,1}\Sigma_{1,1}^{-1}\Sigma_{1,2})
\label{eq:conditionalGaussian}
\end{equation}

A stochastic process $\{X_t\}_{t\in\mathcal{T}}$ is a Gaussian process if each of its finite-dimensional distributions is Gaussian. Let us introduce the mean function such as $m : t\in\mathcal{T} \rightarrow m(t)=\mathbb{E}[X_t]$ and the correlation function such as $ K : (t,t')\in\mathcal{T}\times \mathcal{T}\rightarrow K(t,t')=corr(X_t,X_{t'})$. A Gaussian process with a scale parameter noted $\sigma^2$ will be defined as the equation (\ref{eq:GaussianDef}).

\begin{equation}
X(.) \sim \mathcal{PG}(m(.),\sigma^2K(.,.))
\label{eq:GaussianDef}
\end{equation}

Gaussian processes are used in this article in two cases. In the fist one, $f$ is a code function long to run and the Gaussian process emulates its behavior. The Gaussian process is called the surrogate of the code. The second case is when we want to estimate the error made by the code (called code error or discrepancy in this article). For the former, we want to create a surrogate $\tilde{f}$ of a deterministic function $f$. In a Bayesian framework, the Gaussian process is a "functional" \textit{a priori} on $f$ \citep{currin1991}.\newline

Let us note:
\begin{equation}
f(.) \sim \mathcal{PG}(h(.)^T\boldsymbol{\beta}_f,\sigma_f^2K_{\boldsymbol{\psi}_f}(.,.))
\end{equation}
where $\boldsymbol{\beta}_f$, $\sigma_f^2$, $\boldsymbol{\psi}_f$ are the parameters specifying the mean and the variance-covariance structure of the process and $h(t)=(h_1(t),\dots,h_n(t))$ is a vector of regressors. For $(t,t')\in \mathcal{T}\times\mathcal{T}$:
\begin{equation}
cov(f(t),f(t'))=\sigma_f^2K_{\boldsymbol{\psi}_f}(t,t')
\end{equation}

Let us consider the code have been tested on $N$ points \textit{i.e.} on $N$ different vectors $\bm{t}$. The design of experiments (DOE) is noted $D=(t_1,\dots,t_N)^T$ and the outputs of $D$ by $f$ will be defined as $y=(f(t_1),\dots,f(t_N))^T$. The correlation matrix  induced by $y$ can be defined by the correlation function $K_{\boldsymbol{\psi}_f}(.,.)$ and can be written as $\Sigma_{\boldsymbol{\psi}_f}=\Sigma_{\boldsymbol{\psi}_f}(D,D)$ such as $\forall (i,j) \in [1,\dots,n] \  \Sigma_{\boldsymbol{\psi}_f}(D)(i,j)=K(t_i,t_j)$.

\begin{equation}
\begin{pmatrix}
f(t)\\
f(D)
\end{pmatrix} \sim \mathcal{N}\Big( \begin{pmatrix}
h(t)^T\boldsymbol{\beta}_f\\
h(D)^T\boldsymbol{\beta}_f
\end{pmatrix}, \sigma_f^2\begin{pmatrix}
\Sigma_{\boldsymbol{\psi}_f}(t) &  \Sigma_{\boldsymbol{\psi}_f}(t,D) \\
\Sigma_{\boldsymbol{\psi}_f}(t,D)^T & \Sigma_{\boldsymbol{\psi}_f}(D)
\end{pmatrix}\Big)
\label{eq:Conditionnal}
\end{equation}

From the equation (\ref{eq:conditionalGaussian}), it comes straightforwardly that $f(t)|f(D)\sim\mathcal{PG}(\mu_p(t),\Sigma_p(t))$. This conditional is called \textit{posterior} distribution with :
\begin{equation*}
\mu_p(t)=h(t)^T\boldsymbol{\beta}_f+\Sigma_{\boldsymbol{\psi}_f}(t,D)\Sigma_{\boldsymbol{\psi}_f}(D)^{-1}(f(D)-h(D)^T\boldsymbol{\beta}_f)
\end{equation*}
\begin{equation*}
\Sigma_p(t,t')=\sigma_f^2\Big(\Sigma_{\boldsymbol{\psi}_f}(t,t')-\Sigma_{\boldsymbol{\psi}_f}(t,D)^T\Sigma_{\boldsymbol{\psi}_f}(D)^{-1}\Sigma_{\boldsymbol{\psi}_f}(t',D)\Big)
\end{equation*}

The mean obtained \textit{a posteriori} is called the Best Linear Unbiased Predictor (BLUP) which the linear predictor without bias $\tilde{f}$ of $f$ which minimize the Mean Square Error (MSE) :
\begin{equation}
MSE(\tilde{f})=\mathbb{E}[(f-\tilde{f})^2]
\end{equation}

In this appendix, we will not discuss the choice of $K_{\boldsymbol{\psi}_f}$, the parameter estimation, nor the validation of the Gaussian process.

\newpage
\section{Estimation algorithm implemented}
\label{MCMC}

\begin{algorithm}[H]
\footnotesize
  // Beginning of the Metropolis within Gibbs algorithm \;
 $\boldsymbol{\theta}_{\text{\tiny{MHWG}}}^1=\boldsymbol{\theta}_{init}$\;
 $p = dim(\boldsymbol{\theta}_{init})$\;
 $\tau_{\text{\tiny{MHWG}}} \leftarrow (0,\dots,0)^T$ //p size vector \;
 \For{$i$ in $1:N_{\text{\tiny{Gibbs}}}$}{
  \For{$j$ in $1:p$}{
    $\boldsymbol{\theta}[j]_{\text{\tiny{MHWG}}}^{*} \sim \mathcal{N}(\boldsymbol{\theta}[j]_{\text{\tiny{MHWG}}}^i,\boldsymbol{k}\sigma[j,j])$\;
    $r=\frac{\pi(\boldsymbol{\theta}[j]_{\text{\tiny{MHWG}}}^{*}|\boldsymbol{\theta}[-j]_{\text{\tiny{MHWG}}}^{*},\boldsymbol{x},\boldsymbol{y}_e)}{\pi(\boldsymbol{\theta}[j]_{\text{\tiny{MHWG}}}^{i}|\boldsymbol{\theta}[-j]_{\text{\tiny{MHWG}}}^{i},\boldsymbol{x},\boldsymbol{y}_e)}=\frac{\pi(\boldsymbol{\theta}[j]_{\text{\tiny{MHWG}}}^*)\mathcal{L}(\boldsymbol{\theta}[j]_{\text{\tiny{MHWG}}}^*|\boldsymbol{\theta}[-j]_{\text{\tiny{MHWG}}}^{*},\boldsymbol{x},\boldsymbol{y}_e)}{\pi(\boldsymbol{\theta}[j]_{\text{\tiny{MHWG}}}^i)\mathcal{L}(\boldsymbol{\theta}[j]_{\text{\tiny{MHWG}}}^i|\boldsymbol{\theta}[-j]_{\text{\tiny{MHWG}}}^{i},\boldsymbol{x},\boldsymbol{y}_e)}$\;
   \eIf{$r > u$, with $u\sim \mathcal{U}(0,1)$}{
   $\boldsymbol{\theta}[j]_{\text{\tiny{MHWG}}}^{i+1} \leftarrow \boldsymbol{\theta}^*$\;
   $\tau_{\text{\tiny{MHWG}}}[j] \leftarrow \tau_{\text{\tiny{MHWG}}}[j+1]$\;
   }{
   $\boldsymbol{\theta}[j]{\text{\tiny{MHWG}}}^{i+1} \leftarrow \boldsymbol{\theta}{\text{\tiny{MHWG}}}^i$\;
  }
  \If{$i \equiv 0[100]$}{
    \If{$\tau_{\text{\tiny{MHWG}}}[j]/i<0.25$}{$k \leftarrow k(1-r[1])$}
    \If{$\tau_{\text{\tiny{MHWG}}}[j]/i>0.5$}{$k \leftarrow k(1+r[1])$}
  }
  }
 }
 $S \leftarrow cov(\boldsymbol{\theta}_{\text{\tiny{MHWG}}})$\;
 // Beginning of the Metropolis Hasting Algorithm\;
 $\boldsymbol{\theta}_{\text{\tiny{MH}}}^1=\boldsymbol{\theta}_{init}$\;
 $\tau_{\text{\tiny{MH}}} \leftarrow 0$\;
 $t \leftarrow 1$\;
  \For{$i$ in $1:N_{\text{\tiny{MH}}}$}{
    $\boldsymbol{\theta}_{\text{\tiny{MH}}}^{*} \sim \mathcal{N}(\boldsymbol{\theta}_{\text{\tiny{MH}}}^i,tS)$\;
    $r=\frac{\pi(\boldsymbol{\theta}_{\text{\tiny{MH}}}^{*}|\boldsymbol{\theta}_{\text{\tiny{MH}}}^{*},\boldsymbol{x},\boldsymbol{y}_e)}{\pi(\boldsymbol{\theta}_{\text{\tiny{MH}}}^{i}|\boldsymbol{\theta}_{\text{\tiny{MH}}}^{i},\boldsymbol{x},\boldsymbol{y}_e)}=\frac{\pi(\boldsymbol{\theta}_{\text{\tiny{MH}}}^*)\mathcal{L}(\boldsymbol{\theta}_{\text{\tiny{MH}}}^*|\boldsymbol{\theta}_{\text{\tiny{MH}}}^{*},\boldsymbol{x},\boldsymbol{y}_e)}{\pi(\boldsymbol{\theta}_{\text{\tiny{MH}}}^i)\mathcal{L}(\boldsymbol{\theta}_{\text{\tiny{MH}}}^i|\boldsymbol{\theta}_{\text{\tiny{MH}}}^{i},\boldsymbol{x},\boldsymbol{y}_e)}$\;
  \eIf{$r > u$, with $u\sim \mathcal{U}(0,1)$}{
    $\boldsymbol{\theta}[j]_{\text{\tiny{MH}}}^{i+1} \leftarrow \boldsymbol{\theta}^*$\;
    $\tau_{\text{\tiny{MH}}}[j] \leftarrow \tau_{\text{\tiny{MH}}}[j+1]$\;
  }{
    $\boldsymbol{\theta}_{\text{\tiny{MH}}}^{i+1} \leftarrow \boldsymbol{\theta}_{\text{\tiny{MH}}}^i$\;
  }
  \If{$i \equiv 0[100]$}{
    \If{$\tau_{\text{\tiny{MH}}}/i<0.25$}{$t \leftarrow t(1-r[2])$}
    \If{$\tau_{\text{\tiny{MH}}}/i>0.5$}{$ t \leftarrow t(1+r[2])$}
  }
 }
\end{algorithm}

\end{appendix}

\end{document}